%% file: paper_QEC_TV.tex
\documentclass[prx,showpacs,twocolumn,superscriptaddress,longbibliography,floatfix,notitlepage,nofootinbib]{revtex4-2}
\usepackage{natbib}

\usepackage{times}
\usepackage{amsmath}
\usepackage{amsfonts}
\usepackage{amssymb}
\usepackage{dsfont}
\usepackage{graphicx}
\usepackage{physics}
\usepackage{braket}

\usepackage{epsfig}
\usepackage{verbatim}
\usepackage{bm}
\usepackage{mathrsfs}
\usepackage{hyperref}
\hypersetup{hidelinks}
\usepackage[labelformat=simple]{subcaption}

\usepackage[utf8]{inputenc}
\usepackage[english]{babel}
\usepackage{csquotes}

\usepackage{pst-all}
\usepackage{leftidx}

\graphicspath{./fig/}

\newcommand{\1}{\mathbf{1}}
\newcommand{\e}{\mathrm{e}} 
\newcommand{\ii}{\mathrm{i}} 
\renewcommand{\H}{\mathcal{H}}
\newcommand{\D}{\mathcal{D}}
\newcommand{\C}{\mathcal{C}}
\newcommand{\cP}{\mathcal{P}}

\newcommand{\RR}{\mathbb{R}}
\newcommand{\NN}{\mathbb{N}}

\newcommand{\CC}{\mathbb{C}}

\newcommand{\figref}[1]{Fig.~\ref{#1}}

\newcommand{\be}{\begin{equation}}
\newcommand{\ee}{\end{equation}}
\newcommand{\bc}{\begin{center}}
\newcommand{\ec}{\end{center}}
\newcommand{\nin}{\noindent}

\renewcommand{\degrees}{^\circ}

\makeatletter
\def\l@subsubsection#1#2{}
\makeatother

\begin{document}
\title{Quantum error correction thresholds for the universal Fibonacci Turaev-Viro code}
\author{Alexis Schotte}
\email{alexis.schotte@ugent.be}
\affiliation{Department of Physics and Astronomy, Ghent University, Krijgslaan 281, 9000 Gent, Belgium}

\author{Guanyu Zhu}
\email{guanyu.zhu@ibm.com}
\affiliation{IBM Quantum, IBM T.J. Watson Research Center, Yorktown Heights, NY 10598, and IBM Almaden Research Center, San Jose, CA 95120, USA}

\author{Lander Burgelman}
\affiliation{Department of Physics and Astronomy, Ghent University, Krijgslaan 281, 9000 Gent, Belgium}
\author{Frank Verstraete}
\affiliation{Department of Physics and Astronomy, Ghent University, Krijgslaan 281, 9000 Gent, Belgium}

	\input{sections/abstract.tex}

\maketitle

\tableofcontents
\clearpage
	\input{sections/introduction.tex}

	\input{sections/extended_levin-wen.tex}	
	
	\input{sections/measurement_scheme.tex}	
		
	\input{sections/threshold_simulation.tex}

	\input{sections/decoders.tex}		
	
	\input{sections/results.tex}

	\input{sections/conclusion.tex}
	
    \input{sections/acknowledgment.tex}	
	
\appendix

	\input{sections/ribbon_graphs}

	\input{sections/tensor_networks.tex}		
	
	\input{sections/result_scaling.tex}

	\input{sections/relation_iid_noise.tex}

	\input{sections/paperclip_list.tex}	

\bibliography{references.bib}

\end{document}

%% file: sections/abstract.tex
\begin{abstract}
We consider a two-dimensional quantum memory of qubits on a torus which encode the extended Fibonacci string-net code, and devise strategies for error correction when those qubits are subjected to depolarizing noise. Building on the concept of tube algebras, we construct a set of measurements and of quantum gates which map arbitrary qubit errors to the string-net subspace and allow for the characterization of the resulting error syndrome in terms of doubled Fibonacci anyons. Tensor network techniques then allow to quantitatively study the action of Pauli noise on the string-net subspace. We perform Monte Carlo simulations of error correction in this Fibonacci code, and compare the performance of several decoders.  For the case of a fixed-rate sampling depolarizing noise model, we find an error correction threshold of 4.7\% using a clustering decoder. 
To the best of our knowledge, this is the first time that a threshold has been estimated for a two-dimensional error correcting code for which universal quantum computation can be performed within its code space via braiding anyons.

\end{abstract}

%% file: sections/introduction.tex
\section{Introduction}

The biggest theoretical challenge in achieving scalable quantum computation is the construction of more efficient schemes for quantum error correction and fault-tolerance \cite{terhal_quantum_2015, campbell_roads_2017}. Topological quantum error correcting codes (QECC), with the most famous representative being the surface code \cite{kitaev2003fault, Dennis:2002ds, bravyi_quantum_1998, raussendorf2007fault, Fowler:2012fi}, are among the most promising candidates for near-term implementation due to their geometric locality which makes them well suited for practical implementation using state-of-the-art hardware technology  such as superconducting or semiconducting qubits, ion traps, silicon photonics, NV centers, cold atoms, just to name a few. Surface codes allow for high quantum error correction thresholds, but have the drawback that one needs a very large overhead of magic states to make the scheme universal for quantum computation \cite{bravyi2005, Fowler:2012fi, campbell_roads_2017}.

In order to overcome this limitation, one has to consider topological codes which allow for non-Clifford logical gates. One approach in this direction is to consider stabilizer codes in three and higher dimensions, which allow for non-Clifford transversal gates \cite{Bravyi:2013dx, Bombin:2015hia, vasmer_three-dimensional_2019, JochymOConnor:2018is, JochymOConnor:2021ih, Bombin:2018wj, Browneaay4929}.  This approach also includes schemes based on 3D color codes or 3D surface codes which simulate the third spatial dimension using time within a 2D measurement-based quantum computing architecture \cite{Bombin:2018wj, Browneaay4929}.  A perpendicular direction is to look beyond the stabilizer formalism and consider non-Abelian codes in 2D which are universal for quantum computation without the need for magic-state distillation \cite{Nayak:2008dp}. In this paper, we follow this second path, as we believe that this is a more natural setting for hardware platforms currently being pursued.

Most conventional topological error correcting codes fall within the framework of the stabilizer formalism \cite{gottesman_stabilizer_1997}, and admit quasiparticle excitations which can be characterized as Abelian anyons. However, braiding of these excitations does not allow for a universal gate set. Similarly, double semion topological codes \cite{freedman2016, Dauphinais_2019}, while going beyond the stabilizer code class, exhibit Abelian topological order and are not universal. In order to achieve universal topological quantum computation, a necessary condition is to use systems with a more intricate topological order allowing for non-Abelian anyonic excitations, which have the property that the fusion of two anyons can yield several outcomes \cite{Nayak:2008dp}. In particular, braiding of Fibonacci anyons can be used to realize a fault-tolerant universal gate set \cite{freedman2002modular}. Several lattice models supporting this non-Abelian topological order have been proposed, such as Kitaev's quantum double models \cite{kitaev2003fault} or the string-net models of Levin and Wen \cite{levin2005string}. While their ground space is still defined as the simultaneous eigenspace of a set of mutually commuting local check operators, their description falls beyond the stabilizer formalism.

In recent years, there has been significant progress in the study of quantum error correction and decoding for non-Abelian topological codes with a non-stabilizer structure, including numerical estimates of their error thresholds \cite{brell2014thermalization, wootton2014error, wootton2016active, dauphinais2017fault, burton2017classical}. These works however assume the existence of a protected anyonic fusion space and only consider phenomenological noise models on this space, while not specifying a concrete underlying microscopic quantum mechanical spin model suitable for the description of realistic quantum computer implementations.

In this work, we remedy this shortcoming by considering a non-Abelian error correcting code consisting of generic qubits subjected to depolarizing noise. We focus on one of the simplest of those codes, the Turaev-Viro code constructed from the Fibonacci string-net model \cite{levin2005string}. Building on the pioneering work of K\"onig, Kuperberg and Reichardt \cite{konig2010quantum} and of Bonesteel and DiVincenzo \cite{bonesteel2012quantum}, we construct a set of measurements and quantum gates which map arbitrary qubit errors to the Turaev-Viro subspace, and make crucial use of the framework of tensor networks for simulating the error correction process, giving rise to surprisingly high quantum error correction thresholds. Using a clustering decoder and a fixed-rate sampling noise model, we have obtained a $4.7\%$ threshold for the code subjected to depolarizing noise, and a $7.3\%$ threshold for pure dephasing noise. These numbers are comparable to the code-capacity error threshold of the surface code, which is around $10\%$ for i.i.d.~bit-flip or phase-flip noise \cite{Dennis:2002ds, bravyi2014efficient, yoder2017surface}.  

Before giving a summary of the paper, let us provide a brief review of the history and current status of the Turaev-Viro codes. Following the pioneering work by Jones in 1984 discovering the Jones Polynomials \cite{Jones_1985}, Reshetikhin and Turaev generalized the Jones Polynomials of links by introducing the concept of ribbon graphs and the corresponding invariants derived from quantum groups in their influential work in 1990 \cite{Reshetikhin_Turaev}. Around the same time, Witten \cite{Witten1989_CMP} and Atiyah \cite{atiyah} introduced the formalism of topological quantum field theory (TQFT). Turaev and Viro introduced a path integral in terms of a discrete state-sum describing a wide class of 2+1D topological quantum field theories in 1992 , leading to new quantum invariants of 3-manifold \cite{turaev1992}.  Around 1997, Kitaev had the crucial insight that the problem of constructing quantum error correcting codes is effectively equivalent to the one of constructing quantum spin systems providing a Hamiltonian realization of such topological field theories. He introduced a class of quantum double models \cite{kitaev2003fault}, of which the simplest Abelian version $\mathcal{D}(\mathbb{Z}_2)$ is the well-known toric code. The non-Abelian codes in the quantum double family can be used to implement universal fault-tolerant logical gate sets without magic state distillation. In 2005, Levin and Wen generalized Kitaev's quantum doubles by introducing string-net models \cite{levin2005string}, which provide a local Hamiltonian realization of all the unitary Turaev-Viro TQFTs. Their original  motivation of introducing the string-net condensation picture was to provide a microsopic mechanism for spin liquid in the context of condensed matter physics. The string-nets can be viewed as planar special case of ribbon graphs introduced by Reshetikhin and Turaev. The string-net models capture all non-chiral topological orders (Abelian and non-Abelian) in 2D, including topological orders of the toric code and Kitaev's quantum double model as specific examples. In 2010, K\"onig, Kuperberg and Reichardt \cite{konig2010quantum} studied a class of Levin-Wen models with a modular input category from the point of view of error correction and called them \emph{Turaev-Viro codes}. They developed the tube algebra for this modular case and defined a complete basis of the anyonic excitations. In 2012, Bonesteel and DiVincenzo proposed the quantum circuits to measure the vertex and plaquette projectors in the Fibonacci Levin-Wen model \cite{bonesteel2012quantum} and hence made the first step towards practical implementation of Turaev-Viro codes with ordinary qubits by devising an error detection scheme. 

Several technical difficulties had to be solved however to turn their error detecting scheme into an error correcting one.  First, the original string-net model proposed by Levin and Wen \cite{Levin:2006ij} does not admit an easy microscopic description of all types of anyonic excitations in the corresponding topological phases, but only the fluxons (plaquette excitations). As we will see, a single vertex error in this model can bring the system out of the string-net subspace such that the created excitations are no longer anyons as in the case of phenomenological anyon models \cite{brell2014thermalization, burton2017classical,  dauphinais2017fault}, making error correction and decoding quite challenging. For that purpose, an extended string-net model was defined on a tailed lattice ~\cite{feng2015non, bonesteel_inpreparation} (see also Ref.~\cite{hu2018full} where tail qubits were introduced for the purpose of incorporating charged and dyonic excitations in topological phases). In the same work ~\cite{feng2015non, bonesteel_inpreparation}, a scheme to trap vertex errors and a tadpole swapping scheme to trap plaquette errors were introduced. 

In this work, we  adopt this tailed-lattice construction, and use a similar strategy to trap local vertex errors. We introduce a measurement scheme in terms of tube algebras or \textit{tube operators} \cite{ocneanu2001operator, bultinck2017anyons} whose outcomes contain more syndrome information than the ones reported in Refs.~\cite{feng2015non} and \cite{bonesteel_inpreparation}. This tube algebra enables the definition of an anyonic fusion basis, which can be used to  effectively describe the system evolution in the simpler language of an anyon model.  We then use the tensor network description of those tube algebras to convert microscopic noise processes such as  Pauli errors into anyon-creation processes. The last step needed to calculate error correcting thresholds then consists of simulating the error correcting process.

It is tempting to think that an efficient classical simulation of the error correction process for the Fibonacci Turaev-Viro code is impossible, since braiding Fibonacci anyons is universal for quantum computation. This issue has been addressed in Ref.~\cite{burton2017classical} for a phenomenological Fibonacci anyon model (in which the physical degrees of freedom are anyons as opposed to qubits), and it was demonstrated that it is possible to simulate the error correction threshold with a polynomial complexity. This is because, unlike the quantum computation process where the computational anyons are braided along topologically nontrivial worldlines, the worldlines of noise-created anyons in the error correction process are topologically trivial most of the time. As long as the system is below the percolation threshold corresponding to anyon generation, this classical simulation can be performed in an  efficient way. Our work extends the applicability of that result to the case where the physical degrees of freedom are qubits subject to arbitrary noise processes, and hence allows us to determine error thresholds through classical simulations.

\subsection*{Summary}

Before proceeding with the main exposition, we first dedicate some space to convey the central ideas presented in this paper, free of superfluous technical details. There are in essence two main achievements detailed in this work. The first is the construction of a non-Abelian topological quantum error correcting code consisting of regular qubits, and the design of a complete protocol for error detection and correction in this code. The second is the classical simulation of this error correction procedure using tensor network techniques resulting in an estimate for the associated error correction threshold for a microscopic noise model of Pauli errors.

We begin by introducing the central object in our discussion, the extended string-net code. Our starting point is the Fibonacci Levin-Wen string-net model \cite{levin2005string} of qubits arranged on a hexagonal lattice, defined from the algebraic data of the Fibonacci unitary fusion category (UFC). More specifically, we build on the work of König et al.~\cite{konig2010quantum}, who illustrated that this model may be used as a scheme for universal topological quantum computation, giving rise to the concept of Turaev-Viro codes. By adopting a continuum formulation of the model in terms of trivalent ribbon graphs whose properties are defined by the input category, it was shown that the excitations in a Turaev-Viro code can be identified with the central idempotents of the tube algebra \cite{ocneanu2001operator, bultinck2017anyons},  which correspond to topological sectors labeled by the doubled category. This tube algebra, and the associated characterization of excitations as doubled anyons, form the guiding principle throughout the remainder of our discussion. In this work we adopt an extension of the string-net model that serves as the basis for an error correcting code. This extension has a twofold motivation. On the one hand, we need a way of correcting violations of the ribbon graph branching rules that can be induced by generic errors at the level of the lattice qubits. Moreover, we also require a concise way of characterizing the excitation spectrum in terms of anyonic charges, by defining the action of the tube algebra idempotents in the bulk of the lattice model. Both of these requirements can be met by adding an additional \enquote{tail edge} to each plaquette, inspired by the constructions introduced in Refs.~\cite{feng2015non}, \cite{bonesteel_inpreparation} and \cite{hu2018full} for similar reasons. These considerations then lead to a model of qubits arranged on the edges of a tailed hexagonal lattice on the torus whose fourfold degenerate ground space serves as a topological quantum memory and effectively encodes two logical qubits, and whose excited states can be interpreted as fusion states of doubled Fibonacci anyons. By generalizing the torus setup (genus~=~1) to a higher-genus surface, one can scale up the number of logical qubits, which grows approximately linearly with the genus. One can hence perform universal quantum computation via topological operations corresponding to the elements of the mapping class group of the high-genus surface,  which can be generated by Dehn twists \cite{freedman2002modular, konig2010quantum, PhysRevB.102.075105}.  Alternatively, one can encode the logical  information in the fusion basis of computational anyons, and perform universal quantum computation via braiding these  computational anyons  \cite{freedman2002modular, konig2010quantum, PhysRevLett.125.050502}.  Both Dehn twists and braiding can be implemented by local quantum circuit via F-moves \cite{konig2010quantum} or via code deformation.

With this code definition, we proceed with defining the protocols for error detection and correction. Generic errors on the lattice qubits can cause violations of the string-net (ribbon graph) branching rules. Such a violation can be interpreted as a string ending in a vertex of the lattice, resulting in a qubit state that lies outside of the string-net subspace, which can therefore not be captured as an anyonic fusion state. Adapting the circuits for detecting these vertex violations first introduced in Ref.~\cite{bonesteel2012quantum}, we define local unitary circuits that can correct an arbitrary combination of vertex errors, by pulling the corresponding string end onto the tail edge of the associated plaquette. After returning the system to the string-net subspace in this way, we then define circuits for syndrome extraction, again guided by the concept of the tube algebra. Measuring the idempotents of the tube algebra in each plaquette reveals the location and charge of all anyonic excitations in the system, yielding the error syndrome. Utilizing the ribbon graph formalism, we naturally arrive at a local unitary circuit which can perform these charge measurements. Equipped with this protocol for syndrome extraction, we are left with the task of recovery, which consists of moving excitations on the lattice and fusing them to back to the anyonic vacuum, thereby returning the system to the code space. Similar to the Abelian case, a logical error occurs when an anyon is wound along a nontrivial cycle of the torus in this process. Building on and extending previous works \cite{konig2010quantum, feng2015non, bonesteel_inpreparation, PhysRevLett.125.050502, PhysRevB.102.075105, Lavasani:2019}, we design protocols for the necessary recovery operations at the level of the qubits. For the decoding procedure itself, which entails deciding which recovery operations should be carried out given an error syndrome, we rely on recent advances in error correction for non-Abelian anyon models \cite{brell2014thermalization, wootton2014error, burton2017classical} and tailor the decoders introduced there to our specific code, as well as further design new decoders for our purpose. The main difficulty that arises for the non-Abelian case is the fact that error correction has to proceed in an iterative fashion because of indeterminacy of fusion outcomes for non-Abelian anyons. Specifically, we adapt a clustering decoder to our setting and further design a fusion-aware iterative matching decoder.
      
These considerations conclude our discussion of the code definition and associated error correction protocol, giving a complete scheme for error correction in an extended Fibonacci Turaev-Viro code. We then move on to our second main result:  the classical simulation of the error correction procedure and the estimate of the error correction threshold.

The main problem to be tackled here is the question of how to determine what distribution of anyonic excitations is generated by Pauli errors acting at the level of the lattice qubits. The complex description of the excited anyonic fusion states in terms of qubit states makes this a highly nontrivial task however. Again relying on the concept of the tube algebra, we extend the tensor network anyon ansatz known from the matrix operator description of topological order \cite{bultinck2017anyons} to construct a projected entangled pair state (PEPS) representation for the anyonic fusion states that appear as excited states in our model. Armed with these PEPS, we utilize tensor network methods to analyze the effect of Pauli noise on the code, which effectively allows to translate a physical error rate at the level of the qubits to an anyon generation rate. We can hence simplify the classical simulation of the decoding problem to the simulation of noise-driven dynamics of anyonic fusion states, which is infinitely more feasible than directly simulating the full microscopic model itself in the qubit basis. Having overcome this main difficulty, we adapt recently developed techniques for simulating non-Abelian error correction \cite{burton2017classical} to the hexagonal geometry and doubled Fibonacci excitations relevant to our model. We proceed to illustrate a scheme for the classical simulation of error correction in our code, and conclude with an estimate of the error correction threshold using different decoders.

To the best of our knowledge, our results provide the first threshold estimate for a two-dimensional error correcting code of qubits which is universal for topological  quantum computation, without the need for additional non-topological operations.

%% file: sections/extended_levin-wen.tex
\section{The extended string-net code}\label{sec:model}
\subsection{Definition of the code}
\begin{figure}
		\centering
		\includegraphics[scale=.6]{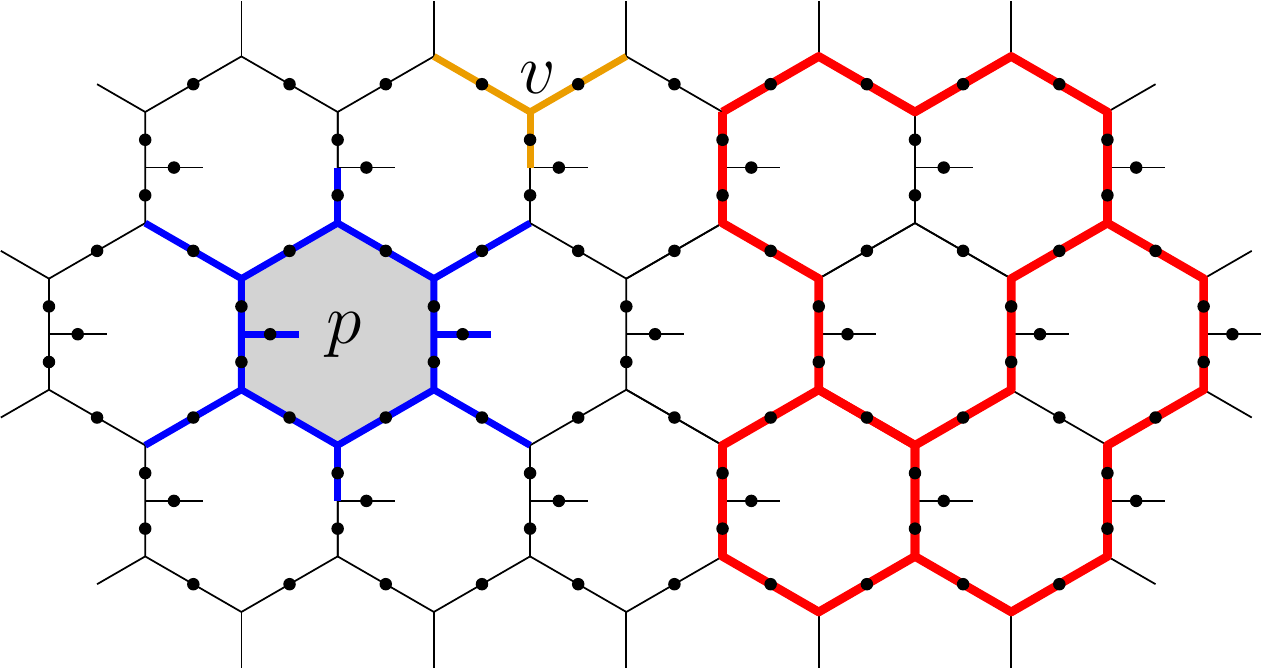}
		\caption{Qudits arranged on the tailed honeycomb lattice. The support of $ B_p $ and $ Q_v $ are indicated in blue and orange, respectively. The red edges represent a valid string-net configuration of qubits in the $ \ket{1} $ state when using the Fibonacci input category.}
		\label{fig:tailed_lattice_support}
	\end{figure}

	The extended string-net code is a microscopic realization of a Turaev-Viro code \cite{konig2010quantum}. 
	Its code space is defined as the ground space of the \emph{extended} Levin-Wen string-net model \cite{levin2005string, hu2018full}.
	This is a microscopic model of qudits situated on the edges of a tailed trivalent lattice $ \Lambda $,  obtained by modifying the Levin-Wen Hamiltonian \cite{levin2005string} to accommodate one additional ``tail edge'' in every plaquette as shown in \figref{fig:tailed_lattice_support}. The code space is hence denoted by $\mathcal{H}_\Lambda$.  
	Such a modification was first proposed in Ref.~\cite{hu2018full}, with the original goal of incorporating charged and dyonic excitations in (doubled) topological order. 
	Below, we give a brief summary of its definition and of its most important properties. 
	We discuss this model and its relation to topological quantum field theory in much greater detail in App.~\ref{sec:extended_LW_and_TQFT}. 
	
	The model is defined starting from the algebraic data of a unitary fusion category $ \C $. For simplicity, we limit ourselves to multiplicity-free self-dual categories. The generalization to generic unitary fusion categories is straightforward but quite tedious. 
	Since we will work with the Fibonacci category later (which is self-dual), we will not be needing the general case. 
	The algebraic data of such an object consists of:
	\begin{enumerate}
		\item \textbf{String types}: A set of all possible string types $ \{ \1, i_2, \dots, i_N \} $. 
		The label $ \1 $ is referred to as the \emph{vacuum label}, and represents the absence of a string on a particular edge. 
		
		\item \textbf{Branching rules}: The set of all triplets $ \{ i,j,k\} $ that are allowed to meet at a vertex (also known as \textit{fusion rules}). 
		We introduce the symbol $ \delta_{ijk} $ defined by the branching rules as
		\begin{equation}\label{eq:strnet_branching_rules}
			\delta_{ijk} =
			\begin{cases}
				1, \quad \text{if the triplet $\{i,j,k\}$ is allowed},\\
				0, \quad \text{otherwise}.
			\end{cases}
		\end{equation}
		For every label $ i $, there is unique dual label $ i^* $ satisfying $ \delta_{i i^* \1}  = 1$, and $ (i^*)^*=i $.  Note that we are considering self-dual categories\footnote{For generic unitary fusion categories, one must pick an orientation for every edge. An edge with label $ i^* $ is equivalent to an edge with label $ i $ and the opposite orientation.}, which satisfy $ i^* = i $ for every string type $ i $. 
		
		\item \textbf{Numerical data}: For each string type $ i $, a real constant $ d_i $, called the \emph{quantum dimension}, satisfying
		\begin{equation}\label{eq:quantdim_cond_short}
			d_i d_j = \sum_k \delta_{ijk} d_k \,, \quad 	d_1 = 1\,, \quad \text{and} \quad d_{i^*} = d_{i}\,,
		\end{equation}
		and a six-index symbol $F^{ijm}_{kln}$, which is a complex constant dependent on 6 string types $ i, j, k, l, m, n $. These quantities are required to satisfy the following consistency conditions:
		\begin{align}
			\text{physicality}&: \quad		 F^{ijm}_{kln} \delta_{ijm} \delta_{klm^*} = F^{ijm}_{kln} \delta_{iln} \delta_{jkn^*}  \label{eq:strnet_physicality} \\
			\text{pentagon equation}&:		 \sum_{n=1}^N F^{mlq}_{kpn} F^{jip^*}_{mns} F^{jsn}_{lkr} = F^{jip^*}_{q^*kr} F^{r^*iq^*}_{mls} \label{eq:strnet_pentagon} \\
			\text{unitarity}&:	\quad	\left(F^{ijm}_{kln}\right)^* = F^{lin}_{jkm^*} \label{eq:strnet_unitarity} \\
			\text{tetrahedral symmetry}&: \quad	 F^{ijm}_{kln} = F^{jim}_{lkn^*} = F^{lkm^*}_{jin} \nonumber \\ 
			& \qquad \qquad = F^{imj}_{k^*nl}\frac{v_m v_n}{v_j v_l} \label{eq:strnet_tetrahedral} \\
			\text{normalization}&:	\quad	 F^{ii^*1}_{j^*jk} = \frac{v_k}{v_i v_j}\, \delta_{ijk} \label{eq:strnet_normalization}
		\end{align}
		where $ v_i = \sqrt{d_i}$.
	\end{enumerate}
	Note that for self-dual categories, the $F$-symbols are real-valued. 
	
	We associate the string types to the elements of an orthonormal basis of the qudit Hilbert space $ \CC^N $ at each edge.
	The extended Levin-Wen Hamiltonian is then defined as: 
	\begin{equation}\label{eq:code_hamiltonian}
		H_\Lambda = - \sum_v Q_v - \sum_p B_p \,,	
	\end{equation}
	where $ v $ and $ p $ label the vertices and plaquettes of the trivalent lattice $ \Lambda $, and $ \{Q_v,  B_p\} $ are a set of commuting projectors whose support is shown in \figref{fig:tailed_lattice_support}.
	
	For every vertex $ v $, the three-body projector $ Q_v $ imposes the branching rules:
	\begin{equation}\label{eq:Qv_short}
		 Q_v 
		\big |	\raisebox{-.15cm}{\includegraphics[scale=.7]{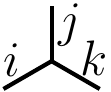}} \big >
		=  \delta_{ijk} \big |	\raisebox{-.15cm}{\includegraphics[scale=.7]{fig/vertex_small.pdf}} \big > \,.
	\end{equation}
	The subspace $ \H_{\text{s.n.}} $ of states that satisfy all vertex projectors is known as the \emph{string-net subspace}. States in $ \H_{\text{s.n.}} $ can be understood as superpositions of string-nets, which are defined as string configurations that obey the branching rules.
	
	We will work with the tailed honeycomb lattice shown in \figref{fig:tailed_lattice_support}, for which the plaquette projector $ B_p $ is a 16-body operator.
	On a generic trivalent tailed lattice, is defined as
	\begin{equation}\label{eq:Bp_short}
			B_p = \frac{1}{\D^2} \sum_s d_s O_p^s\,,
	\end{equation}
	where $ \D = \sqrt{\sum_i d_i^2} $, and 
	\begin{multline} \label{eq:Ops}
		 O_p^s \, \Bigg |
		\raisebox{-1.1cm}{\includegraphics[scale=.32]{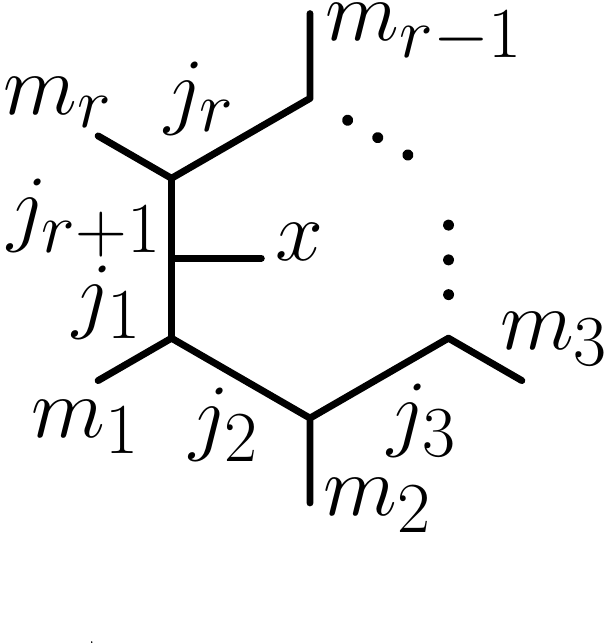}}
		\Bigg \rangle 
		= \delta_{x,\1} \!\!
		\sum_{k_1,\ldots,k_{r+1}} 
		\!\!\!\! \delta_{k_1, k_{r+1}} \\ \cdot \bigg( \! \prod_{\nu = 1}^{r} F^{m_\nu j_{\nu} j_{\nu+1}}_{s k_{\nu+1} k_{\nu} } \! \bigg) \,
		\Bigg |
		\raisebox{-1.1cm}{\includegraphics[scale=.32]{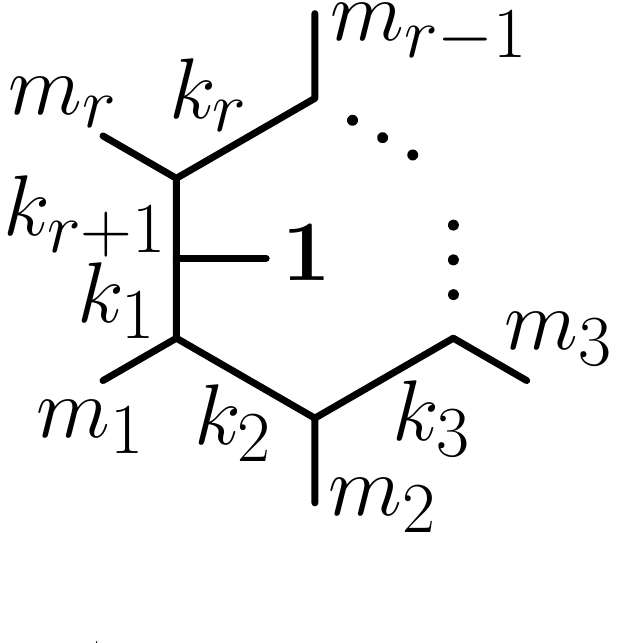}}
		\Bigg \rangle \,.
	\end{multline}
	
	The error correction scheme and numerical simulations described in Sects.~\ref{sec:error_correction} and \ref{sec:simulation} were designed specifically for the Fibonacci input category ($ \C = $ FIB), which contains only two string types, $ \1 $ and $ \tau $. 
	Hence the model we consider in the remainder of this paper is a system of qubits. 
	We choose to relate the string types to the standard computational basis states: $ \1 \to \ket{0} $ and $ \tau \to \ket{1} $.
	The nontrivial fusion rule is $ \tau \times \tau = \1 + \tau $, which leads to the following branching rules:
	\begin{equation}\label{eq:branching_fib}
		\delta_{ijk} = 
		\begin{cases}
			1, \quad \text{if}\; (ijk) \in \{ \1\1\1,\, \tau\tau\1,\, \1\tau\tau,\, \tau\1\tau, \, \tau\tau\tau \},\\
			0, \quad \text{otherwise}.
		\end{cases}
	\end{equation}
	The quantum dimensions are
	\begin{equation}\label{eq:fib_qd_short}
		d_\mathbf{1} = 1\,, \qquad d_\tau = \phi\,,
	\end{equation}
	where $\phi=\frac{1+\sqrt{5}}{2}$ is the golden ratio.
	The only nontrivial $F$-matrix is
	\begin{equation}\label{eq:fib_F_short}
		[F^{\tau\tau}_{\tau\tau}] = 
		\begin{pmatrix}
			\phi^{-1} & \phi^{-\frac{1}{2}}  \\
			\phi^{-\frac{1}{2}}&  -\phi^{-1}
		\end{pmatrix}. \vspace{1pt}
	\end{equation}
	For all other combinations of indices, $ F^{ijm}_{kln} $ is either 1 or 0, depending on whether or not the corresponding indices in Eq.~\eqref{eq:strnet_physicality} satisfy the branching rules. 
	
	The ground space of Hamiltonian \eqref{eq:code_hamiltonian} has a degeneracy that depends on the genus of the surface on which the model is defined. 
	On a torus, and with the Fibonacci input category, the code space is four-dimensional, which enables one to encode the state of two logical qubits. 
	
\subsection{The fattened lattice picture}\label{sec:fattened}
	It is convenient to think about the string-net Hilbert space as the lattice realization of the ribbon graph Hilbert space on a punctured surface.
	We discuss the ribbon graph Hilbert space, and its relation to the Levin-Wen model extensively in App.~\ref{sec:extended_LW_and_TQFT}. 
	For our purpose, it is sufficient to state that this is the space of formal linear combinations of labeled trivalent graphs which satisfy the branching rules in Eq.~\eqref{eq:strnet_branching_rules}, modulo continuous deformations and the following relations:
	\begin{align}
		\raisebox{-0.3cm}{\includegraphics[scale=.40]{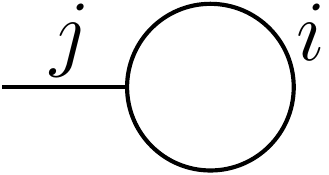}} \;
		&= \delta_{j1} d_i \label{eq:tadpole} \,,\\
		\raisebox{-0.5cm}{\includegraphics[scale=.40]{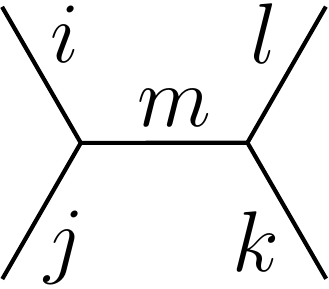}} \;\;
		&= \sum_{n} \; F^{ijm}_{kln} \;\;
		\raisebox{-.5cm}{\includegraphics[scale=.40]{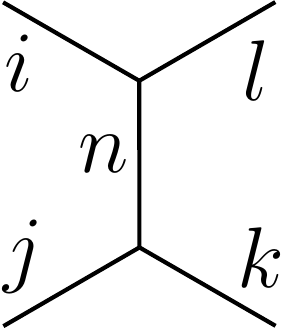}} \,. \label{eq:F} 
	\end{align}
	The second relation is known as an $F$-move or 2-2 Pachner move. 
		
\begin{figure}[h]
		\centering
		\begin{subfigure}{.5\linewidth}
			\includegraphics[scale=.50]{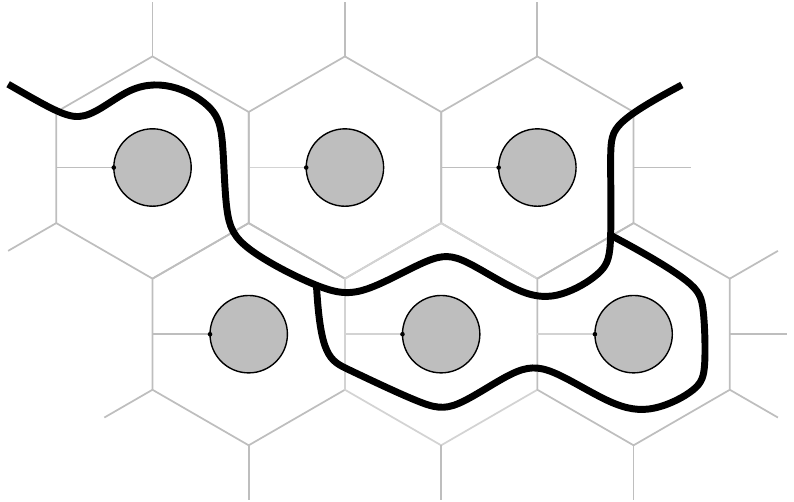}
			\caption{}
			\label{fig:fattened_lattice_1}
		\end{subfigure}
		\begin{subfigure}{.49\linewidth}
			\includegraphics[scale=.50]{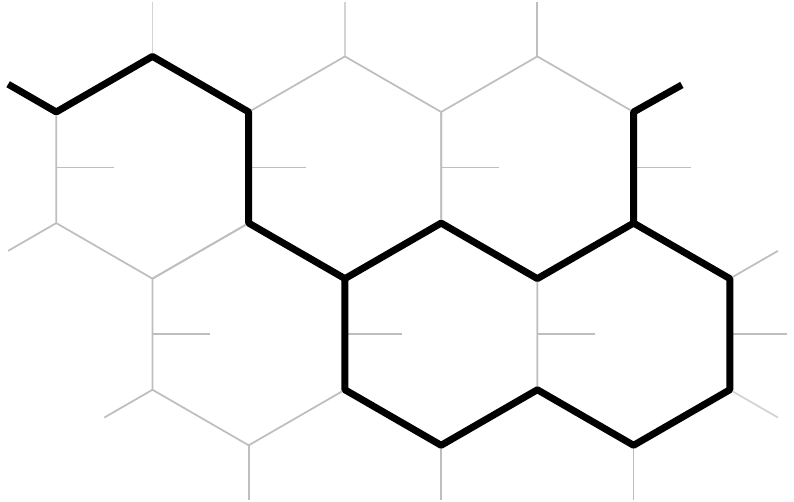}
			\caption{}
			\label{fig:fattened_lattice_2}
		\end{subfigure}
		\caption{(a) A ribbon graph on the fattened lattice.
			(b) The corresponding string-net configuration on the lattice. The grey edges correspond to qudits in the $ \ket{0} $ state, while the state of the black edges is given by the string type of the corresponding piece of the ribbon graph in (a).}
	\end{figure}

	These \emph{ribbon graphs} are defined on a compact, orientable, surface $ \Delta $ containing one puncture for every plaquette in the lattice. Each boundary component has a unique marked boundary point, and ribbons are only allowed to end on these marked boundary points.
	We can relate ribbon graphs on the surface $ \Delta $  to string-nets using the ``fattened lattice'' picture, which represents an embedding of the lattice $ \Lambda $ in the surface $ \Delta $ as shown in \figref{fig:fattened_lattice_1}.
	Whenever ribbons have a more complicated shape that can't be smoothly deformed to the shape of the embedded lattice, one can first deform them using 2-2 Pachner moves, and 1-3 Pachner moves.
	The latter are defined as
	\begin{equation}\label{eq:G_def}
		\raisebox{-0.62cm}{\includegraphics[scale=.40]{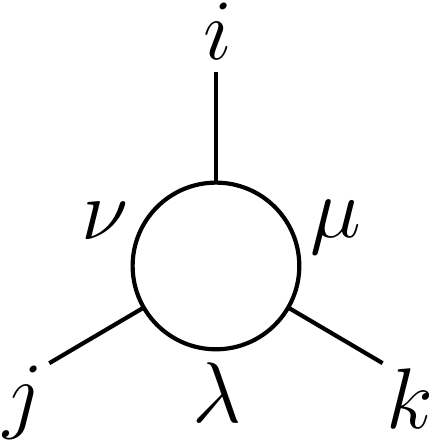}} = v_\lambda v_\mu v_\nu G^{i j k}_{\lambda \mu \nu} \raisebox{-0.42cm}{\includegraphics[scale=.40]{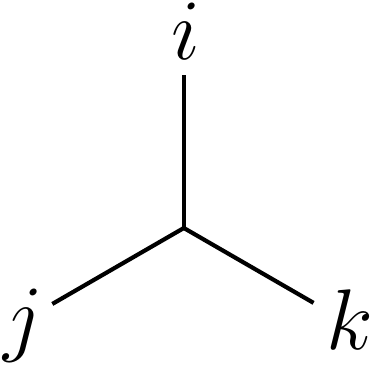}}\,,
	\end{equation}
	where 
	\begin{equation}\label{eq:G_value}
		G^{ijk}_{\lambda \mu \nu}  = \dfrac{1}{v_i v_\lambda} F^{\nu j \lambda}_{k \mu i}  = \dfrac{1}{v_\nu v_k} F^{ijk}_{\lambda \mu \nu} \,.
	\end{equation}
	
	The action of $ O_p^s $, defined in Eq.~\eqref{eq:Ops}, can be represented in the fattened lattice picture as the inclusion of a loop with label $ s $ around the puncture in plaquette $ p $. 
	Hence, the action of the plaquette projector $ B_p $ can be represented on the fattened lattice as follows:
	\begin{equation}\label{eq:Bp_fattenedlattice}
		B_p : \quad \raisebox{-.8cm}{\includegraphics[scale=.55]{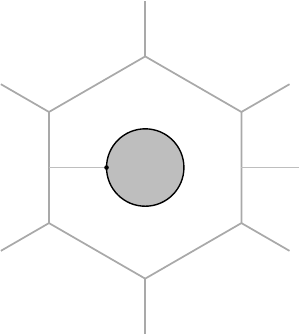}} \; \mapsto \; \dfrac{1}{\D} \raisebox{-.8cm}{\includegraphics[scale=.55]{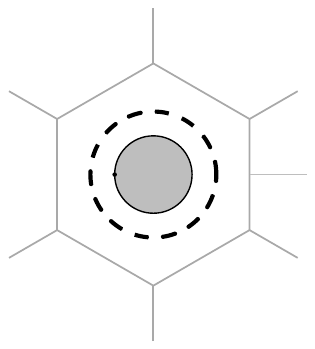}}\,,
	\end{equation}
	where
	\begin{equation}\label{eq:vacuum_line_strnet}
		\raisebox{-0.4cm}{\includegraphics[scale=.40]{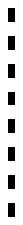}} \quad
		= 
		\frac{1}{\D} \sum_{i} d_i  \,\,\,
		\raisebox{-.4 cm}{\includegraphics[scale=.40]{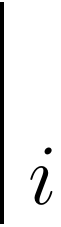}} \;.
	\end{equation}
	The dashed loop is referred to as a \emph{vacuum loop}.
	Note that we have omitted the tail edge on the right hand side of Eq.~\eqref{eq:Bp_fattenedlattice}. After resolving the loop into the lattice using a sequence of $F$-moves, a trivial tail edge should be included. Keep in mind that $ B_p = 0 $ whenever there is a (nontrivial) ribbon ending in the puncture $ p $, which corresponds to a nontrivial tail edge.

\begin{figure}[h]
		\centering
		\begin{subfigure}{.5\linewidth}
			\includegraphics[scale=.5]{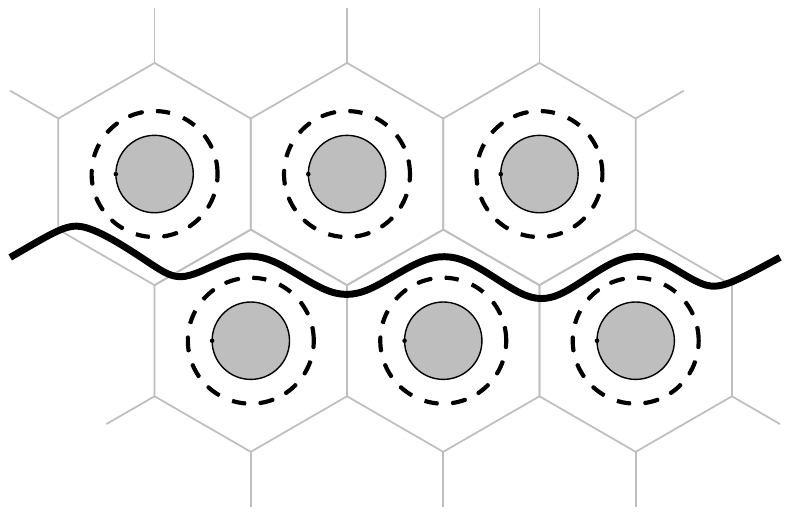}
		\end{subfigure}
		\begin{subfigure}{.49\linewidth}
			\includegraphics[scale=.5]{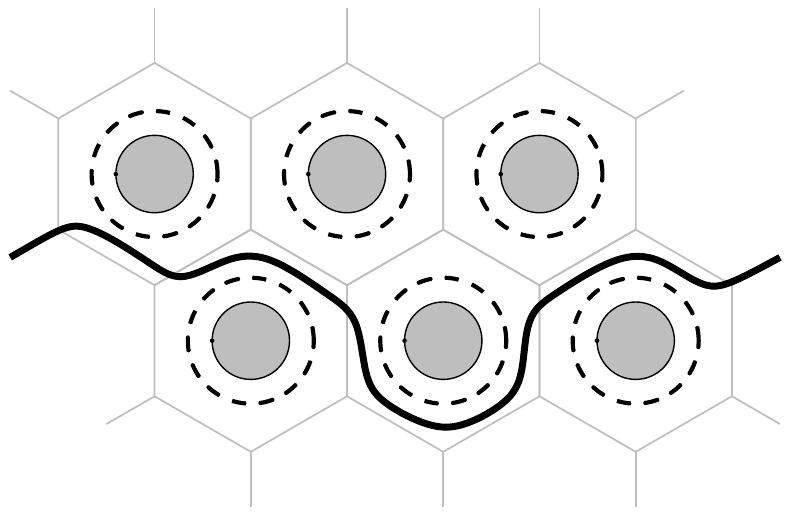}
		\end{subfigure}
		\caption{Two ribbon graphs on the fattened lattice representing the same string-net ground state.}
		\label{fig:fattened_lattice_groundstate}
	\end{figure}

	Ground states (up to a normalization factor) then correspond to ribbon graph configurations without any ribbons ending in punctures, and where a vacuum loop is added around every puncture as shown in \figref{fig:fattened_lattice_groundstate}.
	A short calculation shows that ribbons can be ``pulled across'' vacuum loops without changing the corresponding string-net state, meaning that one can obtain the same ground state by applying the $ \prod_p B_p $ to different string-net states. 
	
	\subsection{Code deformation using Pachner moves}
\begin{figure}
		\centering
		\begin{subfigure}{\linewidth}
			\centering
			\includegraphics[width=\columnwidth]{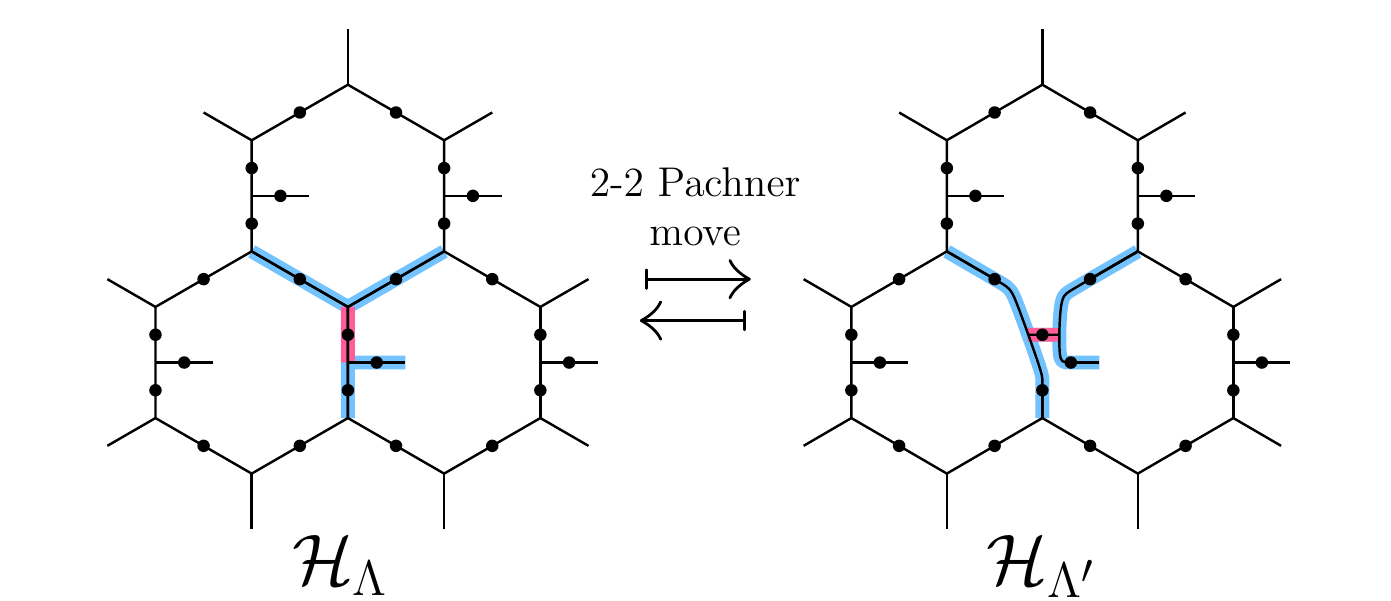} 
		\end{subfigure}
		\begin{subfigure}{\linewidth}
			\centering
			\includegraphics[width=\columnwidth]{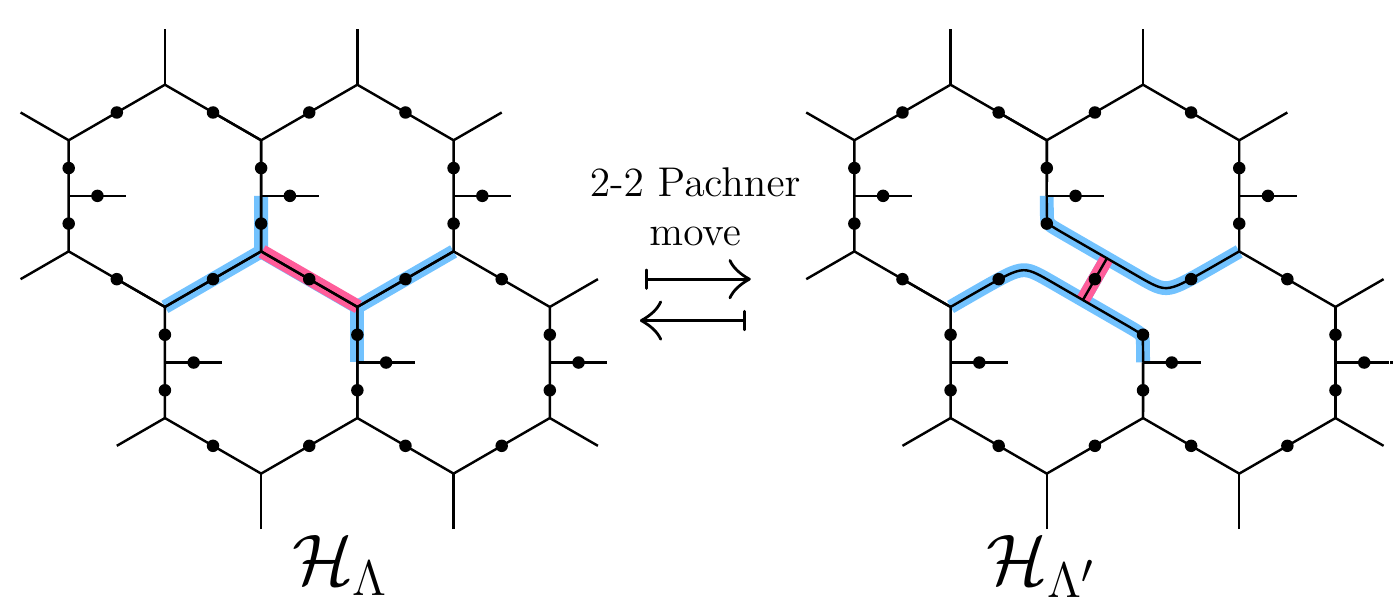}
		\end{subfigure}
		\caption{Lattice deformations by 2-2 Pachner moves on different edges. The affected edge is highlighted in pink, the other edges contained in the $F$-matrix is are highlighted in blue.}
		\label{fig:lattice_deformation}
	\end{figure}

\begin{figure}
		\centering
	    \includegraphics[width=.9\columnwidth]{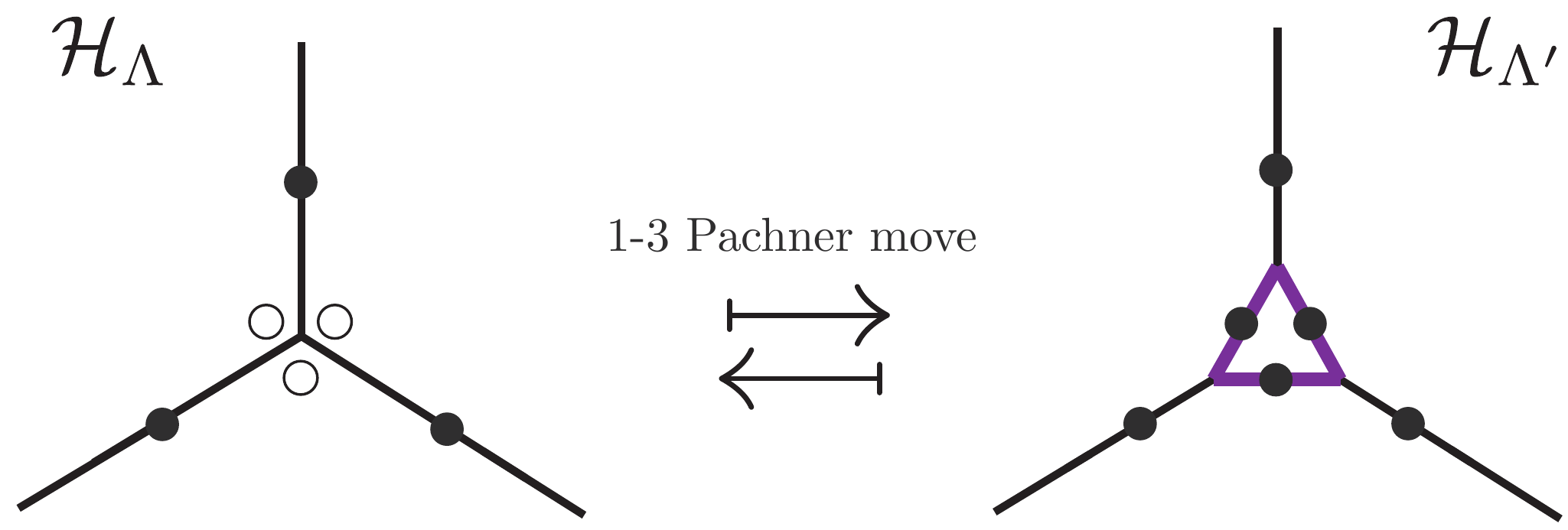}
		\caption{Lattice deformation by 1-3 Pachner move. The white dots on the left represent ancilla qudits. The fine-graining process (left to right) entangles the three ancilla qudits into the code space. 
		The coarse-graining process (right to left) disentangles the three central qudits out of the code space.}
		\label{fig:1-3_Pachner}
	\end{figure}

	The Pachner moves introduced above can be used to relate string-net states defined on different lattice geometries. 
	In particular, when transforming the lattice $ \Lambda $ to $ \Lambda' $, Eqs.~\eqref{eq:F} and \eqref{eq:G_def} give the appropriate map between the corresponding code spaces $\mathcal{H}_\Lambda$ and $\mathcal{H}_{\Lambda'}$, defined as the ground spaces of Hamiltonians $ H_\Lambda $ and $ H_{\Lambda'} $, respectively. 
	For instance, applying the unitary $F$-move (see \figref{fig:F-move_circuit}) to the qudits on certain edges of a ground state, transforms this state to a ground state of the string-net Hamiltonian defined on a new lattice obtained by recoupling these edges in the original lattice.
	The recoupling of lattice edges by a 2-2 Pachner move is shown in Fig.~\ref{fig:lattice_deformation}.
	The lattice deformation corresponding to a 1-3 Pachner move is shown in Fig.~\ref{fig:1-3_Pachner}, where one adds a triangular loop on the original vertex.  The 1-3 Pachner move can be thought as a fine/coarse-graining process. 
	In Fig.~\ref{fig:1-3_Pachner} one entangles three ancilla qudits (white dots) from left to right and add them into the code space, which effectively fine-grain the lattice. The inverse process from right to left disentangles the qudits in the center out of the code space, which effectively coarse-grains the lattice.  Both 2-2 and 1-3 Pachner moves can be implemented via unitary circuits as will be shown in Sec.~\ref{sec:measure_charge}. 
	

	\subsection{Anyonic excitations} \label{sec:excitations}
	Localized excitations in the string-net model exhibit anyonic statistics, described by the quantum double $ \D\C $ of the input category $ \C $ \cite{levin2005string}.
	It is important to note that the categorical double of a unitary fusion category is always braided. Hence, the input category $ \C $ does not need to include any braiding structure for the excitations to have well defined anyonic statistics. 
	When the input category $ \C $ is modular (implying it is braided), such as for $ \C = \text{FIB} $, the doubled category has a special structure $ \D\C \cong \C \otimes \bar{\C} $, and its string types can be labeled by pairs $ a_+\overline{a_-} $ with $ a_+, a_- \in \C $ (see App.~\ref{sec:extended_LW_and_TQFT} for more details). 
	For notational simplicity, we will drop the bar notation for labels in $ \bar{\C} $, 
	and we will indicate the string-types of the doubled category with bold labels: $ \bm{a} =  a_+a_- \in \D\C$.  
	
\begin{figure}
		\centering
		\includegraphics[width=1\columnwidth]{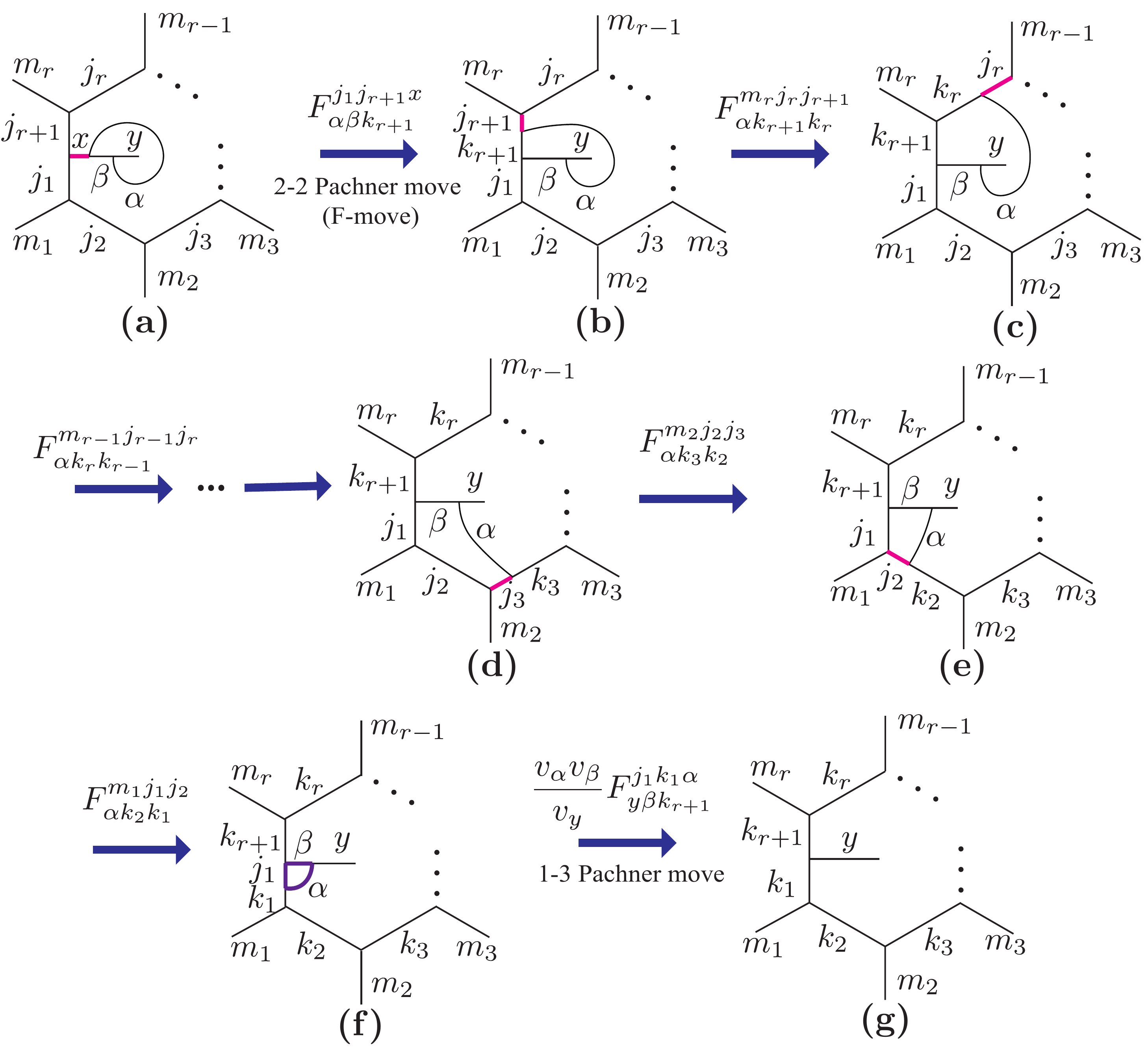}
		\caption{Derivation of the expression for the tube operator.}
		\label{fig:tube_derivation}
	\end{figure}

	The motivation for introducing the tail qudit in every plaquette is that they allow us to define the action of an operator algebra known as Ocneanu's tube algebra \cite{ocneanu2001operator} in TQFT (see Sec.~\ref{sec:tube_algebra} in App.~\ref{sec:extended_LW_and_TQFT}), at the level of the lattice qudits.
	The central idempotents of the tube algebra form projectors onto the different superselection sectors of the theory. 
	By defining their action on an individual plaquette, one obtains a set of projectors corresponding to the different possible values of the doubled anyonic charge contained in that plaquette .
	The generators of this algebra act on an individual plaquette as
	\vspace{-1cm}
	\begin{widetext}
		\begin{align} \label{eq:tube_operator_lattice_short}
			O_{x y \alpha \beta } \, \Bigg |
			\raisebox{-1.1cm}{\includegraphics[scale=.32]{fig/plaquette_stick_1}}
			\Bigg \rangle
			 = \delta_{x,x'} \Bigg |
			\raisebox{-1.1cm}{\includegraphics[scale=.32]{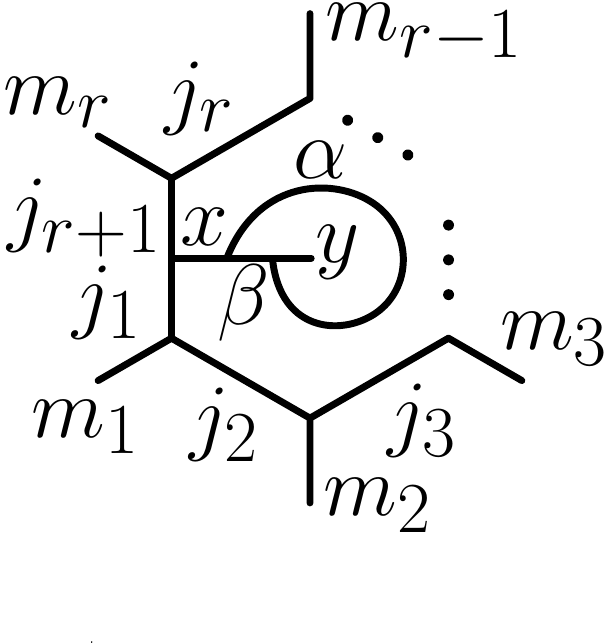}}
			\Bigg \rangle 
			 = \delta_{x,x'} \dfrac{v_\alpha v_\beta}{v_y}\!\!
			\sum_{k_1,\ldots,k_{r+1}} 
			\!\!\!\! F^{j_1 j_{r+1} x}_{\alpha \beta k_{r+1}} \bigg( \! \prod_{\nu = 1}^{r} F^{m_\nu j_{\nu} j_{\nu+1}}_{\alpha k_{\nu+1} k_{\nu} } \! \bigg) F^{ k_{r+1} k_1 y}_{\alpha j_1 \beta}
			\Bigg |
			\raisebox{-1.1cm}{\includegraphics[scale=.32]{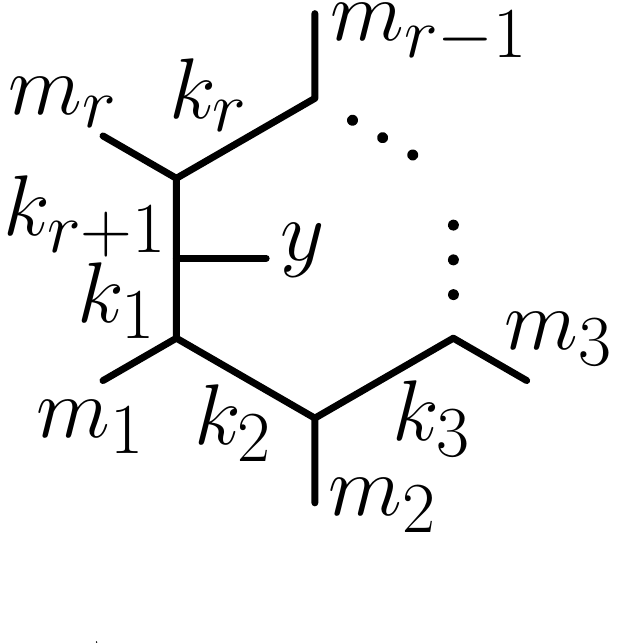}}
			\Bigg \rangle \,.   
		\end{align}
	\end{widetext}
	This corresponds to gluing the ``tube'' 
	\begin{align}\label{eq:tube}
		\raisebox{-.7cm}{\includegraphics[scale=.4]{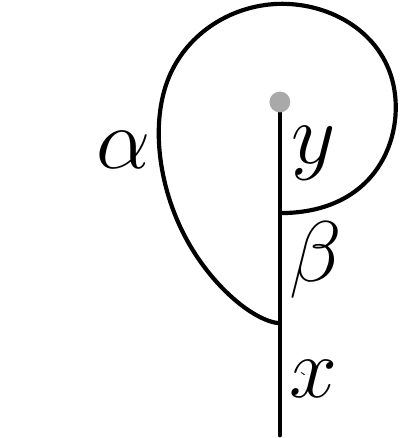}}
	\end{align} 
	onto the tail edge and resolving it into the lattice using a sequence of 2-2 Pachner moves followed by a 1-3 Pachner move, as shown in \figref{fig:tube_derivation}.

	For the Fibonacci input category, the doubled category (DFIB) contains the string types $ \{ \1\1, \1\tau, \tau \1, \tau\tau\} $, which label the different corresponding projectors
	\begin{align}
		\cP^{\1\1} &= \dfrac{1}{\D^2} \left(O_{\1\1\1\1} + \phi O_{\1 \1 \tau \tau} \right ), \label{eq:P11_short} \\
		\cP^{\1\tau} &= \dfrac{1}{\D^2} \left(O_{\tau \tau  \1 \tau}  + \e^{4\pi \ii/5} O_{\tau\tau \tau \1} + \sqrt{\phi} \e^{-3\pi \ii/5} O_{\tau\tau\tau\tau} \right ), \label{eq:P12_short} \\
		\cP^{\tau \1} &= \dfrac{1}{\D^2} \left (O_{\tau \tau \1 \tau}  + \e^{-4\pi \ii/5} O_{\tau\tau\tau \1} + \sqrt{\phi} \e^{3\pi \ii/5} O_{\tau\tau\tau\tau} \right ), \label{eq:P21_short}\\
		\cP^{\tau\tau} &= \dfrac{1}{\D^2} \bigg(\phi^2 O_{\1\1\1\1} - \phi O_{\1\1 \tau\tau} +   \phi O_{\tau \tau \1 \tau} + \phi O_{\tau\tau\tau \1} \nonumber \\ 
		& \qquad + \dfrac{1}{\sqrt{\phi}} O_{\tau\tau\tau\tau} \bigg) . \label{eq:P22_short}
	\end{align}
	Note that $ B_p  = \cP^{\1\1}$ i.e., the ground space is precisely the anyonic vacuum.
	The entry in Eq.~\ref{eq:P22_short} decomposes into two simple idempotents: $ \cP^{\tau \tau} = \cP^{\tau \tau}_\1 + \cP^{\tau \tau}_\tau $, with	
	\begin{align}
		\cP^{\tau\tau}_\1 &= \dfrac{1}{\D^2} (\phi^2 O_{\1\1\1\1} - \phi O_{\1\1\tau \tau}), \label{eq:P221_short}\\
		\cP^{\tau\tau}_\tau &= \dfrac{1}{\D^2}  \left(\phi O_{\tau \tau \1  \tau} + \phi O_{\tau\tau\tau \1} + \dfrac{1}{\sqrt{\phi}} O_{\tau\tau\tau\tau}  \right). \label{eq:P222_short}
	\end{align}
	This decomposition of the central $ \tau\tau $ idempotent into two simple idempotents should be interpreted as the fact that a $ \tau\tau $ anyon charge in a plaquette does not fix the state of its tail qubit. 
	The corresponding projector has a block-diagonal form with the two blocks corresponding to a $ \ket{0} $ or a $ \ket{1} $ state for the tail qubit. We label these two cases as $ \tau\tau_\1 $ and $ \tau\tau_\tau $, respectively.
	Both are in the $ \tau\tau $ anyon sector and their respective +1 eigenspaces are related as follows:
	\begin{equation}
		\cP^{\tau\tau}_{\1\tau} \cP^{\tau\tau}_\1 = \cP^{\tau\tau}_\tau ,
		\qquad \quad 
		\cP^{\tau\tau}_{\tau\1} \cP^{\tau\tau}_\tau = \cP^{\tau\tau}_\1 ,
	\end{equation}
	where 
	\begin{align} \label{eq:nilpotents_fib}
		\cP^{\tau\tau}_{\1\tau}  =  \e^{-3\pi \ii/10} \dfrac{\phi}{\D} O_{\1\tau\tau\tau}\,,  \\
		\cP^{\tau\tau}_{\tau\1} = \e^{3\pi \ii/10} \dfrac{\sqrt{\phi}}{\D} O_{\tau\1\tau\tau} \label{eq:nilpotents_fib_2}\,,
	\end{align}
	are nilpotent operators.
	
	\subsection{The anyonic fusion basis}\label{sec:anyonic_fusion_basis_short}
    An important property of the extended Levin-Wen model is the fact that one can construct a basis of the string-net subspace $ \H_{\text{s.n.}} $, whose elements are labeled by fusion states of $ |P| $ anyons (of the doubled category $ \D\C $), where $ |P| $ is the number of plaquettes. This basis is called the \emph{anyonic fusion basis}.
	Below, we will give the expressions for these bases in terms of ribbon configurations in the fattened lattice picture, for modular input categories. More details are given in Apps.~\ref{sec:anyonic_fusion_basis} and \ref{sec:anyonic_fusion_basis_lattice}.
	
	For simplicity we first show anyonic fusion basis states on a sphere.
	Since we are considering the fusion space of anyonic excitations in plaquettes, the corresponding fusion diagram are labeled by the string types of the doubled category $ \D\C$ (indicated by bold labels):
	\begin{equation}\label{eq:anyonic_fusion_basis_comp_short}
		\ket{\vec{\bm{a}}, \vec{\bm{b}}} =	\raisebox{-1cm}{\includegraphics[scale=.5]{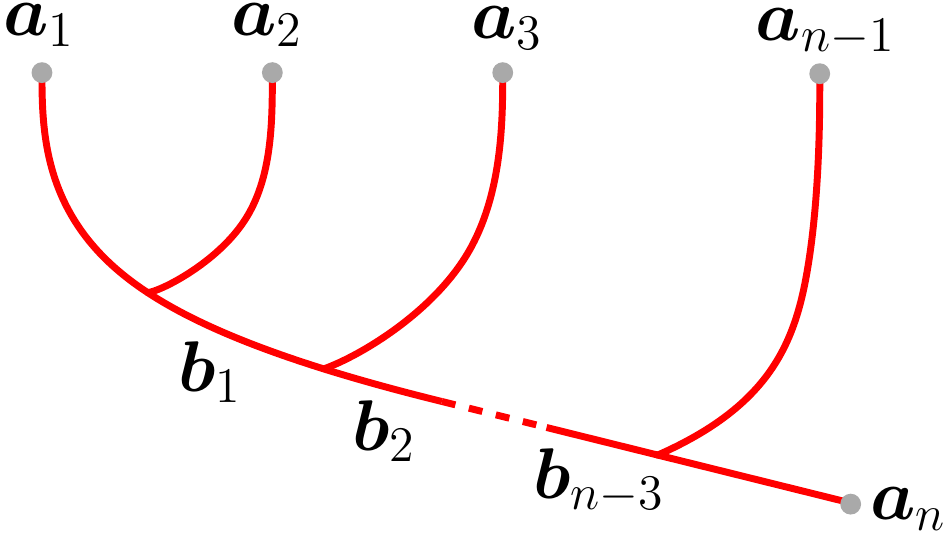}}  .
	\end{equation}

	As mentioned above, the anyon label of a plaquette alone does not always fix the state of the tail qudit. Hence, to fully specify the state, we must also fix the tail labels 
	\begin{equation*}
		  \vec{\ell} = [\ell_1,\, \ell_2,\, \ldots, \, \ell_{|P|} ], 
	\end{equation*}
	where $ \vec{\ell} $ must be consistent with the anyon labels of all plaquettes.
	For $ \C = \text{FIB} $, the allowed combinations of plaquette anyon (DFIB) labels $ \bm{a} = a_+ a_- $ and tail labels $ \ell $ are $ (a_+a_-)_\ell \in \{ \1\1_\1, \1\tau_\tau, \tau\1_\tau, \tau\tau_\1, \tau\tau_\tau \} $, which are simply all combinations satisfying $ \delta_{a_+a_-\ell} = 1 $.
	
	Throughout the remainder of this work, we will often adopt a slight abuse of notation by also using a bold label to indicate the joined labels of the plaquette anyon and tail labels: $ \bm{a} =   (a_+a_-)_\ell$. 
	It will always be clear from the context whether or not a bold label includes the tail label. In particular, only leaf labels can contain a tail label. Internal branch labels of a doubled anyonic fusion tree never include them, since they do not correspond to plaquettes.
	
	An anyonic fusion basis is determined by fixing the branching structure of the corresponding fusion tree and its embedding in the fattened lattice. 
	The basis states are then labeled as $ \ket{\vec{\ell}, \vec{\bm{a}}, \vec{\bm{b}}} $, where $ \vec{\ell} $ are the tail labels, $ \vec{\bm{a}} $ are the leaf labels (corresponding to the anyon charge of each plaquette), and $ \vec{\bm{b}} $ are the internal branch labels.
	Before embedding them into the fattened lattice, the corresponding ribbon configurations are
	\begin{flalign}
		\ket{\vec{\ell},\vec{\bm{a}}, \vec{\bm{b}}} \,
		=  \hspace{-.2cm} \raisebox{-1.8cm}{\includegraphics[scale=.46]{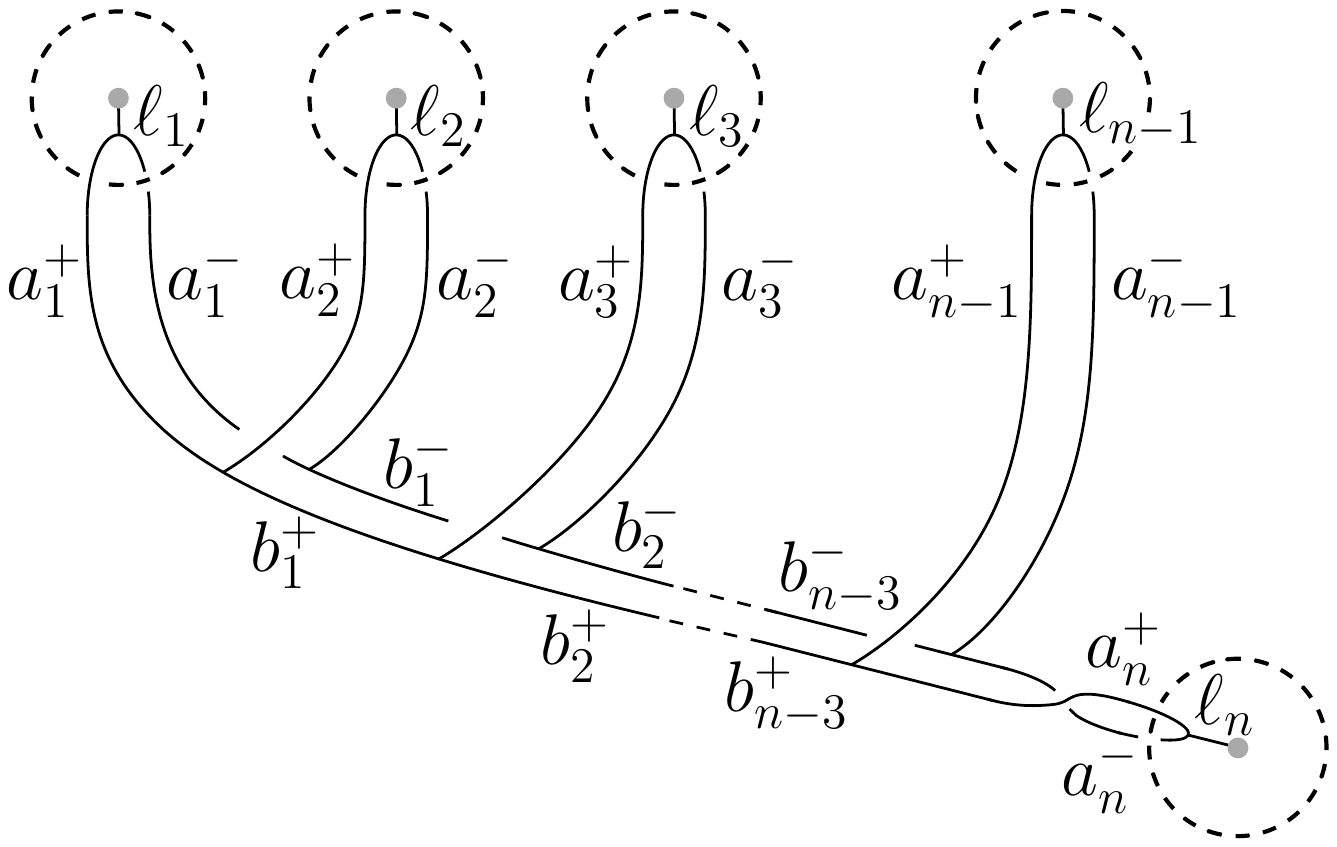}}  \label{eq:anyonic_fusion_basis_states_n_short} \\ %
		\vspace{.4cm}
		  = 	\sum_{\vec{\alpha}, \vec{\beta}, \vec{k}}  X^{\vec{\ell}, \vec{\bm{a}}, \vec{\bm{b}}}_{\vec{\alpha}, \vec{\beta}, \vec{k}, \vec{l}} 
		\hspace{-1.5cm}
		\raisebox{-2.5cm}{\includegraphics[scale=.46]{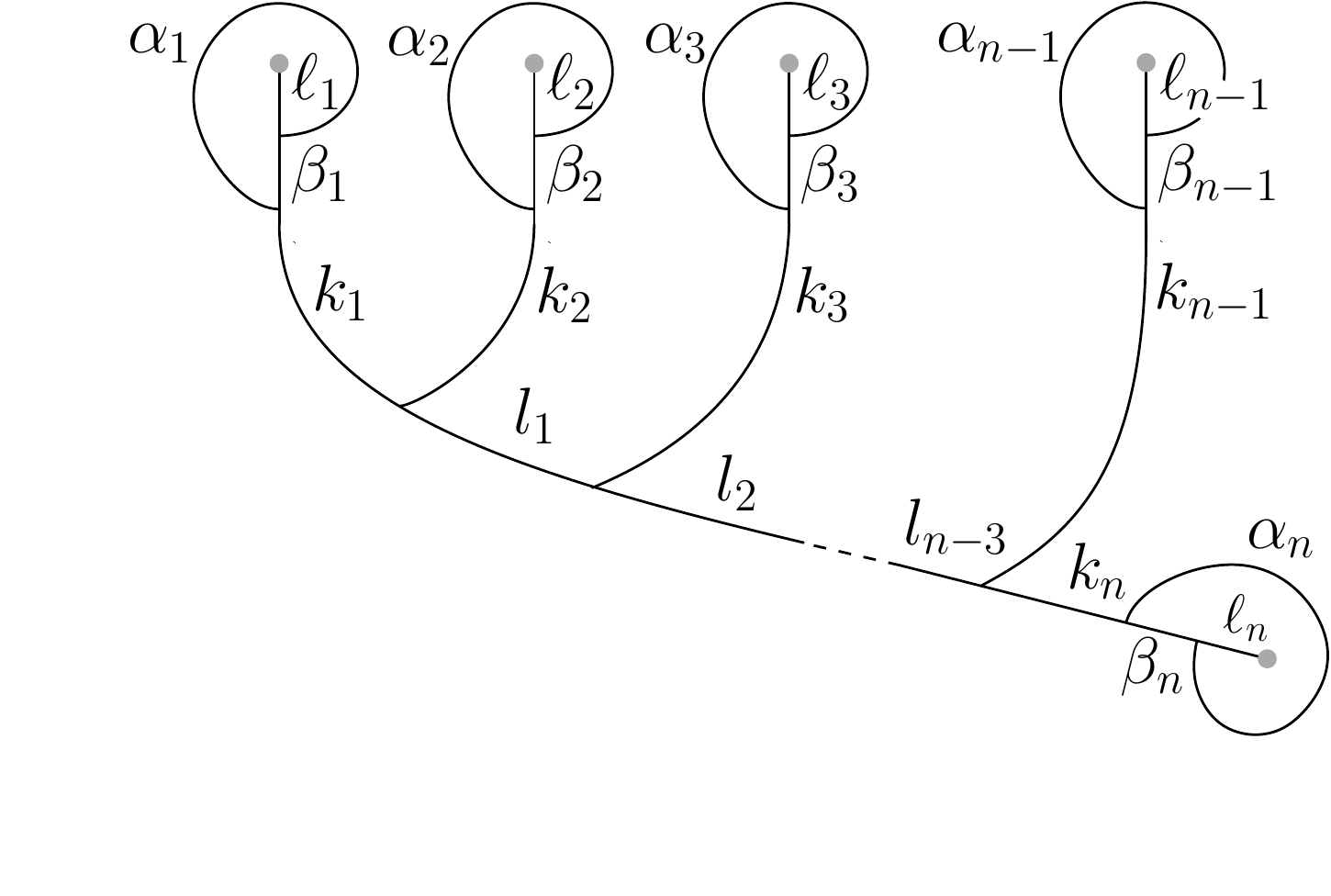}} \hspace{-.3cm}, & \label{eq:anyonic_fusion_basis_red_short}
	\end{flalign}
	where the coefficients $ X^{\vec{\ell}, \vec{\bm{a}}, \vec{\bm{b}}}_{\vec{\alpha}, \vec{\beta}, \vec{k}, \vec{l}}  $ are found by resolving the crossings in Eq.~\ref{eq:anyonic_fusion_basis_states_n_short} using
	\begin{align}\label{eq:resolve_crossing_short}
		\raisebox{-.5cm}{\includegraphics[scale=.40]{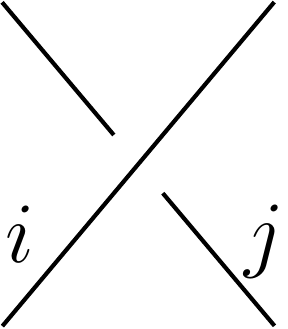}}
		\;\; = \sum_{k} \frac{v_k}{v_i v_j} \, R^{ij}_k \quad
		\raisebox{-.5cm}{\includegraphics[scale=.40]{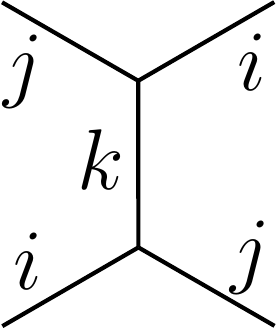}}\;,
	\end{align}
	followed by a sequence of $F$-moves and 1-3 Pachner moves.
	The object $ R^{ij}_k $ above is known as the $ R $-matrix of the input category $ \C $, and defines its braiding properties. 
	It must satisfy certain consistency equations, which are listed in App.~\ref{sec:category}.
	For the Fibonacci category, the only nonzero entries are
	\begin{equation}\label{eq:fib_R_short}
		R^{\tau\tau}_{\mathbf{1}} = \e^{\frac{4\pi\ii}{5}} \,, \quad\; R^{\tau\tau}_{\tau} = \e^{\frac{-3\pi\ii}{5}} \,, \quad\;  R^{\mathbf{1} a}_a = R^{a \mathbf{1}}_a = 1  \,, 
	\end{equation}
	where $ a \in \{\1,\,\tau\} $.
	The necessary calculations to obtain $ X^{\vec{\ell}, \vec{\bm{a}}, \vec{\bm{b}}}_{\vec{\alpha}, \vec{\beta}, \vec{k}, \vec{l}} \,$ were performed explicitly in App.~\ref{sec:anyonic_fusion_basis}.
	
	After picking some embedding in the fattened lattice, we find
	\begin{flalign} \label{eq:anyonic_fusion_basis_lattice_short}
		\ket{\vec{\ell},\vec{\bm{a}}, \vec{\bm{b}}}  = \; \raisebox{-1.1cm}{\includegraphics[scale=.32]{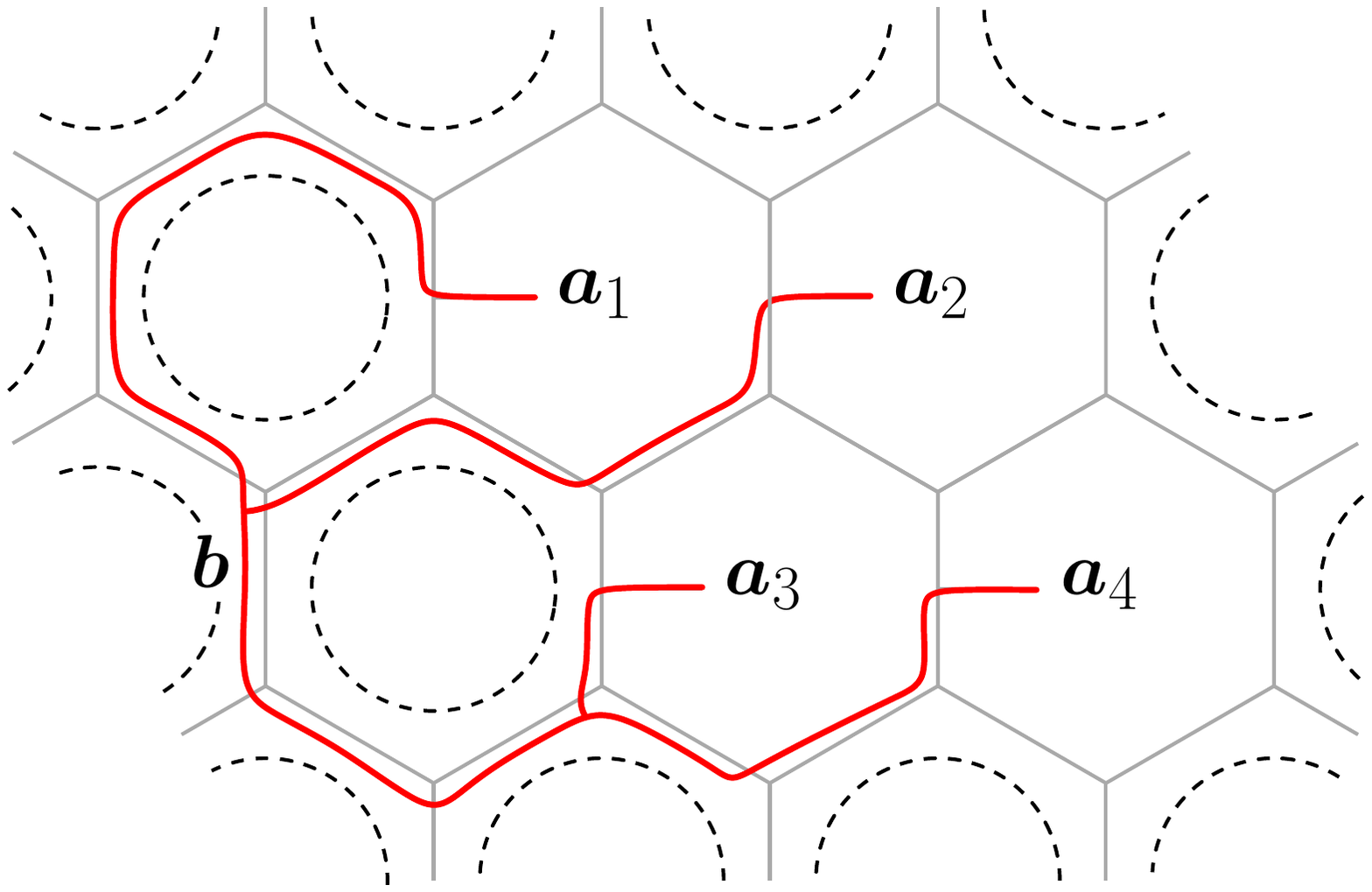}} & \\  
		 =  \sum_{\vec{\alpha}, \vec{\beta}, \vec{k}, \vec{l}} X^{\vec{\ell}, \vec{\bm{a}}, \vec{\bm{b}}}_{\vec{\alpha}, \vec{\beta}, \vec{k}, \vec{l}}   \hspace{-.1cm} \raisebox{-1.1cm}{\includegraphics[scale=.32]{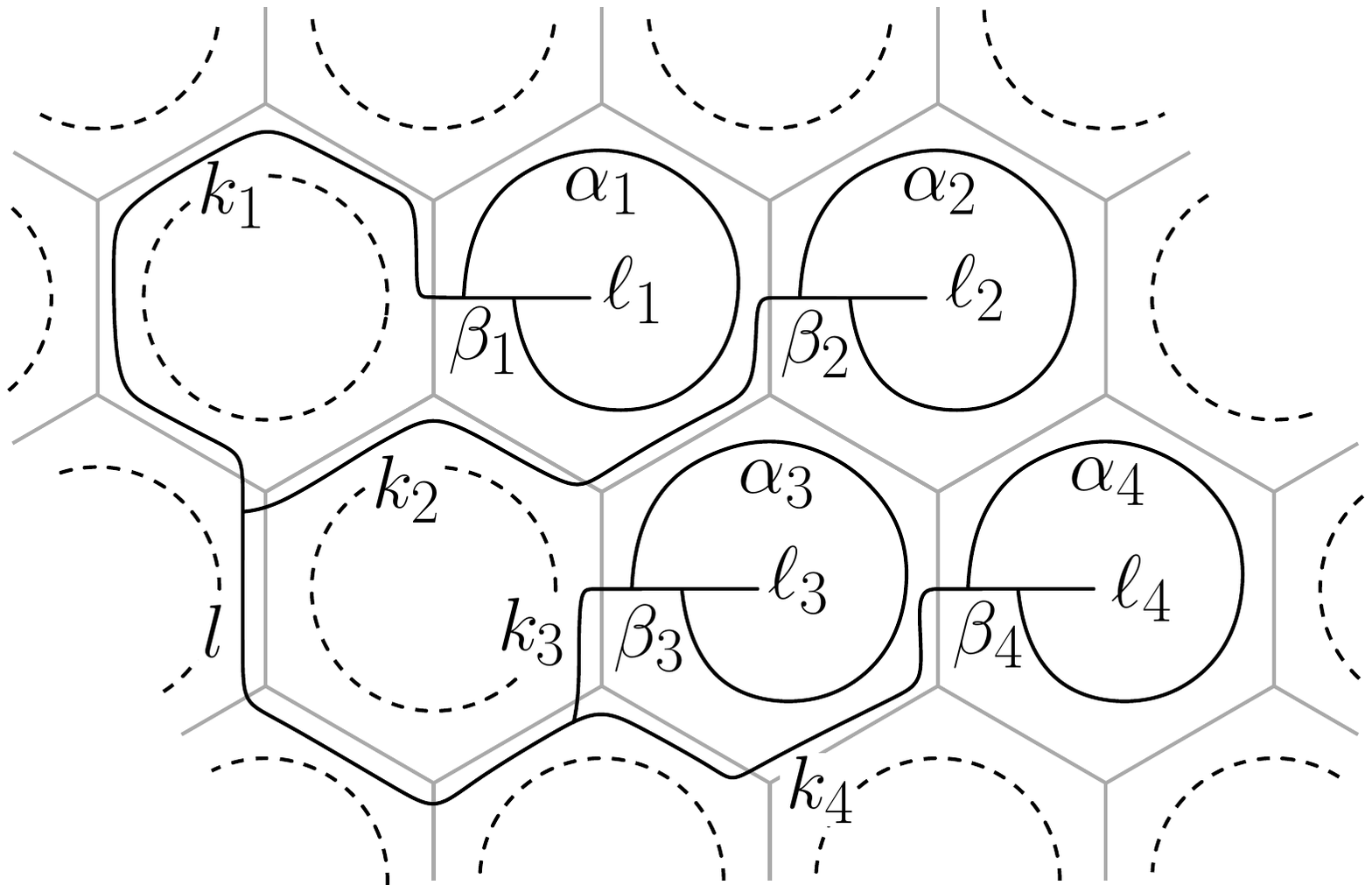}}  & \; , 
	\end{flalign}
	where we chose not to draw the leaves with vacuum labels explicitly, but included vacuum loops in the corresponding plaquettes instead.
	The fattened lattice state above corresponds to the case where only 4 plaquettes carry a nontrivial anyonic charge, other cases are analogous.
	The final expression for the anyonic fusion basis states (as a state of qudits) can then be found by resolving these ribbon graph configurations into the lattice.
	
	Different anyonic fusion bases (that is, bases corresponding to trees with different branching structures, or different embeddings in the fattened lattice), can be related using $F$-moves, braid moves and Dehn twists defined by the categorical data of $ \D\C $:
	\begin{equation}\label{key}
		\raisebox{-.8cm}{\includegraphics[scale=.45]{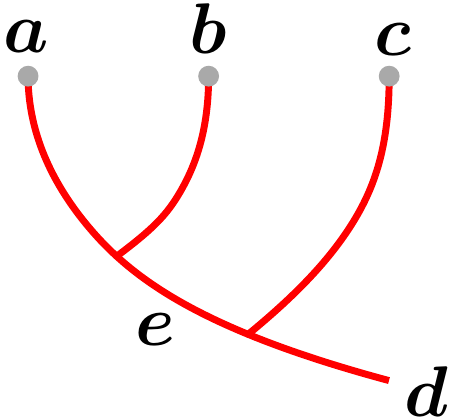}} = \sum_{\bm{f}}F^{\bm{a}\bm{b}\bm{e}}_{\bm{c}\bm{d}\bm{f}} \raisebox{-.8cm}{\includegraphics[scale=.45]{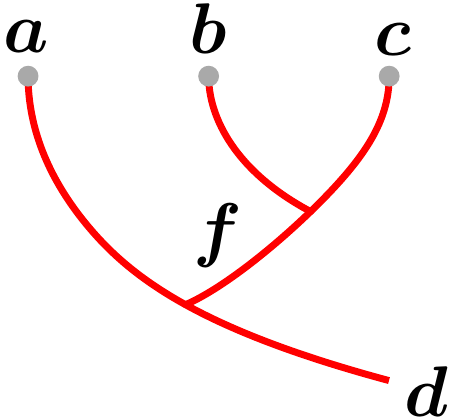}}\,,
	\end{equation}
	\begin{align} 
		\raisebox{-.8cm}{\includegraphics[scale=.45]{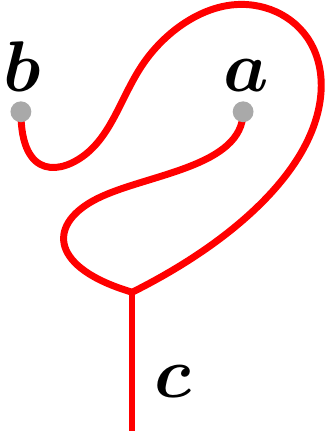}} &= R^{\bm{a}\bm{b}}_{\bm{c}} \raisebox{-.8cm}{\includegraphics[scale=.45]{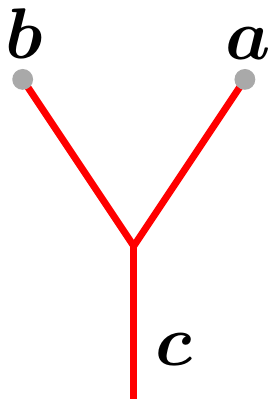}}\,, \\
		\raisebox{-.8cm}{\includegraphics[scale=.45]{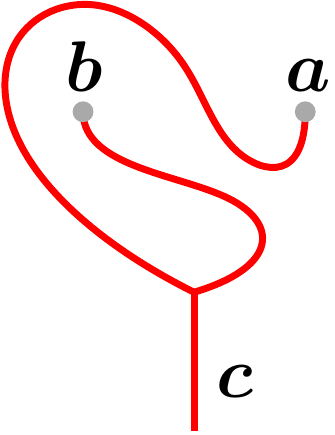}} &= \left( R^{\bm{b}\bm{a}}_{\bm{c}}\right)^* \raisebox{-.8cm}{\includegraphics[scale=.45]{fig/R-move_comp_2}}\,,
	\end{align}
	\begin{align}\label{key}
			\raisebox{-.55cm}{\includegraphics[scale=.45]{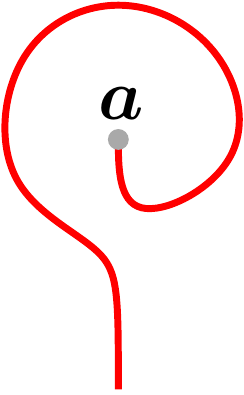}} &= \theta_{\bm{a}}\;\, \raisebox{-.55cm}{\includegraphics[scale=.45]{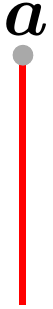}}\,,\\
			\raisebox{-.55cm}{\includegraphics[scale=.45]{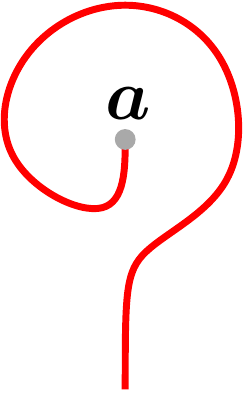}} &= (\theta_{\bm{a}})^*\;\, \raisebox{-.55cm}{\includegraphics[scale=.45]{fig/Dehn-twist_comp_2}}\,.
	\end{align}
	Due to the particular structure of the doubled category, $ \D\C \simeq \C \otimes \bar{\C} $, its numerical data can be deducted from that of the input category $ \C $. 
	In particular, one has 
	\begin{align} 
		F^{\bm{a}\bm{b}\bm{e}}_{\bm{c}\bm{d}\bm{f}} &= F^{a_+b_+e_+}_{c_+d_+f_+} F^{a_-b_-e_-}_{c_-d_-f_-}\,,\\
		R^{\bm{a}\bm{b}}_{\bm{c}} &= R^{a_+b_+}_{c_+} (R^{b_- a_-}_{c_-})^*\,,\\
		\theta_{\bm{a}} &= \theta_{a_+}(\theta_{a_-})^*\,.
	\end{align}
	We discussed how these are obtained for modular input categories in more detail in App.~\ref{sec:anyonic_fusion_basis}.
	
	\begin{figure}[b]
		\centering
		\includegraphics[scale=.55]{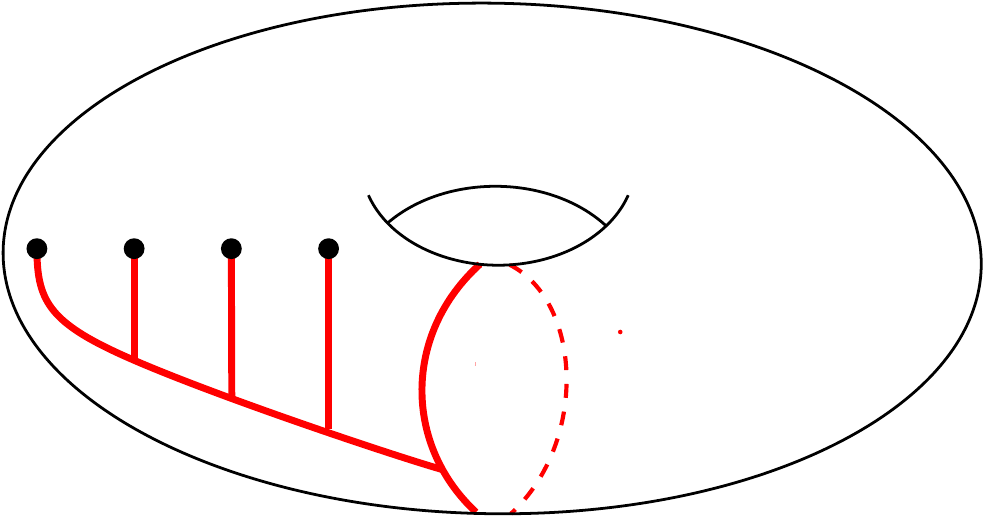}
		\caption{Fusion diagram of a system of anyons defined on a torus. Note the line wrapping around the torus. }
		\label{fig:anyons_torus}
	\end{figure}
	
	The construction of anyonic fusion basis states on a torus is similar.
	An important difference is that fusion states of anyons on a torus requires us to specify a \emph{handle label} (see Sec.~\ref{sec:higher-genus} for more details), which determines how the state transforms when an anyon is moved along a non-contractible loop \cite{pfeifer2012translation}.
	An example of such a fusion state is shown in \figref{fig:anyons_torus}.
	Anyonic fusion basis states are then labeled as  $ \ket{\vec{\ell}, \vec{\bm{a}}, \vec{\bm{b}}, \bm{c} } $, where $ \vec{\ell}$, $\vec{\bm{a}}$ and $ \vec{\bm{b}} $ are again the tail, leaf and internal branch labels, respectively, and $ \bm{c} $ is the handle label.
	The corresponding ribbon configurations are
	\begin{widetext}
	\vspace{-.5cm}
		\begin{align}\label{eq:anyonic_fusion_basis_states_torus_short}
			\ket{\vec{\ell}, \vec{\bm{a}}, \vec{\bm{b}}, \bm{c}}  = 
			\;\;
			\raisebox{-1.6cm}{\includegraphics[scale=.5]{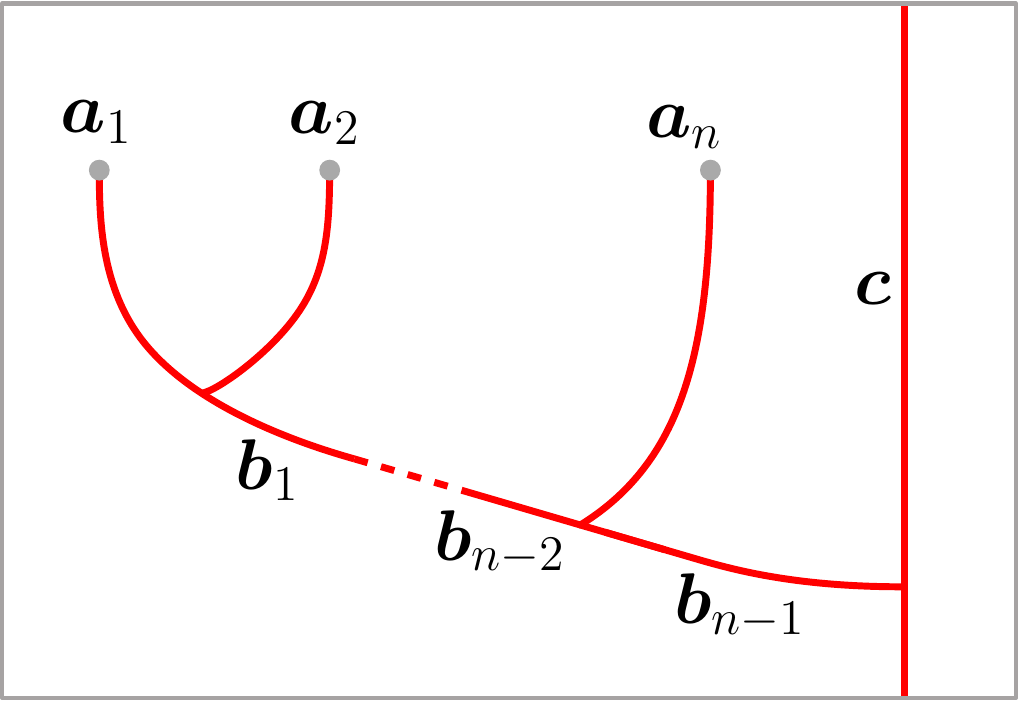}}  \quad = \;\;
			\raisebox{-2.2cm}{\includegraphics[scale=.5]{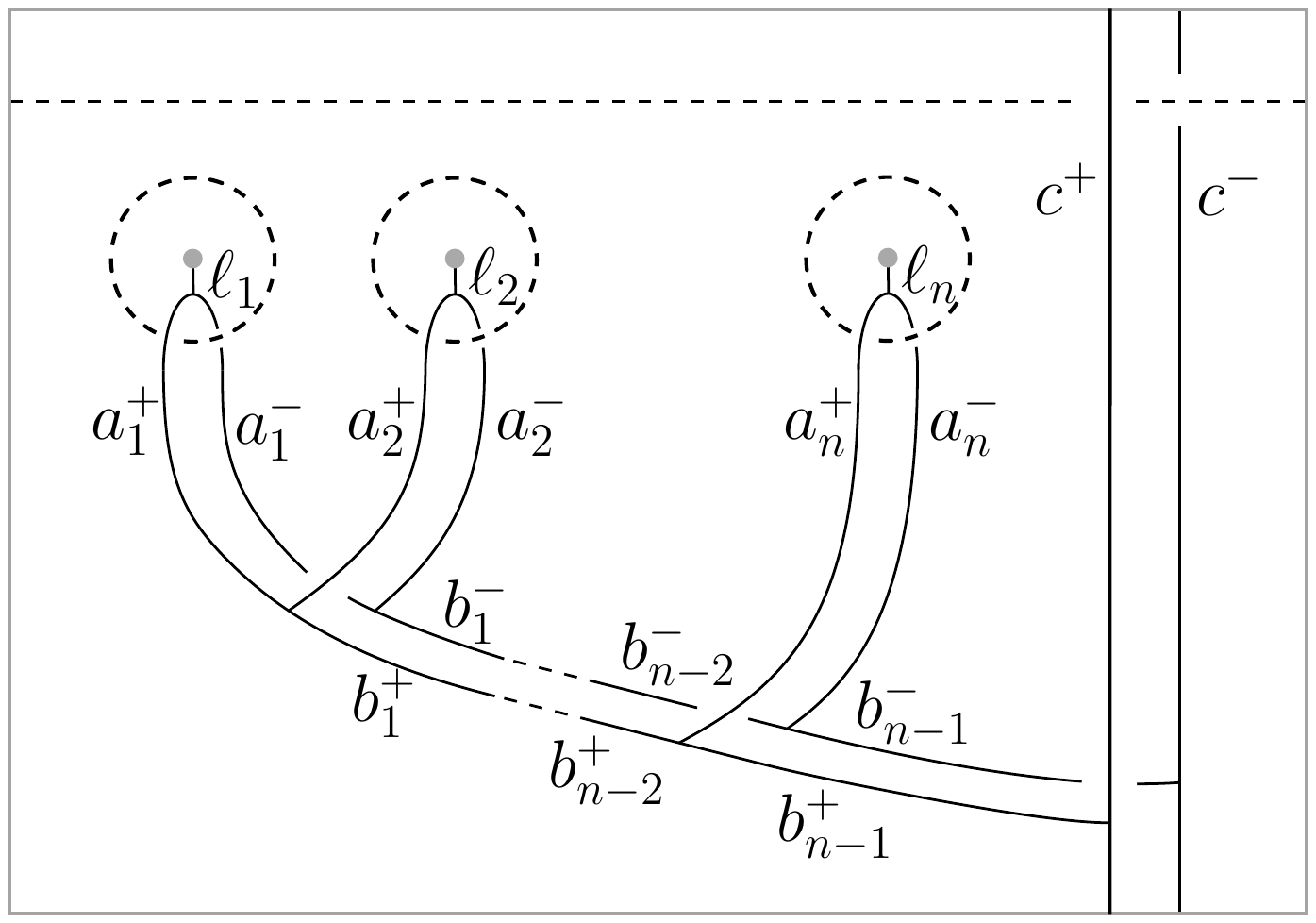}} \; ,
		\end{align}
	\end{widetext}
	where the gray box represents the periodic boundary conditions of a torus. 
	Note that the handle label must satisfy $ \delta_{\bm{b}_{n-1} \bm{c} \bm{c}} $ or, equivalently, $ \delta_{b^+_{n-1} c^+ c^+} $ and $ \delta_{b^-_{n-1} c^- c^-} $, for the corresponding ribbon graph to obey the branching rules.
	The crossings on the right-hand side of Eq.~\eqref{eq:anyonic_fusion_basis_states_torus_short} must again be resolved using Eq.~\ref{eq:resolve_crossing_short}, which will lead to superposition of ribbon configurations similar to Eq.~\eqref{eq:anyonic_fusion_basis_red_short}. These ribbons must then be embedded in the fattened lattice like in Eq.~\eqref{eq:anyonic_fusion_basis_lattice_short}, and resolved into the lattice using $ F $-moves.

\begin{figure}[h]
		\centering
		\begin{subfigure}{\linewidth}
		    \includegraphics[scale=.55]{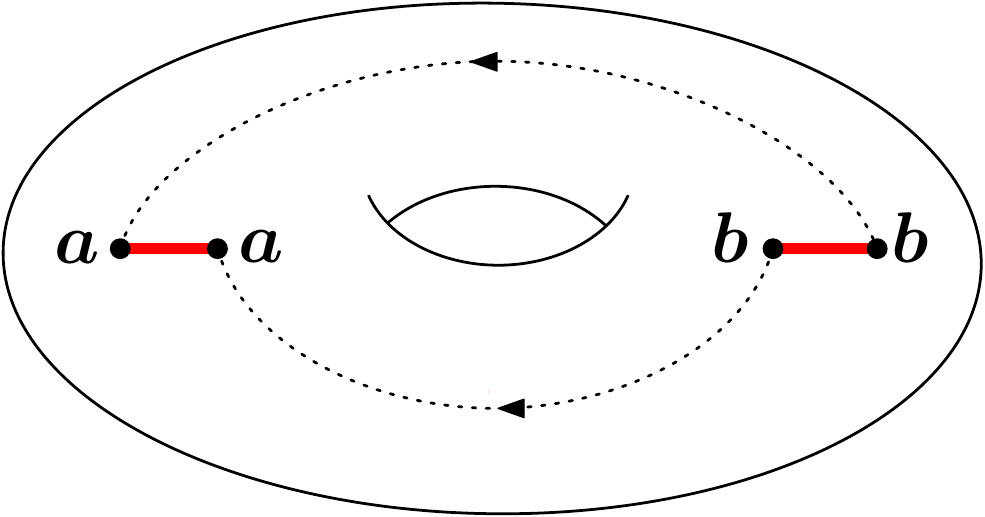}
		    \caption{}
		\end{subfigure}
		\begin{subfigure}{\linewidth}
		    \includegraphics[scale=.55]{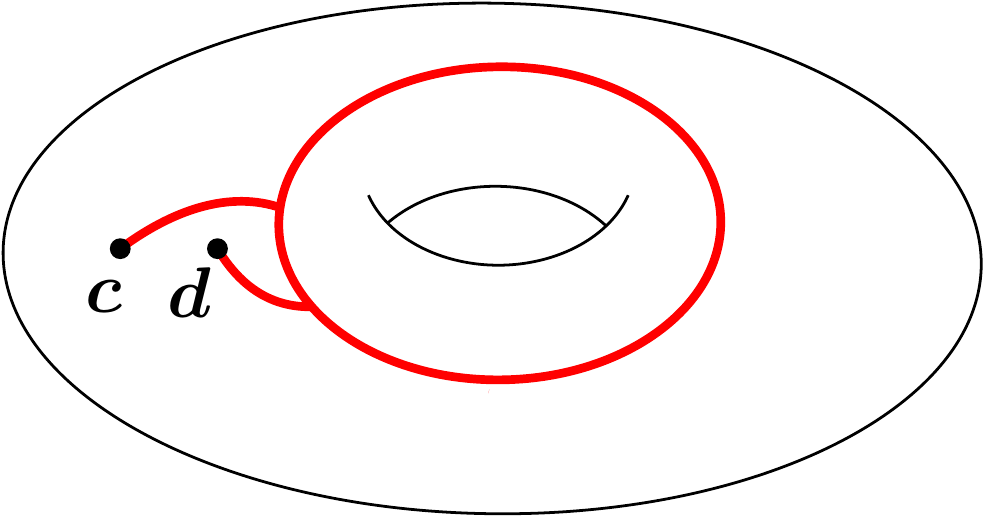}
		    \caption{}
		\end{subfigure}
		\caption{(a) Topologically nontrivial process which results in a logical error: two pairs of anyons $ (\bm{a}, \bm{a}) $ and $(\bm{b}, \bm{b}) $ are created by local qudit errors. The anyons are then fused along the dotted black paths with outcomes $\bm{c}$ and $\bm{d}$. (b) The fusion diagram of the resulting state winds around the torus.}
		\label{fig:logical_error}
	\end{figure}

	Ground states of the model correspond to (linear combinations of) configurations in which all plaquettes carry a trivial charge ($ \bm{a}_i = \1\1 $, $  \bm{b}_j = \1\1 $, $\forall\, i, j $). 
	On a torus, the degenerate ground space is spanned by the states $ \ket{\vec{\1}, \vec{\1\1}, \vec{\1\1}, \bm{c}} $, where $ \vec{\1} $ and $ \vec{\1\1} $ represent arrays containing only trivial entries. One can show that these states are indeed orthonormal.
	Hence, when storing the state of two logical qubits (for $ \C = \text{FIB} $), this information is encoded in the handle label. 
	The operations that affect the handle, are precisely those in which anyons interact along a non-contractible path such as the process depicted in \figref{fig:logical_error}.
	As long as no such operations are performed, the encoded information is preserved.
	Local qudit errors will create pairs, triplets or quadruplets of nontrivial anyons in neighboring plaquettes. 
	The initial ground state can then be recovered by fusing these nontrivial anyons pairwise until none are left, without creating any non-contractible loops in the process. 
	

	
	Clearly, all anyonic fusion basis states correspond to exceedingly complicated superpositions of qudit states. It would seem that this makes them highly impractical for any actual computation. 
	Fortunately, however, one can formulate a tensor network representation for such states, which enables us to use them for practical applications (in particular, see Sec.~\ref{sec:matrix_elements}). This tensor network representation is derived in App.~\ref{sec:TN}.
	
%% file: sections/measurement_scheme.tex
\section{Error correction scheme}\label{sec:error_correction}

We now present the circuits to measure and fix arbitrary local errors. 
From here on, we will work exclusively with the Fibonacci input category, hence we are working with a system of qubits on a lattice.
Our overall error correction procedure is composed of two major steps: 
\begin{enumerate}
	\item
		Measure all the vertex operators $Q_v$ in the extended Levin-Wen model Eq.~\eqref{eq:code_hamiltonian}, and apply a correction which fixes the vertex errors through unitaries $U_V$ conditioned by a measurement projection $P_V$.  This measurement and correction processes projects the many-body state onto the string-net subspace $\mathcal{H}_\text{s.n.}$.
	\item 
		After projecting to the string-net subspace $\mathcal{H}_\text{s.n.}$, we apply additional measurement circuits to measure the simple idempotents of the tube algebra [Eqs.~\eqref{eq:P11_short}, \eqref{eq:P12_short}, \eqref{eq:P21_short}, \eqref{eq:P221_short}, and \eqref{eq:P222_short}] and extract the error syndromes, i.e., the anyon charges and tail labels of all plaquettes.  Based on these syndromes, we use our decoders to identify the error location (up to equivalence classes) and apply the corresponding recovery maps to project the state back to the code (ground) space  $\mathcal{H}_\Lambda$. We note that the code space is a subspace of the string-net subspace: $\mathcal{H}_\Lambda \subset \mathcal{H}_\text{s.n.}$.   
\end{enumerate}
In this section, we discuss all measurement and recovery operators required to implement these steps. 
In particular, we discuss the vertex measurement and correction processes in Sec.~\ref{sec:vertex_measurement}, the anyon charge measurements in Sec.~\ref{sec:measure_charge}, and the recovery operations in Sec.~\ref{sec:recovery}.

\subsection{Vertex measurements and correction} \label{sec:vertex_measurement}
\begin{figure}[t]
	\centering
	\includegraphics[scale=0.7]{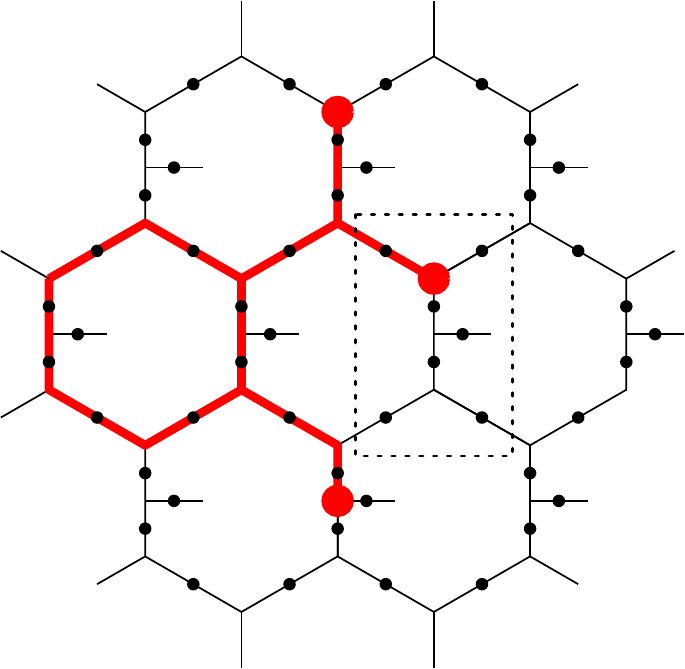}
	\caption{A configuration that violates the branching rules by having string-endings in 3 vertices, indicated  the red dots.}
	\label{fig:tailed_lattice_violated_vertices}
\end{figure}

Certain types of errors, such as a single bit-flip error $\sigma_x^e$ or a coherent error generated by the Pauli-X operator, i.e., $\e^{\ii \theta \sigma_x^e}$, can cause a violation of the vertex projector $Q_v$ for a vertex $ v $ adjacent to edge $ e $.
As illustrated in \figref{fig:tailed_lattice_violated_vertices}, a vertex error corresponds to a broken string ending at that vertex (in the string-net configurations of the system wave function).
Note that a generic error such as $\e^{\ii \theta \sigma_x^e}$ will put the system wave function in a superposition of violating and not violating the vertex condition at any adjacent vertex $ v $.  When we measure the vertex operator $Q_v$, we project to either of the two situations. 
Vertex violations can be resolved by local unitary operators, which take the system back to the string-net subspace $\mathcal{H}_\text{s.n.}$. 
Intuitively, one can think of the action of these operators as pulling string ends into the tail edge of a neighboring plaquette: 
\begin{equation}\label{eq:vertex_correction_intuitive}
	\raisebox{-.5cm}{\includegraphics[scale=0.64]{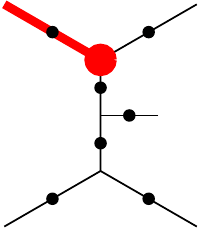}}
	\;\mapsto\;
	\raisebox{-.5cm}{\includegraphics[scale=0.64]{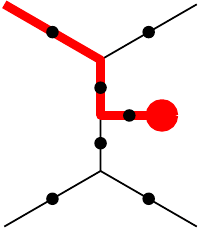}}\,.
\end{equation}

\begin{figure*}[h]
\centering
\includegraphics[width=\textwidth]{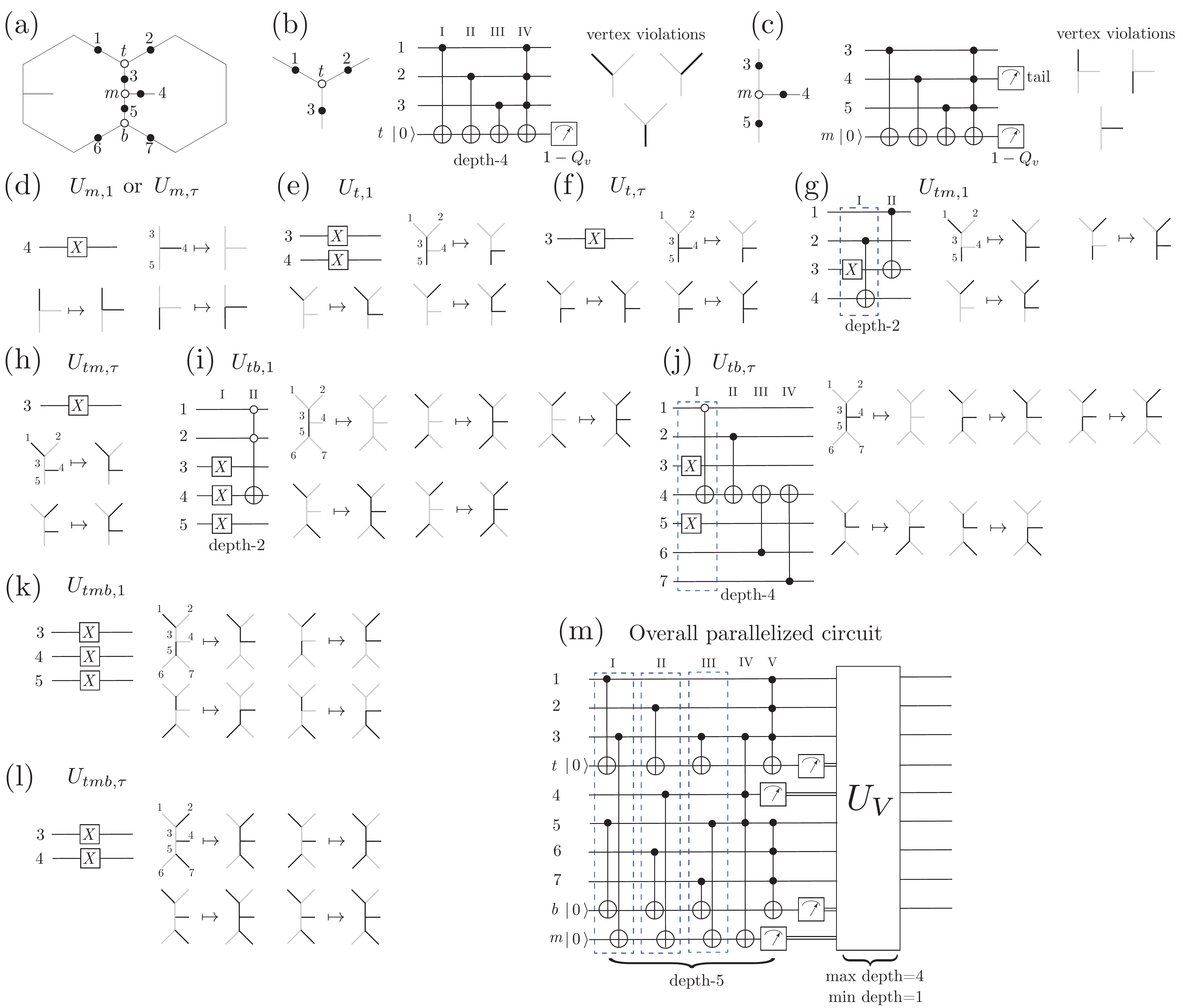}
\caption{Measurement and correction circuit for the vertex errors. (a) The qubit labeling.  The black dots represent data qubits, while the white dots represent ancilla qubits for measurements.  (b,c) The measurement circuit for the top and middle vertex projectors.  The circuit for measuring the bottom vertex can be inferred by symmetry.  (d-l) The circuits for pulling an open string into the tail edge when certain vertices are violated. (m) The overall circuit for measuring and correcting vertex errors. The first part of the circuit measures the three vertex projectors. The second part is the unitary $U_V$ conditioned on the measurement results which is summarized in (d-l).}
\label{fig:vertex_correction}
\end{figure*}

Previously, the circuit for measuring the vertex errors has been discovered in Ref.~\cite{bonesteel2012quantum}. Here we adopt this circuit to measure the top, middle and bottom vertices, with three ancilla qubits (white dots) denoted by $t$, $m$ and $b$, respectively, as shown in Fig.~\ref{fig:vertex_correction}(a-c). The circuit for measuring the top and bottom are symmetric, so we only show the top one in Fig.~\ref{fig:vertex_correction}(a). For the middle vertex $m$, we also apply a measurement of the tail qubit at the end of the circuit. To respect the usual convention of quantum circuits,  we represent the unoccupied edge as $\ket{0}$, which corresponds to the vacuum string label $\mathbf{1}$, and the occupied edge as $\ket{1}$ corresponding to the string label $\tau$.  The ancilla qubits are all initialized in the state $\ket{0}$.  

We apply a correction $U_V$, conditioned by the measurement results of the three vertex operators and the tail qubit, to fix the vertex error.
This correction is selected out of 14 possible unitaries:
\begin{center}
		\begin{minipage}[t]{0.4\linewidth}
            \begin{flalign*}  
				&U_{m, 1} : & \raisebox{-.2cm}{\includegraphics[scale=.35]{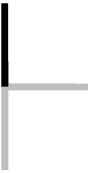}} \;
				\mapsto
				\;\raisebox{-.2cm}{\includegraphics[scale=.35]{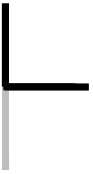}} &\\
				& & \raisebox{-.2cm}{\includegraphics[scale=.35]{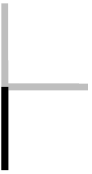}} \;
				\mapsto
				\;\raisebox{-.2cm}{\includegraphics[scale=.35]{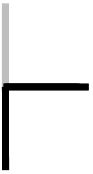}}&
			\end{flalign*}
		\end{minipage}
		\hspace{1.1cm}
		\begin{minipage}[t]{0.4\linewidth}
			\begin{flalign*} 
				 &U_{m,\tau} : & \raisebox{-.2cm}{\includegraphics[scale=.35]{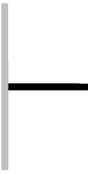}} \;
				\mapsto
				\;\raisebox{-.2cm}{\includegraphics[scale=.35]{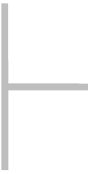}}& 
			\end{flalign*}
		\end{minipage}
		\\

		\begin{minipage}[t]{0.4\linewidth}
			\begin{flalign*}  
				&U_{t, 1} : 
				& \raisebox{-.25cm}{\includegraphics[scale=.35]{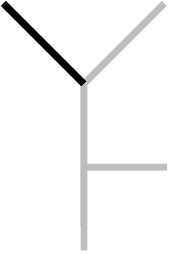}} \;
				\mapsto
				\;\raisebox{-.25cm}{\includegraphics[scale=.35]{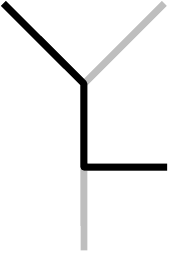}}&\\
				& & \raisebox{-.25cm}{\includegraphics[scale=.35]{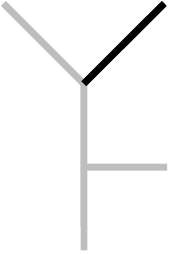}} \;
				\mapsto
				\;\raisebox{-.25cm}{\includegraphics[scale=.35]{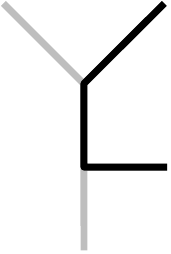}}&\\
				& & \raisebox{-.25cm}{\includegraphics[scale=.35]{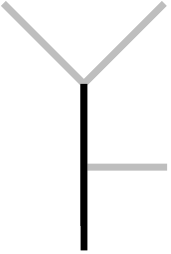}} \;
				\mapsto
				\;\raisebox{-.25cm}{\includegraphics[scale=.35]{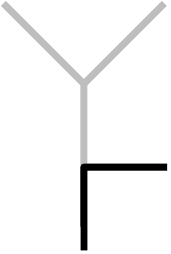}}&
			\end{flalign*}
		\end{minipage}
		\hspace{1.1cm}
		\begin{minipage}[t]{0.4\linewidth}
			\begin{flalign*} 
    			&U_{t, \tau} : 
				& \raisebox{-.25cm}{\includegraphics[scale=.35]{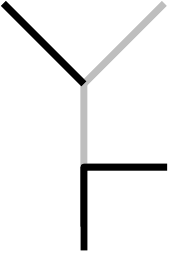}} \;
				\mapsto
				\;\raisebox{-.25cm}{\includegraphics[scale=.35]{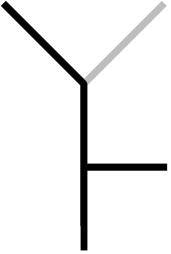}}&\\
				& & \raisebox{-.25cm}{\includegraphics[scale=.35]{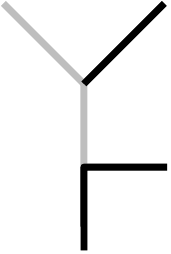}} \;
				\mapsto
				\;\raisebox{-.25cm}{\includegraphics[scale=.35]{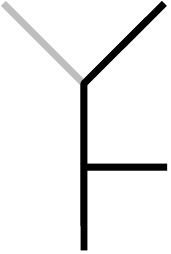}}&\\
				& & \raisebox{-.25cm}{\includegraphics[scale=.35]{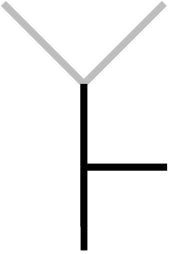}} \;
				\mapsto
				\;\raisebox{-.25cm}{\includegraphics[scale=.35]{fig/segment_top_tb.pdf}}&
			\end{flalign*}
		\end{minipage}
		\\
		
		\begin{minipage}[t]{0.4\linewidth}
			\begin{flalign*}
				&U_{tm, 1} : 
				& \raisebox{-.25cm}{\includegraphics[scale=.35]{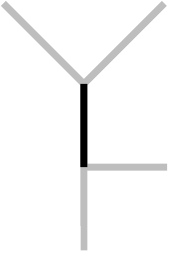}} \;
				\mapsto
				\;\raisebox{-.25cm}{\includegraphics[scale=.35]{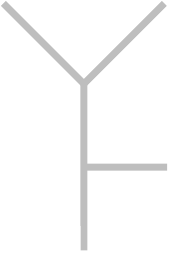}}&\\
				& & \raisebox{-.25cm}{\includegraphics[scale=.35]{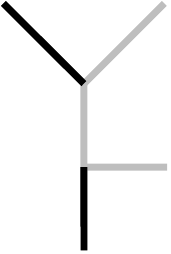}} \;
				\mapsto
				\;\raisebox{-.25cm}{\includegraphics[scale=.35]{fig/segment_top_lmtb.pdf}}&\\
				& & \raisebox{-.25cm}{\includegraphics[scale=.35]{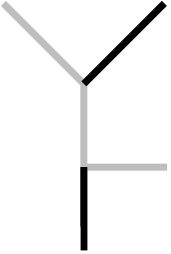}} \;
				\mapsto
				\;\raisebox{-.25cm}{\includegraphics[scale=.35]{fig/segment_top_rmtb.pdf}}&
			\end{flalign*}
		\end{minipage}
		\hspace{1.1cm}
		\begin{minipage}[t]{0.4\linewidth}
			\begin{flalign*} 
				&U_{tm, \tau} :
				& \raisebox{-.25cm}{\includegraphics[scale=.35]{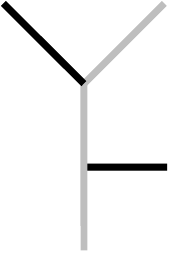}} \;
				\mapsto
				\;\raisebox{-.25cm}{\includegraphics[scale=.35]{fig/segment_top_lmt.pdf}}&\\
				& & \raisebox{-.25cm}{\includegraphics[scale=.35]{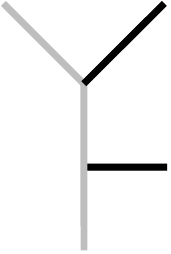}} \;
				\mapsto
				\;\raisebox{-.25cm}{\includegraphics[scale=.35]{fig/segment_top_rmt.pdf}}&
			\end{flalign*}
		\end{minipage}
		\\
		
		\begin{minipage}[t]{0.4\linewidth}
			\begin{flalign*} 
				&U_{tb, 1} : 
				& \raisebox{-.49cm}{\includegraphics[scale=.35]{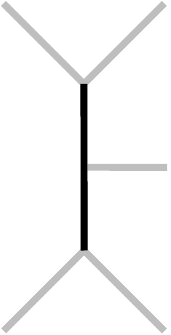}} \;
				\mapsto
				\;\raisebox{-.49cm}{\includegraphics[scale=.35]{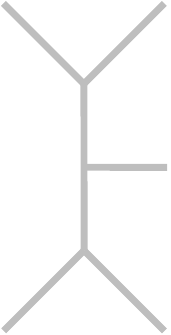}}&\\
                & & \raisebox{-.49cm}{\includegraphics[scale=.35]{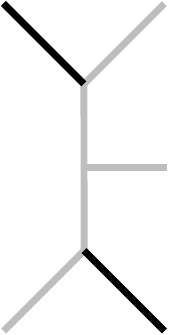}} \;
				\mapsto
				\;\raisebox{-.49cm}{\includegraphics[scale=.35]{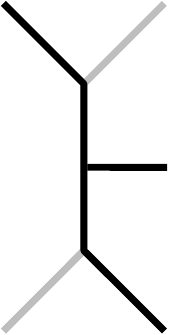}}&\\
				& & \raisebox{-.49cm}{\includegraphics[scale=.35]{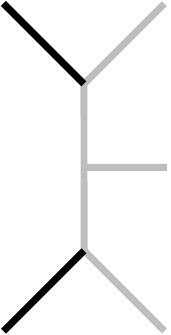}} \;
				\mapsto
				\;\raisebox{-.49cm}{\includegraphics[scale=.35]{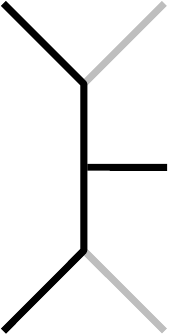}}& \\
				& & \raisebox{-.49cm}{\includegraphics[scale=.35]{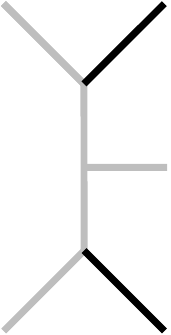}} \;
				\mapsto
				\;\raisebox{-.49cm}{\includegraphics[scale=.35]{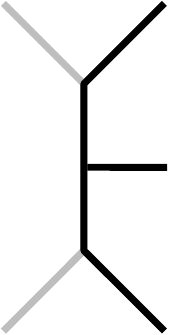}}&\\
                & & \raisebox{-.49cm}{\includegraphics[scale=.35]{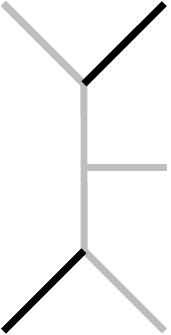}} \;
				\mapsto
				\;\raisebox{-.49cm}{\includegraphics[scale=.35]{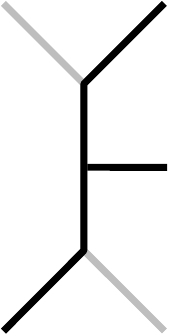}}&
			\end{flalign*}
		\end{minipage}
		\hspace{1.1cm}
		\begin{minipage}[t]{0.4\linewidth}
			\begin{flalign*} 
				&U_{tb, \tau} : 
				& \raisebox{-.49cm}{\includegraphics[scale=.35]{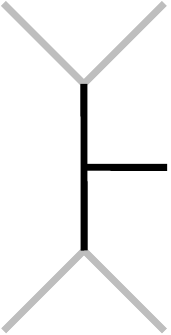}} \;
				\mapsto
				\;\raisebox{-.49cm}{\includegraphics[scale=.35]{fig/segment_empty.pdf}}&\\
				& & \raisebox{-.49cm}{\includegraphics[scale=.35]{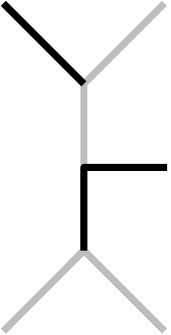}} \;
				\mapsto
				\;\raisebox{-.49cm}{\includegraphics[scale=.35]{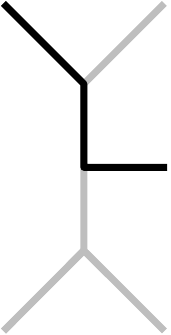}}&\\
				& & \raisebox{-.49cm}{\includegraphics[scale=.35]{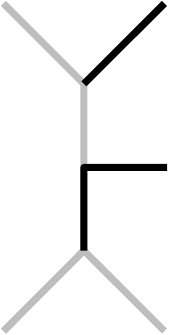}} \;
				\mapsto
				\;\raisebox{-.49cm}{\includegraphics[scale=.35]{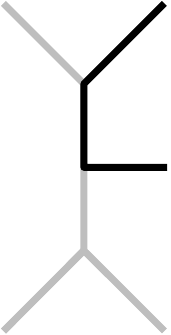}}&\\
				& &\raisebox{-.49cm}{\includegraphics[scale=.35]{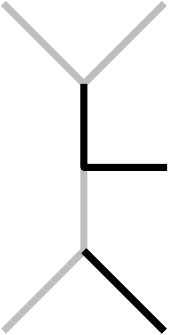}} \;
				\mapsto
				\;\raisebox{-.49cm}{\includegraphics[scale=.35]{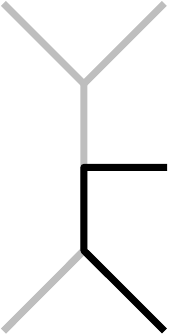}}&\\
				& &\raisebox{-.49cm}{\includegraphics[scale=.35]{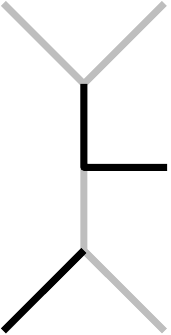}} \;
				\mapsto
				\;\raisebox{-.49cm}{\includegraphics[scale=.35]{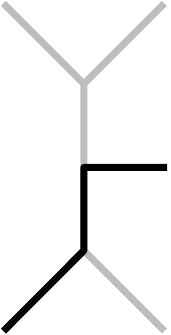}}&
			\end{flalign*}
		\end{minipage}
		\\
        
		\begin{minipage}[t]{0.4\linewidth}
			\begin{flalign*} 
				&U_{tmb, 1} : 
				& \raisebox{-.49cm}{\includegraphics[scale=.35]{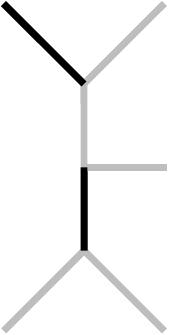}} \;
				\mapsto
				\;\raisebox{-.49cm}{\includegraphics[scale=.35]{fig/segment_157.pdf}}&\\
				& & \raisebox{-.49cm}{\includegraphics[scale=.35]{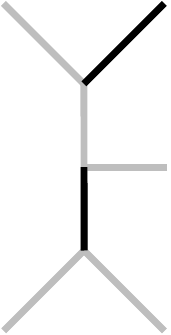}} \;
				\mapsto
				\;\raisebox{-.49cm}{\includegraphics[scale=.35]{fig/segment_257.pdf}}&\\
				& & \raisebox{-.49cm}{\includegraphics[scale=.35]{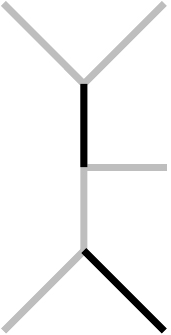}} \;
				\mapsto
				\;\raisebox{-.49cm}{\includegraphics[scale=.35]{fig/segment_367.pdf}}&\\
				& & \raisebox{-.49cm}{\includegraphics[scale=.35]{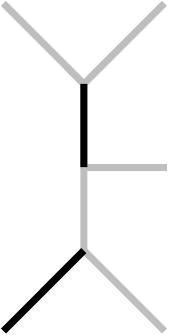}} \;
				\mapsto
				\;\raisebox{-.49cm}{\includegraphics[scale=.35]{fig/segment_467.pdf}}&
			\end{flalign*}
		\end{minipage}
		\hspace{1.1cm}
		\begin{minipage}[t]{0.4\linewidth}
			\begin{flalign*} 
				&U_{tmb, \tau} : 
				& \raisebox{-.49cm}{\includegraphics[scale=.35]{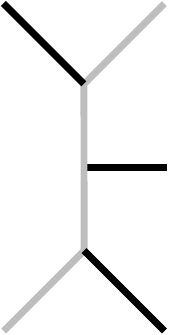}} \;
				\mapsto
				\;\raisebox{-.49cm}{\includegraphics[scale=.35]{fig/segment_13567.pdf}}&\\
				& & \raisebox{-.49cm}{\includegraphics[scale=.35]{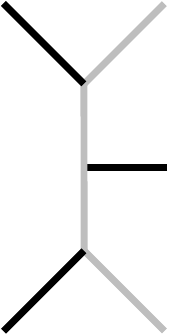}} \;
				\mapsto
				\;\raisebox{-.49cm}{\includegraphics[scale=.35]{fig/segment_14567.pdf}}&\\
				& & \raisebox{-.49cm}{\includegraphics[scale=.35]{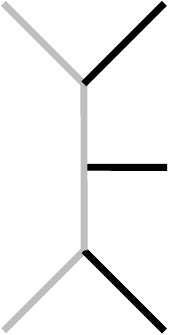}} \;
				\mapsto
				\;\raisebox{-.49cm}{\includegraphics[scale=.35]{fig/segment_23567.pdf}}&\\
                & & \raisebox{-.49cm}{\includegraphics[scale=.35]{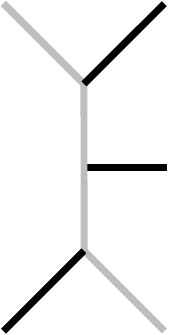}} \;
				\mapsto
				\;\raisebox{-.49cm}{\includegraphics[scale=.35]{fig/segment_24567.pdf}}&
			\end{flalign*}
		\end{minipage}
		\\
	\end{center}
\vspace{.2cm}	   
\noindent Note that we have omitted the mappings which are  mirror symmetric ($t \leftrightarrow b$) to the listed ones.

The corresponding quantum circuits of the above unitaries are listed in Fig.~\ref{fig:vertex_correction}(d-l). For gates which do not have overlap in qubit support, we can parallelize them in a single time step, as indicated by the dashed boxes. As we can see, most of the unitary circuits have depth 1 or 2, while  only one of them, $U_{tb, \tau}$ in Fig.~\ref{fig:vertex_correction}(i), has depth 4.  The overall measurement and correction circuit is summarized in Fig.~\ref{fig:vertex_correction}(m) where we have parallelized the measurement of the three vertex operators into a depth-5 circuit (in terms of the unitary gates). When taking into account the readout of the ancilla qubits before applying $U_V$, the depth is 6.   Note that we have neglected the step of state preparation of the ancilla qubits in the beginning, because in the situation of repetitive syndrome measurements, the ancilla qubits can always be prepared during the application of the correction unitary $U_V$.   Overall, the depth of the measurement circuits ranges from 5 to 9, or from 6 to 10 when taking into account the measurement step. 

Note that a different scheme of fixing vertex errors has been previously proposed in Refs.~\cite{feng2015non} and \cite{bonesteel_inpreparation}, which also uses tail qubits and hence has a similar spirit.

\subsection{Anyon charge measurements}\label{sec:measure_charge}
After measuring the vertex operators and applying the corresponding corrections on the extended string-net code, we have transformed the many-body state to the string-net subspace $\mathcal{H}_{s.n.}$, where it can be described in terms of anyonic fusion states. 
The local qubit errors, including both the Pauli-$X$ and $Z$ types, now result in the creation of anyonic excitations inside $\mathcal{H}_{s.n.}$.  

\begin{figure*}[h]
    \centering
    \includegraphics[width=.9\textwidth]{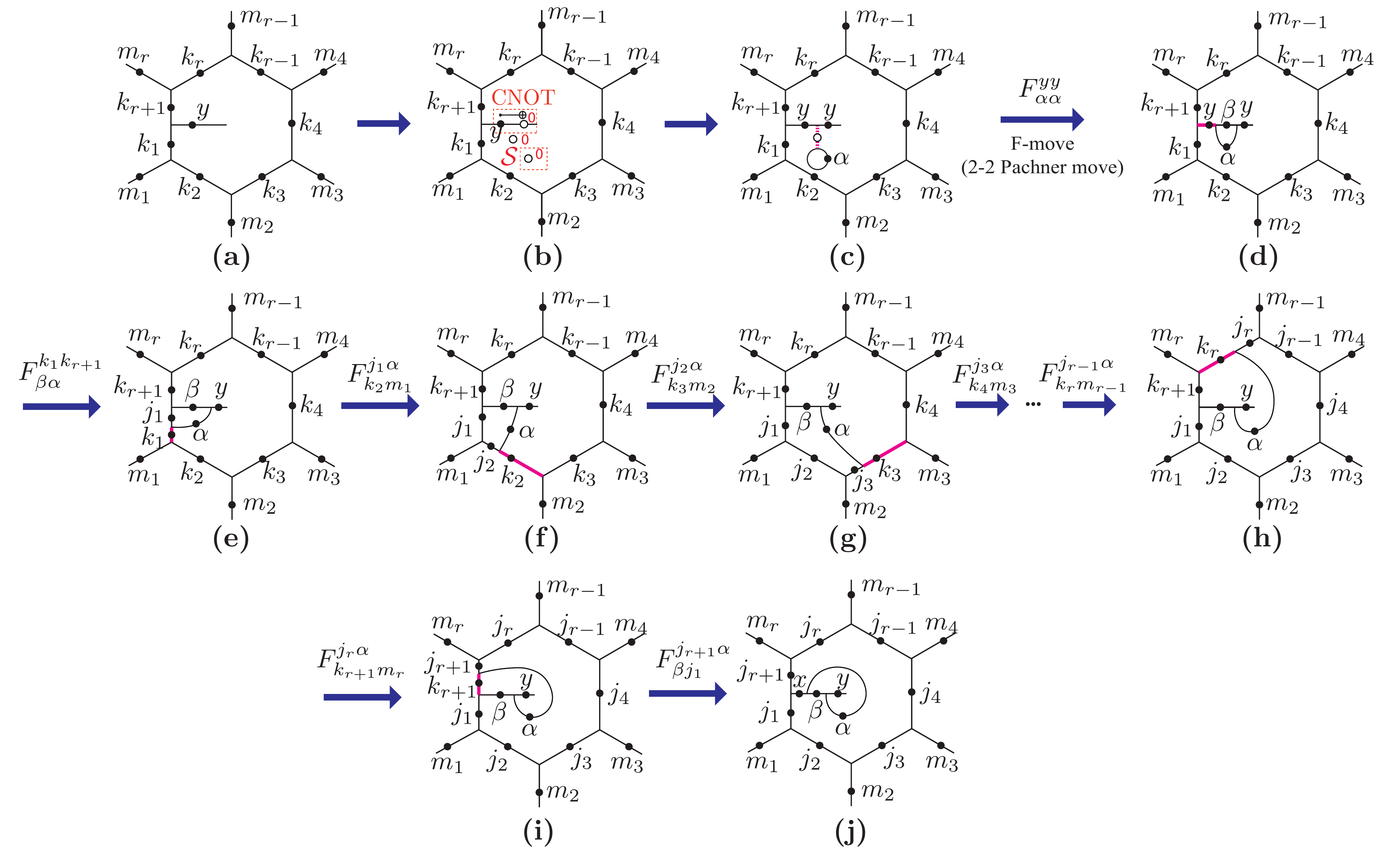}
    \caption{Protocol and circuit of growing a tube onto a puncture in a plaquette on a tailed lattice via a sequence of local gates and Pachner moves. }
    \label{fig:grow_circuit}
\end{figure*}

As explained in Sec.~\ref{sec:excitations}, one can measure the anyonic charge of a plaquette using the central idempotents of the tube algebra. 
To fully characterize an excitation, we must also measure its tail label.
A joint measurement of the anyonic charge and the tail label is achieved by measuring the \emph{irreducible} idempotents of the tube algebra, listed in Eqs.~(\ref{eq:P11_short}-\ref{eq:P21_short}), \eqref{eq:P221_short}, and \eqref{eq:P222_short}.
Measuring these irreducible idempotents is done in 3 steps:
\begin{enumerate}
    \item Grow a tube inside the plaquette by introducing ancilla qubits and performing the appropriate quantum circuit, as shown in \figref{fig:grow_circuit}.
    \item Measure the tube qubits in the appropriate basis. 
    \item Either trace out the tube qubits immediately, or first resolve the tube back into the lattice before tracing out the ancillas.  
\end{enumerate}

\begin{figure}
    \includegraphics[width=1\columnwidth]{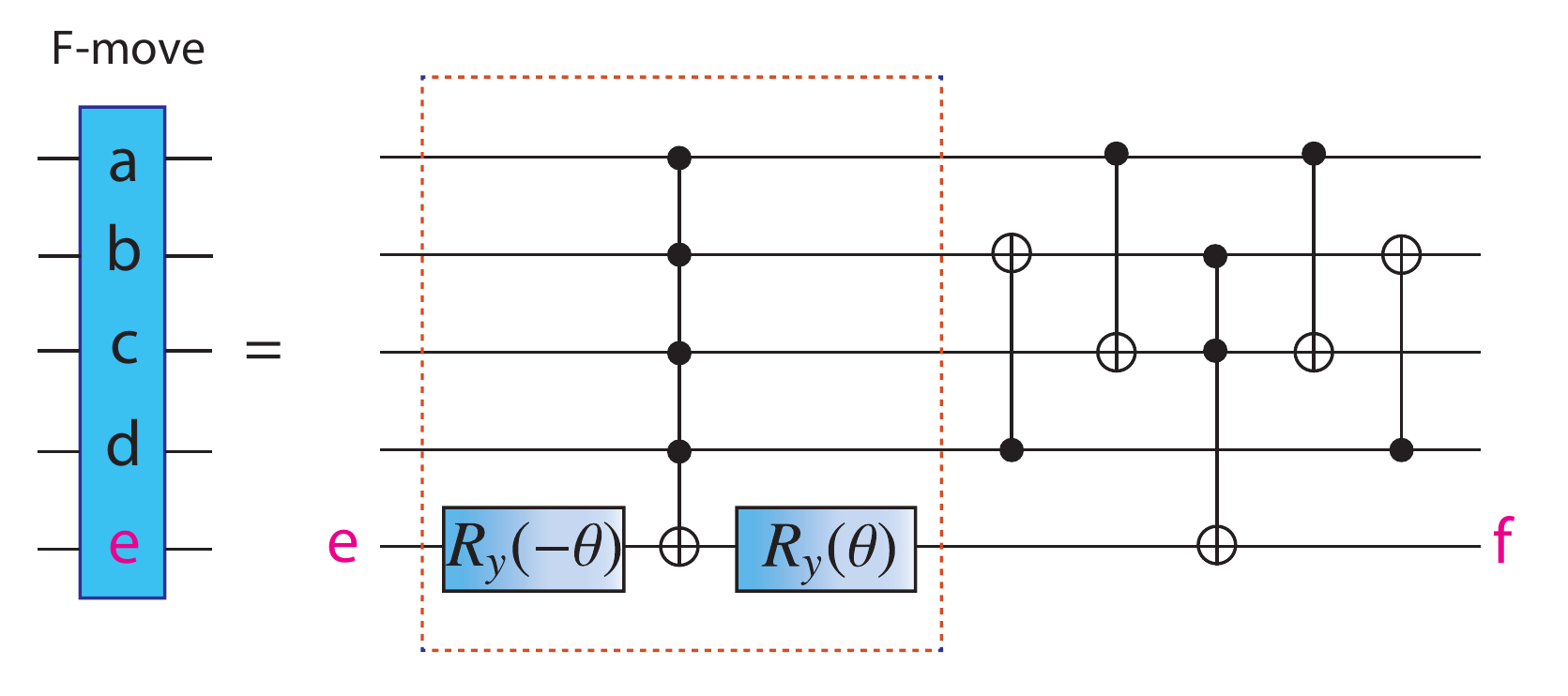}
    \caption{Quantum circuit to implement the F-move (2-2 Pachner move) operation $F^{abe}_{cdf}$ for the Fibonacci Turaev-Viro code. }
    \label{fig:F-move_circuit}
\end{figure}

As a basic ingredient, the quantum circuit to implement the $F$-move (2-2 Pachner move) operation $F^{abe}_{cdf}$ in the Fibonacci Turaev-Viro code is shown in Fig.~\ref{fig:F-move_circuit}. This circuit was first proposed in Ref.~\cite{bonesteel2012quantum}. The $F$-move operation can be viewed as a controlled unitary operation, where the external legs $a,b,c,d$ are control qubits determining the resulting unitary $F^{ab}_{cd}$, with the matrix elements being $[F^{ab}_{cd}]_{ef}$. 
For the Fibonacci Turaev-Viro code, the F-matrix is given in Eq.~\eqref{eq:fib_F_short}.
The circuit inside the red dashed box, composed of a 5-qubit Toffoli gate in between
 two single-qubit rotations, applies the conditional unitary corresponding to the F-matrix $F^{\tau \tau}_{\tau \tau}$,  where $R_y(\pm \theta)=\e^{\pm \ii \theta \sigma_y/2}$
are single-qubit rotations about the y-axis with angle $\theta = \tan^{-1} (\phi^{-\frac{1}{2}})$. Note that this conditional unitary is only activated if the control qubits $a$, $b$, $c$ and $d$ are all in the $\ket{1}$ state corresponding to the string label $\tau$. All the other conditional unitaries are implemented by the rest of the quantum circuit.

\begin{figure}
    \includegraphics[width=1\columnwidth]{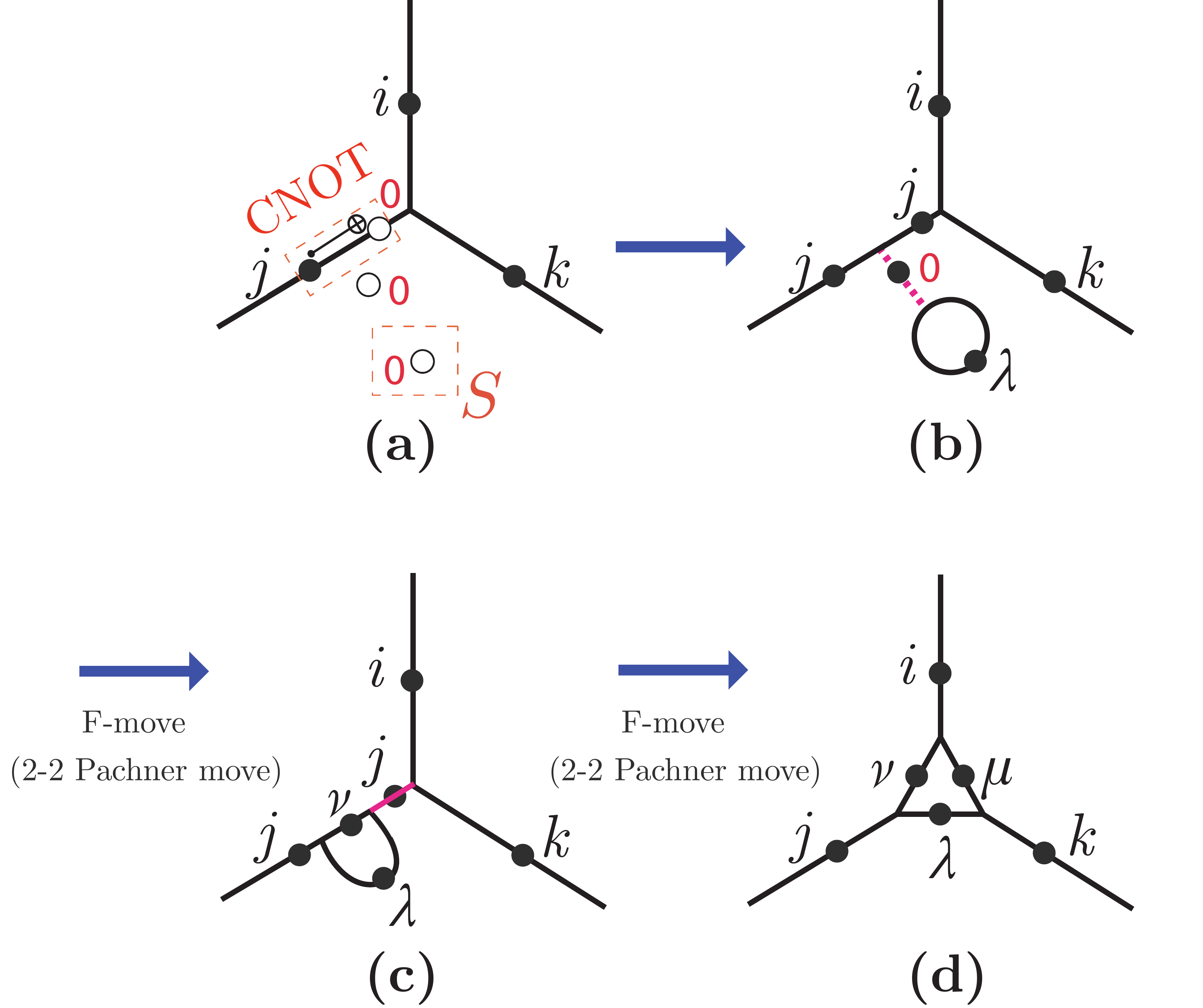}
    \caption{Quantum circuit to implement the 1-3 Pachner move for the Fibonacci Turaev-Viro code.}
    \label{fig:1-3_Pachner_circuit}
\end{figure}

Based on the circuit for 2-2 Pachner move ($F$-move), one can also implement the 1-3 Pachner move with unitary circuit, as shown in Fig.~\ref{fig:1-3_Pachner_circuit}. The protocol consists of the following steps: 
(1) Initialize three ancilla qubits (white dots) in state $\ket{0}$.  (2) Apply a CNOT gate which entangles the data qubit labeled $j$ to the new ancilla qubit on the same edge,  CNOT:~$\ket{j} \ket{0} $$\mapsto$$\ket{j} \ket{j}$, as shown in (a) and (b). (3) Apply a modular-$\mathcal{S}$ gate  on one ancilla to create a tadpole diagram, as shown in (b). The modular--$\mathcal{S}$ does the following transformation: $\mathcal{S}: |0\rangle $$\mapsto $$ \sum_\lambda \frac{d_\lambda}{D} |\lambda \rangle$.  For the Fibonacci Turaev-Viro code, the modular $\mathcal{S}$-matrix is  
\begin{equation}
	\mathcal{S}=\frac{1}{\sqrt{2+\phi}}
	\begin{pmatrix}
		1 & \phi  \\
		\phi & -1
	\end{pmatrix} \,.
\end{equation}
(4) Apply an $F$-move to absorb the tadpole onto the edge, as shown in (b) and (c).  (5) Apply another $F$-move to sweep edge $\lambda$ to attach the right leg (with label $k$), as shown in (c) and (d).  From left to right, the circuit effectively fine-grain the lattice by entangling the three ancilla qubits into the code space.  One can also reverse the circuit (from right to left) which corresponds to a coarse-graining process disentangling  three qubits out of the code space.

%
	
\begin{figure*}[h]
    \centering
    \includegraphics[scale=.35]{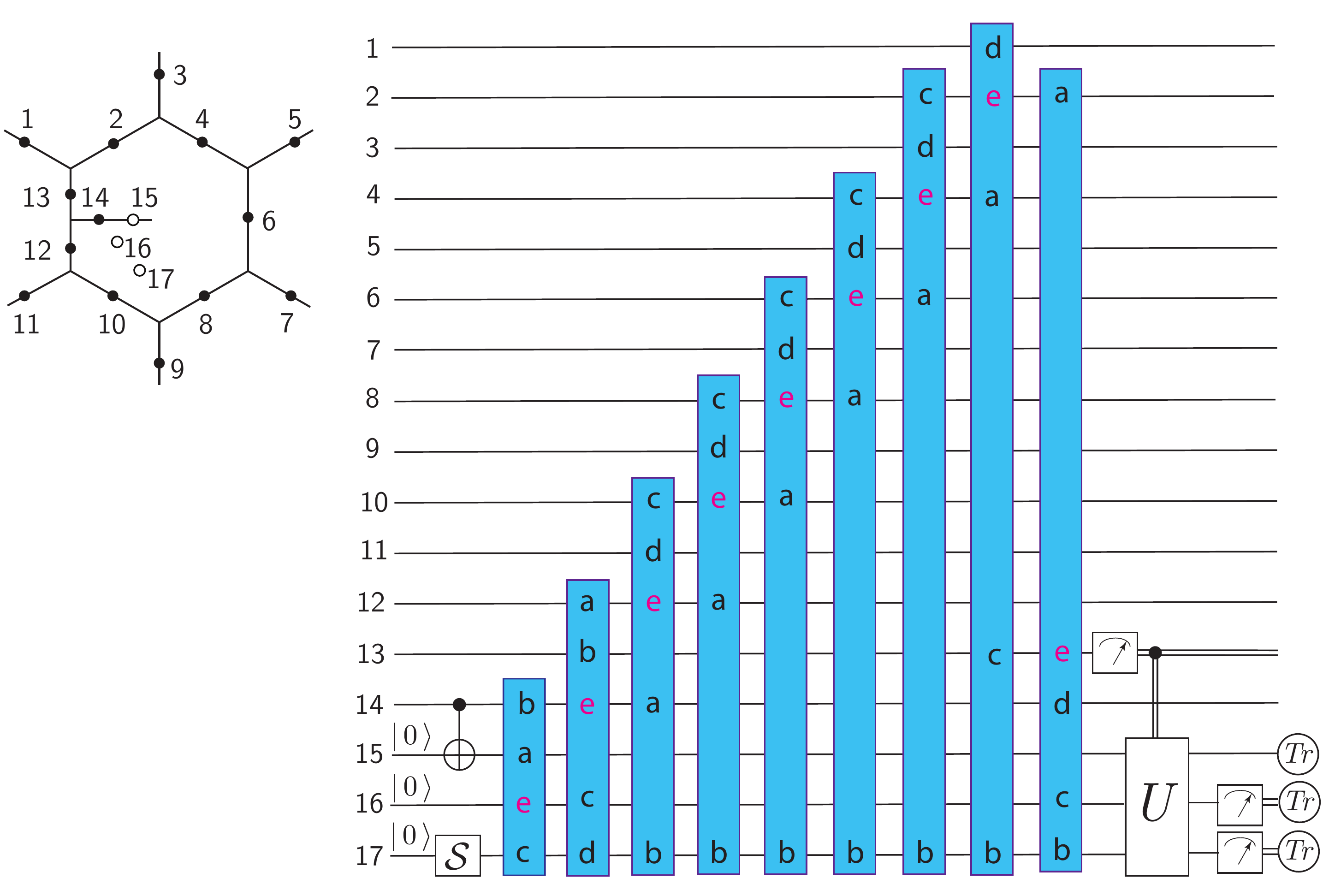}
	\caption{The complete quantum circuit for the joint measurement of the anyon charge and tail label of a plaquette. Note that after the grow circuit, qubit 13 corresponds to the tail edge.}
	\label{fig:charge_measurement_circuit_1}
    \clearpage
\end{figure*}

The ``growing'' of a tube onto the tailed lattice can then be implemented by a quantum circuit, as shown in Figs.~\ref{fig:grow_circuit} and \ref{fig:charge_measurement_circuit_1}. We start with the tailed lattice in Fig.~\ref{fig:grow_circuit}(a) with data qubits (black dots) residing on every edge. We then introduce three ancilla qubits (white dots) in Fig.~\ref{fig:grow_circuit}(b) initialized at $\ket{0}$.  From panel (b) to (e), we apply a series of operations to achieve a 1-3 Pachner move to add a triangle loop below the tail:  (1) Apply a CNOT gate which entangles the data qubit on the tail to the new ancilla qubit on the tail,  CNOT: $\ket{y} \ket{0} \mapsto \ket{y} \ket{y}$, as shown in (b) and (c). (2) Apply a modular-$\mathcal{S}$ gate  on one ancilla to create a tadpole diagram, as shown in (b), i.e., $\mathcal{S}: |0\rangle $$\mapsto $$ \sum_\alpha \frac{d_\alpha}{D} |\alpha \rangle$.  (3) Apply an $F$-move to absorb the tadpole onto the tail, as shown in (c) and (d).  (4) Apply another $F$-move to sweep edge $\alpha$ to attach the left edge, as shown in (d) and (e).  Afterwards, we apply a sequence of $F$-moves to sweep edge $\alpha$ around the whole plaquette, after which it ends up in the upper side of the tail.   
In this way, we have grown a tube.  
Note that as a result, the qubits on the plaquette and tail edges (with labels $ k_i $ and $ y $ in \figref{fig:grow_circuit}(a)) get rotated one position counterclockwise.
The details of the quantum gates in this circuit are  shown in Fig.~\ref{fig:charge_measurement_circuit_1}.  As we can see, in total 9 $F$-moves have been applied.

After growing the tube, the anyon charge can be inferred by measuring the four qubits on the tube.
In order to find the appropriate basis for this measurement, we first note that the growing procedure can be thought of as creating a vacuum bubble (in \figref{fig:grow_circuit}~(c)), stretching it out along the boundary of the plaquette, and finally resolving only half of it into the lattice. The remaining half then constitutes the tube in \figref{fig:grow_circuit} (j).
In terms of ribbon diagrams the stretched out vacuum bubble inside the plaquette can be written as
\begin{align} \label{eq:grow_tube}
	\sum_{\alpha} \dfrac{d_\alpha}{\D} \;\; \raisebox{-1.3cm}{\includegraphics[scale=.44]{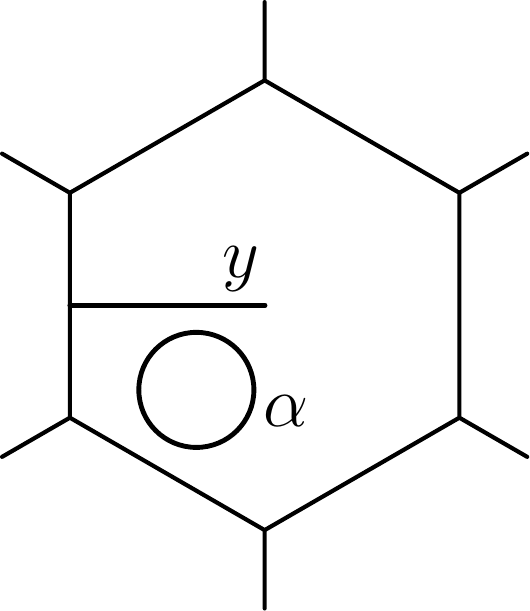}} \;
	& = 
	\sum_{\alpha} \dfrac{d_\alpha}{\D}\;\; \raisebox{-1.3cm}{\includegraphics[scale=.44]{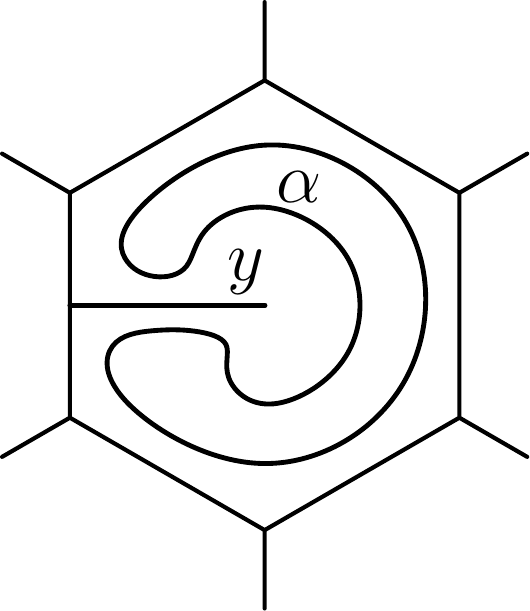}} \nonumber \\
	&  \hspace{-.8cm} =	\sum_{\alpha, \beta} \dfrac{d_\alpha}{\D} \dfrac{v_\beta}{v_\alpha v_y}\;\; \raisebox{-1.3cm}{\includegraphics[scale=.44]{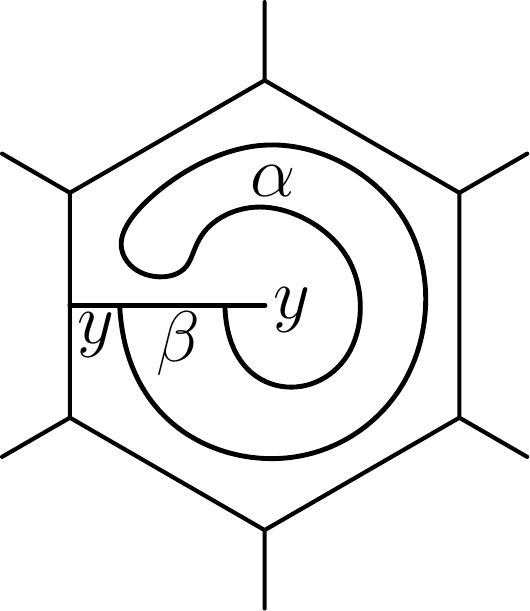}} \;\nonumber
	\\
	& \hspace{-.8cm}= \sum_{\alpha, \beta, x } \dfrac{1}{\D} \dfrac{v_x}{v_y}\;\; \raisebox{-1.3cm}{\includegraphics[scale=.44]{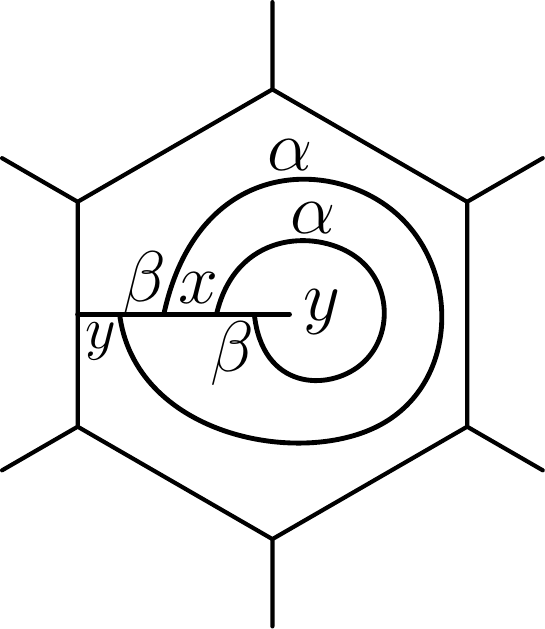}} \,.
\end{align}

The sequence of $F$-moves appearing in the grow circuit corresponds exactly to resolving only the outer tube into the lattice\footnote{Note that resolving both the inner and the outer tube into the lattice would yield a trivial operation, since they constitute the vacuum bubble together.}.
If we denote the initial state of the lattice qubits as $ \ket{\Psi_0} = \sum_y \varepsilon_y \ket{\phi_y} \otimes \ket{y} $ and the ancillas are initially in the $ \ket{000} $ state, then the action of the grow circuit is
\begin{multline}\label{eq:grow_action}
	 \ket{\Psi_0} \otimes \ket{000} \\
	 \mapsto \, \sum_y \varepsilon_y \sum_{\alpha,\beta,x} \dfrac{1}{\D} \dfrac{v_x}{v_y} 
	 \tilde{O}_{y x \alpha \beta} \, (\ket{\phi_y} \otimes \ket{y}) \otimes \ket{y\alpha\beta} ,
\end{multline}
where we used the following abbreviation for the tube state vectors: 
\begin{align}
	\nonumber \Bigg | \!\!\!\! \raisebox{-0.5cm}{\includegraphics[scale=.3]{fig/tube_1.pdf}} \,
	\Bigg \rangle \equiv \ket{x} \otimes \ket{y \alpha \beta },
\end{align}
and where
\begin{equation}\label{eq:Otilde}
	\tilde{O}_{y x \alpha \beta} \equiv \sum_\gamma F^{\alpha x \beta}_{\alpha y \gamma}\, O_{y x \alpha \gamma}
\end{equation}
is the operator which corresponds to resolving a outer tube\footnote{Note that it has a different shape than our convention Eq.~\eqref{eq:tube}, hence the F-matrix in Eq.~\eqref{eq:Otilde}} into the lattice.
If a measurement projects the tube qubits onto the state  
\begin{equation} \label{eq:outcome_psi}
	\ket{\psi} = \sum_{\alpha,\beta} A_{\alpha \beta} \ket{x}\otimes\ket{y\alpha\beta} ,
\end{equation}
then the full state gets projected onto
\begin{widetext}
\begin{equation} \label{eq:post-measurement}
	\ket{\Phi}\otimes \ket{\psi} = 
	 \dfrac{1}{\mathcal{N}} 
	\Bigg( 
	\sum_{\alpha,\beta} \dfrac{1}{\D} \dfrac{v_x}{v_y} \left( A_{\alpha \beta} \right)^* (\mathds{1} \otimes \bra{x})\, \tilde{O}_{y x \alpha \beta} (\ket{\phi_y} \otimes \ket{y})
	\Bigg)  \otimes \ket{\psi} ,
\end{equation}
where $ \mathcal{N} $ is a normalization factor. 
Hence, by selecting the basis for the measurement of the tube qubits carefully, we can effectively apply the idempotent projectors Eqs.~(\ref{eq:P11_short}-\ref{eq:P222_short}) or the nilpotent operators Eqs.~\eqref{eq:nilpotents_fib} and \eqref{eq:nilpotents_fib_2}.

We choose the following basis\footnote{There are seven ways to label a tube according to the Fibonacci fusion rules. Hence, the 7 vectors below span the string-net subspace for the 4 qubits on the tube.} for the measurement of the tube qubits:
\begin{align}
	\ket{\psi_{\mathbf{1}\mathbf{1}}} 		\label{eq:psi11} 
	& =\frac{1}{\D}  \Bigg |\!\!\!\! \raisebox{-0.5cm}{\includegraphics[scale=.3]{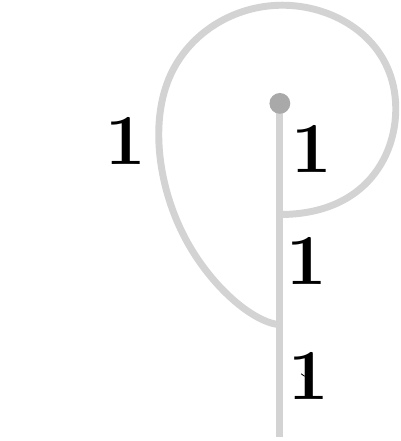}} \, \Bigg \rangle
	+ \frac{\phi}{\D}\Bigg | \!\!\!\! \raisebox{-0.5cm}{\includegraphics[scale=.3]{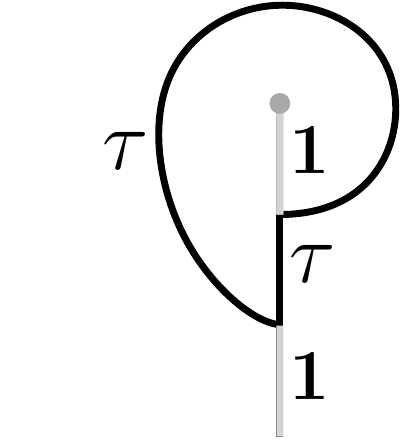}} \, \Bigg \rangle   
	 =  \ket{0} \otimes  \frac{1}{\D} \Big( \ket{000}+\phi \ket{011} \Big)
	\equiv   \ket{0} \otimes
	\ket{\tilde{\psi}_{\mathbf{1}\mathbf{1}}} , 
\end{align}
\begin{align} 
	\ket{\psi_{\1\tau}} 			\label{eq:psi1t}	
	& = \frac{1}{\D}
	\Bigg | \!\!\!\! \raisebox{-0.5cm}{\includegraphics[scale=.3]{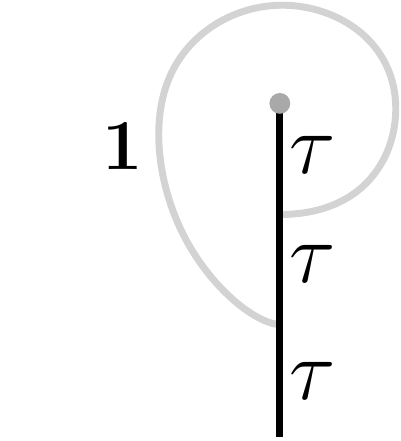}} \, \Bigg \rangle 
	+ \frac{\e^{4 \pi \ii/5}}{\D} \Bigg | \!\!\!\! \raisebox{-0.5cm}{\includegraphics[scale=.3]{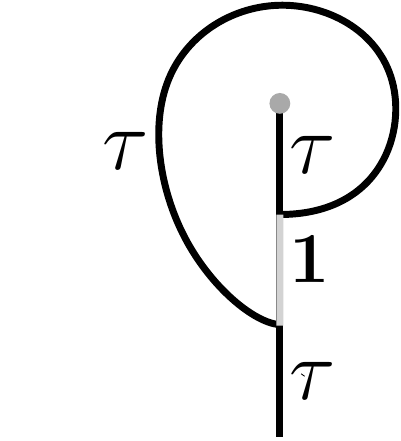}} \, \Bigg \rangle 
	+ \sqrt{\phi} \frac{\e^{-3\pi \ii/5}}{\D} \Bigg | \!\!\!\! \raisebox{-0.5cm}{\includegraphics[scale=.3]{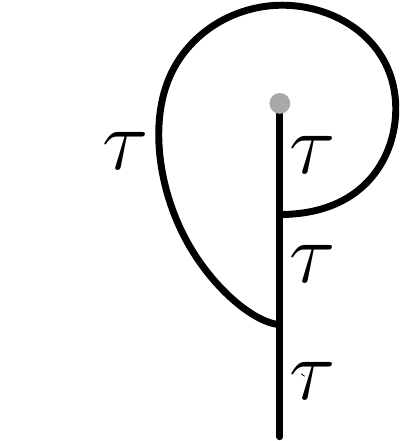}} \, \Bigg \rangle   \\
	& 
	=   \ket{1} \otimes \frac{1}{\D} \Big( \ket{101} + \e^{4 \pi \ii/5} \ket{110}  + \sqrt{\phi} \e^{-3\pi \ii/5}  \ket{111} \Big)  \nonumber 
	 \equiv  \ket{1} \otimes \ket{\tilde{\psi}_{\mathbf{1}\tau}} ,  \nonumber\\
	\nonumber \\
	\ket{\psi_{\tau \1}} 				\label{eq:psit1}
	& = \frac{1}{\D} \Bigg | 
	\!\!\!\! \raisebox{-0.5cm}{\includegraphics[scale=.3]{fig/tube_t1tt.pdf}} \, \Bigg \rangle 
	+  \frac{\e^{-4 \pi \ii/5}}{\D} \Bigg | \!\!\!\! \raisebox{-0.5cm}{\includegraphics[scale=.3]{fig/tube_ttt1.pdf}} \, \Bigg \rangle   
	+ \sqrt{\phi} \frac{\e^{3\pi \ii/5}}{\D} \Bigg | \!\!\!\! \raisebox{-0.5cm}{\includegraphics[scale=.3]{fig/tube_tttt.pdf}} \, \Bigg \rangle  \\
	& 
	=   \ket{1} \otimes \frac{1}{\D} \Big( \ket{101} + \e^{-4 \pi \ii/5} \ket{110}  + \sqrt{\phi} \e^{3\pi \ii/5}  \ket{111} \Big)     \nonumber 
	 \equiv  \ket{1} \otimes \ket{\tilde{\psi}_{\tau \1}}  , \nonumber
\end{align}
\begin{align}
	\ket{\psi_{\tau\tau,\1}} 			\label{eq:psitt_1}
	& = \frac{\phi }{\D}\Bigg | \!\!\!\! \raisebox{-0.5cm}{\includegraphics[scale=.3]{fig/tube_1111.pdf}} \, \Bigg \rangle
	-  \dfrac{1}{\D} \Bigg | \!\!\!\! \raisebox{-0.5cm}{\includegraphics[scale=.3]{fig/tube_1t1t.pdf}} \, \Bigg \rangle 
	 =  \ket{0} \otimes \frac{1}{\D} \Big( \phi\ket{000} -  \ket{011}  \Big)  
	\equiv  \ket{0} \otimes 	 
	\ket{\tilde{\psi}_{\tau\tau,\1}}   ,     \\
	\nonumber \\ 
	\ket{\psi_{\tau\tau,\tau} }			\label{eq:psitt_t}	
	& = \frac{\sqrt{\phi}}{\D} \Bigg | \!\!\!\! \raisebox{-0.5cm}{\includegraphics[scale=.3]{fig/tube_t1tt.pdf}} \, \Bigg \rangle
	+  \frac{\sqrt{\phi}}{\D} \Bigg | \!\!\!\! \raisebox{-0.5cm}{\includegraphics[scale=.3]{fig/tube_ttt1.pdf}} \Bigg \rangle 
	+  \frac{1}{\phi \D}  \Bigg |  \!\!\!\! \raisebox{-0.5cm}{\includegraphics[scale=.3]{fig/tube_tttt.pdf}} \, \Bigg \rangle \\	 
	& =  \ket{1} \otimes \frac{1}{\D} \Big( \sqrt{\phi} \ket{101} + \sqrt{\phi} \ket{110} + \frac{1}{\phi} \ket{111} 
	\Big) \nonumber
	 \equiv   \ket{1} \otimes \ket{\tilde{\psi}_{\tau\tau,\tau}},  \nonumber
\end{align}
\end{widetext}
\begin{align}
    \ket{\psi_{\tau\tau,\,\1,\tau} } 		 \label{eq:psitt_1t}
    & =  \Bigg | \!\!\!\! \raisebox{-0.5cm}{\includegraphics[scale=.3]{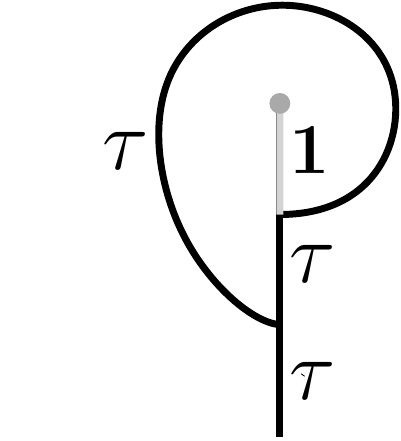}} \, \Bigg \rangle 
     =  \ket{1}  \otimes \ket{011}
     \equiv   \ket{1}  \otimes
    \ket{\tilde{\psi}_{\tau\tau,\, \1,\tau}} ,  
    \\
	\nonumber \\
    \ket{\psi_{\tau\tau,\, \tau,\1}}  	 \label{eq:psitt_t1}
    & =  \Bigg | \!\!\!\! \raisebox{-0.5cm}{\includegraphics[scale=.3]{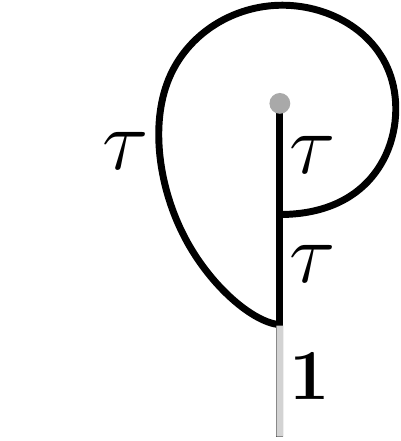}} \, \Bigg \rangle
    =   \ket{0} \otimes \ket{111}   
    \equiv  \ket{0} \otimes \ket{\tilde{\psi}_{\tau\tau,\,\tau, \1}}  . 
\end{align}
Note that the coefficients appearing in the different states are proportional to those in the corresponding irreducible idempotents in Eqs.~\eqref{eq:P11_short}-\eqref{eq:P21_short},  
\eqref{eq:P221_short} and \eqref{eq:P222_short}, or nilpotents in Eq.~\eqref{eq:nilpotents_fib} respectively. In fact, these states are precisely the anyonic fusion basis states on $ \Sigma_2 $, as described in Eq.~\eqref{eq:doubled_leaf_segment_red}.

The measurement of the tube qubits in \figref{fig:grow_circuit}(j) is done in three steps:
\begin{enumerate}	
    \item
	Measure the tail qubit (label $x$). 
	
	\item 
	Apply one of the following unitaries\footnote{\label{note1} The operators below must of course be completed to true unitary operators. The missing terms were left out to improve readability. } conditioned on the measurement of the tail qubit:
    
    (a) if the tail qubit is in state $\ket{0}$ (string label $\mathbf{1}$):
    \begin{flalign*}
        && U_\1= \ket{0}\left(
		\ket{00}\bra{\tilde{\psi}_{\1\1}} 
		+ \ket{11}\bra{\tilde{\psi}_{\tau\tau,\1}}
		+ \ket{10}\bra{\tilde{\psi}_{\tau\tau,\, \tau, \1}} 
		\right ),
    \end{flalign*}

    (b) if the tail qubit is in state $\ket{1}$ (string label $\tau$): 
    \begin{flalign*}
         && U_\tau = \ket{0}\left(
		 \ket{00}\bra{\tilde{\psi}_{\tau\tau,\tau}} 
		+ \ket{11}\bra{\tilde{\psi}_{\tau\tau,\,\1,\tau}} 
		 \right. \hspace{1.8cm} \\
		&& \left. 
		+ \ket{01}\bra{\tilde{\psi}_{\1\tau}}
		+ \ket{10}\bra{\tilde{\psi}_{\tau \1}}
		\right) .
    \end{flalign*}
    
    \item 
	Measure qubits 16 and 17 in \figref{fig:charge_measurement_circuit_1} in the Z-basis.\\
\end{enumerate}

Once the tube qubits have been measured in the basis Eqs.~\eqref{eq:psi11}-\eqref{eq:psitt_t1} using the procedure above,
we can trace out the three ancilla qubits [15, 16 and 17 in Fig.~\ref{fig:charge_measurement_circuit_1}, 
constituting the inner three edges of the tube in \figref{fig:grow_circuit}(j)] to return to the initial tailed lattice layout with a single tail qubit in each plaquette [i.e., the configuration in Fig.~\ref{fig:grow_circuit}(a)].
This results in the following POVM:
\begin{equation} \label{eq:POVM}
	\Big \{ 
	\cP^{\1\1},\;
	\cP^{\1\tau},\;
	\cP^{\tau\1},\;
	\dfrac{1}{\phi^2} \cP^{\tau\tau}_\1,\;
	\dfrac{1}{\phi} \cP^{\tau\tau}_\tau,\;
	\dfrac{1}{\phi} \cP^{\tau\tau}_\1,\;
	\dfrac{1}{\phi^2} \cP^{\tau\tau}_\tau
	\Big \}\,.
\end{equation}

Note that both a $ \tau\tau_\1 $ and a $ \tau\tau_\tau $ excitation corresponds to two different measurement outcomes.
However, within each of these pairs, the post-measurement states are not identical. 
For instance, a $ \tau\tau_\1 $ excitation can result in measurement outcomes $ \ket{\psi_{\tau\tau, \1}} $ and $ \ket{\psi_{\tau\tau,\, \1,\tau}} $.
Obtaining outcome $ \ket{\psi_{\tau\tau, \1}} $, effectively applies the $ \cP^{\tau\tau}_\1 $ idempotent, meaning the post-measurement state will have trivial tail label $ \1 $. 
On the other hand, the outcome $ \ket{\psi_{\tau\tau,\, \1,\tau}} $ corresponds to the application of the $ \cP^{\tau\tau}_{\1\tau} $ nilpotent. Since $ \cP^{\tau\tau}_{\1\tau}  = \cP^{\tau\tau}_{\tau} \cP^{\tau\tau}_{\1\tau}  \cP^{\tau\tau}_{\1}$, this means the plaquette initially contained a $ \tau\tau_\1 $ excitation, which gets transformed to a $ \tau\tau_\tau $ excitation in the post-measurement state.
The situation for a $ \tau\tau_{\tau} $ excitation is analogous.
Hence these measurements do not preserve the tail label in case of a $ \tau\tau $ anyon, and a subsequent measurement might yield a different outcome. It is important to note that this only affects the tail label \emph{within one anyon sector}, the anyon label itself cannot be altered by subsequent measurements. \\

\begin{figure*}[ht]
	\centering
	\includegraphics[scale=.35]{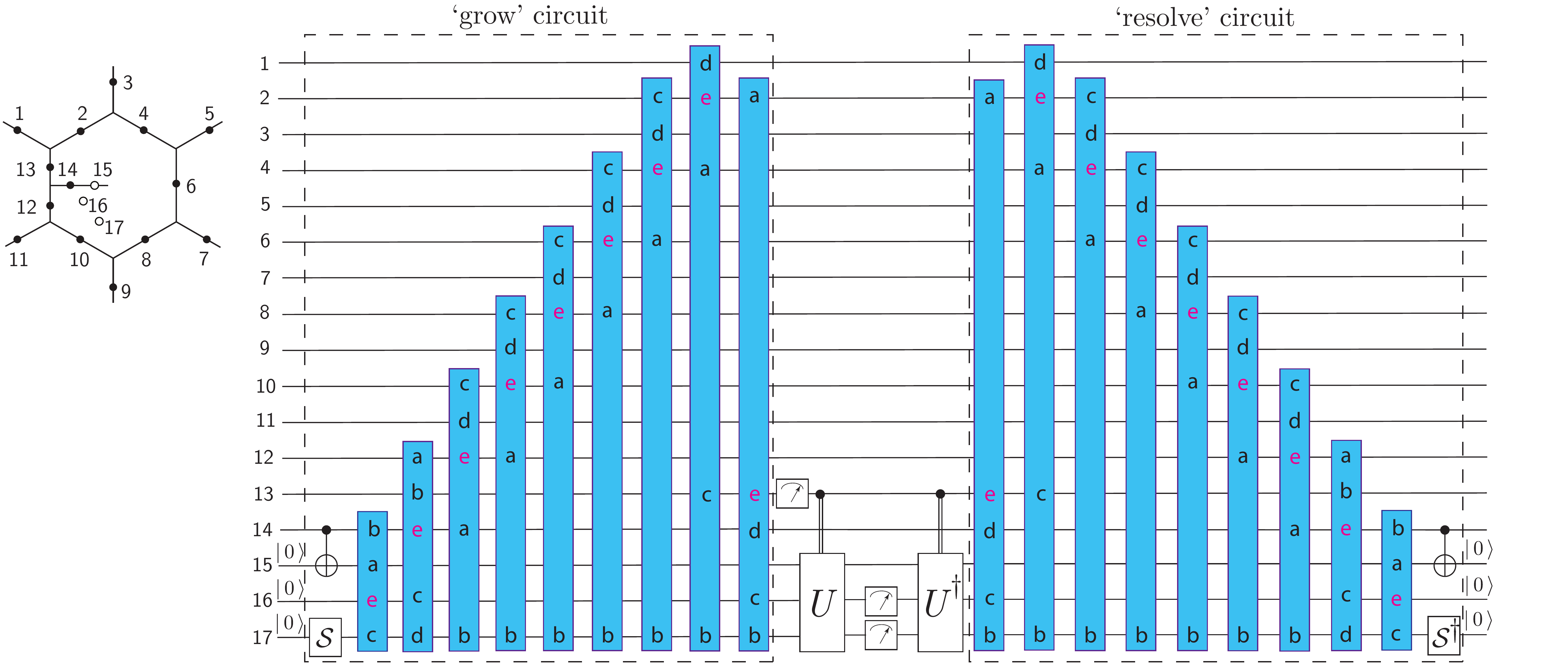}
	\caption{Alternative quantum circuit for the joint measurement of the anyon charge and tail label of a plaquette, which does preserve the tail label.}
	\label{fig:charge_measurement_circuit_2}
\end{figure*}

In case one prefers to preserve the tail label of excitations in subsequent measurements\footnote{Possibly, this could improve the performance of certain decoders. Furthermore, we will assume this for the numerical simulations discussed in Sec~\ref{sec:simulation}.}, an additional step is required before tracing out the three ancillas.
This step consists of resolving the tube into the lattice by applying the gates of the grow circuit in reverse, as indicated in \figref{fig:charge_measurement_circuit_2}.
For measurement outcome $ \ket{\psi} $ defined Eq.~\eqref{eq:outcome_psi}, the resolving process is equivalent to the following transformation on the post-measurement state in Eq.~\eqref{eq:post-measurement}:
\begin{equation} 
	\ket{\Phi} \otimes \ket{\psi} \mapsto \sum_{\alpha,\beta} A_{\alpha\beta} O_{x y\alpha\beta} \left( \ket{\Phi} \otimes \ket{x}\right) \otimes \ket{0 0 0 }.
\end{equation} 
This guarantees that the initial tail label $ y $ will indeed be recovered.
For instance, when obtaining the  $ \ket{\psi_{\tau\tau,\, \1\tau}} $ measurement outcome, the resolving process result in the application of the $  \cP^{\tau\tau}_{\tau\1} $ nilpotent on top of the  $  \cP^{\tau\tau}_{\1\tau} $ nilpotent applied by the measurement, resulting in the combined action $  \cP^{\tau\tau}_{\1} =  \cP^{\tau\tau}_{\tau\1}  \cP^{\tau\tau}_{\1\tau} $, meaning that the tail label is now preserved. 

The same result can be achieved by using repeated measurements with the circuit in \figref{fig:charge_measurement_circuit_1}. In case one measures a $ \tau\tau $ excitation but finds that the tail label gets flipped [corresponding to the tube states in Eqs.~\eqref{eq:psitt_1t} and \eqref{eq:psitt_t1}, and the last two entries in the POVM Eq.~\eqref{eq:POVM}], each subsequent measurement\footnote{Provided that no errors happen on qubits in or adjacent to the plaquette between these repeated measurements.} has a fixed probability of flipping the tail label back to its initial value (as determined by the first measurement). 
For a $ \tau\tau_{\1} $ excitation, it follows from Eq.~\eqref{eq:POVM} that the probability that $ n $ measurements are required before the post-measurement state has a trivial tail label is $ \phi^{-(n+1)} $. Hence, on average, $ \phi^2 $ measurements are required to ensure that the tail label is preserved for a $ \tau\tau_{\1} $ excitation.
Likewise, for a $ \tau\tau_{\tau} $ excitation the probability of needing $ n $ measurements to recover the initial tail label is $ \phi^{-(2n-1)} $, resulting in an average of $ \phi $ measurements.
Whether one should choose for the longer measurement circuit depicted in \figref{fig:charge_measurement_circuit_2} or for repeated measurements with the shorter measurement circuit depicted in \figref{fig:charge_measurement_circuit_1} depends on what types of excitations are more likely to appear.

%

The procedure described above determines the anyon charge of a single plaquette. By repeating it for all plaquettes, we obtain the complete error syndrome (in the form of the anyonic content of every plaquette), which must then be passed to a decoding algorithm to determine the appropriate recovery operation to be performed (see Sec.~\ref{sec:decoding}).

\subsection{Recovery operations} \label{sec:recovery}
After extracting the error syndrome as described above, the decoding algorithm will suggest a sequence of actions to take to fuse pairs of anyonic excitations along specific paths.  
There are two types of fundamental recovery operations: \emph{Fuse}, and \emph{Exchange}.
Here we will introduce the quantum circuits that implement these recovery operations.
The key components of these circuits include the 2-2 Pachner moves and the 1-3 Pachner moves as introduced in Sec.~\ref{sec:model} and the corresponding unitary circuits introduced in Sec.~\ref{sec:measure_charge}.

\subsubsection{Fuse} \label{sec:fuse}
\begin{figure*}
    \centering
    \includegraphics[width=2\columnwidth]{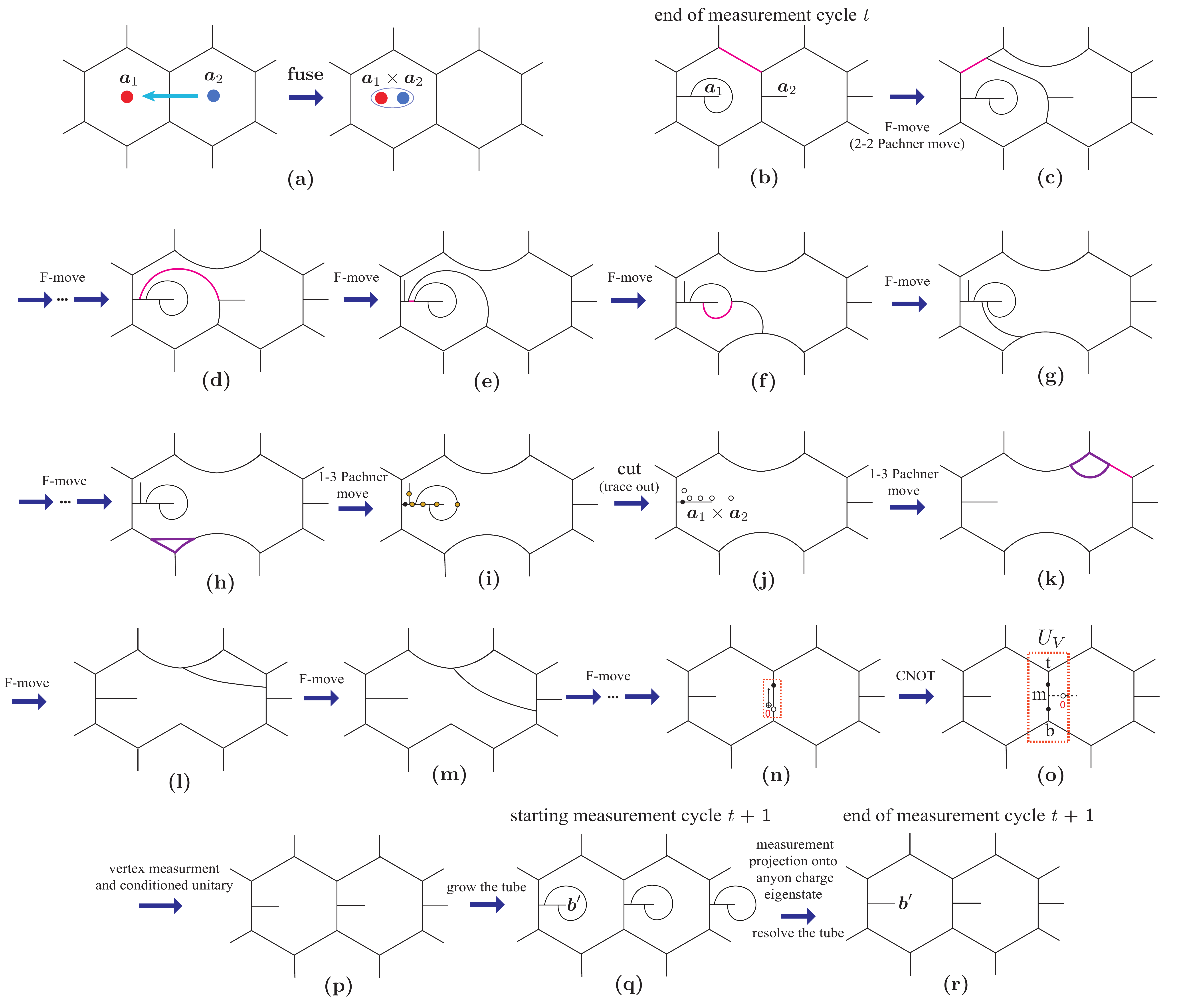}
    \caption{(a) The basic idea of the \textit{fuse} protocol: fuse anyons $\boldsymbol{a}_1$ and $\boldsymbol{a}_2$ by moving $\boldsymbol{a}_2$ on the right plaquette to the left plaquette and then fuse them into a single anyon with charge $\boldsymbol{a}_1 \times \boldsymbol{a}_2$. (b) Starting the fuse protocol in the end of measurement cycle $t$. The tube on the left plaquette is preserved for the convenience of the fuse protocol.  (c) Use an $F$-move to sweep the middle edge towards the left plaquette. (d) Sweep the central edge onto the left tail.  (e) Use an $F$-move to shuttle the right tail onto the left tail. (f-h) Keep sweeping the middle edge until it reaches the bottom vertex of the left plaquette.  (i) Apply a 1-3 Pachner move to shrink the triangular bubble on the bottom vertex. (j) Cut the tubes and extra tail by tracing out all the qubits except the one on the root.  Now the two plaquettes have been merged into one and the two anyons are fused into one with charge $\boldsymbol{a}_1 \times \boldsymbol{a}_2$, which is in general in a superposition state. (k) Apply a 1-3 Pachner move to grow a triangular bubble on the top vertex of the right plaquette. (l-n) Use $F$-moves to keep sweeping the new edge towards the bottom edge and hence grow a new plaquette on the right. Apply a CNOT from the qubit residing on the middle edge to another ancilla qubit initialized at $\ket{0}$, which splits the edge into two.  (o, p) Prepare an ancilla qubit at $\ket{0}$ on the new tail.  Measure the vertices $t$, $m$ and $b$, and apply the unitary $U_V$ conditioned on the measurement result to pull the broken string in and fix all the vertex errors.  We have now rebuilt the tailed lattice in (p). (q,r) In the next measurement cycle $t+1$, start by growing and measuring the tube to project the fused anyon charge $\boldsymbol{a}_1 \times \boldsymbol{a}_2$ to a definite anyon charge $\boldsymbol{b}'$, and then resolve all the tubes back to the tailed lattice.}
    \label{fig:merge_anyons}
\end{figure*}

\begin{figure*}
    \centering
    \includegraphics[width=2\columnwidth]{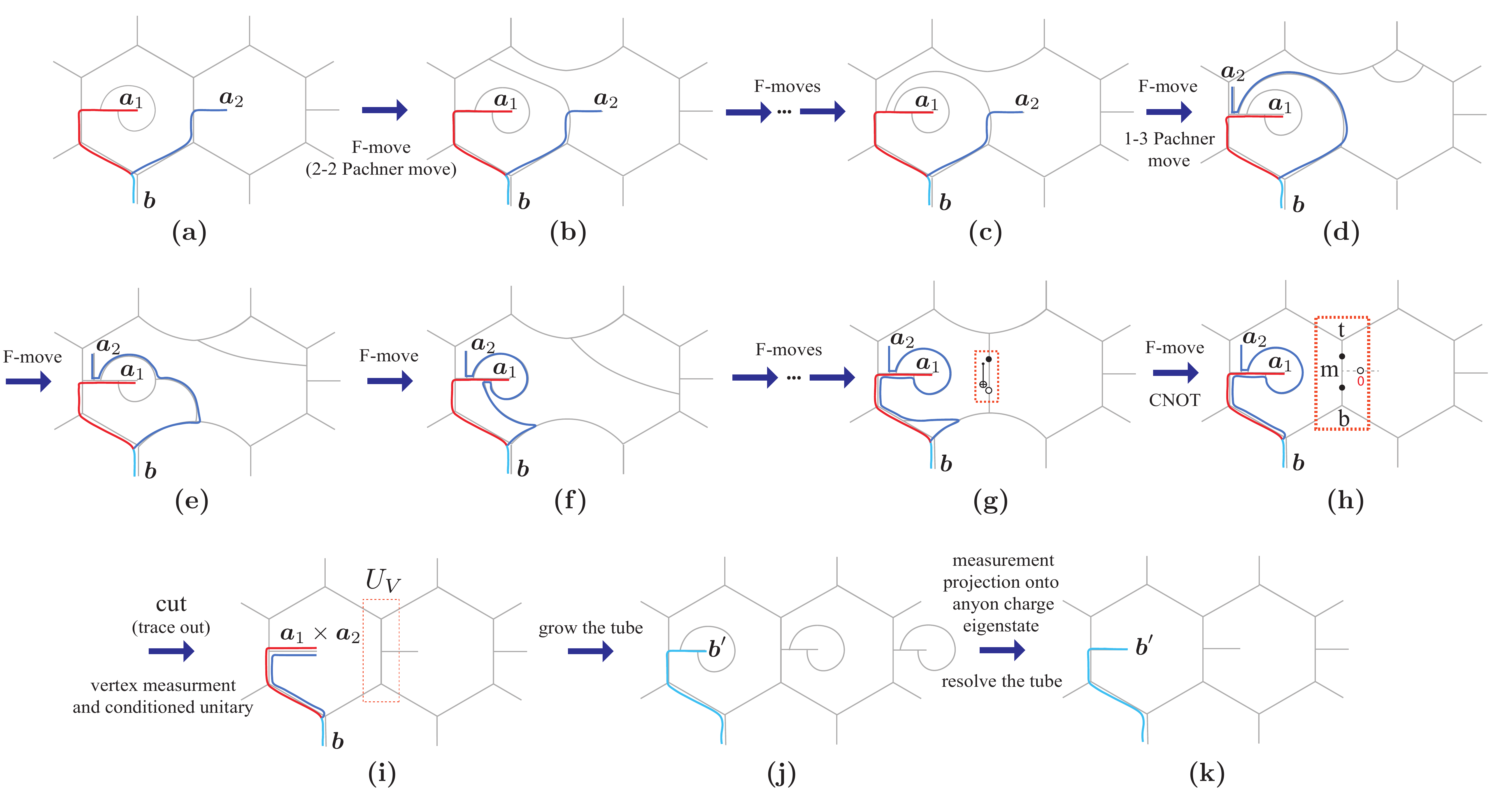}
    \caption{Illustration of the \textit{fuse} protocol with an example of a ribbon graph state with two anyons $\boldsymbol{a}_1$ and $\boldsymbol{a}_2$ and their total branch charge $\boldsymbol{b}$. Note that we have further parallelized the fuse protocol in Fig.~\ref{fig:merge_anyons} by simultaneously doing the plaquette merging and growing a new plaquette on the right.}
    \label{fig:merge_anyons_string}
\end{figure*}

We start by defining \textit{fuse} as a recovery operation that fuses anyons in neighboring plaquettes. The fuse protocol is shown in Fig.~\ref{fig:merge_anyons} for the case where an anyon $\boldsymbol{a}_2$ inside a plaquette is fused with the anyon $\boldsymbol{a}_1$ inside its left neighboring plaquette. 
Other fusing directions will have a very similar process and we omit showing all of them. The application of this protocol to a particular example of a ribbon graph state is illustrated in Fig.~\ref{fig:merge_anyons_string} with further parallelization of the steps in the protocols shown in Fig.~\ref{fig:merge_anyons}.   

Before detailing the quantum circuit to fuse a pair of neighboring anyons, we consider what such an operation means on the level of the ribbon graphs introduced in Sec.~\ref{sec:model}.
We consider two plaquettes $p_1$ and $p_2$, and wish to fuse the anyons contained within them.
For this purpose, we pick an anyonic fusion basis which explicitly shows the total charge $\boldsymbol{b} = b^+ b^-$ of the anyons $\boldsymbol{a}_1 = (a^+_1 a^-_1)_{\ell_1}$ and $\boldsymbol{a}_2 = (a^+_2 a^-_2)_{\ell_2}$ contained in the two plaquettes:
\begin{equation}\label{eq:ribbon_merge}
    \ket{\psi^{\vec{\bm{a}}}_{\bm{b},\vec{\bm{c}} }} =  \!\! \raisebox{-.6cm}{\includegraphics[scale=.4]{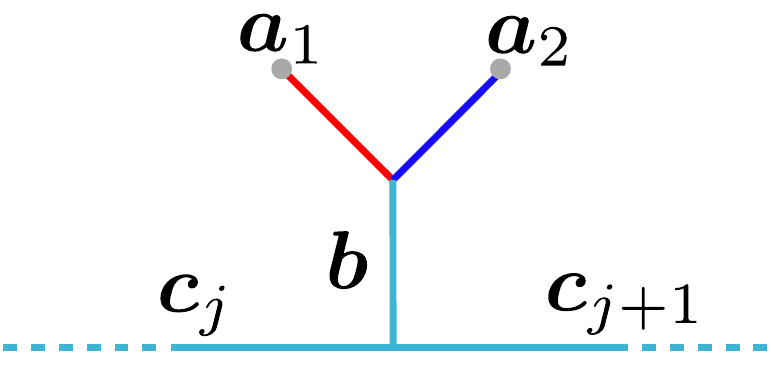}} \!\! \equiv \raisebox{-1.4cm}{\includegraphics[scale=.4]{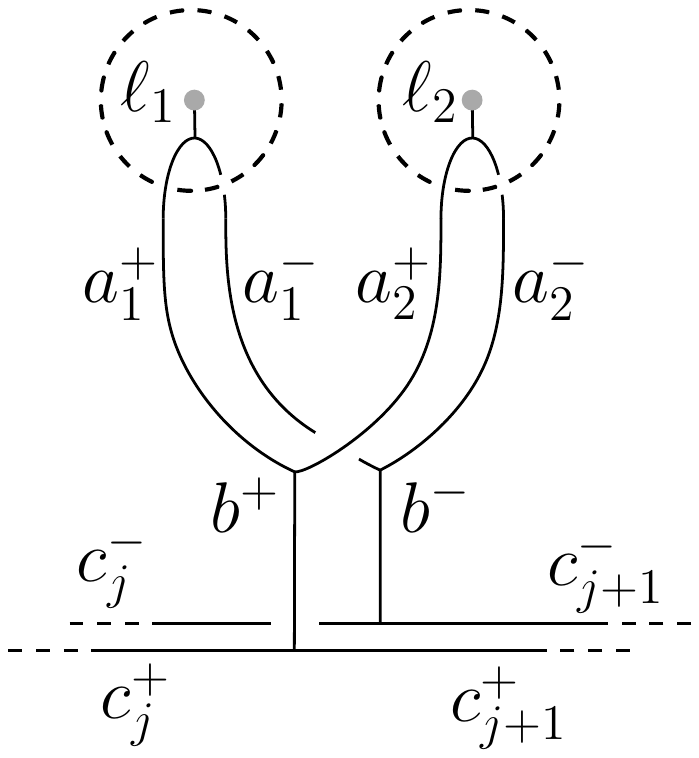}}\,,
\end{equation}
where we have denoted all other branch labels on the fusion tree collectively by $\vec{\bm{c}}$, and have omitted showing all other leaf labels $\{ \bm{a}_3, \bm{a}_4, \ldots \}$.
We will refer to these other leaf labels collectively as $ \vec{\bm{a}}_{\text{ot}}$, writing $\vec{\bm{a}} = [\bm{a}_1, \bm{a}_2, \vec{\bm{a}}_{\text{ot}}]$ .
For simplicity, we only show the corresponding ribbon graphs, and not their embedding in the fattened lattice as described in Sec.~\ref{sec:anyonic_fusion_basis_short}.

After the syndrome measurement detailed in Sec.~\ref{sec:measure_charge}, the system is in definite charge eigenstate for all plaquettes, which means that the leaf labels $\vec{\bm{a}}$ are fixed in the state superposition, while the internal labels $\{\bm{b}, \vec{\bm{c}}\}$ are not.
In general we then have
\begin{equation} \label{eq:fuse_initial_state}
    \ket{\Psi_0} = \sum_{ \bm{b},\vec{\bm{c}}} \alpha_{\bm{b},\vec{\bm{c}}} \ket{\psi^{\vec{\bm{a}}}_{\bm{b},\vec{\bm{c}} }} .
\end{equation}
The fusion basis state $ \ket{\psi^{\vec{\bm{a}}}_{\bm{b},\vec{\bm{c}} }} $ appearing in this decomposition, can be rewritten as follows:
\begin{align} \label{eq:fusion_tail}
	\ket{\psi^{\vec{\bm{a}}}_{\bm{b},\vec{\bm{c}} }} 
	 &=  \frac{1}{\D} \raisebox{-1.6cm}{\includegraphics[scale=.5]{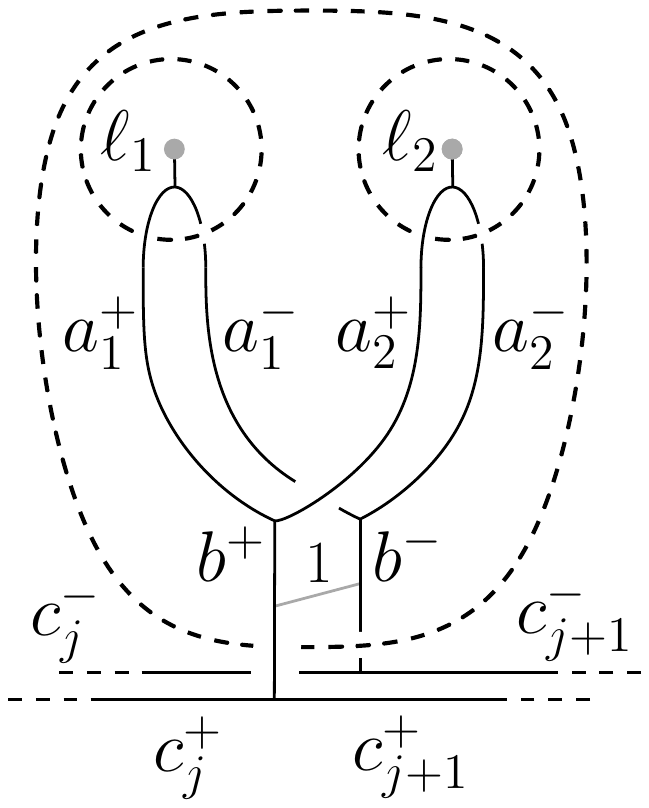}}  \nonumber \\
	 &= \frac{1}{\D} \sum_{\ell=\1,\tau} \! F^{b^- b^+ 1}_{b^+ b^- \ell} \raisebox{-1.6cm}{\includegraphics[scale=.5]{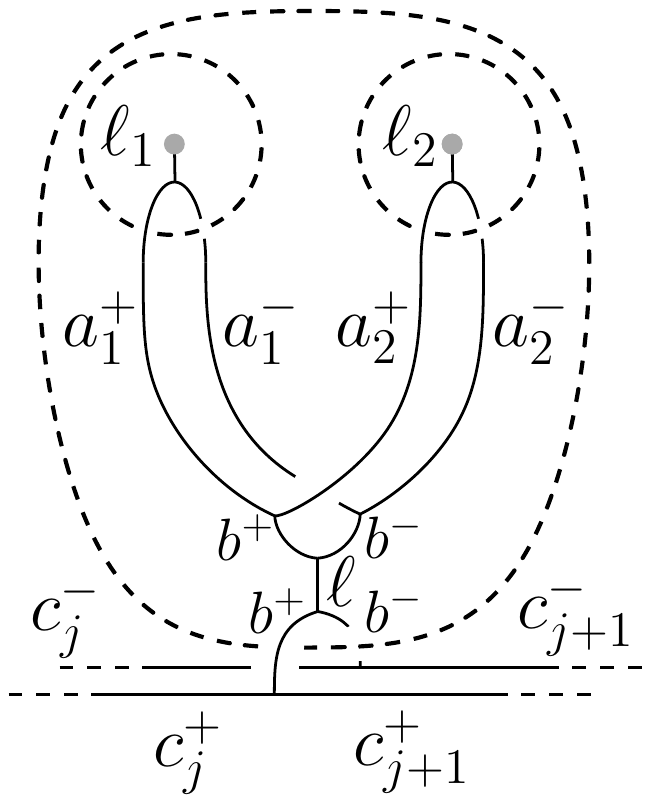}}\,,
\end{align}
where we have used the property that vacuum loops can be ``doubled'' (see Eq.~\ref{eq:vacuum_loop_doubling} in Sec.~\ref{sec:anyonic_fusion_basis}), and have applied an $F$-move on the double ribbon $b^+b^- $ of corresponding to the total charge of plaquettes $p_1$ and $p_2$. 
The fusion of the anyons in these plaquettes corresponds replacing the upper part of the diagram in Eq.~\ref{eq:fusion_tail} by a single puncture as follows
\begin{equation}
    \raisebox{-1.6cm}{\includegraphics[scale=.5]{fig/ribbon_merge_transformed.pdf}} \;\; \mapsto \;\; \raisebox{-1.cm}{\includegraphics[scale=.5]{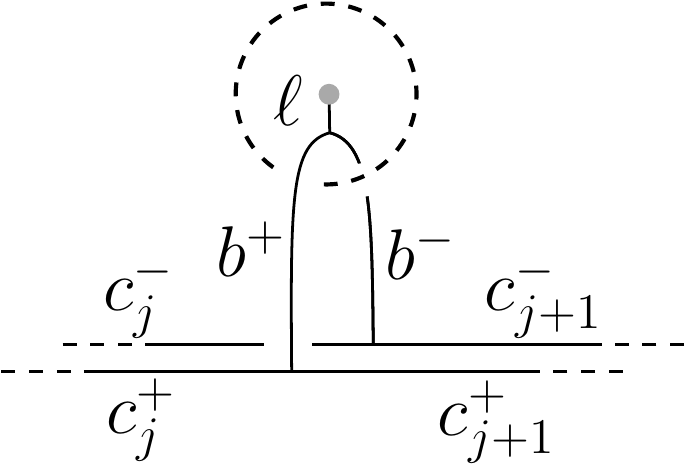}}\,.
\end{equation}
The fusion can then be represented by the following map on anyonic fusion basis states:
\begin{equation}
    \text{fuse : } \quad \ket{\psi^{[\bm{a}_1, \bm{a}_2, \vec{\bm{a}}_\text{ot}]}_{\bm{b},\vec{\bm{c}} }} 
    \mapsto
    \sum_{\ell=\1,\tau} \! F^{b^- b^+ 1}_{b^+ b^- \ell} \ket{\psi^{[\bm{b}_\ell, \vec{\bm{a}}_\text{ot} ]}_{\vec{\bm{c}} }},
\end{equation}
where\footnote{To be precise, we should have written $ \ket{\psi^{[\bm{b}_\ell, \1\1_\1, \vec{\bm{a}}_\text{ot} ]}_{\bm{b}, \vec{\bm{c}} }} $, since plaquette $p_2$ now has a trivial anyonic charge, and the total charge of plaquettes $p_1$ and $p_2$ remains unchanged.}
\begin{equation}
    \ket{\psi^{[\bm{b}_\ell, \vec{\bm{a}}_\text{ot} ]}_{\vec{\bm{c}} }} = \raisebox{-1.cm}{\includegraphics[scale=.5]{fig/ribbon_merge_fused.pdf}} \,.
\end{equation}
The post-measurement state $\ket{\Psi_0}$ then gets mapped to
\begin{equation}
    \ket{\Psi_0} \mapsto \ket{\Psi_1} = \sum_{ \bm{b},\vec{\bm{c}}} \alpha_{\bm{b},\vec{\bm{c}}} \sum_{\ell=\1,\tau} \! F^{b^- b^+ 1}_{b^+ b^- \ell}  \ket{\psi^{[\bm{b}_\ell, \vec{\bm{a}}_\text{ot} ]}_{\vec{\bm{c}} }}.
\end{equation}
Measuring the tail label and the anyonic charge of plaquette $p_1$ after fusing the two anyons, will yield outcomes $\bm{b}_\ell$ with the probabilities 
\begin{equation} \label{eq:probability_fuse}
    p(\bm{b}_\ell) = \sum_{\vec{\bm{c}}} \sum_{\ell=\1,\tau} \left| \alpha_{\bm{b},\vec{\bm{c}}} \, F^{b^- b^+ 1}_{b^+ b^- \ell} \right|^2.
\end{equation}
Note that the probabilities for the outcomes of a tail measurement, depend only on the total charge $\bm{b}$. 
Furthermore, in any measurement scheme where the anyonic charge of a plaquette is measured independently of its tail label, these measurements will commute.


At the end of the charge measurement cycle, we normally trace out the tube qubits or resolve the tube into the lattice. However, before fusing a pair of anyons, we can keep the tube on the left plaquette (instead of growing it again) in the beginning of the fusion protocol, as shown in Fig.~\ref{fig:merge_anyons}(b). The purpose is to carry out the subsequent $F$-moves for an edge to climb around the tube without encountering any broken string as in the case with a single tail qubit on the left plaquette. 
  
The central idea of the protocol is to merge the two plaquettes and hence the corresponding plaquette excitations, while also combining the tail qubits such that the vertex excitations also get merged. Overall, this process fuses the two anyon charges inside the two plaquettes. The protocol starts at the end of measurement cycle $t$. One first  applies an $F$-move (2-2 Pachner move) that sweeps the central edge between the two plaquettes towards the left plaquette, as shown in Fig.~\ref{fig:merge_anyons}(b).  After several $F$-moves, the central edge now climbs onto the tail of the left plaquette as shown in Fig.~\ref{fig:merge_anyons}(d).  Now we apply an $F$-move to move the tail on the right plaquette onto the left tail in Fig.~\ref{fig:merge_anyons}(e). In Fig.~\ref{fig:merge_anyons}(e), we further sweep the central edge onto the tube.  After several steps of further moves, the central edge now forms a triangular bubble with the other two edges in the bottom of the left plaquette, as shown in Fig.~\ref{fig:merge_anyons}(h).   One now applies a 1-3 Pachner move to absorb the bubble into the vertex in  Fig.~\ref{fig:merge_anyons}(i), and now the left and right plaquettes are merged into a single large plaquette.  The corresponding procedures with the example of the ribbon graph state is also illustrated in Fig.~\ref{fig:merge_anyons_string}(a-h).   Now we trace out all the qubits residing on the left tail and only a single tail qubit remains in the lattice, as shown in Fig.~\ref{fig:merge_anyons}(j). We are now left with two ribbons going into the same tail and which correspond to the fused anyon charge $\boldsymbol{a}_1 \times \boldsymbol{a}_2$. Note that due to the non-Abelian nature, the fused anyon can be in a superposition of multiple anyon charge eigenstates, and hence may not have a definite charge yet. For example, if $\boldsymbol{a}_1 = \boldsymbol{a}_2 =\tau \1_\tau  \equiv \tau \1$, then one has  $\boldsymbol{a}_1 \times \boldsymbol{a}_2 = \1\1 + \tau \1 $.
     
Since the two initial plaquettes are now merged into a single one, we need to grow another plaquette to recover the original lattice, which is accomplished in Fig.~\ref{fig:merge_anyons}(k-p). The growing is achieved by first applying a 1-3 Pachner move in Fig.~\ref{fig:merge_anyons}(k) to grow a triangular bubble on the top vertex, followed by a sequence of $F$-moves that sweep the newly created edge to form the original middle edge dividing the two plaquettes, illustrated in Fig.~\ref{fig:merge_anyons}(n).  Now we need to regrow the tail on that middle edge.  We first apply a CNOT from the qubit (black dot) residing on the edge on an ancilla qubit (white dot) initialized in the state $\ket{0}$ in order to split the edge in two, as shown in Fig.~\ref{fig:merge_anyons}(n).  We further prepare another ancilla qubit (white dot) in the state $\ket{0}$ corresponding to the tail qubit (dashed line) in Fig.~\ref{fig:merge_anyons}(o). We then measure the vertex projectors $Q_v$ on the three vertices $t$, $m$ and $b$ and apply the corresponding unitary $U_V$ to pull the potentially broken string into the tail based on the measurement result $V$, using the measurement and unitary circuits previously shown in Fig.~\ref{fig:vertex_correction} in Sec.~\ref{sec:vertex_measurement}. We note that if there is no error introduced in the recovery process, the vertices $ t $, $ m $ and $ b $ should not have any error and the strings on them should not be broken.  However, in reality, additional errors will be introduced during the recovery, and one hence needs to apply the vertex measurements and unitary $U_V$ to pull the string in.  Although we have introduced the plaquette growing process separately, we note it can actually be parallelized with the merging of the two original plaquettes, as illustrated in Fig.~\ref{fig:merge_anyons_string}(d-i).  

After performing the fusion protocol described here, one obtains a state where the second plaquette has a vacuum charge ($B_p = 1$), while the first plaquette contains the total charge of the two excitations that where fused. Note that the first plaquette is not in a definite charge state, instead, it contains a superposition of all possible values of the total charge of the two fused excitations. Likewise, its tail label is also in a superposition of $\1$ and $\tau$. 
In principle, no more actions are required, since the excitations have been fused as expected. However, in the numerical simulation described in Sec.~\ref{sec:simulation}, we will assume that after every fusion process, the fusion outcome is projected to some definite anyon label and tail label. This can simply be achieved by including the charge measurement circuit of Sec.~\ref{sec:measure_charge} in the first plaquette at the end of the fusion procedure.

\subsubsection{Exchange and move} \label{sec:move_operation}         
\begin{figure*}[h]
        \centering
        \includegraphics[width=2\columnwidth]{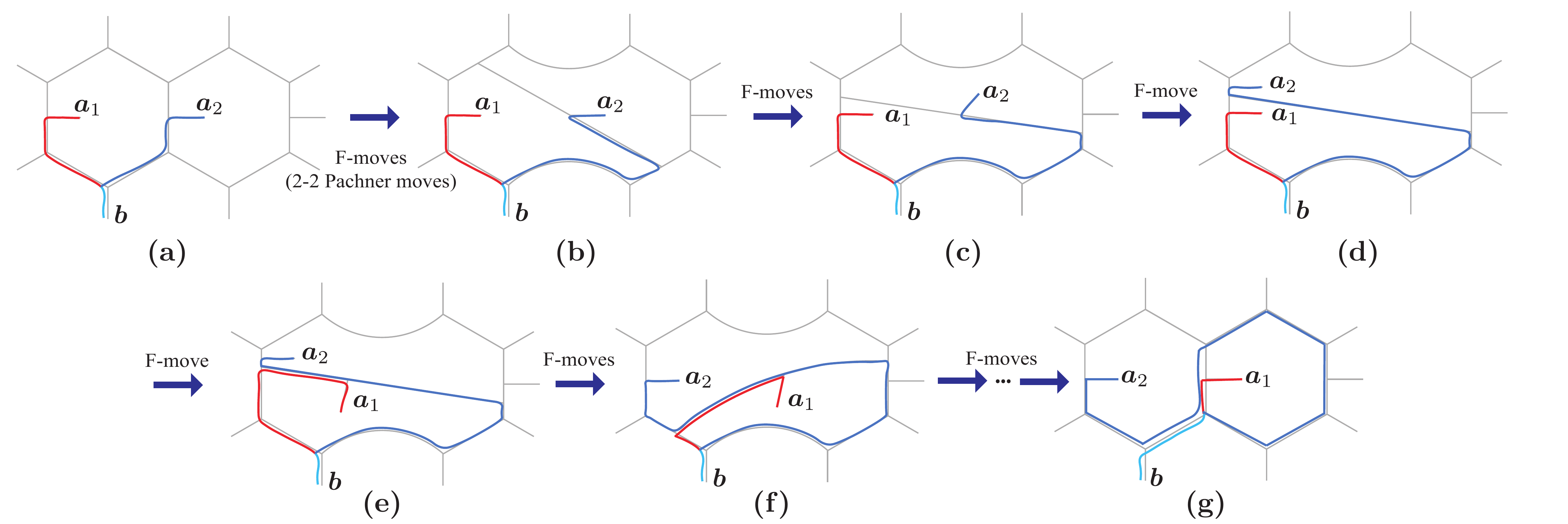}
        \caption{Illustration of the counter-clockwise exchange protocol with an example of ribbon graph state. (a-f) A sequence of F-moves gradually rotates the two plaquettes and the tails accordingly. (g) In the end of the protocol, the position of the two anyons have been exchanged with the ribbons attached to them being braided.  }
        \label{fig:exchange_anyons}
    \end{figure*}

    The second type of recovery operation is a counterclockwise/clockwise \textit{exchange} (braiding) of anyons in neighboring plaquettes. 
    The need for this operation arises when one wants to transport an anyon along a path (which might include plaquettes containing nontrivial excitations).
    In Fig.~\ref{fig:exchange_anyons}, we show the counterclockwise exchange protocol along with the example of the same ribbon graph state shown above in Fig.~\ref{fig:merge_anyons_string}.   The central idea is to rotate the central edge dividing the two plaquette such that the two plaquettes undergo a counterclockwise exchange.  One also needs to shuttle the tail accordingly.  The final ribbon graph state after the exchange operation is shown in Fig.~\ref{fig:merge_anyons_string}(g), and a counterclockwise exchange/braid can be clearly seen on the ribbon graph. 
    Note that, when applied to excitations with a definite anyon and tail labels, these labels are preserved by the exchange procedure.\\
    
\begin{figure*}[ht]
        \centering
        \includegraphics[width=2\columnwidth]{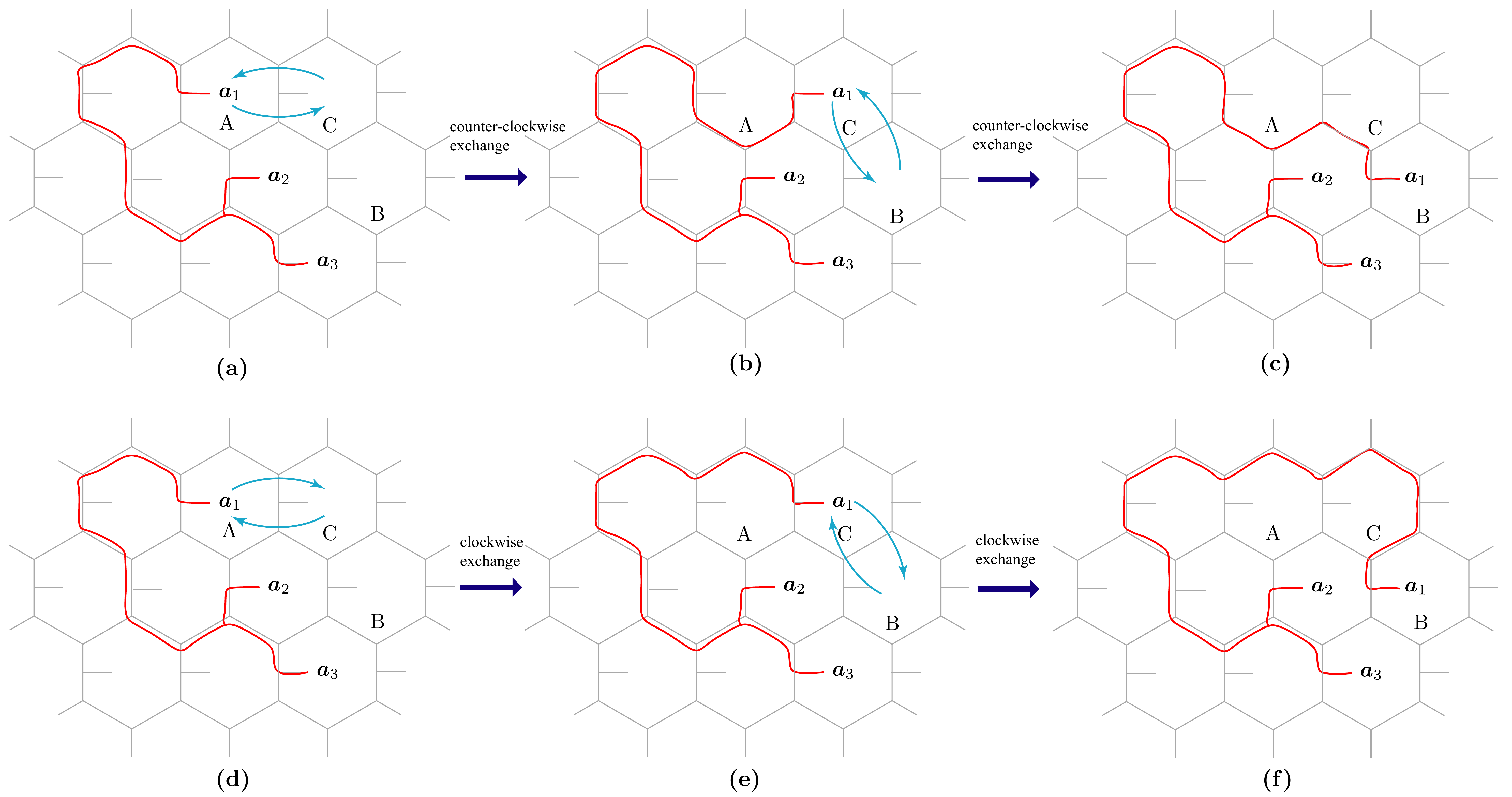}
        \caption{Two equivalent processes to move anyon $\boldsymbol{a}_1$ from plaquette A to plaquette B through the path A$\rightarrow$C$\rightarrow$B.  (a-c) Implement the move through two steps of counterclockwise exchange between plaquettes A and C, and then plaquettes C and B.  The anyon reaches plaquette B in (C).  (d-f) Implement the move through two steps of clockwise exchange of neighboring plaquettes.   
        Note that the ribbon graph states (b) and (e) are equivalent because the exchanges happen with vacuum plaquettes and since ribbons can be pulled across those (see Eq.~\eqref{eq:vacuum_line_pulling_through} in Sec.~\ref{sec:anyonic_fusion_basis}, and also Fig.~\ref{fig:fattened_lattice_groundstate} in Sec.~\ref{sec:fattened}).
        In the same way, the state in (c) is equivalent to the one in (f).
        However, if any plaquette on the movement path contains a nontrivial anyon, the resulting states are not equivalent, since they correspond to different sequences of braid-moves on the fusion state.}
        \label{fig:move_protocol}
        \clearpage
    \end{figure*}

    We define \textit{move} as a process of shuttling anyon $\boldsymbol{a}$ from plaquette $A$ to plaquette $B$ along some specified path, as illustrated in Fig.~\ref{fig:move_protocol} with a path $A\rightarrow C \rightarrow B$.
    This is achieved through a sequence of clockwise or counterclockwise exchanges with all plaquettes along this path.
    Note that if any other anyons are encountered on the path, the results of the move procedures with clockwise and counterclockwise exchanges are not identical, as they differ by a braid in the anyonic fusion space.
    
    In case all plaquettes along the path have a vacuum charge, one could also implement the movement procure as a sequence of fusions, since fusion with a $\1\1$ anyon is trivial.
    However, note that this does not guarantee that the tail label of the excitation that is being moved, is preserved. In particular, when fusing any $\tau\tau$ anyon ( $\tau\tau_\1$ or $\tau\tau_\tau$) with a neighboring trivial (vacuum) anyon, the probabilities for the measurement outcomes of the tail qubit after the fusion are $ 1/\phi^2 $ and $1/\phi$ for the outcomes $\ket{0}$ and $\ket{1}$, respectively, independent of the tail label of the excitation prior to the fusion.
    This is due to the fact that \emph{only} the doubled anyon label of an excitation is topologically protected, its tail label is not and can thus be changed by local interactions if one is not careful. 
    Furthermore, the exchange procedure is considerably simpler than the fusion protocol. 
    
%% file: sections/threshold_simulation.tex
\section{Threshold simulation} \label{sec:simulation}
	Besides formulating an error correction scheme for the extended Fibonacci string-net code, the main achievement of this work consists of obtaining error correction thresholds for this code with a microscopic noise model.	
	This was achieved through Monte Carlo simulations, in which the quantum state of the system is updated to reflect the application of noise, measurement, and recovery operations until either the initial state is recovered successfully or a logical error occurs. 
	Due to the complicated nature of the extended string-net code, and the fact that our simulations require the ability to simulate the dynamics of non-Abelian anyons, this is a very nontrivial task. Below we discuss the various technical details of these simulations. 
	Note that the numerical simulations presented here are independent of the specific circuits used to realize the projective measurements and unitary transformations required during the error correction process (assuming these circuits can be carried out perfectly).
	
	The structure of this section is as follows:
	We start by going over some comments on the classical simulation of non-Abelian QEC, made in Refs.~\cite{brell2014thermalization, wootton2014error, burton2017classical} in Sects.~\ref{sec:simulating_non-Abelian} and \ref{sec:classical_simulatbility}.
	We then introduce a microscopic noise model of Pauli errors in Sec.~\ref{sec:noise_model}, and study the effect of individual qubit errors on anyonic fusion states in Secs.~\ref{sec:Pauli-noise} and \ref{sec:matrix_elements}.
	Sec.~\ref{sec:curve_diagrams} describes the framework of curve diagrams \cite{burton2016short}, which is used to efficiently characterize anyonic fusion basis bases during the simulation. 
	Finally, in Sec.~\ref{sec:outline_of_simulation} we provide a detailed outline of all steps performed for a single Monte Carlo sample.

\subsection{Simulating non-Abelian quantum error correction} \label{sec:simulating_non-Abelian}
	The defining difference between Abelian and non-Abelian anyon models is the fact that the fusion outcome of a pair of anyons is no longer uniquely determined for non-Abelian models.
	This fact makes simulating noise, syndrome measurement and error correction processes for a system exhibiting non-Abelian anyonic excitations considerably more complicated than for Abelian models such as the surface code. 
	
	After anyonic excitations have been created through the application of noise, their locations and anyon labels can be determined through a syndrome measurement. If we think of the state in terms of fusion trees of anyons, the syndrome measurement only projects onto fixed values for the leaf labels. For non-Abelian anyons, the internal (branch) labels of the fusion tree may still be in a superposition of different configurations.
	One of the implications is that braiding processes occurring during the recovery phase do not necessarily result in a global phase.
	Hence, different paths used to physically approach a pair of anyons in order to fuse them might give rise to different fusion outcomes (more precisely, to different probability distributions for the measurement outcome of the total charge of the pair).
	When simulating the error-correction process, we must therefore be very careful in specifying and keeping track of the precise paths followed by the various anyons.
	
	Another implication is that, for non-Abelian models, the decoding process must happen in an iterative fashion.	
	Based on the initial positions and charges of the anyons, a recovery step that consists of a number of fusion processes is suggested. As the result of this recovery cannot be predicted, all fusion processes must be performed in the given order, after which the fusion outcomes must be measured.
	These measured outcomes then give a new error syndrome that serves as the basis for suggesting a next recovery step, consisting of a new series of fusion processes. This cycle continues until all anyons are fused away and decoding is successful, or until some anyon is wound along a nontrivial cycle during a recovery step resulting in a decoding failure. Error correction thus proceeds as a dialogue between syndrome extraction and decoder, where the new syndrome resulting from a given recovery step is used to determine the next recovery step. 
	This is in stark contrast to Abelian models, where a single syndrome measurement provides all the necessary information to determine all fusion processes that must be caried out in order to return the system to the code space.
	
\subsection{Classical simulability} \label{sec:classical_simulatbility}
	The ability to reliably simulate the general dynamics of Fibonacci anyons implies the ability to simulate (and hence perform) universal quantum computation. 
	It is therefore highly unlikely that such simulations are feasible on a classical computer.
	However, typical noise and recovery processes such as those that we will simulate in the remainder of this work exhibit structure that can be exploited to classically simulate them in regimes where we expect successful error correction to be possible.
	
	Individual local error operations either create a distinct connected group of anyons with vacuum total charge, or extend such an existing group (see Sec.~\ref{sec:Pauli-noise}). These groups can be understood as anyons that have interacted at some point during their lifetime.
	Since each disconnected group has a trivial total charge, braiding between separate groups is trivial.
	Hence, the total fusion space factorizes into a tensor product of fusion spaces of individual connected groups, and we are only required to simulate anyon dynamics within each of these groups separately.
	
	With regards to the creation of connected groups of anyons, the noise process behaves as a kind of percolation process.
	Hence, below the percolation threshold, one expects that the size of the largest connected group scales as $ O(\log(L)) $ (with variance $ O(1) $), where $ L $ is the linear system size \cite{bazant2000largest}.
	As this is a probabilistic statement, there will be instances where the largest connected group has a size larger than $ O(\log(L) $, but the probability of such events is suppressed exponentially with the system size $ L $. 
	This logarithmic scaling of the largest cluster size $ s = O(\log(L)) $ counters the exponential scaling of the dimension of the fusion space $ d = O(\exp(s)) $ for individual connected groups. 
	Therefore, the fusion spaces of individual connect groups will have dimension $  \text{dim} = O(\text{poly}(L)) $, meaning that the dynamics within connected groups can be simulated efficiently.
	
		These arguments on the classical simulability of topological quantum error correction with a universal anyon model where first made in Ref.~\cite{burton2017classical}. 
	The behavior of connected clusters of anyons created in the phenomenological model studied there corresponds exactly to the bond percolation model for which the logarithmic scaling mentioned above was verified numerically \cite{bazant2000largest}.
	To ensure that this still holds for the microscopic model studied here, we explicitly verified the logarithmic scaling of the average size of the largest connected group of anyons after subjecting all qubits to a depolarizing noise model, using Monte Carlo simulations.
	The results of these simulations are presented and discussed in App.~\ref{sec:scaling}.
	
	During the iterative decoding procedure, anyons are fused over increasing length scales, thereby potentially joining together connected groups of anyons through this interaction. As long as large connected groups are sparsely distributed (which is the case on average below the percolation threshold) this should not pose a problem, as the dimension of the fusion space is automatically reduced once the fusion process is actually performed, since this then reduces the number of anyons that must be simulated. We can conclude that we expect efficient simulation of the dynamics relevant to error correction to be possible in the regime where the combined action of noise and recovery does not percolate.
	
	If noise is strong enough, connected groups of anyons will percolate and the fusion space will no longer factorize into small disconnected parts. In this case classical simulation of braiding and fusion will become intractable. However, this is precisely the regime where we expect regular transport of anyons along nontrivial cycles during recovery, leading to failed error correction. Therefore, the error correction threshold itself will lie below this regime and estimating its value through classical simulations should be possible.
	
\subsection{Noise model}\label{sec:noise_model}
	Our goal is to stimulate the dynamics of a quantum-computing architecture of qubits, which directly implements our error-correcting code.
	We model noise in this system as individual (either depolarizing or dephasing) Pauli errors acting on random qubits, while the system is constantly being monitored\footnote{Note that under these assumptions, it is not appropriate to characterize the noise in terms of an i.i.d.~noise strength per qubit, since the non-Abelian nature of our code implies that the combined action of individual error and measurement operators do not commute for consecutive errors. We discuss the relation with i.i.d.~noise in App.~\ref{sec:relation_iid_noise}}.
	We treat the occurrence of Pauli errors on individual qubits as independent Poisson processes with a fixed rate $ p $, which characterizes the noise strength. 
	The total number of qubit errors, denoted by $ T $, then follows a Poisson distribution with mean $ T_0 = |E| p  $,  where $ |E| $ is the total number of edges.
	(For a tailed honeycomb lattice with periodic boundary conditions and a total of $ L^2 $ plaquettes, the total number of edges (and hence qubits) is $ |E| = 5L^2 $, resulting in $ T_0 = 5L^2 p$.)
	This fixed-rate sampling noise model is similar to those used for the simulation of non-Abelian quantum error correction with phenomenological models such as in Refs.~\cite{brell2014thermalization, burton2017classical}. 
	For simplicity, we assume that all measurements are perfect and are carried out on timescales which are negligible compared to the average time between individual qubit errors [$ \sim 1/( |E| p ) $].
		
	We simulate the dynamics of the system for $ T $ time steps, each of which corresponds to the occurrence of a single Pauli qubit error, and where $ T $ is drawn from a Poisson distribution with mean $ T_0 $.
	Each individual time step consists of the following: 
	\begin{enumerate}
		\item An edge $ e $ of the lattice is chosen at random and a Pauli operator $ \sigma_i $ is picked according to relative probabilities $\{ \gamma_x,\, \gamma_y,\, \gamma_z\} $. 
		We specifically use depolarizing noise ($ \gamma_x = \gamma_y = \gamma_z = 1/3 $), dephasing noise ($ \gamma_z = 1 $), and bit-flip noise ($ \gamma_x = 1 $).
		The operator $ \sigma_i $ is then applied on the qubit corresponding to edge $ e $.
		\item All vertex stabilizers $ Q_v $, and tail qubits are measured (in the $ Z $-basis). 
		\item The appropriate unitary operator $ U_V $, conditioned on the measurement outcome $ V $ from the measurements above, is applied to fix any violated vertices.
		\item The anyon charge in each plaquette is measured.		
	\end{enumerate}
	The unitary operators used to locally correct violated vertices were introduced in Sec~\ref{sec:error_correction}, their purpose is to move the system back to the string-net subspace $ \H_{\text{s.n.}} $. Inside this subspace states can be described in terms of anyonic fusion states, and plaquette anyon charges are well defined.
	Note that the measurement at the end of each time step means that we will never encounter any superpositions of different anyonic charges in individual plaquettes. (In terms of fusion states, this fixes all leaf labels.)
	
	After the $ T $ error operations have been applied, and if no logical error was induced by the noise process, the syndrome is fed to an iterative (classical) decoding algorithm, which returns a list of anyons to be fused and paths to be followed when doing so. We assume that all recovery operations and additional syndrome measurements are perfect and instantaneous, meaning no additional errors happen during the recovery process.\\
	
	While the connection between percolation and logical errors in topological codes is not exact \cite{hastings2014self} (i.e., not all percolation events cause logical errors), all logical errors are the result of events where a pair of anyons are fused along a non-contractible loop. 
	Our Monte Carlo simulations (detailed below in Sec.~\ref{sec:outline_of_simulation}) classify all such percolation events as failures, and therefore slightly overestimate the logical failure rates. 
	This provides a heuristic argument for the validity of our noise model. 
	The immediate collapse of superpositions in the anyon charge of individual plaquettes does not inhibit the occurrence of percolation processes. We therefore do not expect our assumption of constant syndrome measurement to significantly affect the obtained decoder failure rates.
	Furthermore, in the low noise strength limit, our noise model approximates an i.i.d.~noise model, as the random qubits affected by Pauli errors are unlikely to be in each others vicinity. The existence of an error correction threshold for our fixed-rate sampling noise model therefore implies the existence of a threshold for an i.i.d.~noise model as well.
	
\subsection{Pauli-noise in the anyonic fusion basis}\label{sec:Pauli-noise}
	We will now investigate the effect of the application of noise and subsequent measurement and vertex recovery operations performed in each of the $ T $ time steps of the simulation as described above.
	Because of the syndrome measurement performed at the end of every time step, the quantum state of the system between these steps can always be decomposed as a superposition of anyonic fusion basis states, which all share the same set of leaf labels.
	It is therefore sufficient to understand the action of these different operations on an initial state
	\begin{equation}\label{eq:initial_state_noise}
		\ket{\Psi_0} = \sum_t \alpha_t \ket{\psi_t},
	\end{equation}
	where the $ \ket{\psi_t} $ represent anyonic fusion basis states Eq.~\eqref{eq:anyonic_fusion_basis_lattice_short}, and the states $ \{ \ket{\psi_t} | \alpha_t \neq 0 \} $ all share the same set of leaf labels.
		
	\subsubsection{Pauli-Z errors}\label{sec:Pauli-Z}
	We begin our analysis by studying the case where we act with a $ \sigma_z $ operator on the qubit residing at edge $ e $.
	One can easily see that for any edge $ e $, $ \sigma_z^e $ commutes with all vertex operators  $ Q_v $ defined in Eq.~\eqref{eq:Qv_short}, since both operators are diagonal in the same basis. Hence, a $ \sigma_z $ error does not take the system out of the string-net subspace $ \H_{\text{s.n.}} $, and the resulting state can again be decomposed in the anyonic fusion basis:
	\begin{align}\label{eq:z_noise_res_state}
		\ket{\Psi_1} &= \sigma_z^e \ket{\Psi_0} = \sum_t \alpha_t \sigma_z^e \ket{\psi_t}\\
		& = \sum_{t'} \ket{\psi_{t'}} \left(\sum_t \alpha_t \braket{\psi_{t'}|\sigma_z^e|\psi_t} \right),
	\end{align}
	where we have used that $ \sum_{t'} \ket{\psi_{t'}} \bra{\psi_{t'}} $ acts as the resolution of the identity within $  \H_{\text{s.n.}} $. 
	In particular, if we start from a superposition of basis states $ \ket{\psi_t} $ that all have the same specific handle label (see Sec.~\ref{sec:anyonic_fusion_basis_short}), the states $ \ket{\psi_{t'}} $ in the resulting superposition will also have that same label, unless a logical error has occurred through the interaction of the thermal anyons due to the noise operator. Since the simulation of error correction is immediately aborted in the event of such a logical error, it is sufficient to only consider basis states that all have the same handle label as the initial state.
	
	When the edge $ e $ is a tail edge, acting with $ \sigma_z^e $ results in a global $ (\pm 1) $ factor, which has no physical consequences.
	On a general edge $ e $ different from a tail edge, the action of $  \sigma_z^e $ commutes with the irreducible idempotents of the tube algebra [Eqs.~\eqref{eq:P11_short} \eqref{eq:P12_short}, \eqref{eq:P21_short}, \eqref{eq:P221_short} and \eqref{eq:P222_short}] acting on any plaquette, except for the two that contain the edge $ e $.
	This means that a $ \sigma_z^e $ error on a general edge can modify the anyon charge of at most two plaquettes. Furthermore, the total charge of these two plaquettes cannot be changed by the action of $ \sigma_z^e $, since this is a collective property of the pair that is insensitive to the local operator $ \sigma^z_e $.
	
	With these insights in mind, we pick an anyonic fusion basis of the following form:
	\begin{equation}\label{eq:z_error_basis}
		\ket{\psi^{\vec{\bm{a}}}_{\bm{b}, \vec{\bm{c}}}} = \raisebox{-.2cm}{\includegraphics[scale=.4]{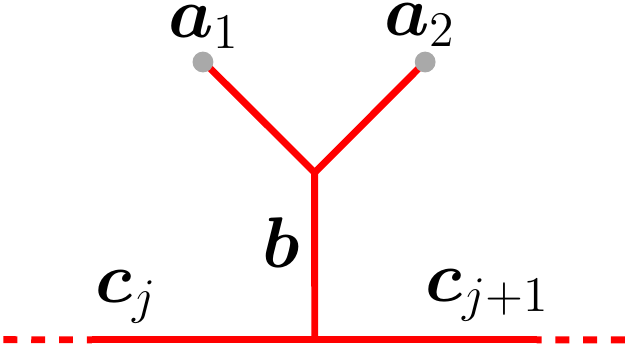}}\;,
	\end{equation}
	where $ \bm{a}_1 $ and $ \bm{a}_2 $ are the charges of the two affected plaquettes, $ \bm{b} $ denotes their total charge, and $\vec{\bm{c}} $ collectively denotes all other leaf, branch and handle labels.
	We can then rewrite the initial state Eq.~\eqref{eq:initial_state_noise} as
	\begin{equation}\label{eq:initial_state_z}
		\ket{\Psi_0} = \sum_{ \bm{b},\vec{\bm{c}}} \alpha_{\bm{b},\vec{\bm{c}}} \ket{\psi^{\vec{\bm{a}}}_{\bm{b},\vec{\bm{c}} }} .
	\end{equation}
	Note that we dropped the summation over $ \vec{\bm{a}} $, which is allowed because the initial state contains no superposition in plaquette charges.
	Since the labels  $ \bm{b} $ and $\vec{\bm{c}} $ are unaffected by $ \sigma_z^e $, the matrix elements appearing on the right hand side of Eq.~\eqref{eq:z_noise_res_state} are block diagonal in these labels, and the expression for the state $\ket{\Psi_1} = \sigma_z \ket{\Psi_0} $ can be reduced to
	\begin{equation}\label{eq:z_noise_res_state_blocks}
		\ket{\Psi_1} = \sum_{ \bm{b},\vec{\bm{c}}} \sum_{\vec{\bm{a}}'} \ket{\psi^{\vec{\bm{a}}'}_{\bm{b}, \vec{\bm{c}}}} \left(\alpha_{\bm{b}, \vec{\bm{c}}} \braket{\psi^{\vec{\bm{a}}'}_{\bm{b}, \vec{\bm{c}}}|\sigma_z^e|\psi^{\vec{\bm{a}}}_{\bm{b}, \vec{\bm{c}}}} \right).
	\end{equation}
	The probability of finding outcome $ \vec{\bm{a}}' = (\bm{a}_1', \bm{a}_2') $ when performing a syndrome measurement in the two affected plaquettes is then given by
	\begin{equation}\label{eq:z_noise_charge_prob}
		p(\vec{\bm{a}}') = \sum_{\bm{b}, \vec{\bm{c}}}  \left|\alpha_{\bm{b}, \vec{\bm{c}}} \braket{\psi^{\vec{\bm{a}}'}_{\bm{b}, \vec{\bm{c}}}|\sigma_z^e|\psi^{\vec{\bm{a}}}_{\bm{b}, \vec{\bm{c}}}} \right|^2.
	\end{equation}
	After performing this measurement, the state of the system collapses to
	\begin{equation}\label{eq:z_noise_final_state}
		\ket{\Psi_2} = \frac{1}{\sqrt{p(\vec{\bm{a}}')}} \sum_{\bm{b}, \vec{\bm{c}}} \ket{\psi^{\vec{\bm{a}}'}_{\bm{b}, \vec{\bm{c}}}} \left(\alpha_{\bm{b}, \vec{\bm{c}}} \braket{\psi^{\vec{\bm{a}}'}_{\bm{b}, \vec{\bm{c}}}|\sigma_z^e|\psi^{\vec{\bm{a}}}_{\bm{b}, \vec{\bm{c}}}} \right),
	\end{equation}
	which is again a superposition of basis states with fixed leaf labels. 
	Note that the matrix elements on the right hand side of Eqs.~\eqref{eq:z_noise_charge_prob} and \eqref{eq:z_noise_final_state} are independent of the unaffected labels $\vec{\bm{c}}$. Hence, we may replace then with matrix elements of the form $ \braket{\psi^{\vec{\bm{a}}'}_{\bm{b}}|\sigma_z^e|\psi^{\vec{\bm{a}}}_{\bm{b}}} $, where $ \ket{\psi^{\vec{\bm{a}}}_{\bm{b}}} $ are states that only contain nontrivial anyon labels for the two affected plaquettes and for their total charge. 
	Also note that since the total charge of the affected plaquettes is a collective property, the precise location of the third plaquette which contains this total charge does not affect the value of the matrix elements, and furthermore, the tail label of that plaquette does not affect the matrix elements.

\begin{figure}[h]
		\centering
		\begin{subfigure}[b]{0.49\columnwidth}
			\centering
			\includegraphics[scale=.31]{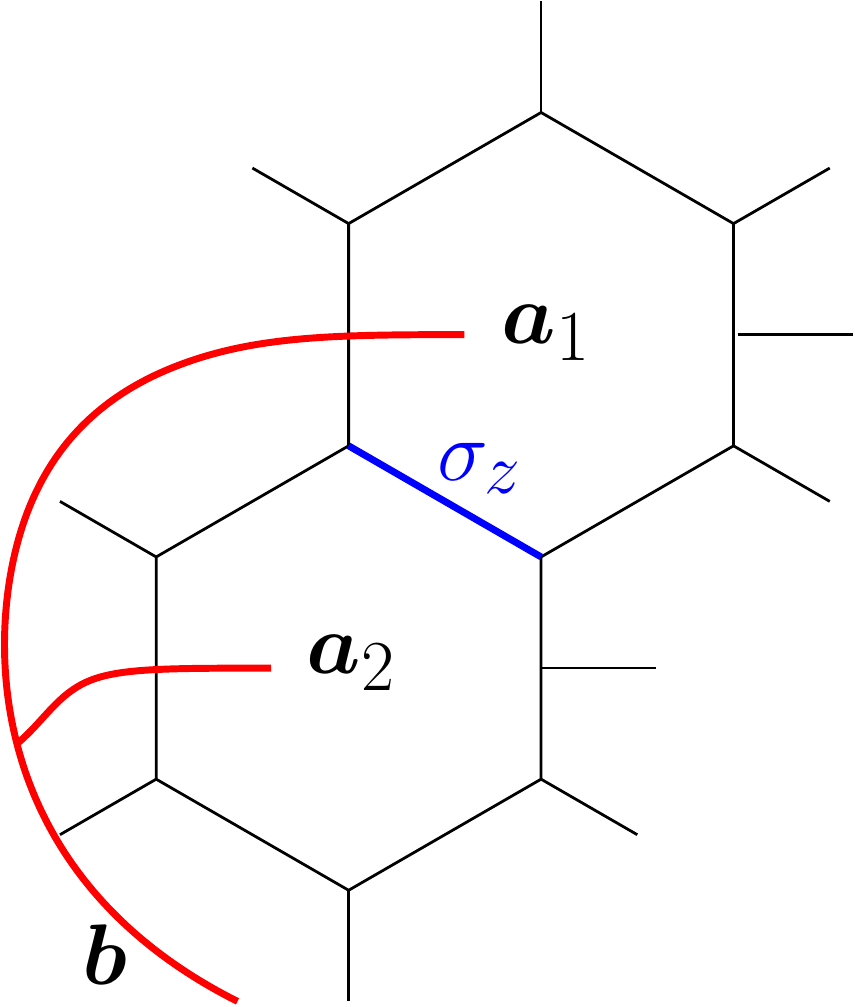}
			\caption{} 
			\label{fig:latticez1}
		\end{subfigure}
		\hfill
		\begin{subfigure}[b]{0.49\columnwidth}
			\centering
			\includegraphics[scale=.31]{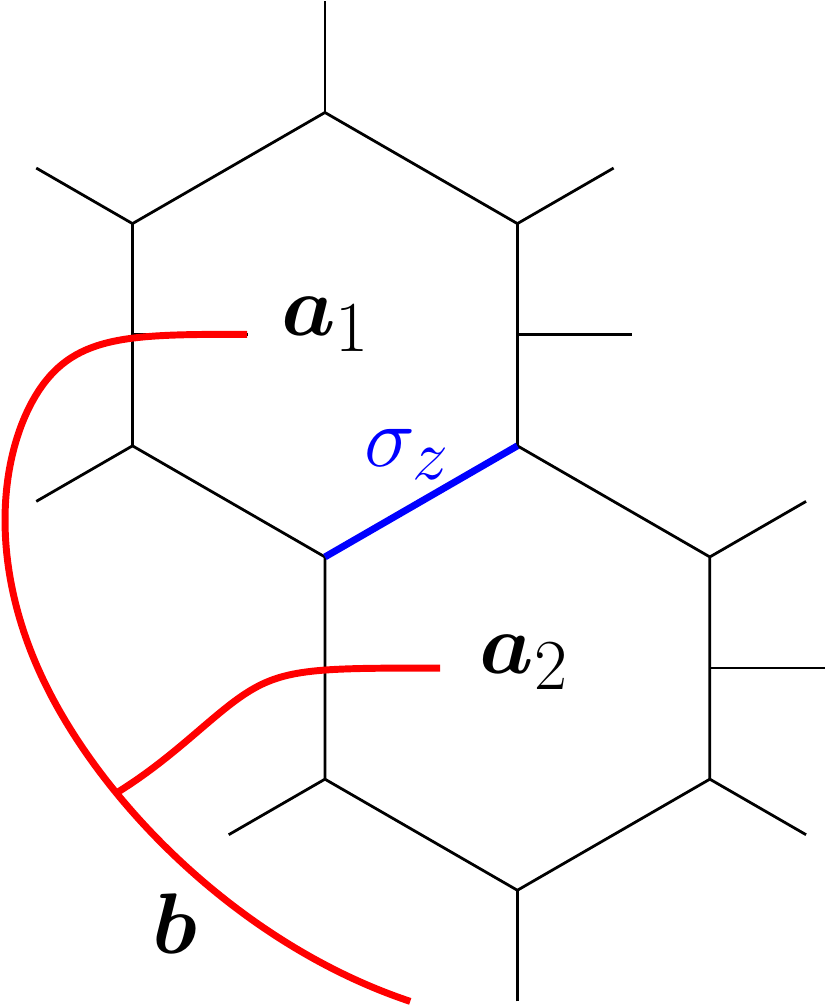}
			\caption{} 
			\label{fig:latticez2}
		\end{subfigure}
		\begin{subfigure}[b]{0.49\columnwidth}
			\centering
			\includegraphics[scale=.31]{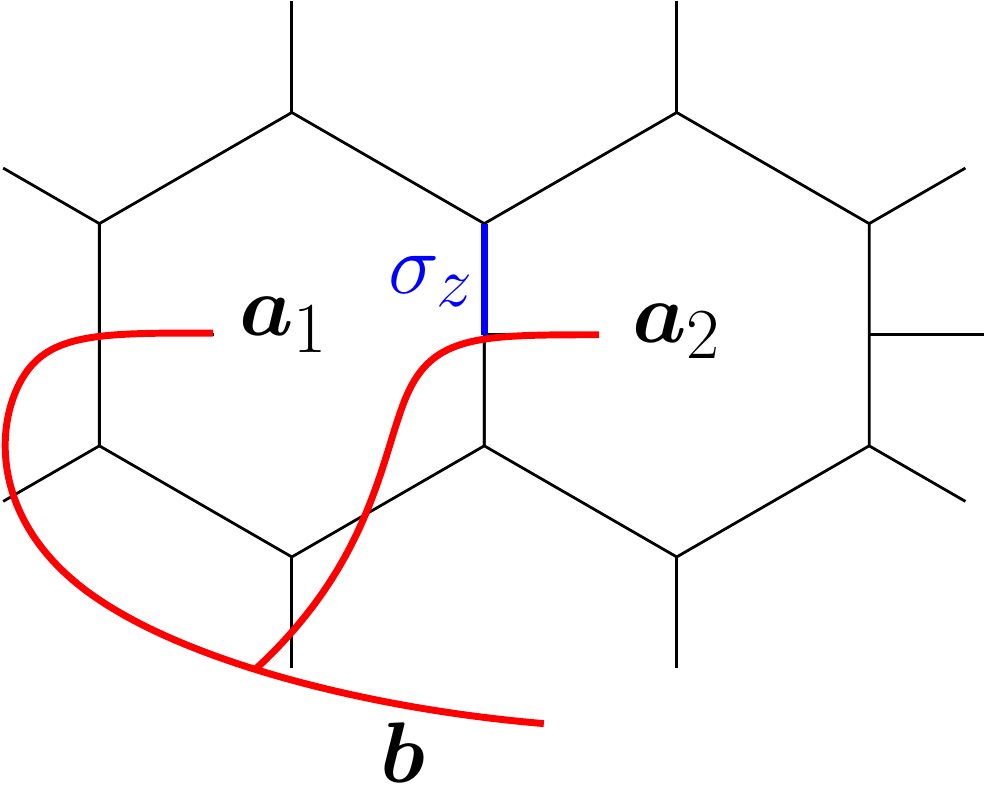}
			\caption{} 
			\label{fig:latticez3}
		\end{subfigure}
		\begin{subfigure}[b]{0.49\columnwidth}
			\centering
			\includegraphics[scale=.31]{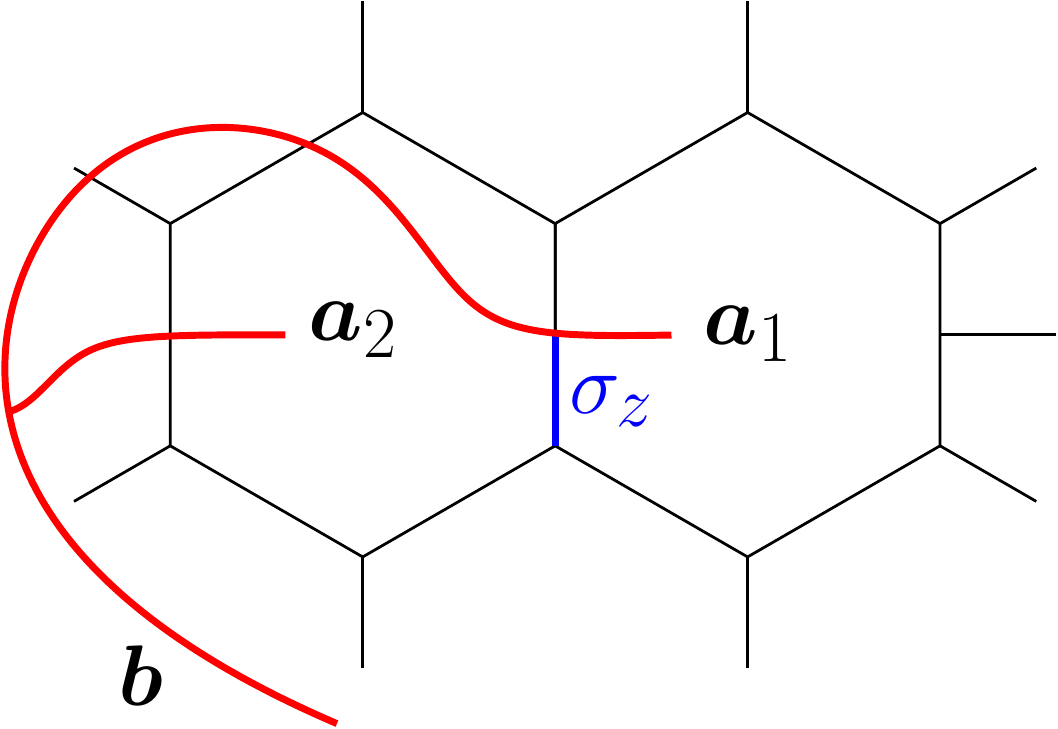}
			\caption{} 
			\label{fig:latticez4}
		\end{subfigure}
		\caption{Basis convention for the affected plaquette charges for all nontrivial orientations of the edge $ e $ (highlighted in blue) in the case of a $\sigma_z$ error. }
		\label{fig:z_error_basis_convention}
	\end{figure}

    When calculating the matrix elements, we must pick some convention for the basis elements  $ \ket{\psi^{\vec{\bm{a}}}_{\bm{b}}} $ [i.e., for the embedding of the fusion tree in Eq.~\eqref{eq:z_error_basis} in the lattice], for each of the possible orientations of $ e $. Since $ \sigma_z^e $ has no physical effect when $ e $ is a tail edge, we only need to consider four orientations for $ e $. 
	Our choice of basis convention for the $ \sigma^z_e $ matrix elements is given in \figref{fig:z_error_basis_convention}. 
	
	\subsubsection{Pauli-X and Y errors}
	The case of a $ \sigma_x^e $ error is similar, but comes with some additional complications. 
	Since a $ \sigma_x^e $ operator does not commute with the vertex operators $ Q_v $ associated to the vertices bounding $ e $, the state
	\begin{equation}\label{eq:x_noise_res_state}
		\ket{\Psi_1} = \sum_t \alpha_t \sigma_x^e \ket{\psi_t}
	\end{equation} 
	does not necessarily belong to the sting-net subspace $ \H_{\text{s.n.}} $ and hence cannot be expressed as a superposition of anyonic fusion basis states.
	In particular, upon measuring the vertex stabilizers for the vertices that bound $ e $, one might find that the ribbon graph branching rules are violated in either of these vertices.
	Each violated vertex will belong to a set of 7 connected qubits on the lattice that we will call a \emph{segment}:
	\begin{equation*} 
		\raisebox{-0.55cm}{\includegraphics[scale=.4]{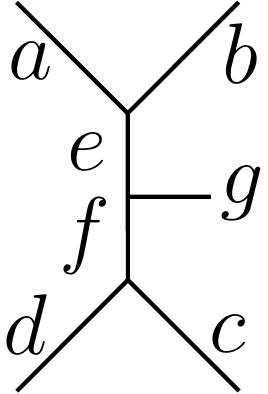}} \;
	\end{equation*}
	The three vertices in each segment will be denoted as $ t $, $ m $ and $ b $, corresponding to the top, middle and bottom vertex, respectively. 
	For each segment with some violated vertices, we also measure the label of the tail qubit (corresponding to edge $ g $ in the diagram above). 
	Depending on these measurement outcomes $ V $, a unitary operator $ U_V $ is applied to the segment in order to take it back to the +1 eigenstate of the three vertex operators $ Q_{t} $, $ Q_{m} $ and $ Q_{b} $ associated to the segment. 
	The definitions and corresponding circuits for these unitaries (conditioned on all possible measurement outcomes) are given in Sec.~\ref{sec:vertex_measurement}.
	
	The probability $ p(V) $ of a combined outcome $ V $ for the vertex and tail qubit measurements in the relevant segments is given by
	\begin{align}\label{eq:x_noise_vert_prob}
		p(V) &= \braket{\Psi_1|P_V|\Psi_1} \nonumber \\
		 &= \sum_{t',t} \bar{\alpha}_{t'} \alpha_t \braket{\psi_{t'}|\sigma_x^e P_V \sigma_x^e |\psi_t},
	\end{align}
	where $ P_V $ is the projector onto the total measurement outcome $ V $. For example, for the case of a $ \sigma_x^e $ acting on the edge $ e $ bounded by vertices $ v_1 $ and $ v_2 $ where only $ v_1 $ is violated and the tail qubit $ q_1 $ of the segment containing $ v_1 $ has label $ \tau $, the associated projector $ P_V $ is given by
	\begin{equation}\label{eq:x_noise_vert_proj_example}
		P_V = (1 - Q_{v_1}) (Q_{v_2}) (\ket{\tau}_{q_1} \bra{\tau}_{q_1})\,.
	\end{equation}
	After performing the vertex and tail qubit measurements with outcome $ V $ and applying the appropriate unitary operator $ U_V $ to bring the system back to the string-net subspace, the state is given by
	\begin{align}\label{eq:x_noise_res_state_post_vert}
		\ket{\Psi_2} &= \frac{1}{\sqrt{p(V)}} \sum_{t} \alpha_t U_V P_V \sigma_x^e \ket{\psi_t} \nonumber \\
		& = 
		\frac{1}{\sqrt{p(V)}} \sum_{t'} \ket{\psi_{t'}} \left(\sum_t \alpha_t \braket{\psi_{t'}| U_V P_V \sigma_x^e |\psi_t}\right),
	\end{align}
	where we have again inserted the resolution of the identity $ \sum_{t'} \ket{\psi_{t'}} \bra{\psi_{t'}} $ in $ \H_{\text{s.n.}} $ on the right hand side in order to explicitly express the state as a superposition of anyonic fusion basis states.
	
	As in the case of a $ \sigma^z_e $ error, the expressions \eqref{eq:x_noise_vert_prob} and \eqref{eq:x_noise_res_state_post_vert} can be simplified by considering which plaquette charges are actually affected by the combined action of the operators $ U_V $, $ P_V $ and $ \sigma_x^e $. Since a $ \sigma_x $ operator acting on an edge $ e $ of the tailed lattice results in at most two violated vertices, the combined action of $ U_V $, $ P_V $ and $ \sigma_x^e $ involves at most two segments. These operators therefore commute with any tube operators acting on plaquettes that have no edges in common with these segments. This means that only the charges associated to the plaquettes in the immediate neighborhood of the error can be affected. The number of affected plaquettes depends on the orientation of the edge $ e $. 
	
	As before, we pick our anyonic fusion basis to reflect this fact. In case of 4 affected plaquettes, the basis states have the form
	\begin{equation}\label{eq:x_error_basis}
		\ket{\psi^{\vec{\bm{a}}, \vec{\bm{d}}}_{\bm{b}, \vec{\bm{c}}}} = \raisebox{-.2cm}{\includegraphics[scale=.4]{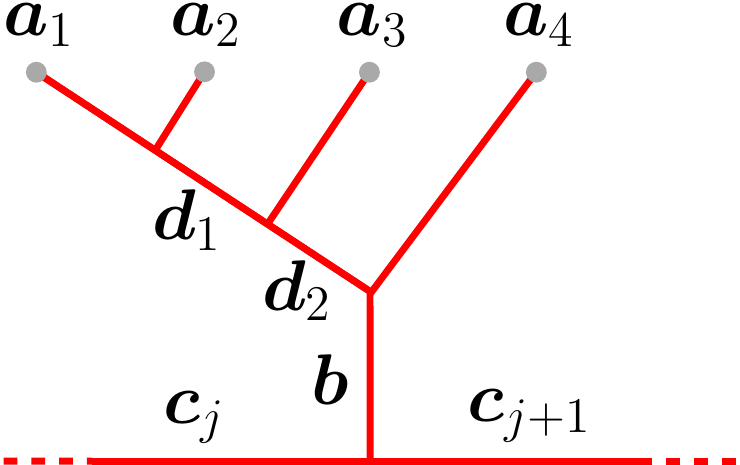}}\;,
	\end{equation}
	where, again, $ \vec{\bm{a}} $ denotes the charges of the affected plaquettes, $ \bm{b} $ denotes their total charge, and $\vec{\bm{c}} $ collectively denotes all other unaffected leaf, branch and handle labels. Note that we now need additional labels $ \vec{\bm{d}} $ to denote the affected internal branch labels.
	With this choice of basis, the sum over $ t $ in 
	Eq.~\eqref{eq:initial_state_noise} is split into a sum over the unaffected labels $ \vec{\bm{c}} $, the total charge $ \bm{b} $, and the branch labels $ \vec{\bm{d}} $ of the affected part:
	\begin{equation}\label{eq:initial_state_x}
		\ket{\Psi_0} = \sum_{\bm{b}, \vec{\bm{c}}, \vec{\bm{d}}} \alpha^{\vec{\bm{d}}}_{\bm{b}, \vec{\bm{c}}} \, \ket{\psi^{\vec{\bm{a}}, \vec{\bm{d}}}_{\bm{b}, \vec{\bm{c}}}}.
	\end{equation}
	
	As all matrix elements are block diagonal in the unaffected labels, the expression for the vertex measurement outcome probability Eq.~\eqref{eq:x_noise_vert_prob} reduces to
	\begin{equation}\label{eq:x_noise_vert_prob_red}
		p(V) = \sum_{\bm{b},\vec{\bm{c}}, \vec{\bm{d}}} \sum_{\vec{\bm{d}}'}  \bar{\alpha}^{\vec{\bm{d}}'}_{\bm{b}, \vec{\bm{c}}} \, \alpha^{\vec{\bm{d}}}_{\bm{b}, \vec{\bm{c}}} 
		\braket{\psi^{\vec{\bm{a}}, \vec{\bm{d}}'}_{\bm{b}, \vec{\bm{c}}} |\sigma_x^e P_V \sigma_x^e|
		\psi^{\vec{\bm{a}}, \vec{\bm{d}}}_{\bm{b}, \vec{\bm{c}}}}.
	\end{equation}
	The sum over $ t' $ in Eq.~\eqref{eq:x_noise_res_state_post_vert} can again be split into a sum over the unaffected labels $ \vec{\bm{c}}' $, the affected leaf labels $ \vec{\bm{a}}' $ and the branch labels $ \vec{\bm{d}}' $ of the affected part, reducing the expression to
	\begin{widetext}
	\begin{equation}\label{eq:x_noise_res_state_post_vert_red}
		\ket{\Psi_2} = 
		\frac{1}{\sqrt{p(V)}} \sum_{\vec{\bm{a}}', \vec{\bm{d}}'}  \sum_{\bm{b},\vec{\bm{c}}} \, 
		\ket{\psi^{\vec{\bm{a}}', \vec{\bm{d}}'}_{\bm{b}, \vec{\bm{c}}}} \left(\sum_{\vec{\bm{d}}} \alpha^{\vec{\bm{d}}}_{\bm{b}, \vec{\bm{c}}}  \braket{\psi^{\vec{\bm{a}}', \vec{\bm{d}}'}_{\bm{b}, \vec{\bm{c}}}
		| U_V P_V \sigma_x^e |
		\psi^{\vec{\bm{a}}, \vec{\bm{d}}}_{\bm{b}, \vec{\bm{c}}}}\right) .
	\end{equation}
	Measurement of the affected plaquette charges will then yield charges $ \vec{\bm{a}}' $ with a probability $ p(\vec{\bm{a}}') $ given by
	\begin{equation}\label{eq:x_noise_charge_prob}
		p(\vec{\bm{a}}') = \frac{1}{p(V)} \sum_{\vec{\bm{d}}'}  \sum_{\bm{b},\vec{\bm{c}}} 
		\left| \sum_{\vec{\bm{d}}} \alpha^{\vec{\bm{d}}}_{\bm{b}, \vec{\bm{c}}}  \braket{\psi^{\vec{\bm{a}}', \vec{\bm{d}}'}_{\bm{b}, \vec{\bm{c}}}|U_V P_V \sigma_x^e|
		\psi^{\vec{\bm{a}}, \vec{\bm{d}}}_{\bm{b}, \vec{\bm{c}}}} \right|^2 .
	\end{equation}
	The resulting state after this charge measurement is
	\begin{equation}\label{eq:x_noise_final_state}
		\ket{\Psi_3} = 
		\frac{1}{\sqrt{p(\vec{\bm{a}}')p(V)}} 
		\sum_{\vec{\bm{d}}'}  \sum_{\bm{b},\vec{\bm{c}}}
		\ket{\psi^{\vec{\bm{a}}', \vec{\bm{d}}'}_{\bm{b}, \vec{\bm{c}}}}
		\left( \sum_{\vec{\bm{d}}} \alpha^{\vec{\bm{d}}}_{\bm{b}, \vec{\bm{c}}}  
		\braket{ \psi^{\vec{\bm{a}}', \vec{\bm{d'}}}_{\bm{b}, \vec{\bm{c}}}
		| U_V P_V \sigma_x^e |
		\psi^{\vec{\bm{a}}, \vec{\bm{d}}}_{\bm{b}, \vec{\bm{c}}}} \right) .
	\end{equation}
	\end{widetext}
	Once again, the matrix elements on the right hand side of Eqs.~\eqref{eq:x_noise_vert_prob_red}, \eqref{eq:x_noise_charge_prob} and \eqref{eq:x_noise_final_state} are independent of the unaffected labels $ \vec{\bm{c}} $, so we only require matrix elements of the form $ \braket{\psi^{\vec{\bm{a}}', \vec{\bm{d}}'}_{\bm{b}}| O |\psi^{\vec{\bm{a}}, \vec{\bm{d}}}_{\vec{\bm{b}}}} $ for fusion basis states involving up to four plaquettes and their total charge.
	
\begin{figure}[ht]
		\centering
		\begin{subfigure}[b]{0.5\columnwidth}
			\centering
			\includegraphics[scale=.31]{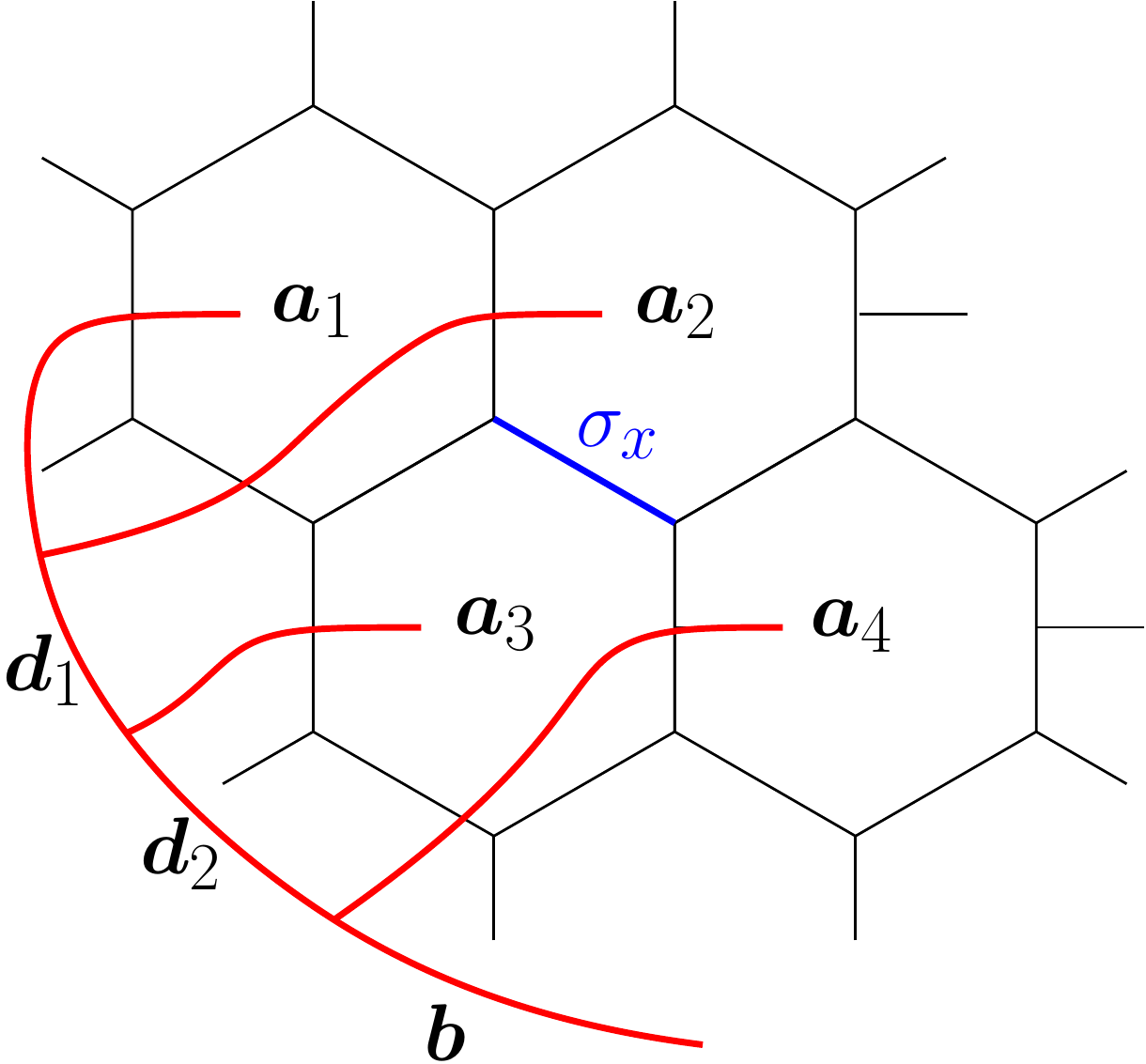}
			\caption{} 
			\label{fig:latticex1}
		\end{subfigure}
		\begin{subfigure}[b]{0.49\columnwidth}
			\centering
			\includegraphics[scale=.31]{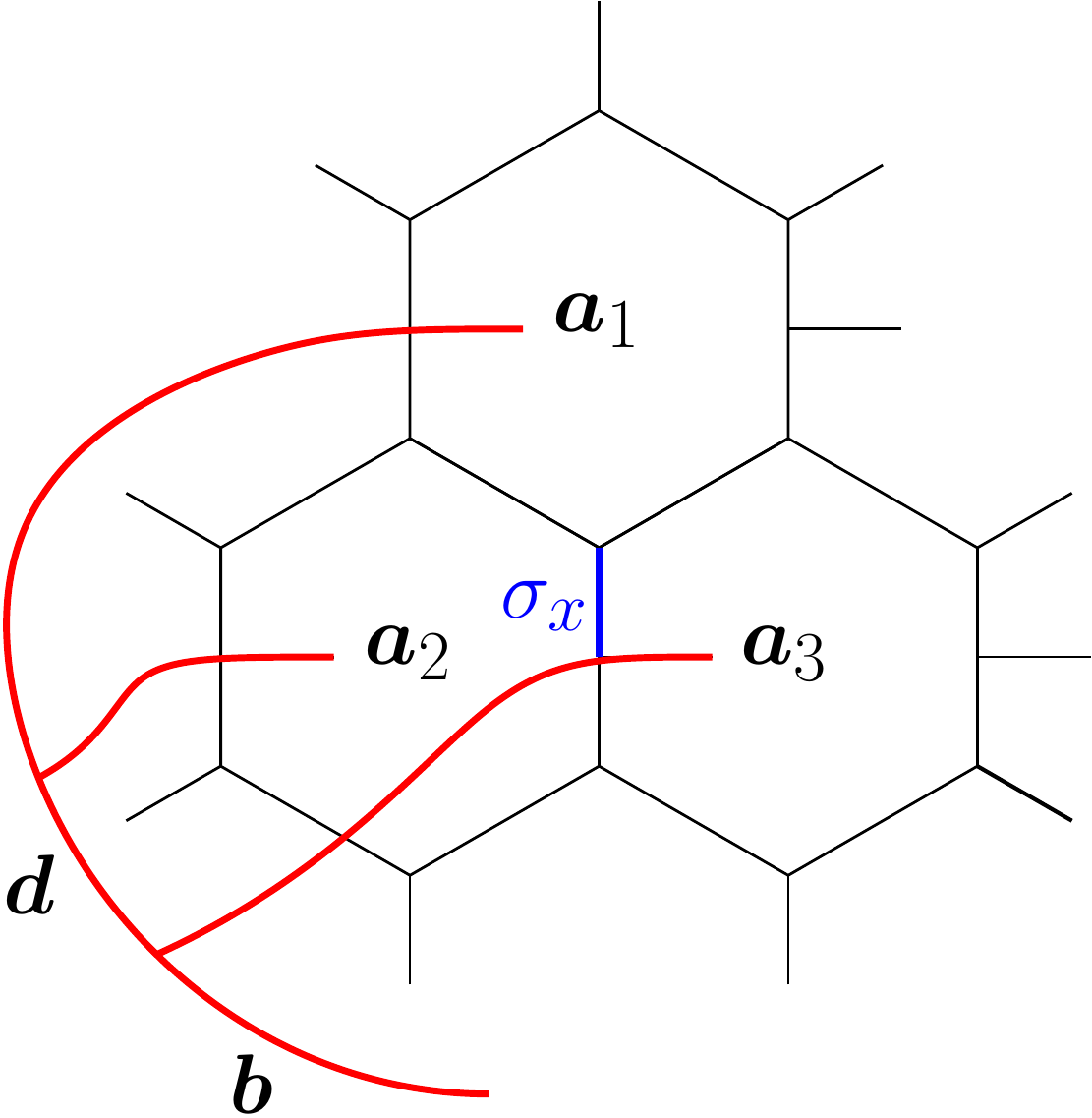}
			\caption{} 
			\label{fig:latticex3}
		\end{subfigure}
		\begin{subfigure}[b]{0.5\columnwidth}
			\centering
			\includegraphics[scale=.3]{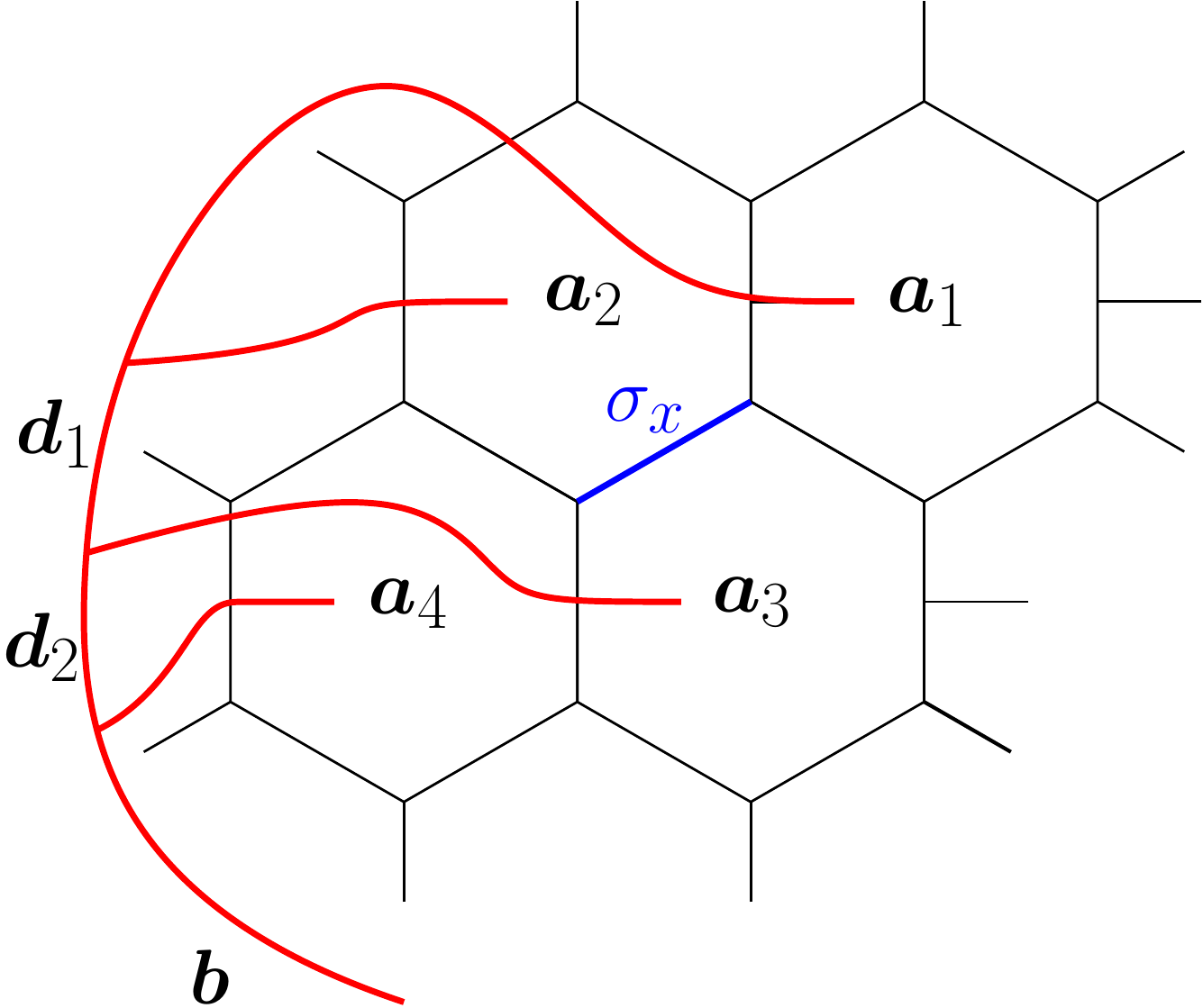}
			\caption{} 
			\label{fig:latticex2}
		\end{subfigure}
		\begin{subfigure}[b]{0.49\columnwidth}
			\centering
			\includegraphics[scale=.31]{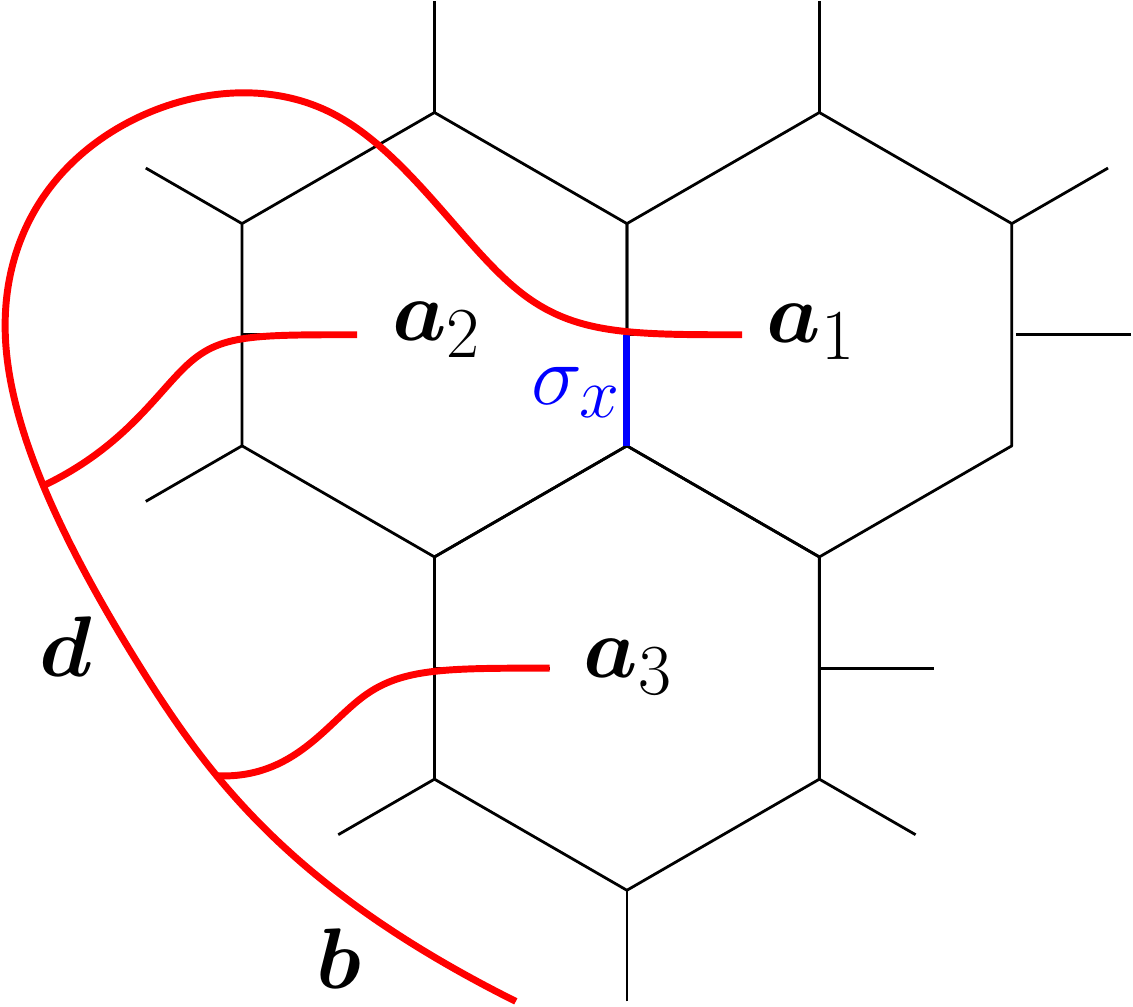}
			\caption{} 
			\label{fig:latticex4}
		\end{subfigure}
		\begin{subfigure}[b]{0.49\columnwidth}
			\centering
			\includegraphics[scale=.31]{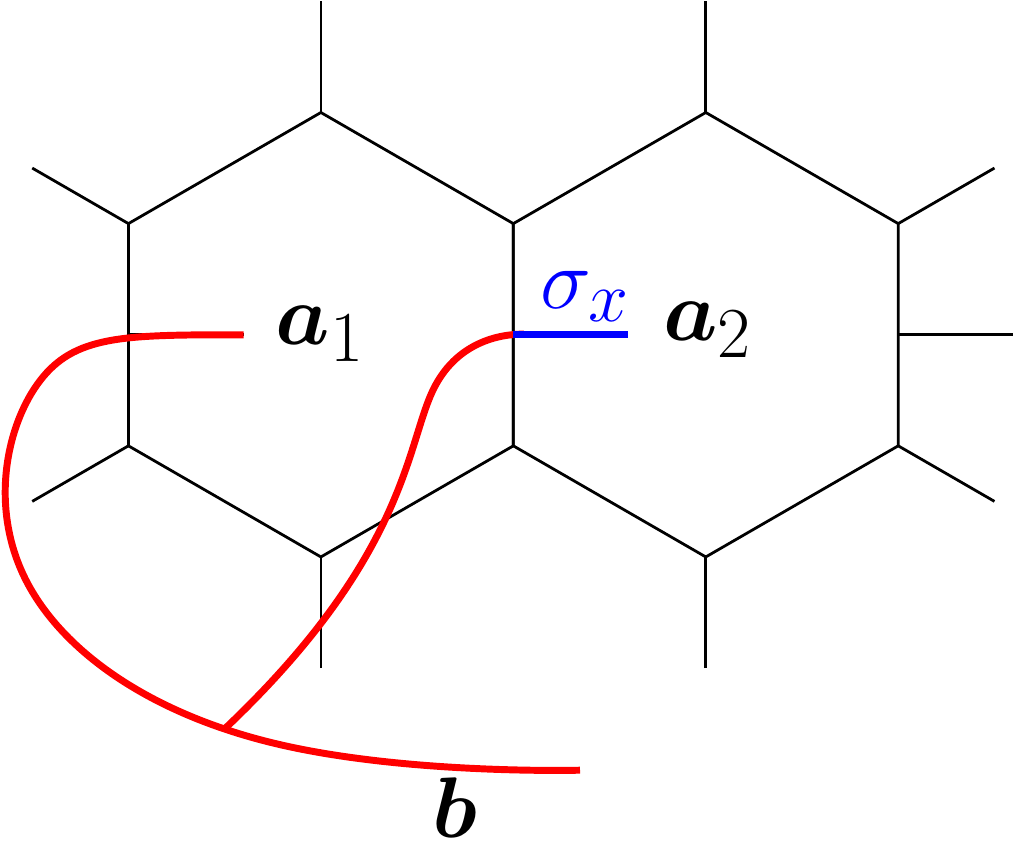}
			\caption{} 
			\label{fig:latticex5}
		\end{subfigure}
		\caption{Basis convention for the affected plaquette charges for the different orientations of $ e $ (highlighted in blue) in the case of a $\sigma_x$ or a $\sigma_y$ error.}
		\label{fig:x_error_basis_convention}
	\end{figure}

	Our choice of basis convention for the charges of the affected plaquettes for all possible orientations of $ e $ is given in \figref{fig:x_error_basis_convention}. In the case of a $ \sigma_x $ error all five possible orientations of $ e $ are nontrivial. 
	By considering the potentially violated vertices and the definition of the associated unitaries given in Sec.~\ref{sec:vertex_measurement} for each orientation, it can easily be verified that the depicted plaquettes are indeed the only affected ones and the combined action of the error, measurements and unitaries commutes with all other tube operators.
	
	The case of a $ \sigma_y $ error is entirely analogous to that of a $ \sigma_x $. The same operators $ P_V $ and $ U_V $ appear, and we use the same basis convention as the one depicted in \figref{fig:x_error_basis_convention}.\\
	
	In summary, all that is required to capture the effect of Pauli-noise on the state of the system are the following matrix elements:
	\begin{gather}
		\braket{\psi^{\vec{\bm{a}}, \vec{\bm{d}}'}_{\bm{b}} | \sigma_x^e P_V \sigma_x^e|\psi^{\vec{\bm{a}}, \vec{\bm{d}}}_{\vec{\bm{b}}}} , \label{eq:noise_mat_el_1} \\
		\braket{\psi^{\vec{\bm{a}}', \vec{\bm{d}}'}_{\bm{b}} | U_V P_V \sigma_x^e |\psi^{\vec{\bm{a}}, \vec{\bm{d}}}_{\vec{\bm{b}}}} , \label{eq:noise_mat_el_2} \\
		\braket{\psi^{\vec{\bm{a}}, \vec{\bm{d}}'}_{\bm{b}} | \sigma_y^e P_V \sigma_y^e|\psi^{\vec{\bm{a}}, \vec{\bm{d}}}_{\vec{\bm{b}}}} , \label{eq:noise_mat_el_3}\\
		\braket{\psi^{\vec{\bm{a}}', \vec{\bm{d}}'}_{\bm{b}} | U_V P_V \sigma_y^e |\psi^{\vec{\bm{a}}, \vec{\bm{d}}}_{\vec{\bm{b}}}} , \label{eq:noise_mat_el_4}\\
		\braket{\psi^{\vec{\bm{a}}'}_{\bm{b}}|\sigma_z^e|\psi^{\vec{\bm{a}}}_{\bm{b}}} .
		\label{eq:noise_mat_el_5}
	\end{gather}
	These matrix elements must be calculated for all possible orientations of the edge $ e $ in Figs.~\ref{fig:z_error_basis_convention} and \ref{fig:x_error_basis_convention}, and for all possible combined outcomes $ V $ of the vertex and tail qubit measurements in the case of a $ \sigma_x $ or $ \sigma_y $ error.
	
\subsection{Computing the relevant matrix elements} \label{sec:matrix_elements}
	At first sight, calculating the matrix elements listed in Eqs.~\eqref{eq:noise_mat_el_1}-\eqref{eq:noise_mat_el_5} seems like an intractable job, due to the highly entangled nature of the anyonic fusion basis states on the tailed lattice.
	Fortunately, as detailed in App.~\ref{sec:TN}, this complex entanglement structure can be captured using tensor network representations for anyonic fusion basis states.
	By using some key insights together with tensor network techniques, we can harness the computational power of tensor networks to compute said matrix elements. 
	Below, we will illustrate this procedure for the matrix element Eq.~\eqref{eq:noise_mat_el_1}, where $ e $ is an edge with the orientation depicted in \figref{fig:latticex1}.
	All other matrix elements can be computed in an analogous manner.
	
	For simplicity, we will work with square PEPS tensors (see App.~\ref{sec:square_PEPS}), which each correspond to one segment of the lattice as shown in the following diagram:
	\vspace{.5cm}
	\begin{center}
		\includegraphics[scale = .6]{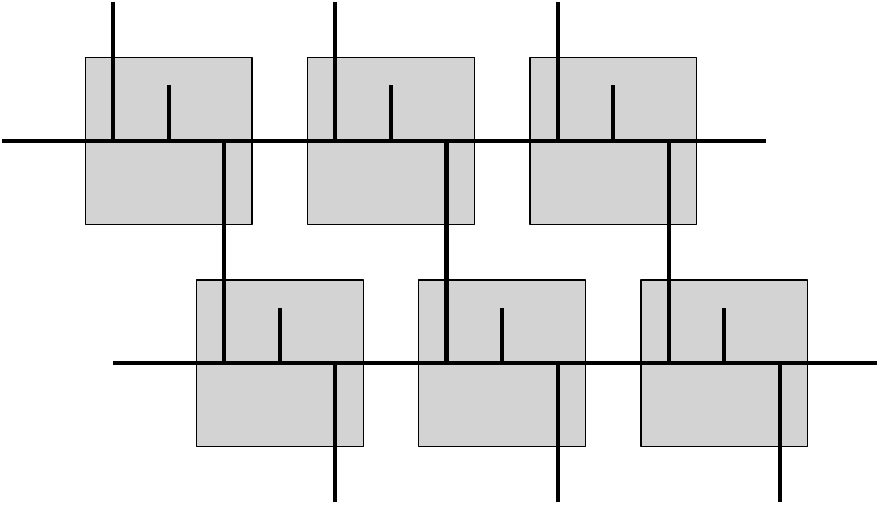}
	\end{center}
	where we have rotated the lattice by $ 60\degrees $ counterclockwise.
	The PEPS tensors will be colored gray and red for segments of which the tails end inside plaquettes containing trivial and nontrivial DFIB charges, respectively.
	With this notation (and after a counterclockwise rotation by $ 60\degrees $), the anyonic fusion state represented in \figref{fig:latticex1} looks as follows:
	\begin{equation}\label{eq:PEPS_squared_basis_state}
		\ket{\psi^{\vec{\bm{a}}, \vec{\bm{d}}}_{\bm{b}}} = \ \raisebox{-1.9cm}{\includegraphics[scale=.4]{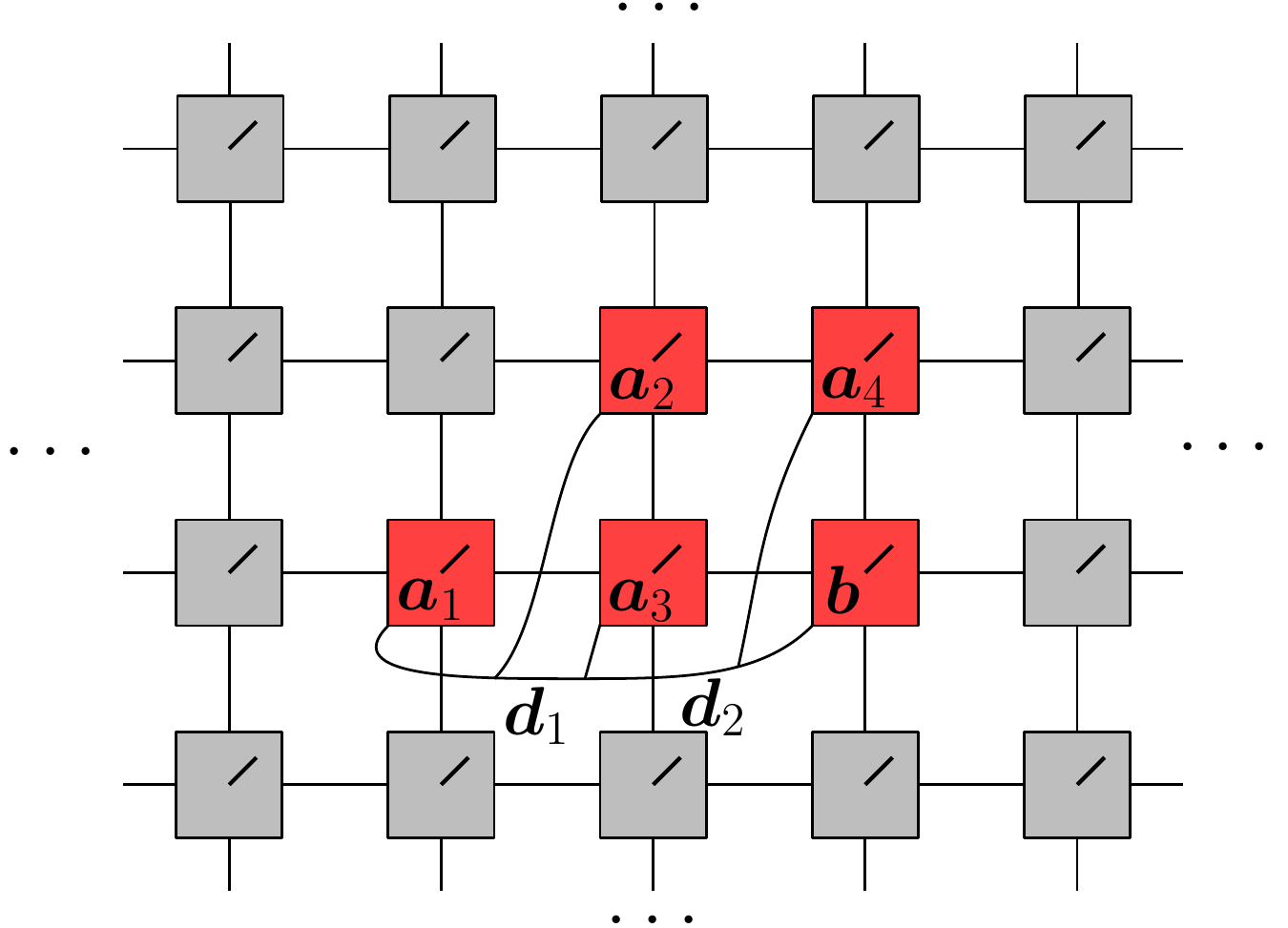}}\;.
	\end{equation}
	The crossings and branchings in this diagram imply the presence of certain crossing and fusion tensors. Details of these are given in App.~\ref{sec:TN}.
	The total charge $ \bm{b} $ of the four leaf charges in \figref{fig:latticex1} was assigned to a neighboring segment. 
	As was mentioned before, since the total charge of the group of anyons is a collective property, the precise location of the excitation tensor encoding this total charge does not affect the computation of the matrix elements itself. Hence, we choose to place it next to the other charges for convenience.
	
	The matrix elements can then be computed by applying the appropriate operators on the physical indices, and then contracting the result with the conjugate PEPS corresponding to the bra vector $ \bra{\psi^{\vec{\bm{a}}', \vec{\bm{d}}'}_{\vec{\bm{b}}}} $. 
	In this specific case, the operator $ P_V $ that projects out the combined vertex and tail qubit measurement outcome $ V $ can be decomposed into two operators $ P_A $ and $ P_B $ that act on the segments of leaf charges $ \bm{a}_2 $ and $ \bm{a}_4 $, and project out the measurement outcomes for the vertices and tail edge in these segments, respectively. 
	The matrix element is then given by the contraction
	\begin{widetext}
	\begin{equation}\label{eq:noise_mat_el_contraction1}
		\braket{\psi^{\vec{\bm{a}}', \vec{\bm{d}}'}_{\bm{b}}|\sigma_x^e P_V \sigma_x^e|\psi^{\vec{\bm{a}}, \vec{\bm{d}}}_{\bm{b}}} = \;
		\raisebox{-1.9cm}{\includegraphics[scale=.4]{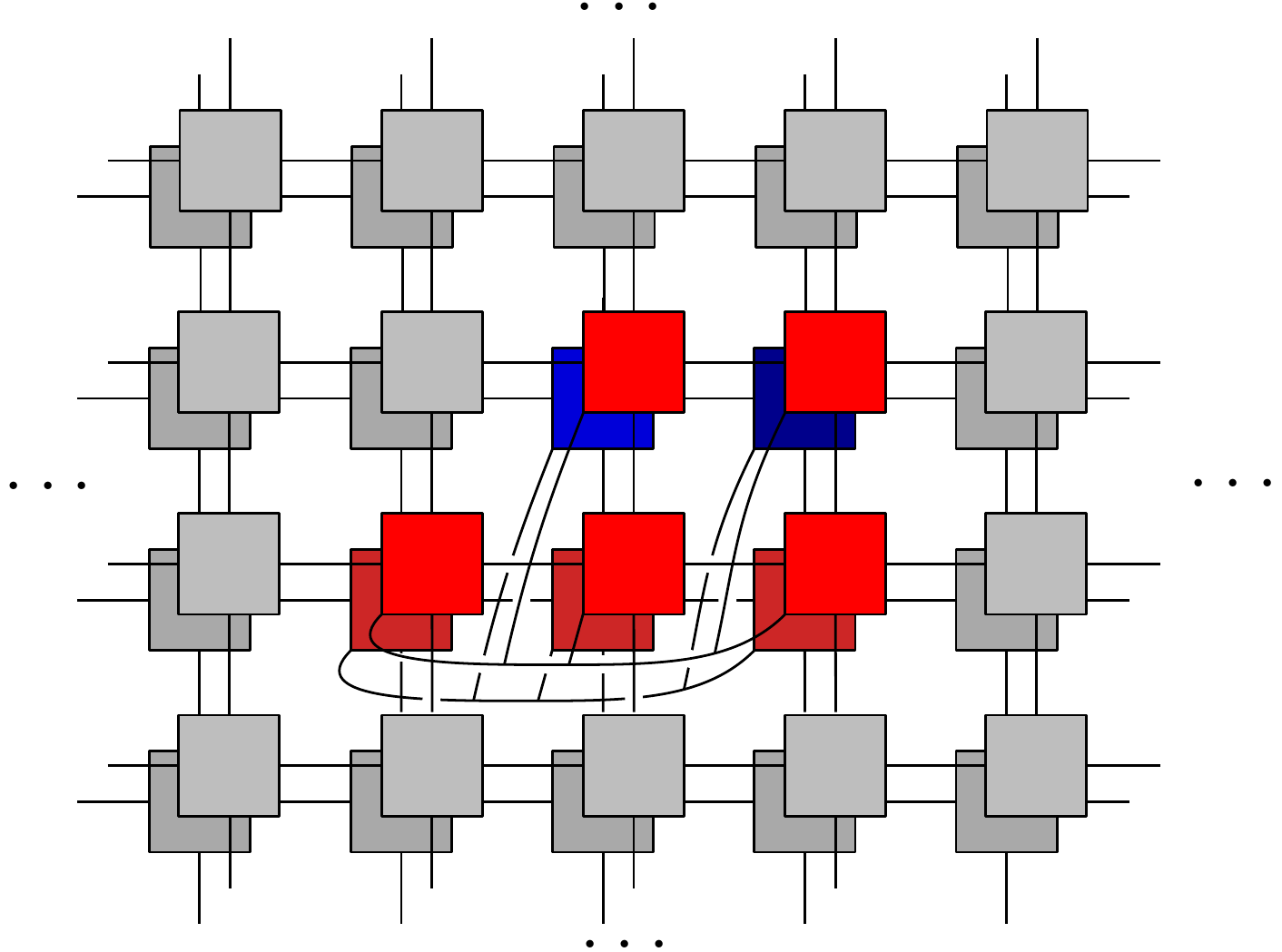}}\;,
	\end{equation}
	\end{widetext}
	where
	\begin{equation}\label{eq:PEPS_square_excitation_operators}
		\raisebox{-.3cm}{\includegraphics[scale=.4]{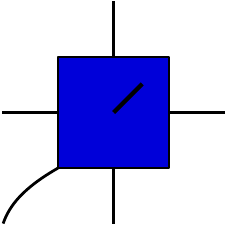}} = \raisebox{-.3cm}{\includegraphics[scale=.4]{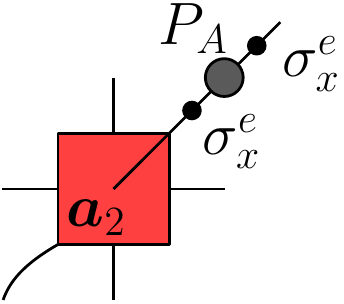}}
		\qquad \text{and}\qquad\;
		\raisebox{-.3cm}{\includegraphics[scale=.4]{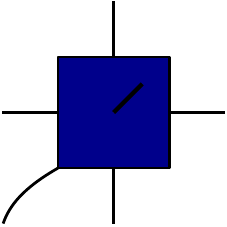}} = \raisebox{-.3cm}{\includegraphics[scale=.4]{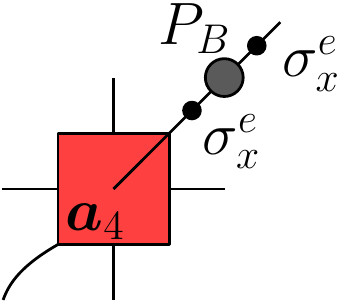}} .
	\end{equation}
	In order to avoid too much clutter, we have omitted the different labels in the graphical notation. The labels $ \vec{\bm{a}}', \bm{b}, \vec{\bm{d}}' $ and $ \vec{\bm{a}}, \bm{b}, \vec{\bm{d}} $ are always implied for the top and bottom layer, respectively.
	Note that the $ \sigma_x^e $ operator was applied on two different tensors. This is nothing more than a side-effect of the fact that the physical degrees of freedom are doubled in the PEPS representation.
	
	The result of the tensor contraction in Eq.~\eqref{eq:noise_mat_el_contraction1} should be independent of the size of the system, and must be entirely determined by  the nontrivial anyonic charges of the colored segment tensors.
	We may therefore assume the system to be infinite, where all tensors except the colored ones depicted in Eq.~\eqref{eq:noise_mat_el_contraction1} correspond to segments with a trivial charge. 
	This infinite contraction can then be simplified by determining the top fixed point MPS of the double layer transfer matrix \cite{zauner2018variational}:
	\begin{multline}\label{eq:PEPS_fixed_point_top}
		\includegraphics[scale=.4]{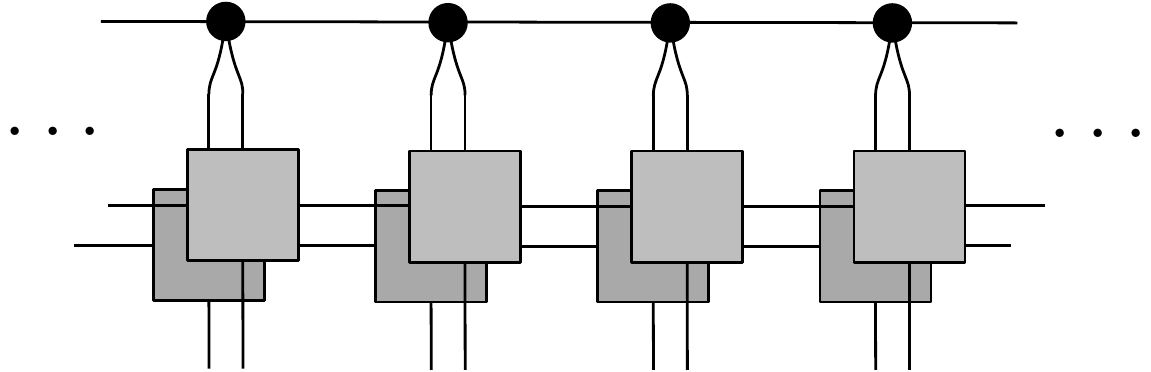} \\
		= \; \raisebox{-.2cm}{\includegraphics[scale=.4]{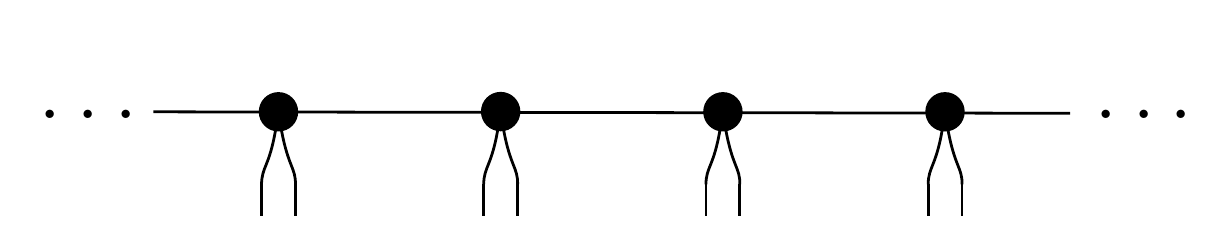}}\;.
	\end{multline}
	After finding a similar bottom fixed point MPS, Eq.~\eqref{eq:noise_mat_el_contraction1} can be reduced to
	\begin{widetext}
	\begin{equation}\label{eq:noise_mat_el_contraction2}
		\braket{\psi^{\vec{\bm{a}}', \vec{\bm{d}}'}_{\bm{b}}|\sigma_x^e P_V \sigma_x^e|\psi^{\vec{\bm{a}}, \vec{\bm{d}}}_{\bm{b}}} = \;
		\raisebox{-1.2cm}{\includegraphics[scale=.4]{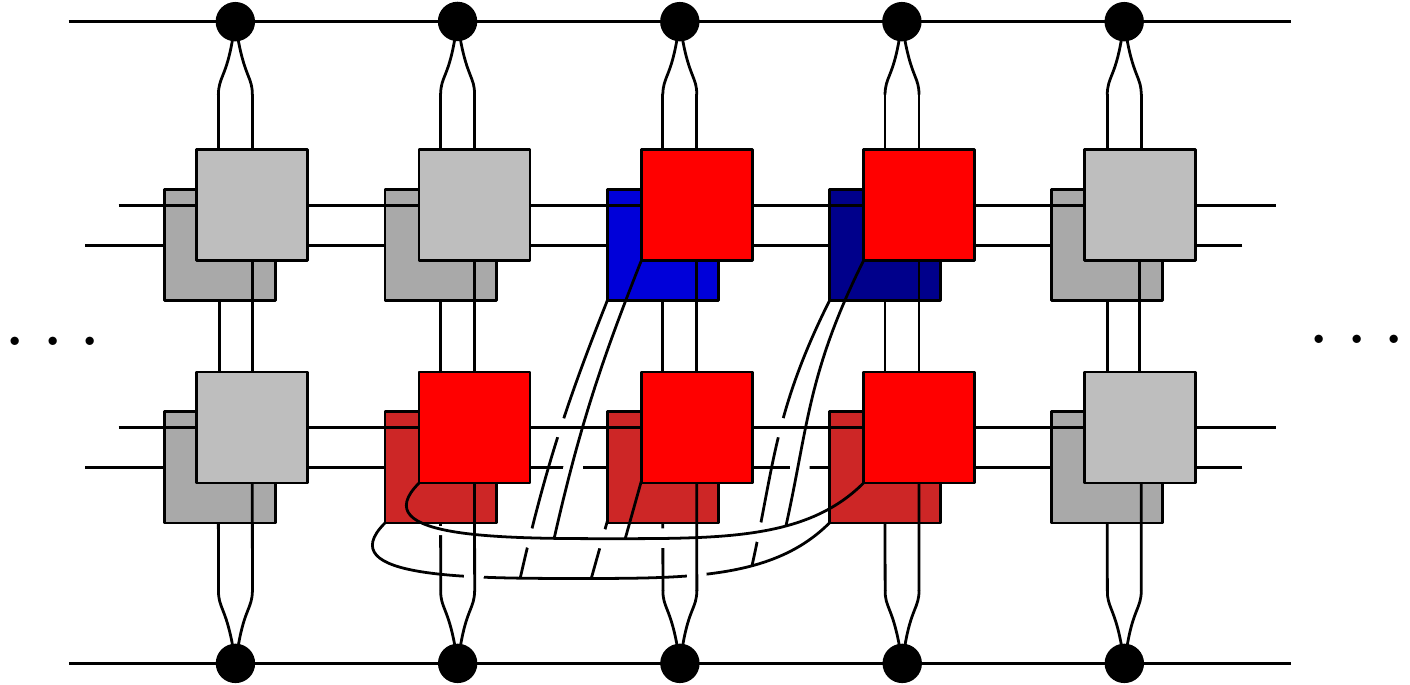}}\;.
	\end{equation}
    \end{widetext}
	In the same way, left and right fixed points can be determined for this expression:
	\begin{align}\label{eq:PEPS_fixed_point_left}
		\raisebox{-1.2cm}{\includegraphics[scale=.4]{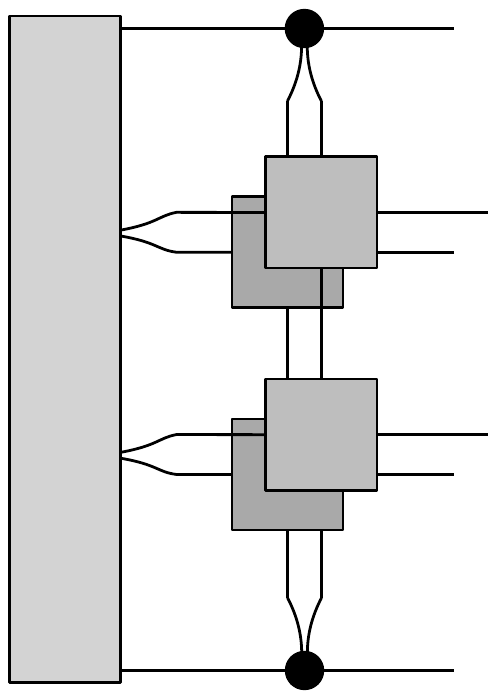}}
		\; = \; \raisebox{-1.2cm}{\includegraphics[scale=.4]{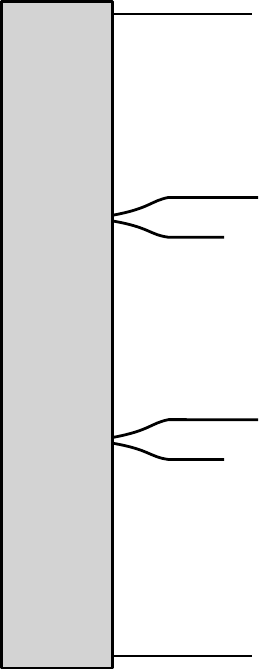}}\;\;,
	\end{align}
	and similarly for the right fixed point. This finally results in the finite contraction
	\begin{equation}\label{eq:noise_mat_el_contraction3}
	\braket{\psi^{\vec{\bm{a}}', \vec{\bm{d}}'}_{\bm{b}}|\sigma_x^e P_V \sigma_x^e|\psi^{\vec{\bm{a}}, \vec{\bm{d}}}_{\bm{b}}} = \;
		\raisebox{-1.2cm}{\includegraphics[scale=.4]{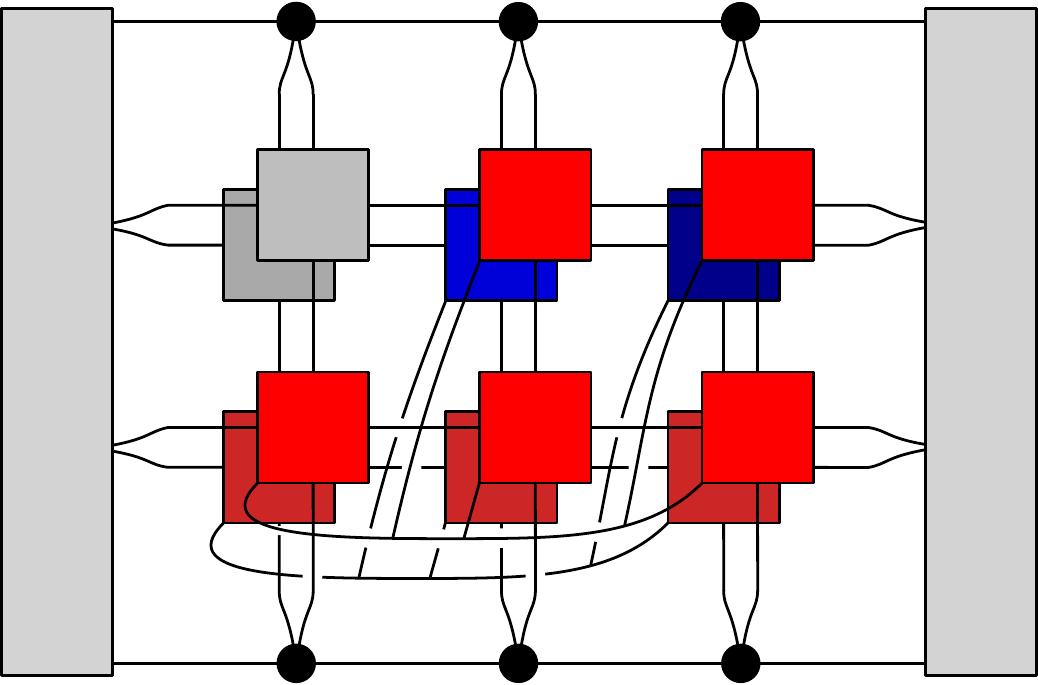}}\;\;,
	\end{equation}
	which can be computed in the regular way. In Eqs.~\eqref{eq:PEPS_fixed_point_top} and \eqref{eq:PEPS_fixed_point_left} we have implicitly assumed that the PEPS tensors were rescaled such that the leading eigenvalue of the relevant transfer matrices in these expressions is equal to 1. Because of these rescalings the PEPS for the fusion basis states will not have a unit norm, and the final expression Eq.~\eqref{eq:noise_mat_el_contraction3} should therefore be divided by $ \sqrt{\braket{\psi^{\vec{\bm{a}}, \vec{\bm{d}}}_{\bm{b}}|\psi^{\vec{\bm{a}}, \vec{\bm{d}}}_{\vec{\bm{b}}}} \braket{\psi^{\vec{\bm{a}}', \vec{\bm{d}}'}_{\vec{\bm{b}}}|\psi^{\vec{\bm{a}}', \vec{\bm{d}}'}_{\bm{b}}} } $ in order to obtain a matrix element that is properly normalized.
	
\subsection{Storing and manipulating anyonic fusion basis states} \label{sec:curve_diagrams}
	As explained in Sec.~\ref{sec:anyonic_fusion_basis_short}, states in $ \H_{\text{s.n.}} $ can be expressed as linear combinations of anyonic fusion basis states. 
	The anyonic fusion basis itself is determined by picking a pants decomposition of the surface and choosing a basis for each handle if the genus is nonzero.
	To keep track of the quantum state $ \ket{\Psi} $ of the system, one could in principle pick a basis $ \{ \ket{\psi_i}\} $ for the entire string-net subspace and then update all coefficients $ \braket{\psi_i | \Psi} $ throughout the different steps in the simulation. 
	Such a naive approach is doomed to fail however	as implementing $ F $-moves and braiding in the exponentially large Hilbert space $ \H_{\text{s.n.}} $ quickly becomes intractable as the system-size grows. 
	Instead, we need to select a basis that reflects the factorization of the fusion space discussed in Sec.~\ref{sec:classical_simulatbility}.
	Hence, we require the ability to dynamically introduce a basis for each of the connected groups of anyons separately, in a way that takes into account the specific structure of the relevant noise processes.	
	
	For each connected group of anyons, we can pick out a linear ordering by drawing a directed curve connecting all the anyons (which we locate at the center of their corresponding plaquette) within the group.	
	Directed curves, corresponding to the same linear ordering and only differing by continuous deformations that keep the anyon positions fixed, form a set of equivalence classes that we shall call \emph{curves} or \emph{curve diagrams}. 
	Each configuration of (non-intersecting) curves on the surface corresponds to an equivalence class of anyonic fusion bases that are related up to local Dehn twists in individual plaquettes. 
	
	A rigorous definition of curve diagrams, using the language of modular functors, can be found in Ref.~\cite{burton2016short}. For our purposes however, the following simplified construction will suffice.	
	For a given curve containing $ n $ anyons, the corresponding fusion basis 
	 (up to local Dehn twists) is found as follows: start by drawing a fusion tree connecting only the first and last anyons, in a way that is homologically equivalent to the curve. Next, following the orientation provided by the curve, add anyons $2, 3, \dots, n-1 $ to this fusion tree one by one, by adding a leaf to the tree on the left hand side.
	This is best illustrated with an example:
	\begin{equation*}
	\raisebox{-2cm}{\includegraphics[scale=.3]{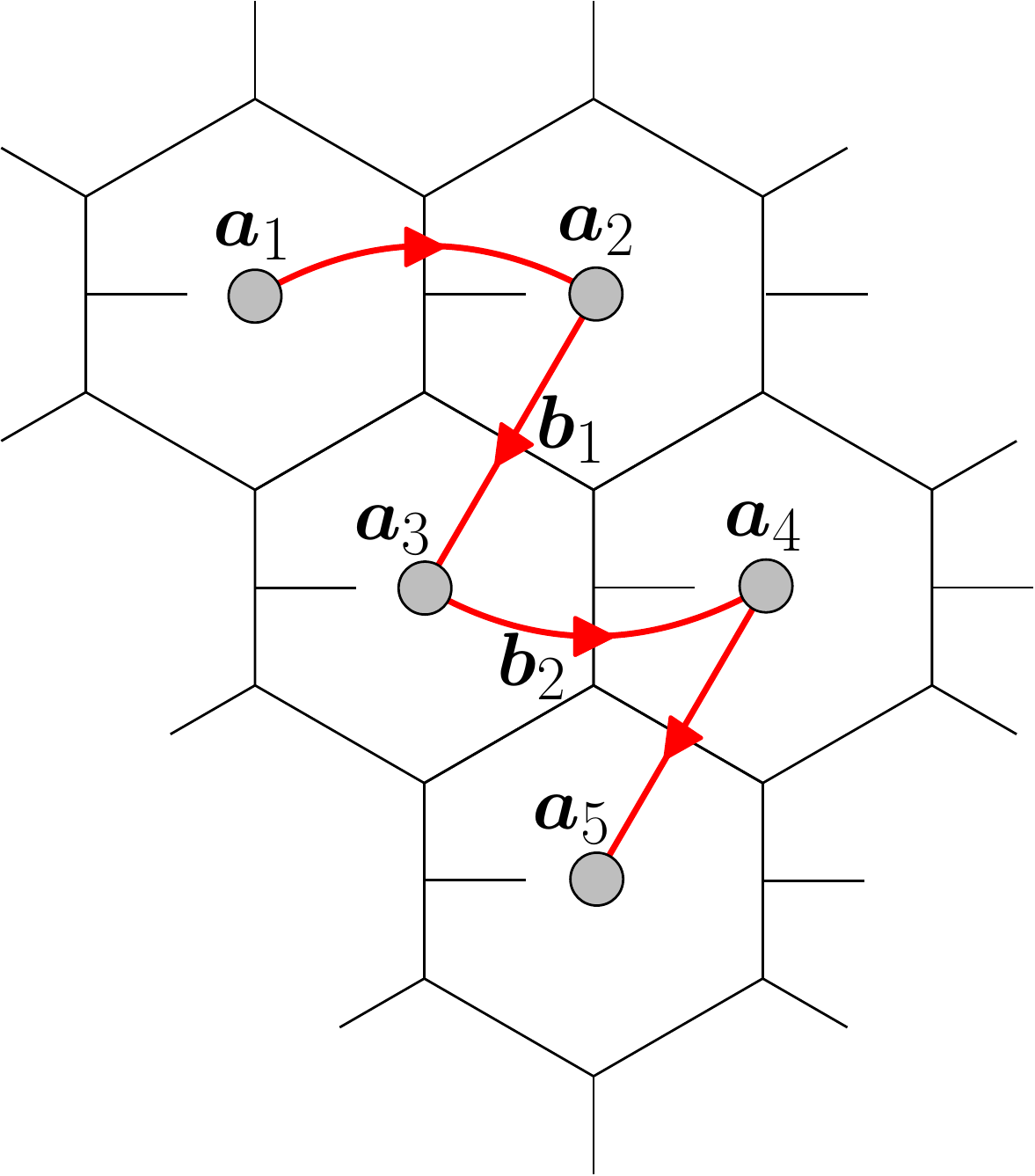}}
	\;\, \leftrightarrow \;\,
	\raisebox{-2cm}{\includegraphics[scale=.3]{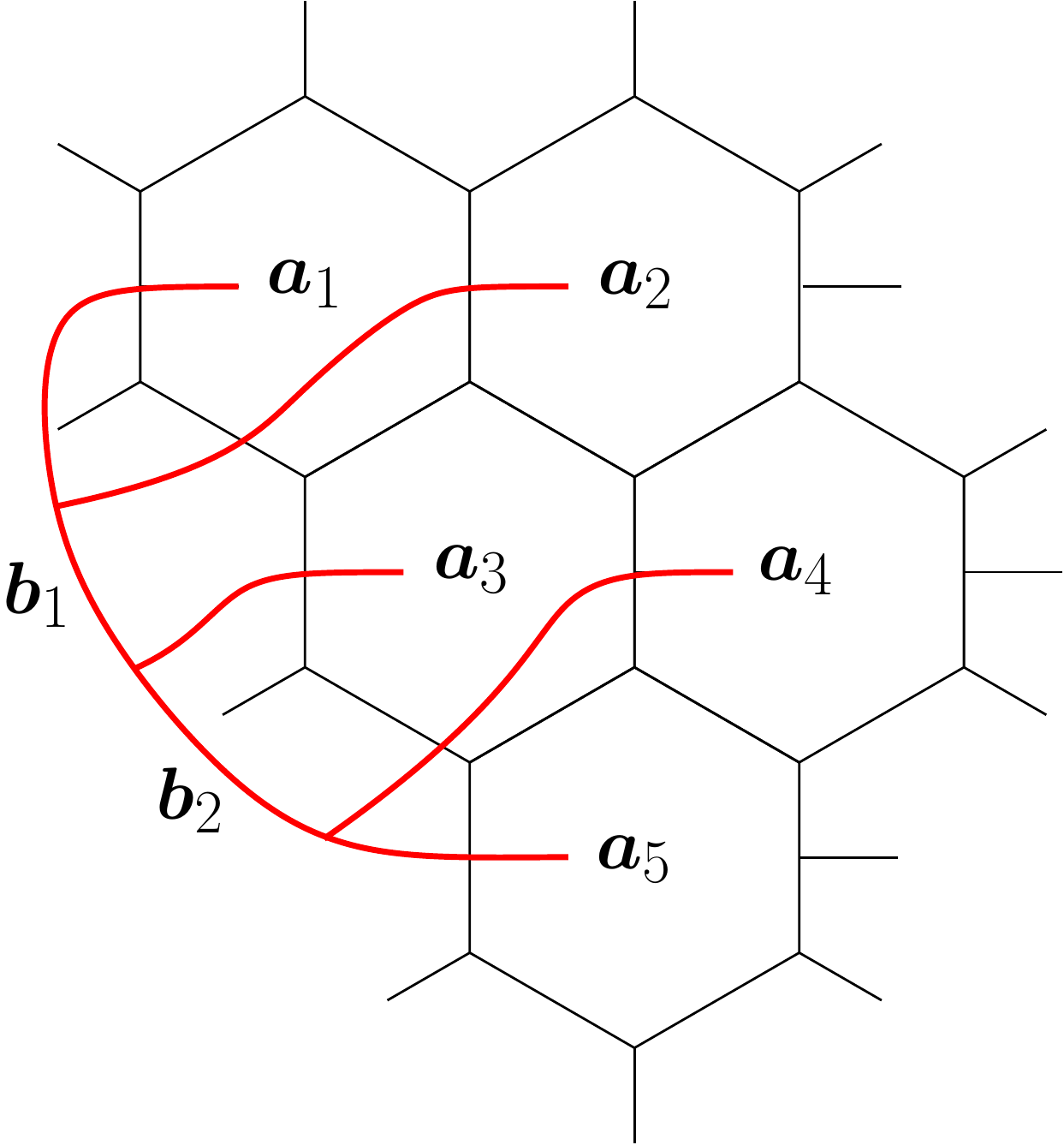}} \;.
	\end{equation*}
	A different basis for the same fusion space corresponding to a different curve diagram would be given by
	\begin{equation*}
	\raisebox{-2cm}{\includegraphics[scale=.3]{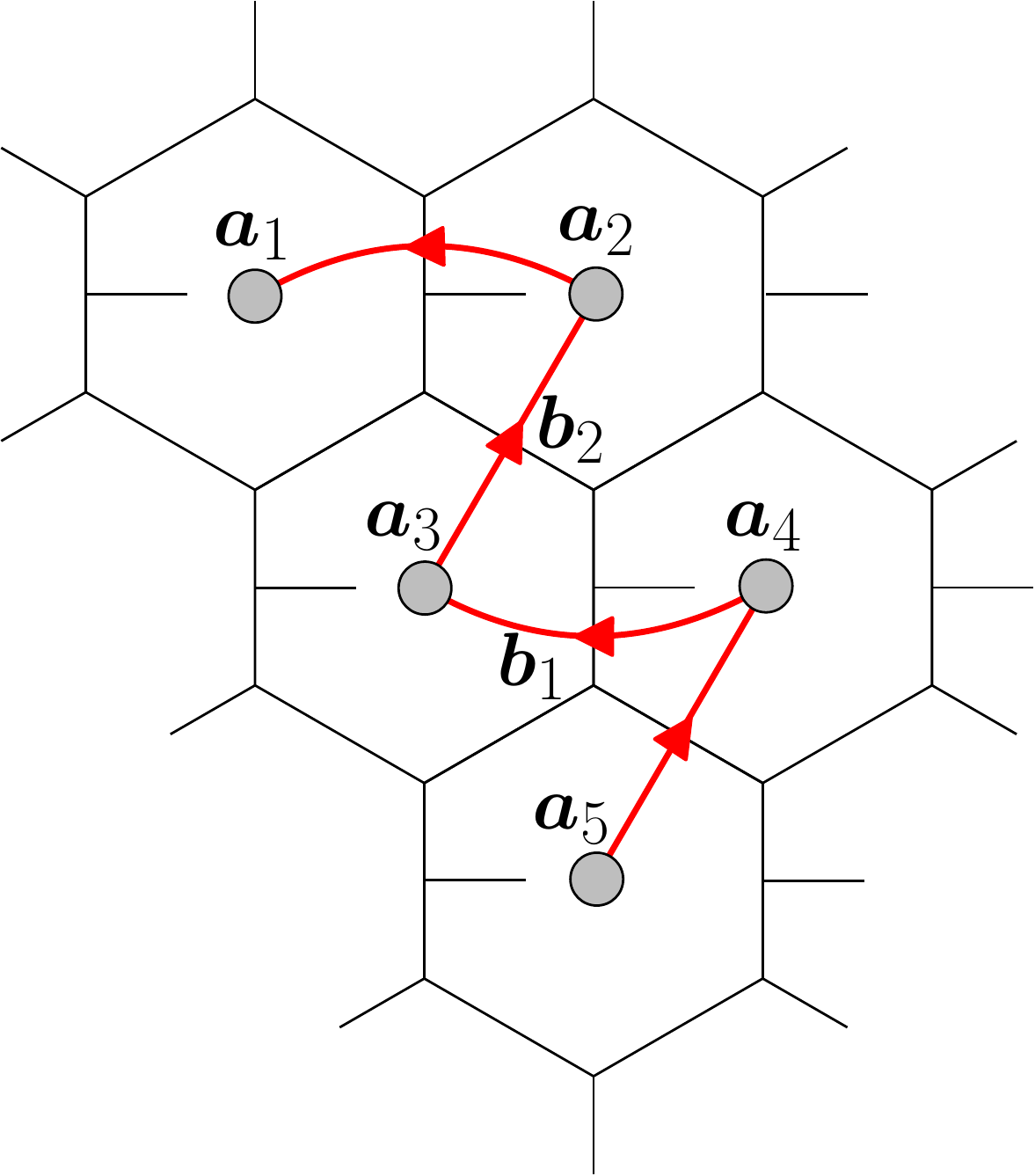}}
	\;\, \leftrightarrow 
	\raisebox{-2cm}{\includegraphics[scale=.3]{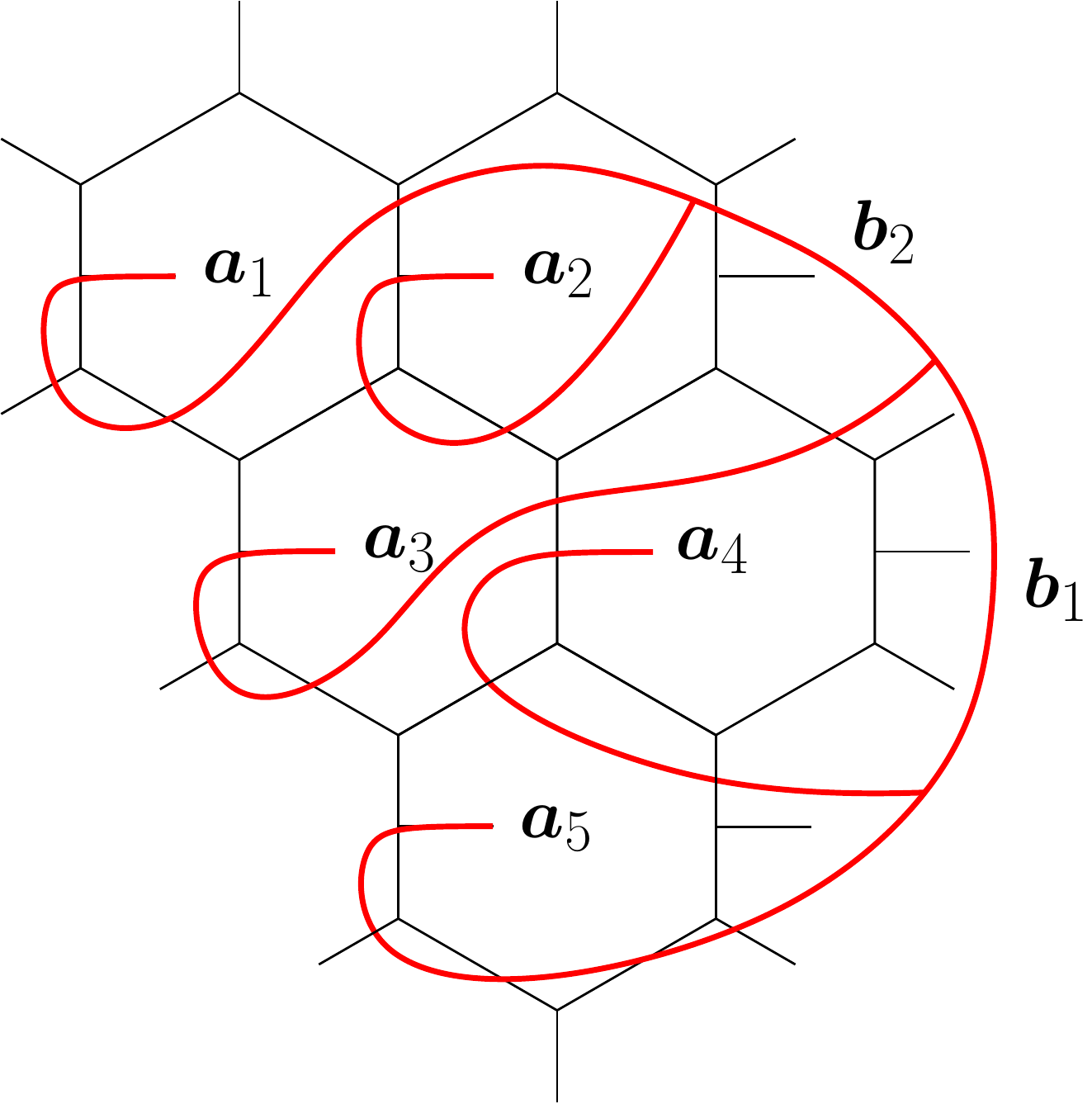}} \;.
	\end{equation*}
	
	Note that the way in which the leaf ribbons wrap around the plaquettes containing anyons is not uniquely determined by this construction. It is precisely this freedom that translates to local Dehn twists in individual plaquettes.
	While the construction could be modified in order to remove this freedom, there is no need for us to bother with such details. Our simulations are set up such that we never need to store any superposition in the anyon labels of individual plaquettes. As such, local Dehn twists give rise to global phases, which have no physical relevance.\\

	
	The relation between curve diagrams and anyonic fusion bases detailed above, does not specify the handle labels. 
	However, the nature of the simulation does not require the ability to store these values.
	The handle label (or superposition of such labels), can only be modified by processes in which a pair of anyons on the same curve interact along a path that is \emph{not} homologically equivalent to the piece of curve between them.
	Since such events are precisely those for which the simulation declares a failure, we do not need the ability to update the handle label.
	Futhermore, since the total charge of all created excitations is trivial, the value of the handle label has no influence on the outcome of any physical process.  
	Hence, all we need to store is the fusion state of anyonic excitations created \emph{on top of} some initial ground state.
	
\subsubsection{Data-structure}		
	In order to simulate the noise and error correction processes, one needs to be able to efficiently store and update the basis in which the quantum state of the system is expressed.
	The curve diagrams, as defined above, provide a convenient way of doing so.
	An efficient method for storing the curve diagrams describing the current basis was introduced in Ref.~\cite{burton2017classical}. 
	We slightly modify this construction in order for it to be better suited for the system at hand.
	
\begin{figure}[h]
		\centering
		\begin{subfigure}[b]{0.46\textwidth}
			\centering
			\includegraphics[scale=.28]{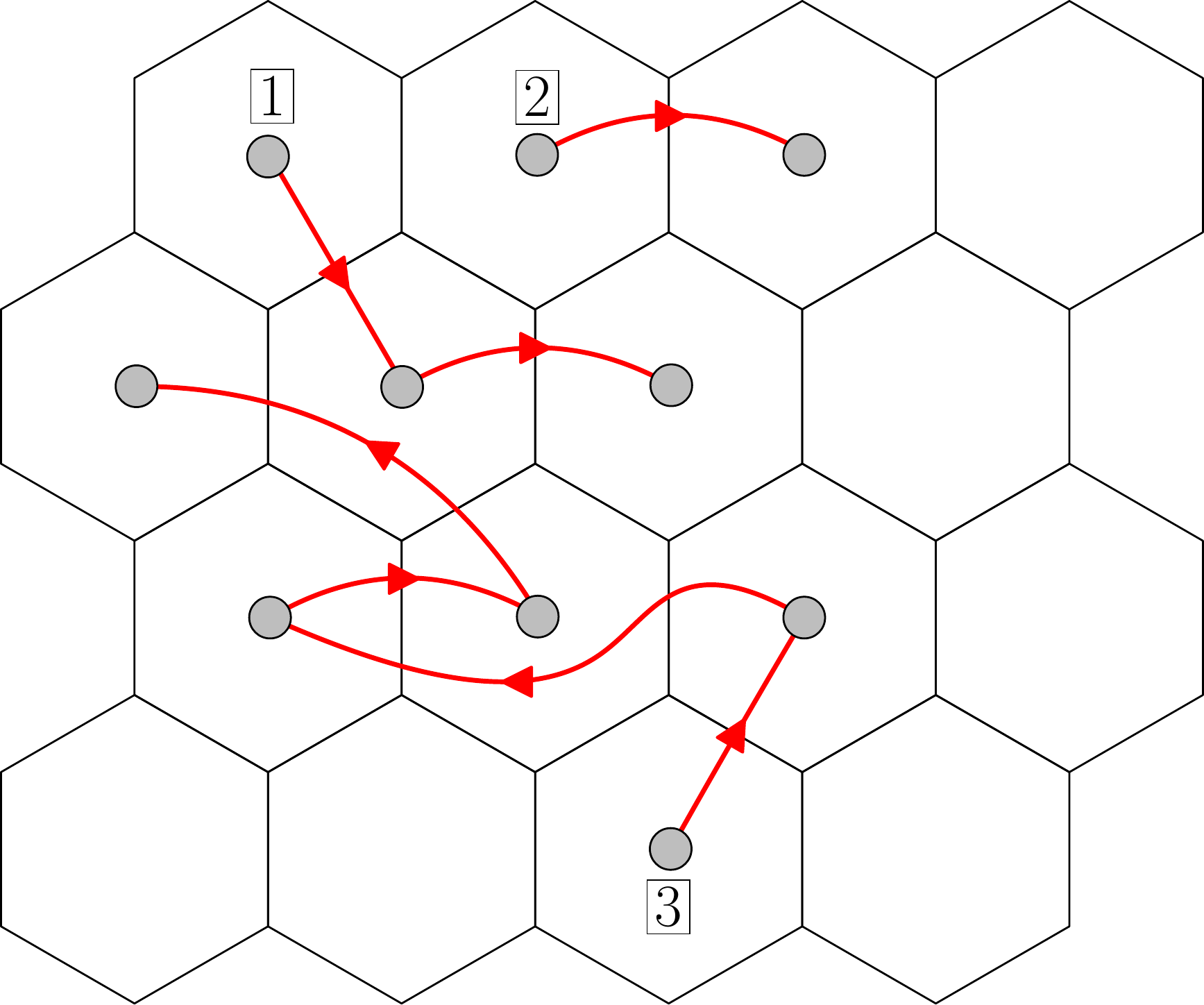}
			\caption{}
			\label{fig:curves_lattice_example_regular}
		\end{subfigure}
		\begin{subfigure}[b]{0.46\textwidth}
			\centering
			\includegraphics[scale=.28]{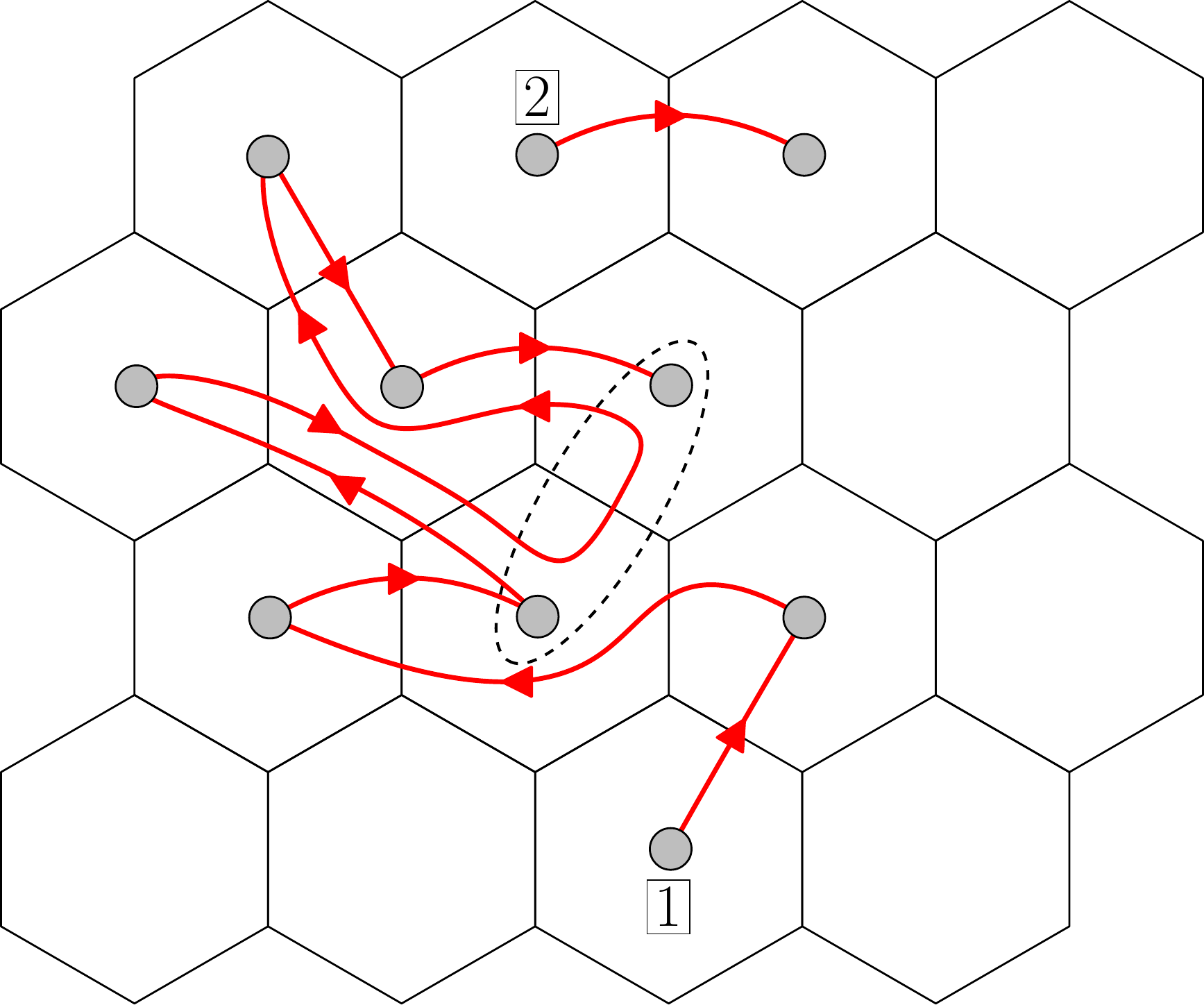}
			\caption{}
			\label{fig:curves_lattice_example_merged}
		\end{subfigure}
		\begin{subfigure}[b]{0.46\textwidth}
			\centering
			\includegraphics[scale=.28]{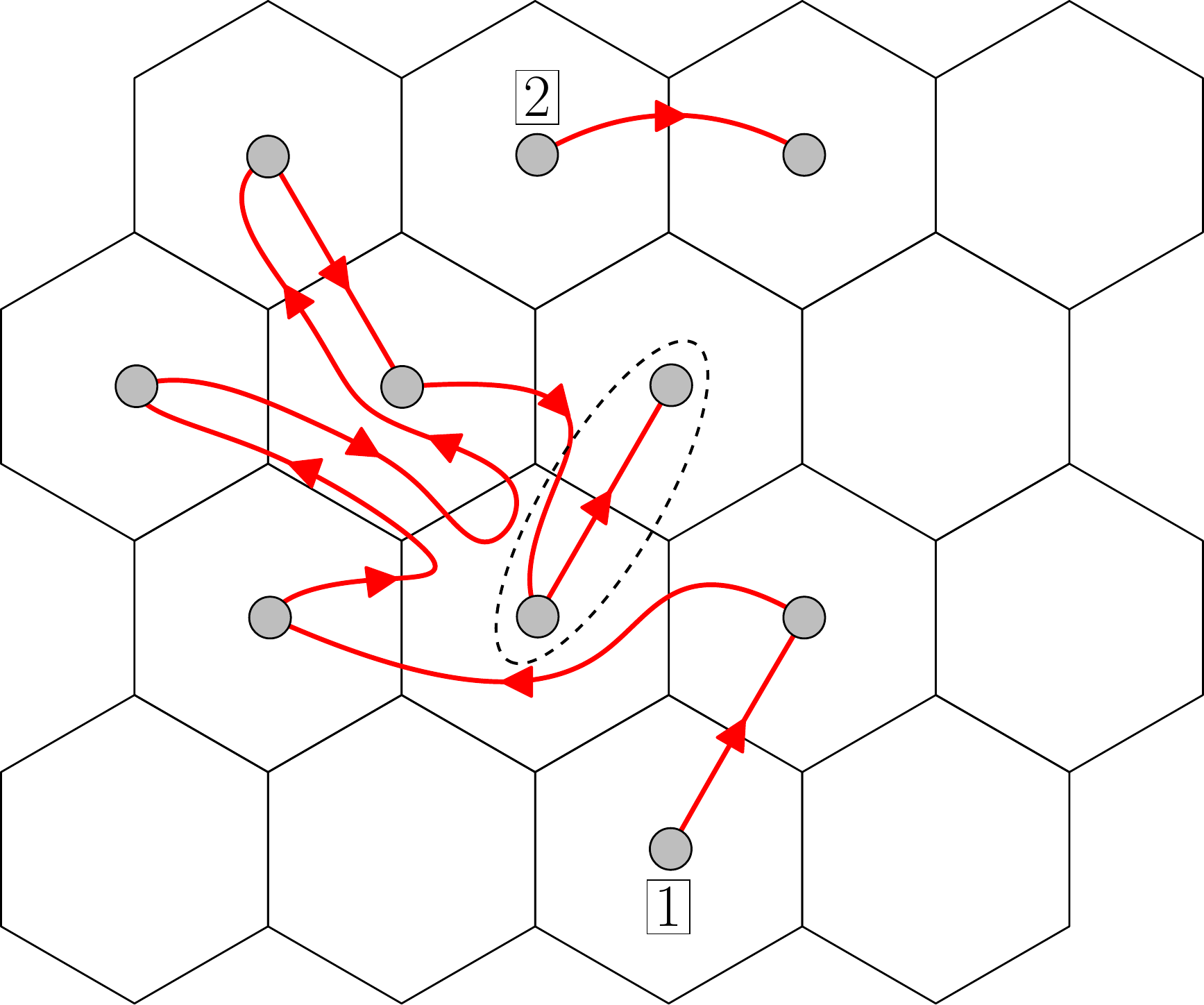}
			\caption{}
			\label{fig:curves_lattice_example_makeneighbors}
		\end{subfigure}
		\caption{(a): Example of a configuration of three connected curves on a $ 4 \times 4 $ periodic hexagonal lattice. Each curve is assigned a unique integer label, depicted here in the square boxes at the start of each curve. (b): Result of a merge procedure in the case the two anyons in the dashed ellipse interact. (c): Result of repeated use of refactoring moves to make the the anyons in the dashed ellipse appear sequentially in the same curve.}
		\label{fig:curves_lattice_example}
	\end{figure}

	We will store the configuration of curves by storing the shape of the curves running through each individual plaquette separately. Since the presence of the tail edges is irrelevant in this context, we will represent the curve diagram layout on hexagonal \emph{tiles} (each of which corresponds to a plaquette).
	The total configuration of the curves can be obtained by piecing all tiles together.
	In order to represent the curves of the connected groups of anyons on the lattice, we assign to each connected curve a unique integer label. An example of three connected curves on a $ 4 \times 4 $ lattice is depicted in \figref{fig:curves_lattice_example_regular}.
	Since they represent the layout of the curve diagrams on the lattice, each tile may contain at most one anyon, and we require that the different curves intersect the edges of tiles transversely.
			
	\begin{figure}[h]
		\centering
		\begin{subfigure}[b]{0.495\linewidth}
			\centering
			\includegraphics[scale=.45]{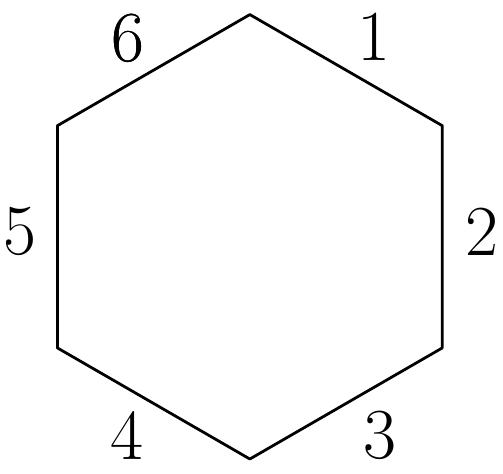}
			\caption{}
			\label{fig:tile_numbering}
		\end{subfigure}
		\begin{subfigure}[b]{0.495\linewidth}
			\centering
			\includegraphics[scale=.45]{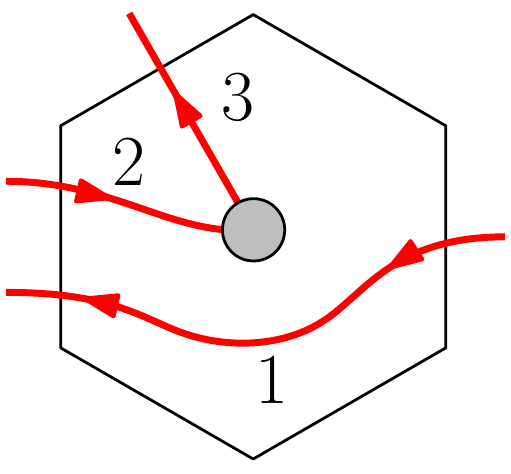}
			\caption{}
			\label{fig:tile_example}
		\end{subfigure}
		\caption{(a): Labeling of the edges of a hexagonal tile in a clockwise direction. (b): Example of a configuration of lines running through tile 10 in \figref{fig:curves_lattice_example_regular} (when counting from left to right and from top to bottom). Each line segment is assigned a unique integer label.}
		\label{fig:tile}
	\end{figure}
	
	In order to save the configuration of curves inside a tile we first assign a unique integer label to each \emph{piece of curve} inside the tile. By a \emph{piece of curve}, we mean a segment of a curve diagram connecting either two edges of the tile, or an edge and an anyon. Note that every piece of curve has an orientation.
	The edges of a tile are numbered clockwise from 1 to 6 as depicted in \figref{fig:tile_numbering}.
	For each edge, we record the label and orientation (+1 for an incoming line, -1 for an outgoing line) of each piece of curve intersecting it, in the order in which they are encountered when going clockwise around the tile.
	In addition, when an anyon is present inside the tile, we also store which pieces of curve are connected to it.
	Finally, for every piece of curve inside the tile, we store the curve label, along with its position inside that curve. If a piece of curve is positioned between the $ n $th and $ n+1 $th anyons on a curve (following the orientation of the curve), its position label is $ n $.
	An example of such a tile is depicted in \figref{fig:tile_example}. The configuration inside this tile would be stored as follows:
	\begin{equation}\label{eq:curve_diagram_tile_example}
		\begin{bmatrix}
		\,
		\end{bmatrix}
		\begin{bmatrix}
		1 & +1
		\end{bmatrix}
		\begin{bmatrix}
		\,
		\end{bmatrix}
		\begin{bmatrix}
		\,
		\end{bmatrix}
		\begin{bmatrix}
		1 & -1\\
		2 & +1
		\end{bmatrix}
		\begin{bmatrix}
		3 & -1
		\end{bmatrix}
		\begin{bmatrix}
		2\\
		3
		\end{bmatrix}
		\begin{bmatrix}
		1 & 3 & 2\\
		2 & 3 & 3\\
		3 & 3 & 4
		\end{bmatrix}.
	\end{equation}
	Here, the first six arrays correspond to the labels and orientations of lines crossing each of the six tile edges, respectively. The next to last array indicates that lines 2 and 3 are connected to the anyon in the tile. The last array indicates the curve and position along the curve of each piece of curve. In this case, all lines belong to curve 3, and the lines with labels 1, 2, 3 appear in the curve after the second, third and fourth anyon along the curve, respectively.
	

\subsubsection{Merge}
	If neighboring anyons on the lattice that lie on different connected curves interact at some point, their corresponding curves must be merged in order to compute the effect of the interaction.
	In terms of fusion trees, one must think of this procedure as connecting two separate fusion diagrams with a line carrying the trivial label.
	
	The curves are merged by connecting the end of one curve to the start of the other, with the condition that the combined path of the resulting curve and the interaction\footnote{By the path of an interaction we mean the shortest path connecting any pair of plaquettes affected by it that initially belong to two different curves.} does \emph{not} contain any homologically nontrivial loop.
	Other than this requirement, the way in which the curves are merged is arbitrary, one simply has to connect the end of one curve to the start of the other in some suitable way.
	 
	In our framework merging is implemented by extending the end of one curve parallel along the curve in its reverse direction until it reaches the tile of one of the interacting anyons. The extended curve is then crossed over to the tile of the other anyon, after which it follows the other curve in its reversed direction and is attached to the start of the latter. An example of this merging procedure is depicted in \figref{fig:curves_lattice_example_merged}. 
	
	Updating the state superposition in the case of a merge operation is trivial, as it simply amounts to taking the tensor product of the state superpositions associated to each curve.

\subsubsection{Passive exchange}
	Calculating the outcome of both individual Pauli errors and fusion operations, requires expressing the state in the appropriate anyonic fusion basis.
	Basis transformations concerning individual curves are performed using passive exchanges or \emph{swaps}, which are represented as follows for the clockwise and the counterclockwise case, respectively:	
	\begin{align}
		S^{ab}:\;\;\;
		\raisebox{-.85cm}{\includegraphics[scale=.42]{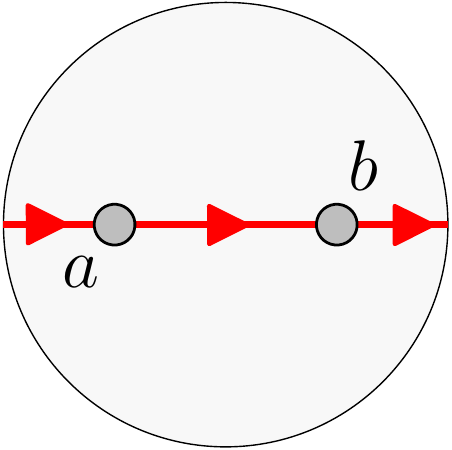}}
		\;\; \mapsto \;\;
		\raisebox{-.85cm}{\includegraphics[scale=.42]{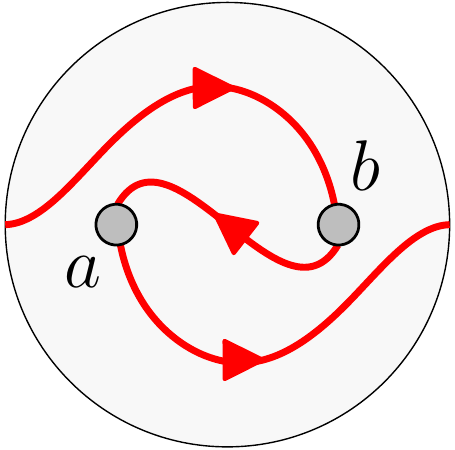}}\;, \label{eq:swap_cw}
	\end{align}
	\begin{equation}
		\left(S^{ab}\right)^{-1}:\;\;\;
		\raisebox{-.85cm}{\includegraphics[scale=.42]{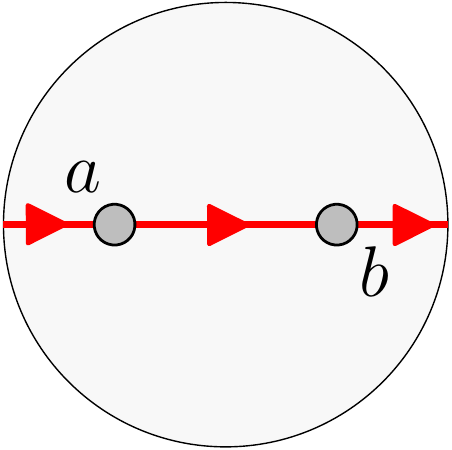}}
		\;\;  \mapsto \;\;
		\raisebox{-.85cm}{\includegraphics[scale=.42]{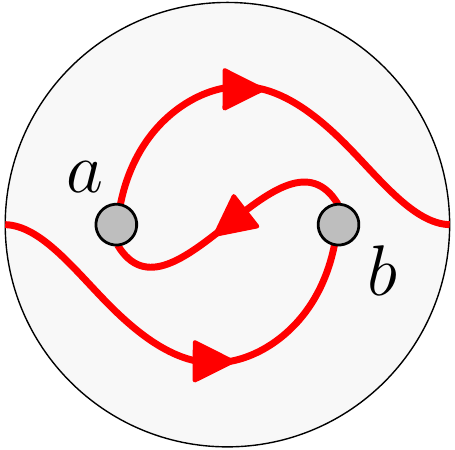}}\;. \label{eq:swap_ccw}
	\end{equation}
	It is important to stress that this does \emph{not} represent an active braid move in which two anyons are exchanged. This is a basis transformation: the quantum state of the system remains unchanged, and the coefficients of the state superposition must be updated to express this state in a new basis. 
	
	The right transformation on the state vector components is found considering the relation between the anyonic fusion bases corresponding to the left and right hand side of these equations.
	For a clockwise swap, represented in Eq.~\eqref{eq:swap_cw}, this relation is	
	\begin{equation}\label{eq:curve_diagr_cw_swap_ribbon}
		\raisebox{-.5cm}{\includegraphics[scale=.38]{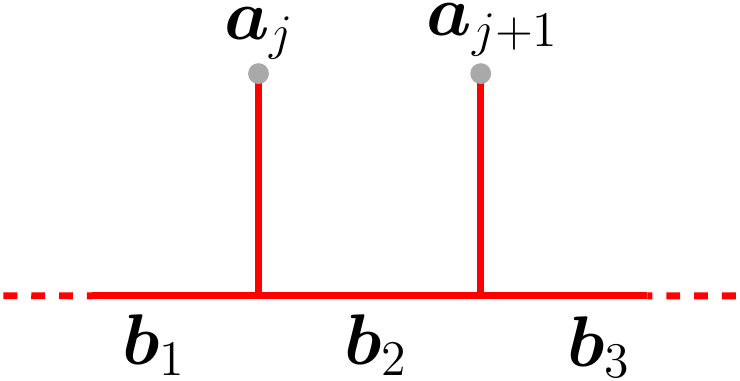}} = 
		\sum_{\bm{b}_2'} B^{\bm{b}_1 \bm{a}_{j} \bm{b}_2}_{\bm{a}_{j+1} \bm{b}_3 \bm{b}_2'} \;\;
		\raisebox{-.5cm}{\includegraphics[scale=.38]{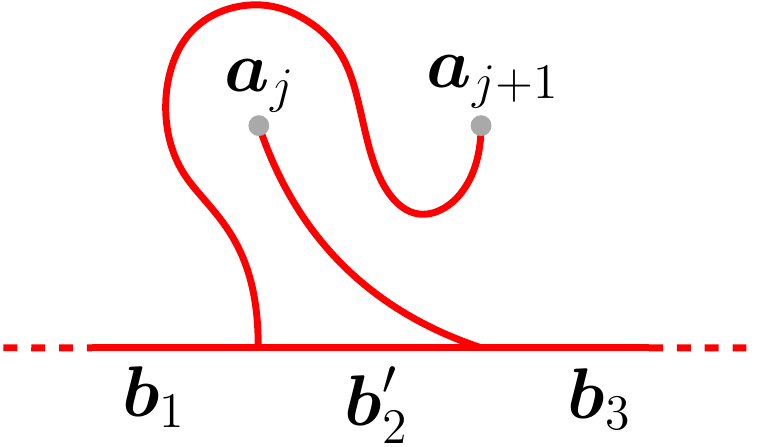}} \;,
	\end{equation}
	where 
	\begin{equation}\label{eq:B}
		B^{\bm{b}_1 \bm{a}_{j} \bm{b}_2}_{\bm{a}_{j+1} \bm{b}_3 \bm{b}_2'} = \sum_{\bm{c}} 
		F^{\bm{a}_{j}  \bm{a}_{j+1} \bm{c}}_{ \bm{b}_3  \bm{b}_1  \bm{b}_2'} 
		R^{\bm{a}_{j} \bm{a}_{j+1}}_{\bm{c}} 
		F^{\bm{b}_1 \bm{a}_{j} \bm{b}_2}_{\bm{a}_{j+1} \bm{b}_3 \bm{c}}\;.
	\end{equation}
	Analogously, for the counter clockwise case, one finds
	\begin{widetext}
	\begin{equation}\label{eq:curve_diagr_ccw_swap_ribbon}
		\raisebox{-.5cm}{\includegraphics[scale=.38]{fig/cw_braid_diagram_initial.pdf}} =
		\sum_{\bm{b}_2'} \left(B^{\bm{b}_1 \bm{a}_{j} \bm{b}_2}_{\bm{a}_{j+1} \bm{b}_3 \bm{b}_2'} \right)^* \;
	\raisebox{-.5cm}{\includegraphics[scale=.38]{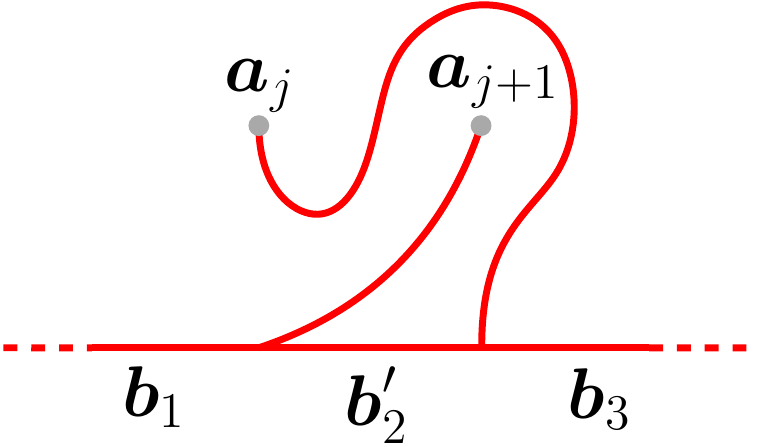}} \,.
	\end{equation}
	\end{widetext}
\subsubsection{The paperclip algorithm} \label{sec:paperclip}
	In order to determine the outcome of an interaction on a set of anyons, be it some local noise operator or the fusion of a pair, we must always transform to a basis where these anyons appear sequentially in the linear ordering determined by their curve diagram.
	Below, we outline how such a basis transformation can be performed using a sequence of swaps for the case where only 2 anyons are involved.
	The general case for $ n $ anyons can then be deduced iteratively. 
	
	In general, any two curve diagrams $ f $ and $ f' $, containing the same anyons, and such that the combined path of $ f $ and $ f' $ does not contain any homologically nontrivial loop, can be related to each other by a sequence of swaps.
	The algorithm for determining this sequence, was introduced in Ref.~\cite{burton2016short} and dubbed the \emph{refactoring algorithm}.
	We will describe a different but entirely equivalent formulation of this algorithm, called the \emph{paperclip algorithm} which is more convenient for our purpose.
	This alternative formulation was introduced in Ref.~\cite{burton2017classical}, and determines the sequence of swaps corresponding to a basis transformation where we ``move'' an anyon (or rather, its position on the curve) to the next piece of curve encountered when moving along the boundary of its tile in a clockwise fashion, as shown in the following diagram: 
	\begin{equation}\label{eq:refactoring_tile}
		\raisebox{-.95cm}{\includegraphics[scale=.4]{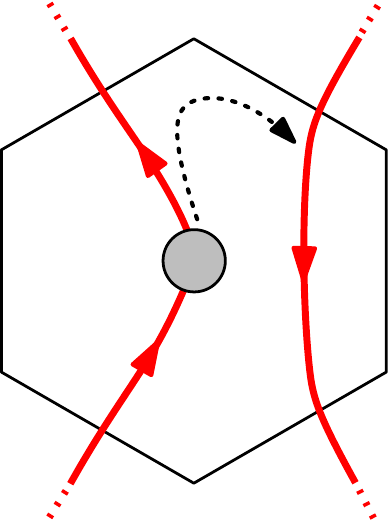}} \quad \mapsto \quad \raisebox{-.95cm}{\includegraphics[scale=.4]{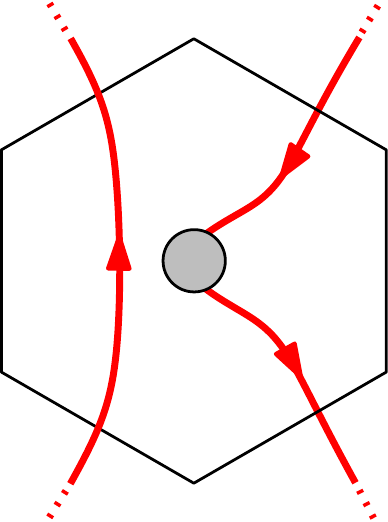}} \;\;.
	\end{equation}
	We will call such transformations \emph{refactoring moves}.
	Schematically, their action on the curve diagram can be represented as
	\begin{equation}\label{eq:refactoring_line}
		\raisebox{-.4cm}{\includegraphics[scale=.48]{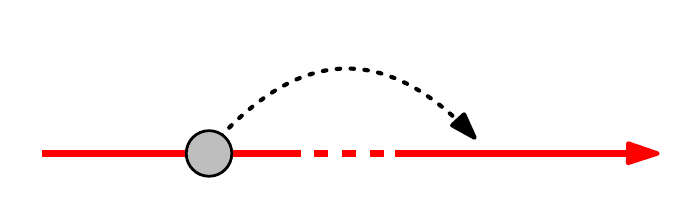}} \; \mapsto  \raisebox{-.4cm}{\includegraphics[scale=.48]{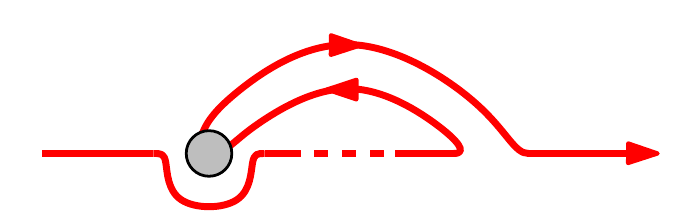}} \,.
	\end{equation}

	The initial position and destination of a refactoring move divide a curve into three disjoint segments which we name \emph{tail} ($ T $), \emph{body} ($ B $) and \emph{head} ($ H $), respectively (following the orientation of the curve). Their content is defined as follows:

	\begin{center}
		\includegraphics[width = 0.8 \linewidth]{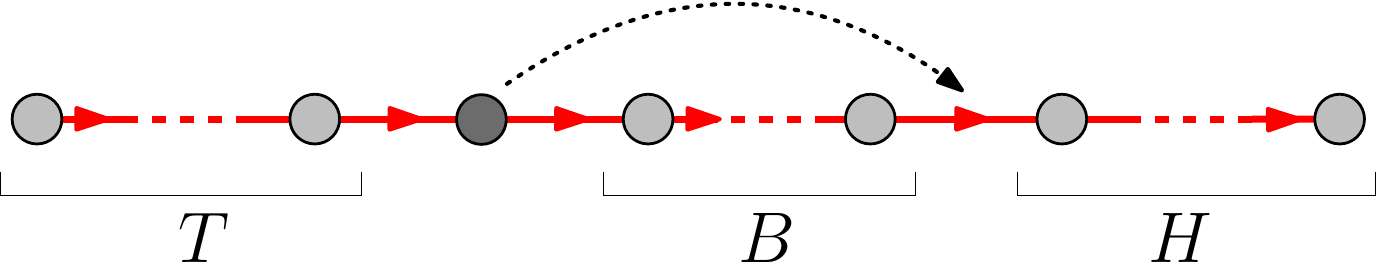}
	\end{center}

	\noindent where the dotted line represents the refactoring move. Note that we do not include the anyon which is being moved (indicated in dark above) in any of these segments.
	

	As interacting anyons are always neighbors on the lattice, repeated use of such moves can be used to obtaining a curve diagram where these anyons appear subsequently on the same curve.
	For example (assuming all encountered lines belong to the same curve): 
	\begin{align*}
		\raisebox{-1cm}{\includegraphics[scale=.4]{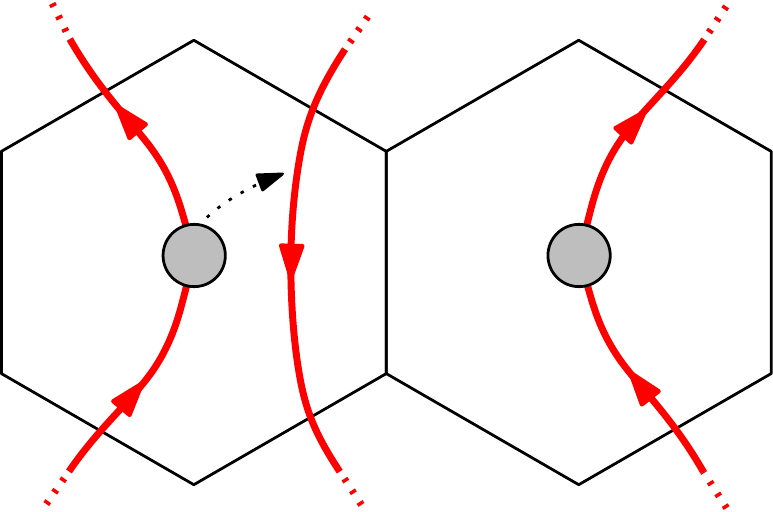}} 
		\quad &\mapsto \quad
		\raisebox{-1cm}{\includegraphics[scale=.4]{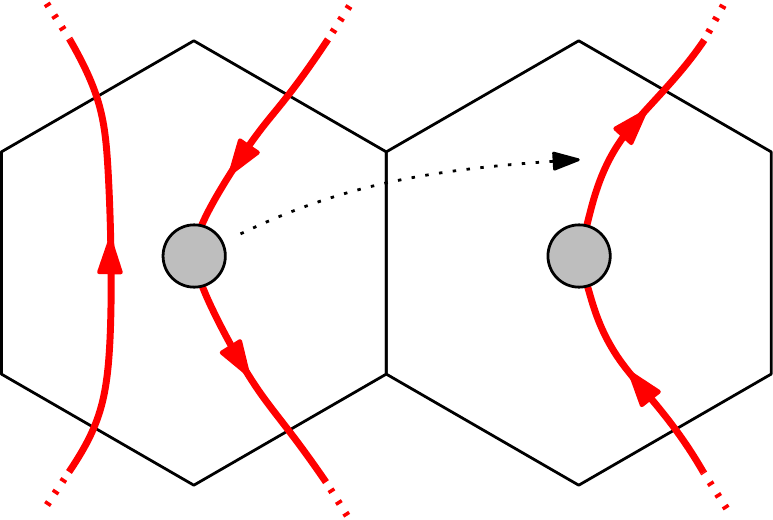}} \\ 
		\quad &\mapsto \quad
		\raisebox{-1cm}{\includegraphics[scale=.4]{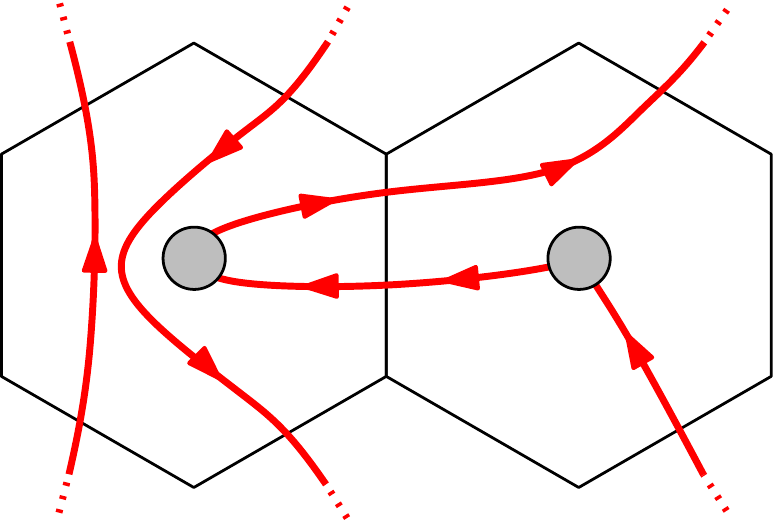}}\;\;.
	\end{align*}
	
	If any lines lie between the two interacting anyons but do not belong to either of their curves, one of the following two actions must be performed: 	
	In case the interacting anyons do not belong to the same curve initially, one can attempt to pull such lines through one of these curves to get it out of the way. Since each curve corresponds to a connected group of anyons with trivial total charge, such a deformation does not affect the state vector and is therefore permitted.
	Whenever this is not possible (i.e., when that does not ``remove'' the obstructing line), the corresponding curve diagram is essentially trapped between the others, and one is forced to merge it with those containing the interacting anyons, before proceeding with the clockwise ``moves''.
	The latter is of course not desirable, since it increases the size of the associated fusion space, but there are situations where this is unavoidable. 
	Of course, in case the pair of anyons are members of different curves initially, these must be merged in the process.
	
	The sequence of swaps corresponding to the refactoring moves described above can be determined with the corresponding \emph{turn number}. 
	This number is found by counting the number of right hand $ 60\degrees $ turns made by the curve between the initial position of the anyon and its destination (that is, the next piece of curve that intersects the tile boundary). 
	Starting at the anyon's initial position, every $ 60\degrees $ right hand turn contributes $ +1 $, while every left hand turn contributes\footnote{Note that we always count the number of right hand turns while following the curve from the start towards the destination of the refactoring move, independently of whether or not the refactoring is along or against the orientation of the curve itself} $ -1 $. 
	In order to ensure that the turn number is independent of where the ``destination piece of curve'' enters the tile, the total value must then be decreased by 1 for every additional edge of the tile boundary we have to move to (in a clockwise fashion) before encountering it (0 if it appears directly after the initial line, $ -1 $ if it appears on the next edge, ...).

	Depending on whether the refactoring move is with or against the orientation of the curve, and depending on whether the piece of curve associated with the initial position of the refactoring move has an incoming or outgoing orientation, it is always isotopic to one of 4 different ``paperclips'', each of which is associated to a specific turn number.
	For example, when the anyon is transported along the curve, and the initial piece of curve has an incoming orientation, the possible turn numbers and corresponding ``paperclips'' are
	\begin{alignat*}{2}
		& \boxed{-3}
		\raisebox{-1cm}{\includegraphics[scale=.42]{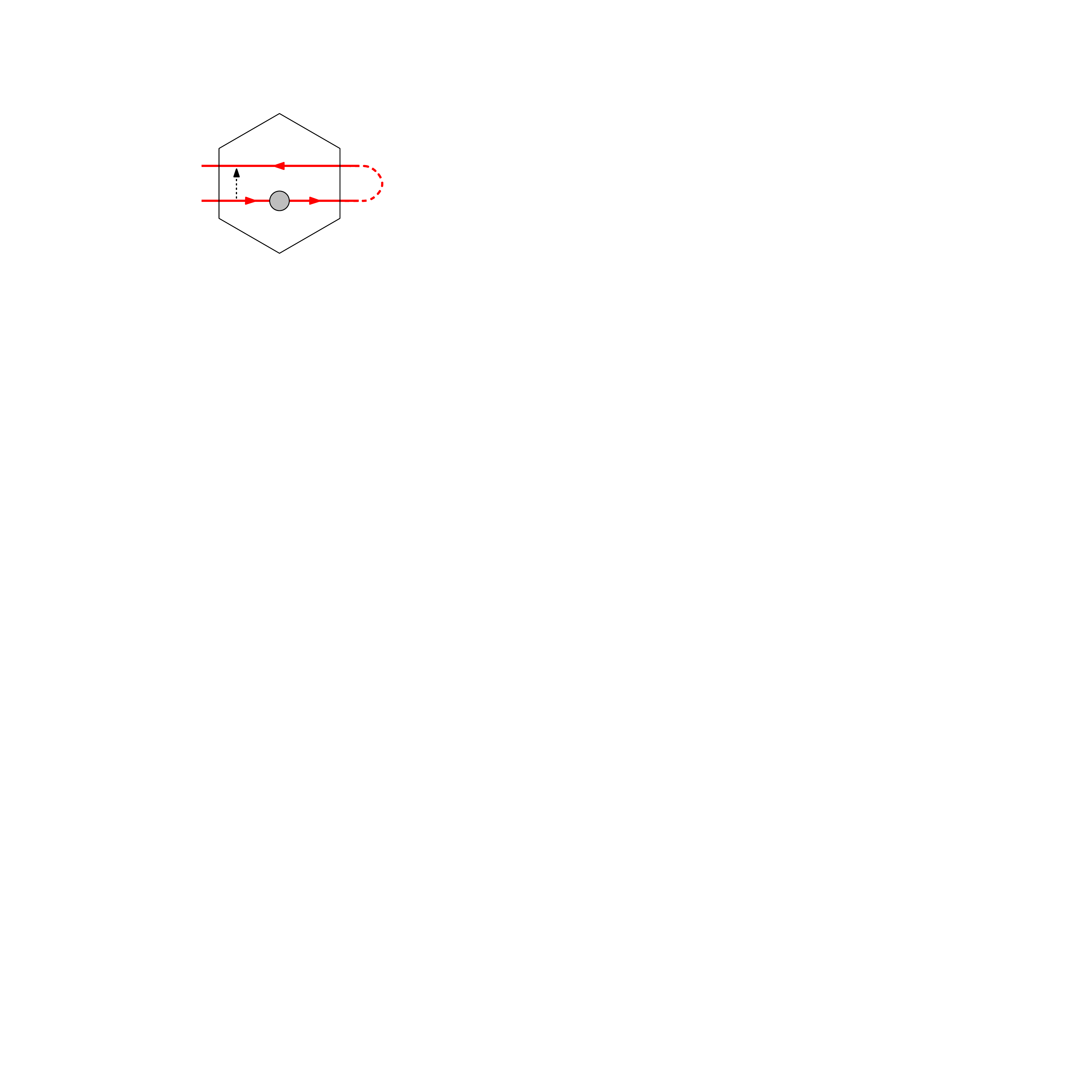}}
		 &&\rightarrow   
		\raisebox{-.5cm}{\includegraphics[scale=.48]{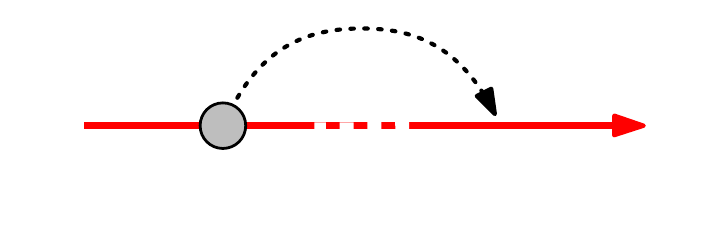}} \\
		& \boxed{-6}   
		\raisebox{-1cm}{\includegraphics[scale=.42]{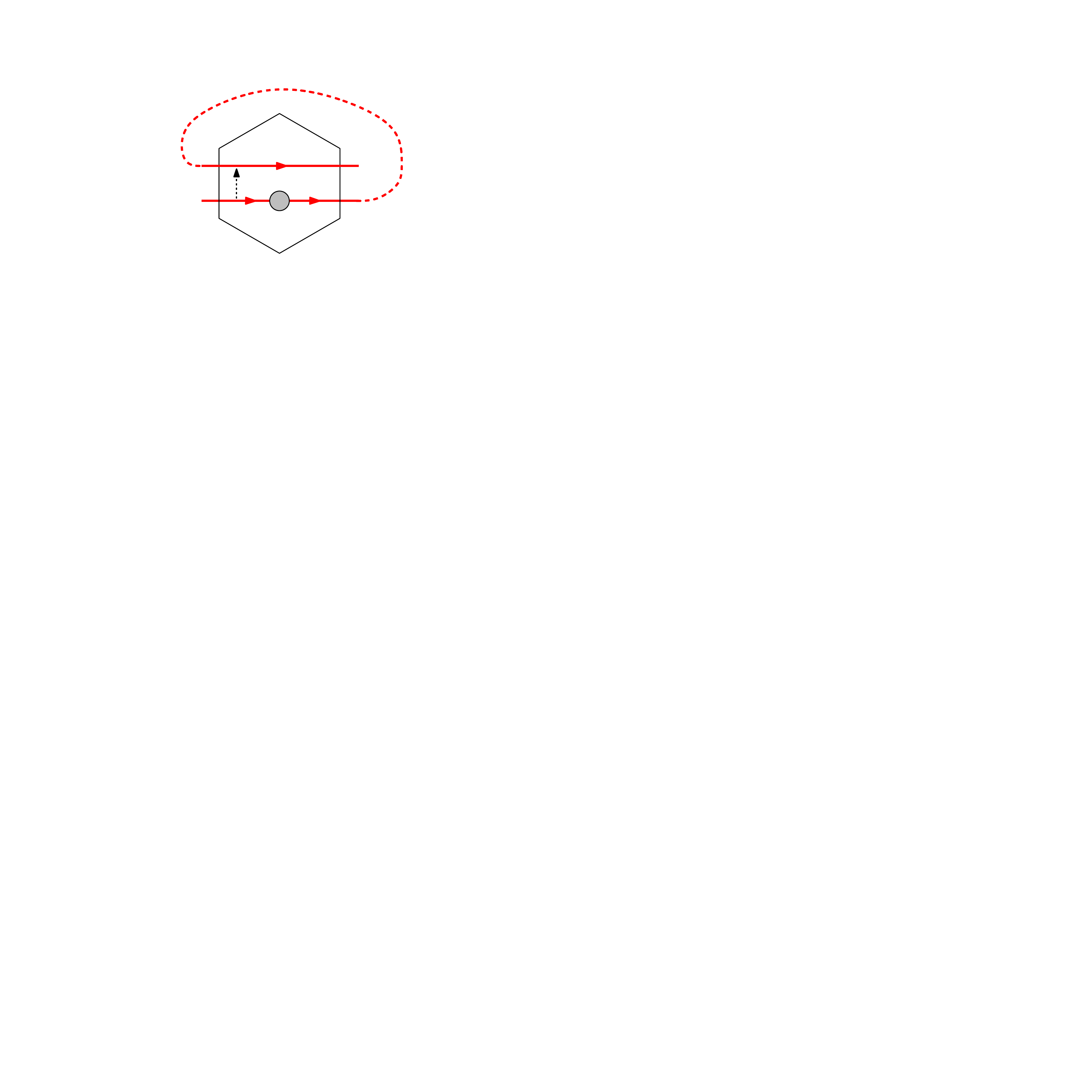}}
		&&\rightarrow 
		\raisebox{-.5cm}{\includegraphics[scale=.48]{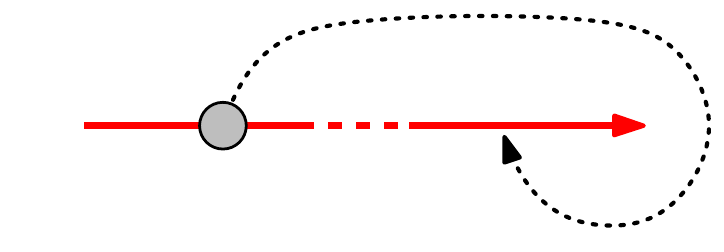}} \\
		& \boxed{+6} 
		\raisebox{-1.3cm}{\includegraphics[scale=.42]{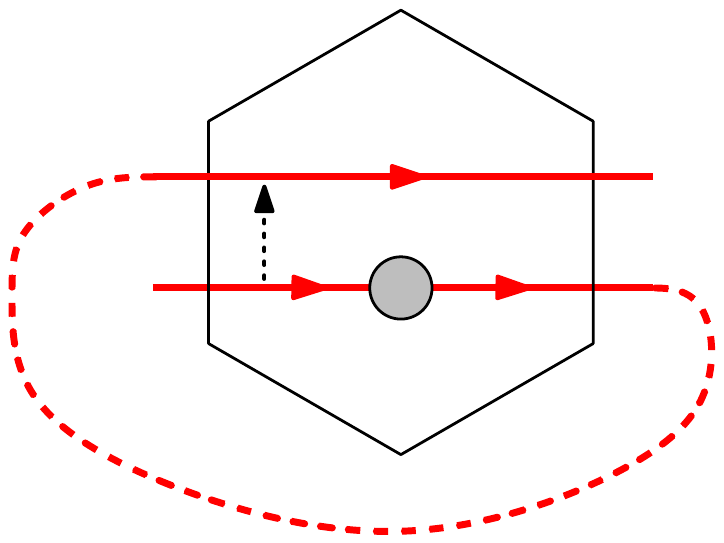}}
		&&\rightarrow 
		\raisebox{-.5cm}{\includegraphics[scale=.48]{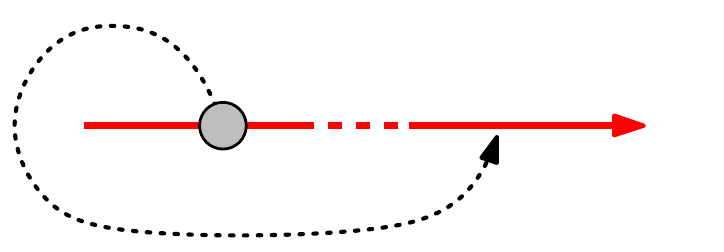}} \\
		& \boxed{+9} 
		\raisebox{-1.3cm}{\includegraphics[scale=.42]{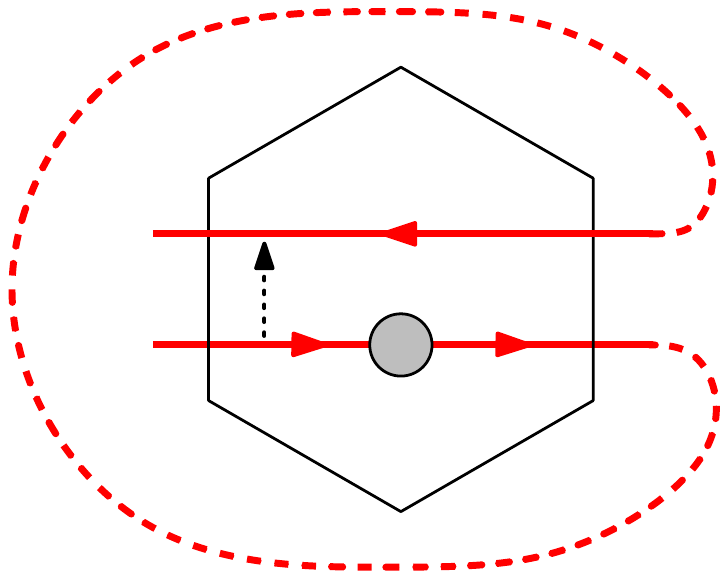}}
		&&\rightarrow 
		\raisebox{-.5cm}{\includegraphics[scale=.48]{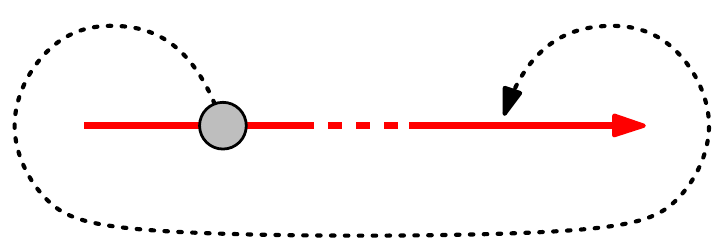}}
	\end{alignat*}
	The appropriate sequences of swaps for these four different situations are
	\begin{align*}
		&-3 \,: \quad S^{-1}[B]\,,\\
		&-6 \,: \quad S^{-1}[B] \, S^{-1}[H] \, S^{-1}[H^r]\,,\\
		&+6 \,: \quad S[T^r] \, S[T] \, S[B] \,,\\
		&+9 \,: \quad S[T^r] \, S[T] \, S[B] \, S[H] \, S[H^r]\,,
	\end{align*}
	where the notation $ S[H] $ stands for sequentially swapping the anyon with all entries in $ H $ in a clockwise fashion. $ H^r $ indicates the sequence of anyons in $ H $, in reversed order. 
	The paperclip configurations for the other cases are similar. All of them are listed together with the corresponding sequences of swaps in App.~\ref{sec:paperclip_list}.
	
\subsubsection{Fusion}
	The fundamental operation during recovery is a pairwise fusion process, where one excitation is moved along a specific path until it neighbors another anyon, followed by the fusion of the pair. 
	The physical implementation of these operations is discussed in Sec.~\ref{sec:recovery}. Here we discuss how they are implemented on the level of curve diagrams during the simulation.
	
	Before fusing neighboring anyons, one must ensure that they appear sequentially on the same curve diagram, which is by achieved using the paperclip algorithm described above.
	To simulate the fusion process, we must first isolate the two anyons from the rest of the fusion tree, which is done using an $ F $-move:	
	\begin{equation}\label{eq:F_diagram}
		\raisebox{-.5cm}{\includegraphics[scale=.38]{fig/cw_braid_diagram_initial.pdf}} = 
		\sum_{\bm{c}} F^{\bm{b}_1 \bm{a}_{j} \bm{b}_2}_{\bm{a}_{j+1} \bm{b}_3 \bm{c}} \;
		\raisebox{-.5cm}{\includegraphics[scale=.38]{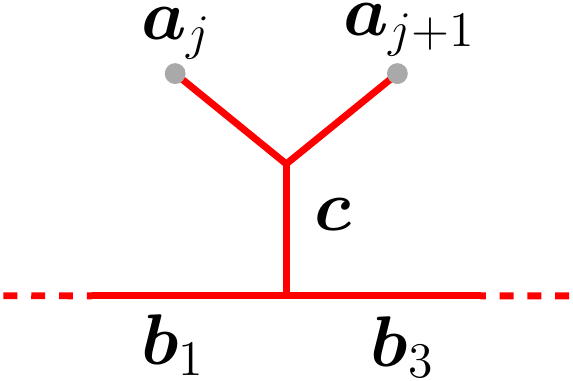}} \,.
	\end{equation}
	The result of the fusion of anyons $ \bm{a}_{j}  $ and $ \bm{a}_{j+1} $ is then picked from the possible values of their total charge $ \bm{c} $ using the probability distribution dictated by the coefficients of the state superposition and the state vector is then projected to the selected outcome. 
	Note that this probability distribution only concerns the resulting anyon charge of the plaquette. 
	In case the outcome $ \boldsymbol{c} = \tau\tau $ was selected, the tail label is picked from the probability distribution $ \{p(\1) = \frac{1}{\phi^2},\, p(\tau) = \frac{1}{\phi} \} $, which follows from Eq.~\eqref{eq:fusion_tail}.
	
	On the level of the curve diagrams, we must then remove one of the anyons as depicted below:
	\begin{equation*}
		\raisebox{-.92cm}{\includegraphics[scale=.4]{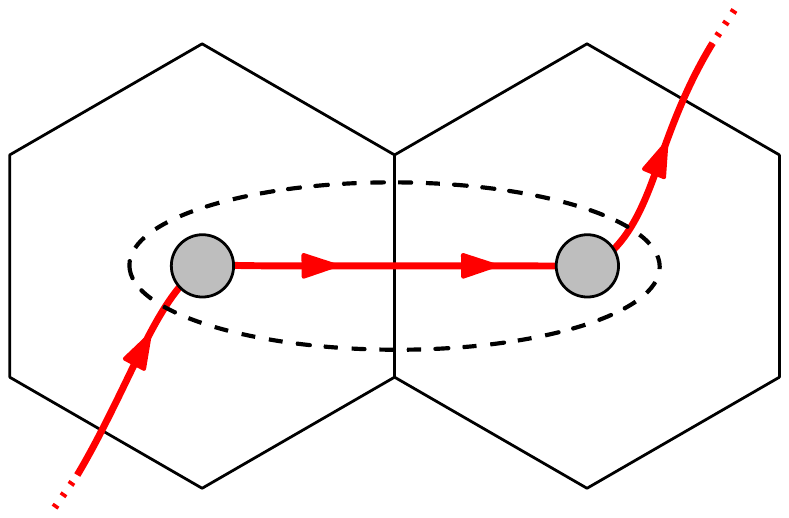}} 
		\quad \mapsto \quad
		\raisebox{-.92cm}{\includegraphics[scale=.4]{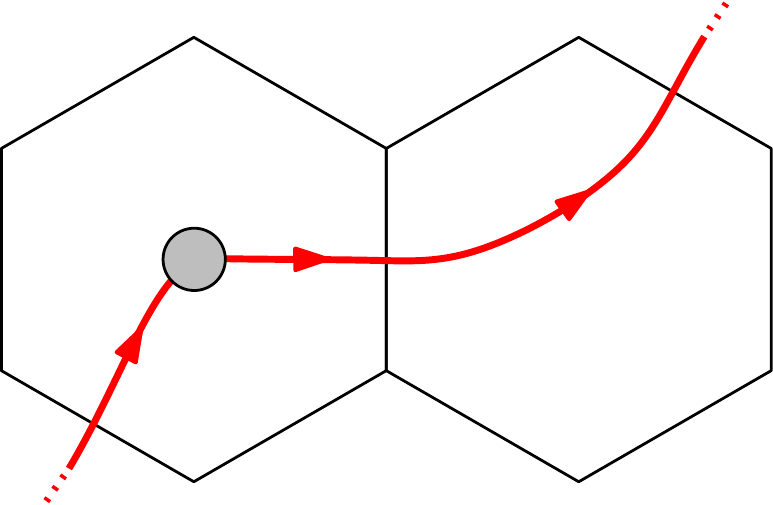}} \;\;.
	\end{equation*}
	Note that we are required to choose a ``target plaquette'' in which the fusion outcome is placed.

\subsubsection{Move and exchange} \label{sec:move_exchange}
	The moving procedure required to bring a pair of anyons to neighboring tiles can be broken down into a sequence of basic steps where an anyon is moved to a neighboring plaquette. 
	If the neighboring plaquette does not contain an anyon, the curve is updated by moving the excitation to the neighboring tile while continuously deforming the curves in the corresponding tiles: 
	\begin{equation}\label{eq:move}
		\raisebox{-.95cm}{\includegraphics[scale=.4]{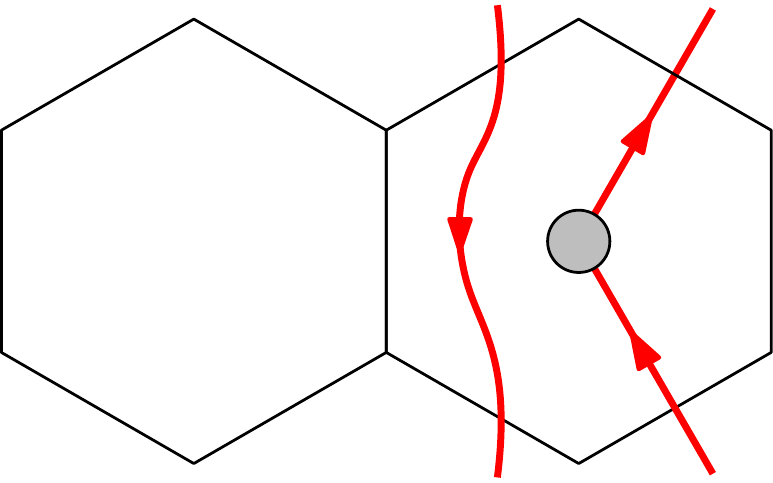}} 
		\quad \mapsto \quad
		\raisebox{-.95cm}{\includegraphics[scale=.4]{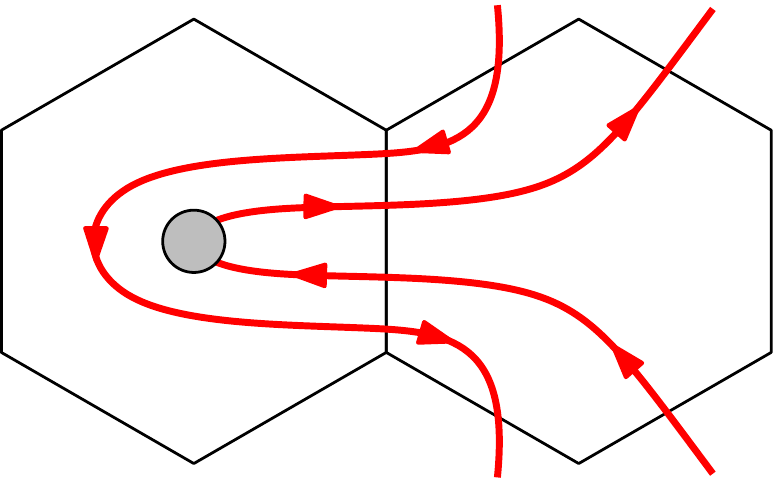}} \;\;.
	\end{equation}
	If the neighboring plaquette contains an anyon, the two anyons are exchanged in a clockwise fashion while deforming the curves in the corresponding tiles accordingly:
	\begin{equation}\label{eq:exchange}
		\raisebox{-.95cm}{\includegraphics[scale=.4]{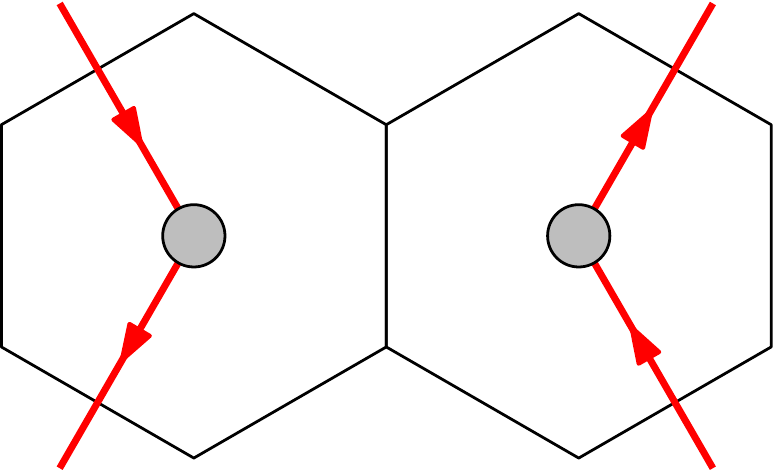}} 
		\quad \mapsto \quad
		\raisebox{-.95cm}{\includegraphics[scale=.4]{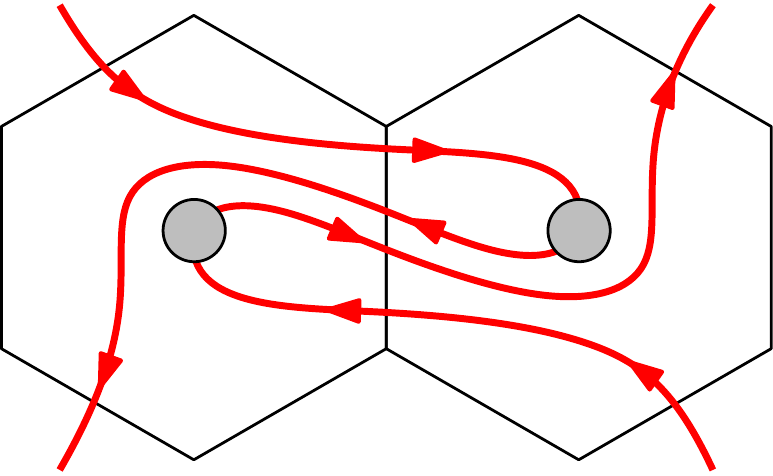}} \;\;.
	\end{equation}
		Note that the choice for a clockwise exchange over a counterclockwise one is arbitrary. A different choice would result in different probabilities for the fusion outcomes after the moving procedure, as remarked in Sec.~\ref{sec:simulating_non-Abelian}. 
	
	Both operations do not affect the coefficients appearing in the state vector. However, they \emph{do} modify the state by changing the corresponding basis elements. 
	Such a transformation can be expressed as 
	\begin{equation}\label{eq:active}
		\ket{\Psi} = \sum_i \alpha_i \ket{\psi_i} \mapsto \ket{\Psi'} = \sum_i \alpha_i \ket{\psi_i'} ,
	\end{equation}
	where the basis state $ \ket{\psi_i'} $ is obtained from the state $ \ket{\psi_i} $ by changing the embedding of the corresponding fusion tree on the surface as dictated by Eq.~\eqref{eq:move} or Eq.~\eqref{eq:exchange}.

\subsection{Outline of the simulation}\label{sec:outline_of_simulation}
		
	With all the groundwork completed, we are now ready to sketch the outline of the entire error correction threshold simulation, performed with the Fibonacci input category on a tailed hexagonal lattice with periodic boundary conditions in both directions (giving the lattice the topology of a torus).
	The simulation consists of a fixed number of Monte Carlo samples, each of which simulates the application of noise and recovery processes to an initial ground state. 
	The quantum state of the system is tracked throughout these processes until either a topologically nontrivial process occurs, in which case \emph{failure} is declared, or all anyonic excitations have been removed (without any logical errors), in which case \emph{success} is declared.
	
	For a given a system size and noise strength, the logical failure rate $ P_L $ is then found by the ratio of failures compared to the total number of Monte Carlo steps.
	Below, we describe in detail all the important aspects of a single Monte Carlo step, with system size  $ L \times L $ and a noise strength characterized by $ p $.\\
	
	\subsubsection{Cutoff parameters}
	In addition to logical errors, failure is also declared whenever any of two cutoff parameters are exceeded.
	The first of these is the maximal allowed tree size $\mathtt{ N_{max} }$. As discussed in Sec.~\ref{sec:classical_simulatbility}, the size of connected groups of anyons can grow very large in some rare occasions. Since the size of the associated fusion space grows exponentially, such situations are extremely costly, both in terms of memory usage and in computation time. 
	Hence we fix some cutoff size $\mathtt{ N_{max} }$, and the simulation is aborted, and a failure is declared, whenever the number of anyons on an individual curve exceeds this value. 	Note that a similar cutoff was used in Ref.~\cite{burton2017classical}.
	
	The second cutoff parameter that we introduce is $\mathtt{ V_{max} }$, which is the maximal number of nonzero coefficients we allow in the vector associated to any individual curve. 
	The motivation behind this cutoff rule is as follows. Since the state vectors appearing during the simulation are generally very sparse, these are stored as sparse arrays. This enables us the keep the value of $\mathtt{ N_{max} }$ higher than one would naively expect (as no memory is allocated to all zero entries, which form the vast majority of the exponentially large state vector).
	To avoid the extreme time and memory cost of the rare cases where any state vector contains a very large number of nonzero entries, such cases are aborted and a failure is declared. 
	
	One must choose these cutoff parameters to be as high as possible, in order to minimize the amount of triggered cutoffs, while still keeping the memory and time cost of the simulations reasonable. 
	Of course, any finite values of these parameters will negatively affect the observed logical failure rates, once we leave the regime in which events with very large connected groups are sufficiently rare.
	However we argue that their influence can only \emph{lower} the obtained threshold, meaning our results will provide a valid lower bound on the actual error correction threshold, independent of the values of $ \mathtt{N_{max}} $ and $ \mathtt{ V_{max} } $. 
	For a fixed value of $ p $, larger system sizes (on average) result in higher values for the size of the largest connected group of anyons (see App.~\ref{sec:scaling}). Hence, it is clear that the cutoffs will be triggered more often for larger system sizes.
	Likewise, for a fixed system size $ L $, larger values of $p $ will also result in more triggered cutoffs.
	When displaying the logical failure rate as a function of $p$, the intersection of the curves corresponding to different system sizes, will be shifted left compared to its true value (had the cutoff parameters been infinite), meaning that our obtained error correction threshold is indeed a valid lower bound to its unknown true value. 
	
	\subsubsection{Noise phase}	
	The system is initialized in a ground state of the code Hamiltonian Eq.~\eqref{eq:code_hamiltonian}, corresponding to the anyonic vacuum with some specific handle labels. 
	However, as explained in Sec.~\ref{sec:curve_diagrams}, there is no need to store these handle labels, since they do not affect any relevant processes. Processes in which the handle labels do affect the outcome are precisely those that result in a logical error. Because the Monte Carlo simulation will automatically declare a failure in those cases, such processes must never be simulated on the level of state-evolution.
	
	The first half of the simulation consists of sequentially applying $ T $ noise processes, described in Sec.~\ref{sec:noise_model}, to this initial ground state, where $ T $ is drawn from a Poisson distribution with mean $ p \, 5 L^2 $.
	%
	For each of these $ T $ steps, an edge $ e $ is chosen at random, and a Pauli operator $ \sigma_i $ is selected according to the relative probabilities $ \{\gamma_x, \gamma_y, \gamma_z\} $. 
	Before the matrix elements computed in Sec.~\ref{sec:matrix_elements} can be used to determine the state after the application of error $ \sigma_i^e $, one must first rewrite the state vector in the appropriate basis.	
	
	Given the orientation of $ e $ and the type of error $ \sigma_i $, the affected plaquettes can be read off from Figs.~\ref{fig:z_error_basis_convention} or \ref{fig:x_error_basis_convention}.
	If none of them contain a nontrivial charge initially, we simply create a new curve diagram with the appropriate shape and with a corresponding trivial fusion state (containing only vacuum charges). 
	If only some of these plaquettes contain a nontrivial charge, we can add vacuum charges to the fusion state of one of the nontrivial charges and modify (``grow'') the curve accordingly such that all affected plaquettes are included in the same cure.
	
\begin{figure}
		\centering
		\includegraphics[width=.7\columnwidth]{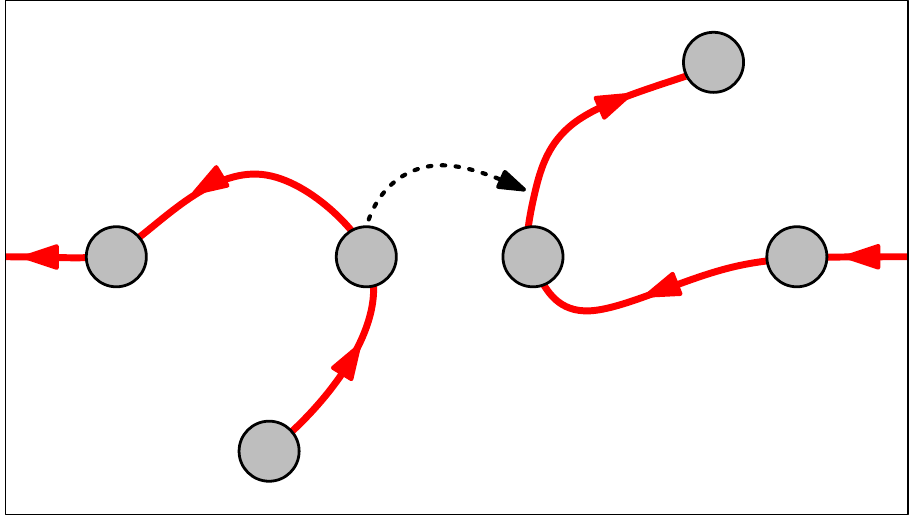}
		\caption{An illegal refactoring move on a torus, which will cause the simulation to declare \emph{failure} and abort.}
		\label{fig:illegal_refactoring}
	\end{figure}

	The desired basis is the one where the affected anyons appear sequentially along the same curve, in the order depicted in Figs.~\ref{fig:z_error_basis_convention} or \ref{fig:x_error_basis_convention}.
	This is obtained by applying a series of refactoring moves as described in Sec.~\ref{sec:paperclip}, 
	and performing the corresponding sequences of swaps and merge operations on the affected state vectors.
	If any of the required refactoring moves is topologically impossible, for instance the one depicted in \figref{fig:illegal_refactoring}, this indicates that a logical error would be caused, as the joint path of the curve diagram and the interaction path form a non-contractible loop. Whenever this happens the simulation declares \emph{failure} and aborts the current Monte Carlo step.
	
	If all refactoring moves above were legal, a sequence of $ F $-moves is performed to isolate the affected anyons from the rest of the fusion tree. 
	This transforms the standard fusion tree shape to the one in Eq.~\eqref{eq:z_error_basis}, Eq.~\eqref{eq:x_error_basis}, or a similar shape in the case of three affected plaquettes.
	In terms of ribbons in the fattened lattice, the basis then locally looks like the ones depicted in \figref{fig:z_error_basis_convention} or \figref{fig:x_error_basis_convention}, depending on the type of error.
	Note that such a basis can no longer be represented using curve diagrams. This is not an issue, since we will return to a standard basis before the full machinery of curve diagrams is required again.
 
	For the case of a $ \sigma_z $ error, the coefficients in the state vector and the matrix elements Eq.~\eqref{eq:noise_mat_el_5} are used to calculate the probabilities Eq.~\eqref{eq:z_noise_charge_prob} of the different possible charge measurement outcomes.
	A result is picked according to this probability distribution, after which the state vector is updated using Eq.~\eqref{eq:z_noise_final_state}. 
	 
	In the case of a $ \sigma_x $ error, the matrix elements \eqref{eq:noise_mat_el_1} are used together with the coefficients in the state vector to compute the probabilities Eq.~\eqref{eq:x_noise_vert_prob_red} for the outcome of the vertex and tail measurements.
	A result is then sampled from this distribution. Next, the matrix elements Eq.~\eqref{eq:noise_mat_el_2} are used to compute the probabilities Eq.~ \eqref{eq:x_noise_charge_prob} for the various outcomes of the charge measurement (that is performed after the appropriate unitary vertex correction was applied). 
	We again pick a result according these probabilities, and update the state vector using Eq.~\eqref{eq:x_noise_final_state}. 
	In case of a  $ \sigma_y $ error, we proceed analogously to the $ \sigma_x $ case, using the matrix elements Eq.~\eqref{eq:noise_mat_el_3} or Eq.~\eqref{eq:noise_mat_el_4}.

	After the state vector has been updated to reflect the collective effect of the noise and the measurements (with the specific outcomes that we picked at random above), we conclude the current ``noise step'' by transforming the fusion basis back to the one that corresponds to the curve diagram we ended up with earlier (where all affected anyons appear sequentially).
	This is done by a sequence of $ F $-moves that reverts the transformation performed by the first series of $ F $-moves.
	
	\subsection{Recovery phase}
	After completing the process above $ T $ times, the error syndrome is given by the locations and charges of all thermal anyons on the lattice. The decoding algorithm is then used to determine an appropriate recovery step. Such a recovery step can be broken down into a sequence of pairwise fusion processes of anyonic excitations. 
	In each such a pairwise fusion process, one member of the pair is moved along a specific path on the lattice until it neighbors the other. 
	The moving procedure consists of basic operations where the anyon is moved to a neighboring tile and the curves are continuously deformed accordingly, as described in Sec.~\ref{sec:move_exchange}. This basic moving step is repeated until the two anyons reside in neighboring tiles. 
	A sequence of refactoring moves (and possibly merges) is then performed to obtain a basis in which they are direct neighbors on the same curve, and the corresponding sequences of swap and merge operations are applied to the affected state vectors.
	As during the noise process, any illegal refactoring moves cause the simulation to abort the current Monte Carlo step and report a decoding failure. Note that this happens precisely when the intended fusion would create a non-contractible loop in terms of the curve diagrams, which in turn corresponds to a logical error.
	
	If necessary, an $ F $-move is applied to transform to a fusion tree shape where the anyons are fused directly.
	Their resulting charge is then projected according to the probabilities dictated by the coefficients in the state superposition, and it is placed in one of the two neighboring plaquettes while the state vector and the curve diagram are updated accordingly.
	
	This basic pairwise fusion process is repeated until the current recovery step is completed, at which point the resulting error syndrome is used to determine the next recovery step. This dialogue is iterated until either all anyonic excitations fused away and decoding is successful, or a logical error occurs during some fusion process and a decoding failure is declared.
	
	Note that throughout the entire Monte Carlo step, the system is constantly monitored for violations of the cutoff parameters. 
	If at any point an individual curve contains more than  $\mathtt{ N_{max} }$ anyons, or a state vector contains more than $\mathtt{ V_{max} }$ nonzero elements, the current Monte Carlo step is automatically reported as a failure.

%% file: sections/decoders.tex
\section{Decoding algorithms}\label{sec:decoding}


After the error syndromes (anyon charges) on all the plaquettes are measured using the circuit discussed in Sec.~\ref{sec:measure_charge}, this syndrome information is passed to a decoder, which in turn outputs the recovery operations required to correct the errors.

In this section we discuss two types of decoders. \textbf{(A)} The first type is the clustering decoder which has been previously applied to decode a phenomenological model of Fibonacci anyons \cite{burton2017classical}. This decoder is based on a hierarchical clustering algorithm \cite{wootton2016active} and shares a similar strategy to the hard-decision renormalization-group decoder \cite{Bravyi:2013ez}. The clustering decoder does not use the detailed syndrome information corresponding to the anyon type. Instead, it just uses the limited syndrome information of the presence of absence of anyon, i.e., whether the anyon charge is nontrivial or trivial (in the doubled vacuum sector $\mathbf{11}$). 
\textbf{(B)} The second type is a fusion-aware iterative minimum-weight perfect matching (MWPM) decoder which modifies the standard MWPM algorithm and incorporates the detailed syndrome information corresponding to anyon type into the decoding strategy.  As a comparison, we also show the results of a ``\textit{blind}'' iterative MWPM decoder which does not use the detailed syndrome information of anyon type but only the presence or absence of anyon. Due to the use of the detailed syndrome information, the logical error rate of the fusion-aware iterative MWPM decoder is lower than the other one.

\begin{figure}
		\centering
		\includegraphics[width=.8\columnwidth]{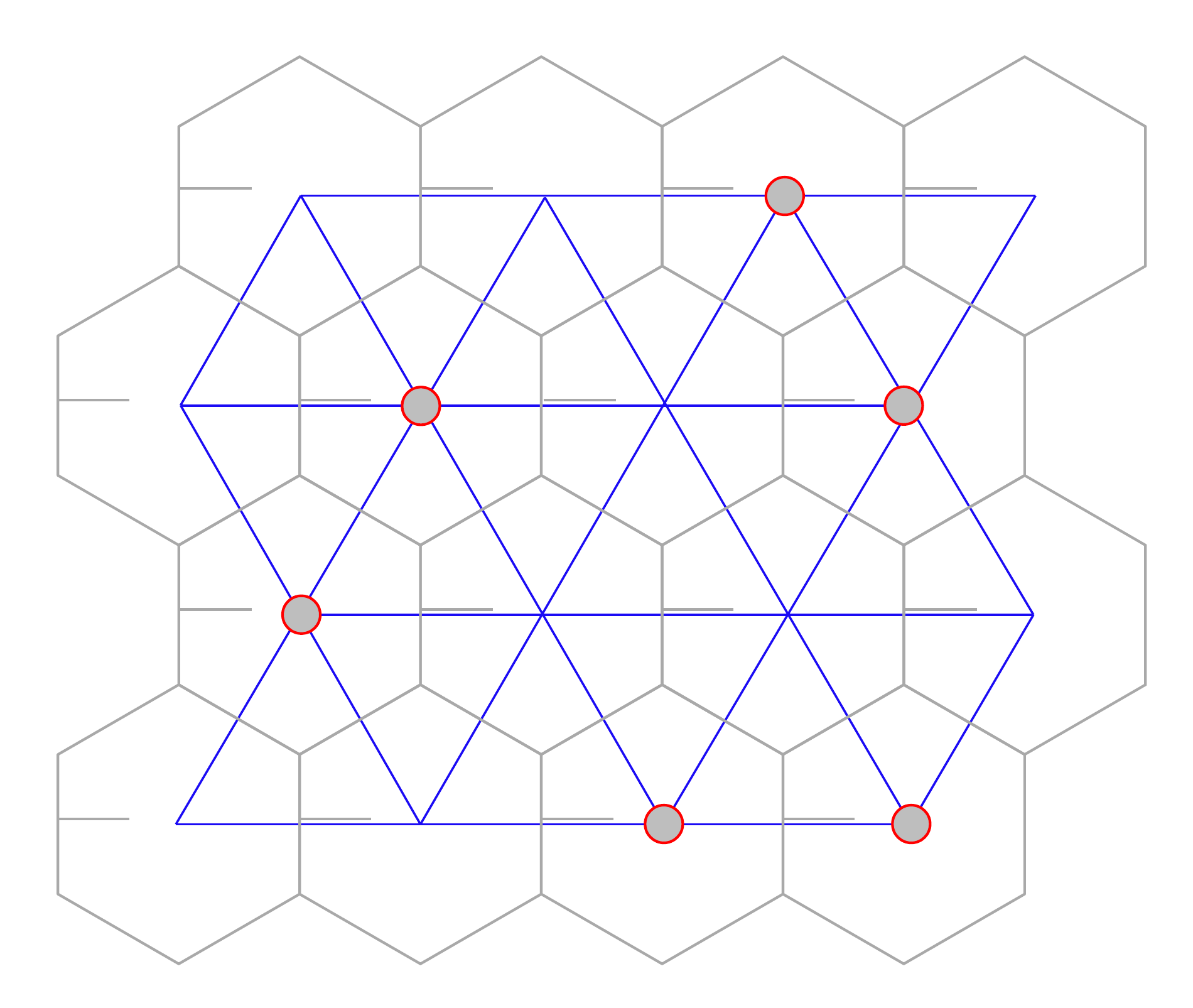}
		\caption{The decoding graph of the extended string-net code. By connecting the center of the plaquettes of the tailed lattice, we obtain a triangular lattice (blue) as the decoding graph.  The anyons charge (grey circles) on the plaquettes of the tailed lattice serve as the syndromes in the error correction scheme and are located on the vertices of the triangular decoding graph.}
		\label{fig:decoding_graph}
	\end{figure}

We will indicate the syndrome using a decoding graph, shown in Fig.~\ref{fig:decoding_graph}.  
This is a triangular lattice which is the dual of the original trivalent graph on which the extended string-net code is defined. 
Anyon charges located in the plaquettes of the trivalent graph, correspond to syndromes on vertices of this triangular decoding graph.  


\subsection{Clustering decoder}\label{sec:clustering_decoder}
The spirit of the clustering decoder is based on the charge conservation of anyons. As discussed in Sec.~\ref{sec:Pauli-noise}, local noise can generate certain pairs or, more generally, clusters of anyons.
Since these anyon clusters are generated from the vacuum sector $\mathbf{11}$, i.e., the ground space of the extended string-net code, the total charge of each anyon cluster should still be trivial ($\mathbf{11}$) due to charge conservation. 
Therefore, if we fuse all the anyons within a particular cluster by moving them towards the same plaquette, we will get a total trivial vacuum charge $\mathbf{11}$, i.e.,  all the anyons are annihilated back to the vacuum sector. 

However, the decoder does not know which cluster each anyons were generated from based on the syndrome information.  Therefore, the decoder may not always apply a correct recovery operation to fuse all the anyons originating from the same cluster. 
Instead, the decoder may fuse anyons from different clusters, effectively joining the two clusters.  Due to the total charge conservation and the fusion rule $\mathbf{11} \times \mathbf{11} = \mathbf{11}$, we know the total anyon charge of the two clusters is still zero.  If we fuse all the anyons in these two clusters together, all the charges will still be annihilated into the vacuum  $\mathbf{11}$. The same argument applies to merging multiple clusters.  As long as the size of the joined cluster is much smaller than the system size,  all the errors can be corrected by merging them into the vacuum $\mathbf{11}$.   
On the other hand, if the joined anyon cluster forms a non-contractible (homologically nontrivial) region, i.e., either wrapping around a cycle of a torus or more generally a high-genus surface in the context of a closed manifold, or connecting two or more gapped boundaries in the context of an open manifold,  the recovery operation with the merging procedure may still annihilate all the anyons but end up applying a nontrivial logical operator along certain homologically nontrivial cycle \footnote{In the case of open manifold with gapped boundaries,  the logical operator corresponds to nontrivial relative homology cycle.} which will cause a logical error and the decoder fails. It is also possible that such a non-contractible anyon cluster will have a nonzero total charge, therefore there is some residual anyon after the merging procedure which cannot be annihilated.   In both cases, we will claim a failure of the clustering decoder.  Therefore, in order to apply a successful recovery operation, we need to make sure that the individual clusters are annihilated before growing to a size comparable to the system size, i.e., with a linear dimension comparable to the code distance. 

The clustering decoding algorithm for the Fibonacci Turaev-Viro code defined on a torus is summarized below by the pseudo-code and will be explained in detail with a concrete example:

\noindent\makebox[\linewidth]{\rule{\columnwidth}{0.4pt}}

\textbf{Algorithm (clustering decoder)}

\noindent\makebox[\linewidth]{\rule{\columnwidth}{0.4pt}}

\begin{small}
\nin \textit{\# Measure the anyon charge of each plaquette on the tailed trivalent lattice; store the nontrivial anyon charge in a list ``anyon\_charge"}

\nin \texttt{anyon\_charge = get\_syndrome(state);}

\

\nin \textit{\# Initialize a cluster for each plaquette with a nontrivial charge}

\nin \texttt{clusters = Cluster(anyon\_charge);}

\

\nin \textit{\# Join any connected clusters}

\nin \texttt{join(clusters, 1);}

\

\nin \textit{\# While there is more than one nontrivial charge}

\nin \textbf{\texttt{while}} \texttt{size(anyon\_charge)>1} 

\

\nin \qquad \textit{\# Fuse all anyons within each cluster and measure }

\nin \qquad \ \ \ \textit{the resulting charge}

\nin \qquad \textbf{\texttt{for}} \texttt{cluster in clusters}

\nin \qquad \qquad  \texttt{fuse\_anyons(cluster);}

\

\nin \qquad \textit{\# Check (in the simulation) whether any anyon path becomes }

\nin \qquad \ \ \ \textit{non-contractible, i.e., homologically nontrivial after the }

\nin \qquad \ \ \ \textit{fusion process}

\nin \qquad \texttt{\textbf{if} anyon\_path == non-contractible}

\nin \qquad \qquad  \textit{\# The decoder claims failure}

\nin \qquad \qquad  \texttt{return} \textbf{\texttt{Failure}}

\

\nin \qquad \textit{\# Discard any empty cluster with trivial vacuum charge $\mathbf{11}$;}

\nin \qquad \texttt{clusters = non\_vac(clusters);}

\ 

\nin \qquad \textit{\# Grow each cluster by a unit length on the triangular }

\nin \qquad \ \ \ \textit{decoding graph}

\nin \qquad \texttt{grow(clusters, 1);}

\

\nin \qquad \textit{\# Join any overlapping clusters}

\nin \qquad \texttt{join(clusters, 0);}

\

\nin \textbf{end}

\

\nin \textit{\# If the list of nontrivial anyon charge is empty, i.e., with no remaining anyon}

\nin \texttt{\textbf{if} anyon\_charge == []}

\nin \qquad \textit{\# The decoder declares success} 

\nin \qquad \texttt{return} \textbf{\texttt{Success}}

\nin \textit{\# If there is a single nontrivial remaining charge}

\nin \texttt{\textbf{else}}

\nin \qquad \textit{\# The decoder claims failure} 

\nin \qquad \texttt{return} \textbf{\texttt{Failure}}
\end{small}

\noindent\makebox[\linewidth]{\rule{\columnwidth}{0.4pt}}

\vspace{0.1 in}

As described above, for a given state of the entire system, we first measure the tube operator within each plaquette of the tailed lattice using the circuit in Figs.~\ref{fig:grow_circuit} and \ref{fig:charge_measurement_circuit_1} and obtain the corresponding anyon charge as the syndromes on the decoding graph. This process is defined as \texttt{get\_syndrome()} in the decoding algorithm, with the \texttt{state} as the input and \texttt{anyon\_charge} as the output. Now we  initialize a cluster on each non-zero anyon charge, defined as \texttt{clusters=Cluster(anyon\_charge)} in the above algorithm.  An example for the decoding graph on a torus with the initial \texttt{clusters} (blue shadow) is shown below:

\begin{center} 
\includegraphics[width=1\columnwidth]{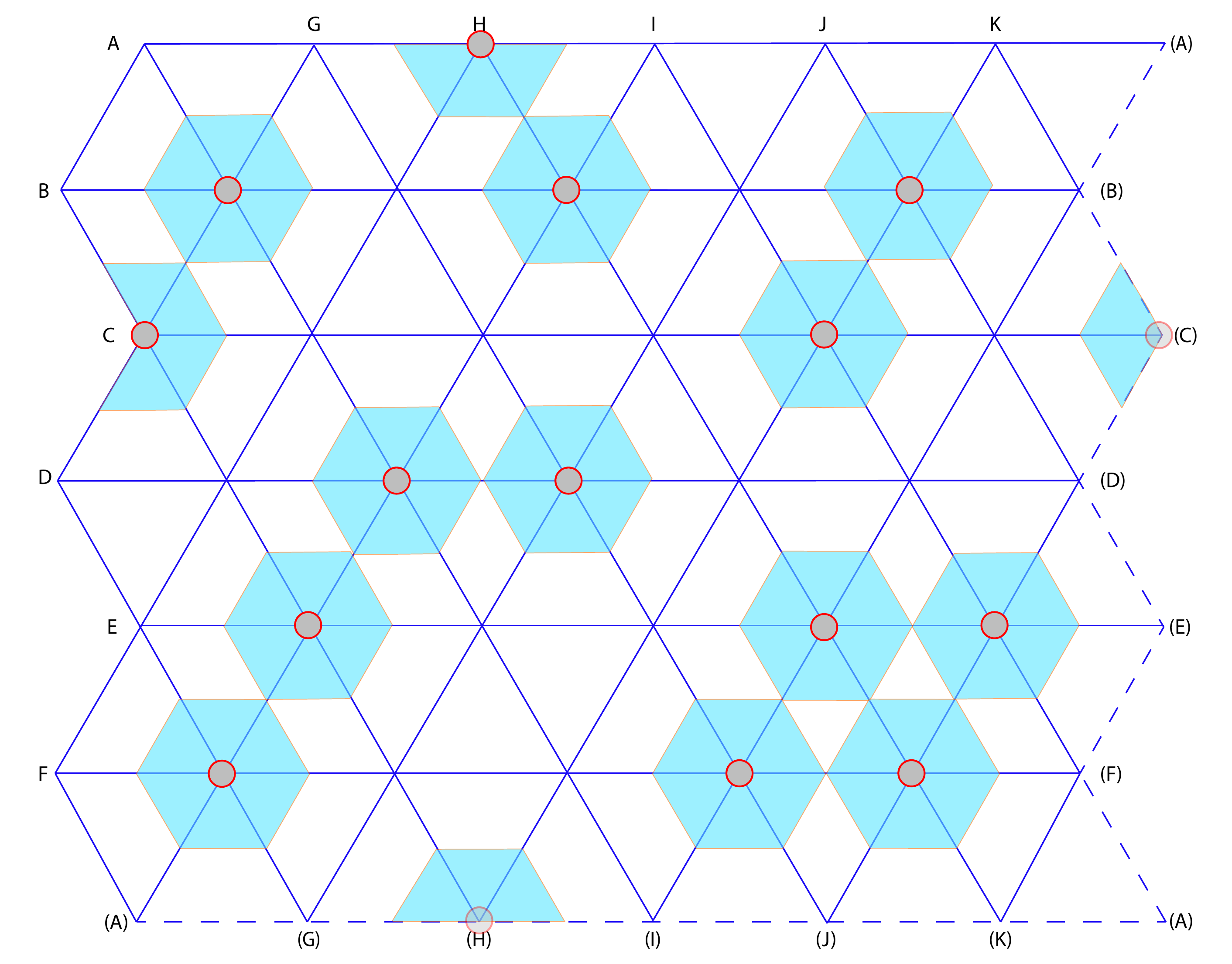}
\end{center}

In the next step,  we call \texttt{join(clusters, 1)} to join neighboring clusters which are separated by only 1 lattice spacing.  Within each cluster we randomly choose a root anyon, and \textit{move} all the other anyons to the root and finally fuse all of them into a single anyon.  The \textit{move operation} is the  recovery operation introduced in Sec.~\ref{sec:move_operation} and Fig.~\ref{fig:move_protocol} and can be simulated classically via modifying the curve diagram as shown in Eqs.~\eqref{eq:move} and \eqref{eq:exchange} in Sec.~\ref{sec:move_exchange}. We note that the order of moving and detailed path do not affect the resulting state after fusing all the anyons within each order, and therefore we could choose an arbitrary order.  The only requirement is that the chosen paths must be inside each cluster.  For simplicity, we choose the shortest paths (with shortest graph distance) towards the root anyons from all the remaining anyons (indicated by the red arrows in the figure below), and start moving the anyons in an order with an increasing graph distance between the remaining and root anyons in our numerical simulation.  

\includegraphics[width=1\columnwidth]{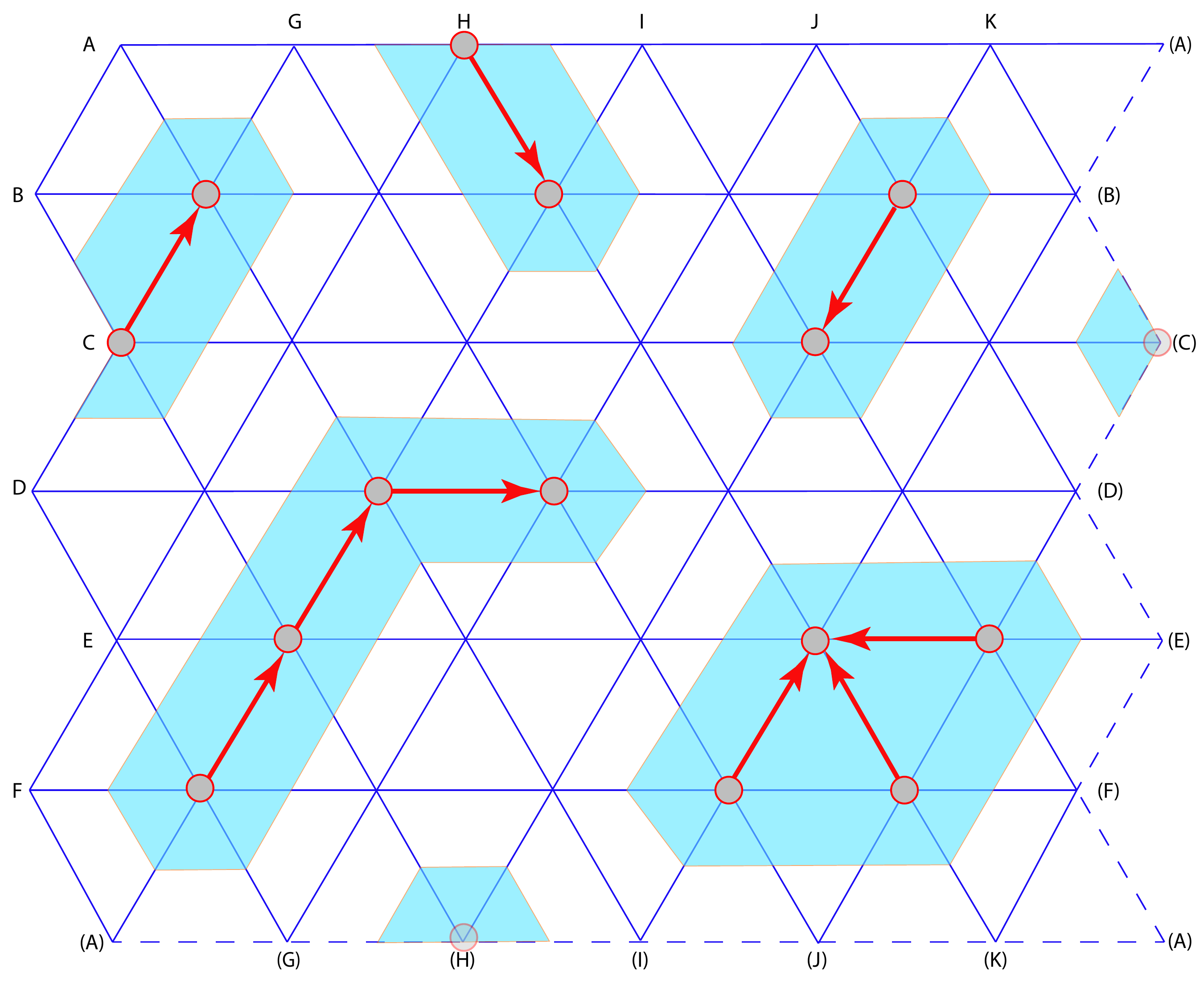}

After the fusion process, we check in the classical simulation whether any anyon path $l$ forms a non-contractible (homologically nontrivial) cycle, which will give rise to a logical error as previously illustrated in Fig.~\ref{fig:logical_error}. The anyon path here is the sum of the error path and the recovery path: $l=l_e+l_r$. We note that, although the decoder itself does not have access to such information, our classical simulation with underlying curve diagrams does. We can hence claim \texttt{failure} of the decoder in our Monte Carlo simulation, if such non-contractible cycle occurs.   

If the decoder does not fail, we continue to measure all the charges of the root anyons.  If the measured charge in a particular cluster is zero, i.e., in the vacuum sector $\mathbf{11}$, we discard the corresponding cluster.  The list of clusters hence gets updated via $clusters $$=$$non\_vac(clusters)$.

\includegraphics[width=1\columnwidth]{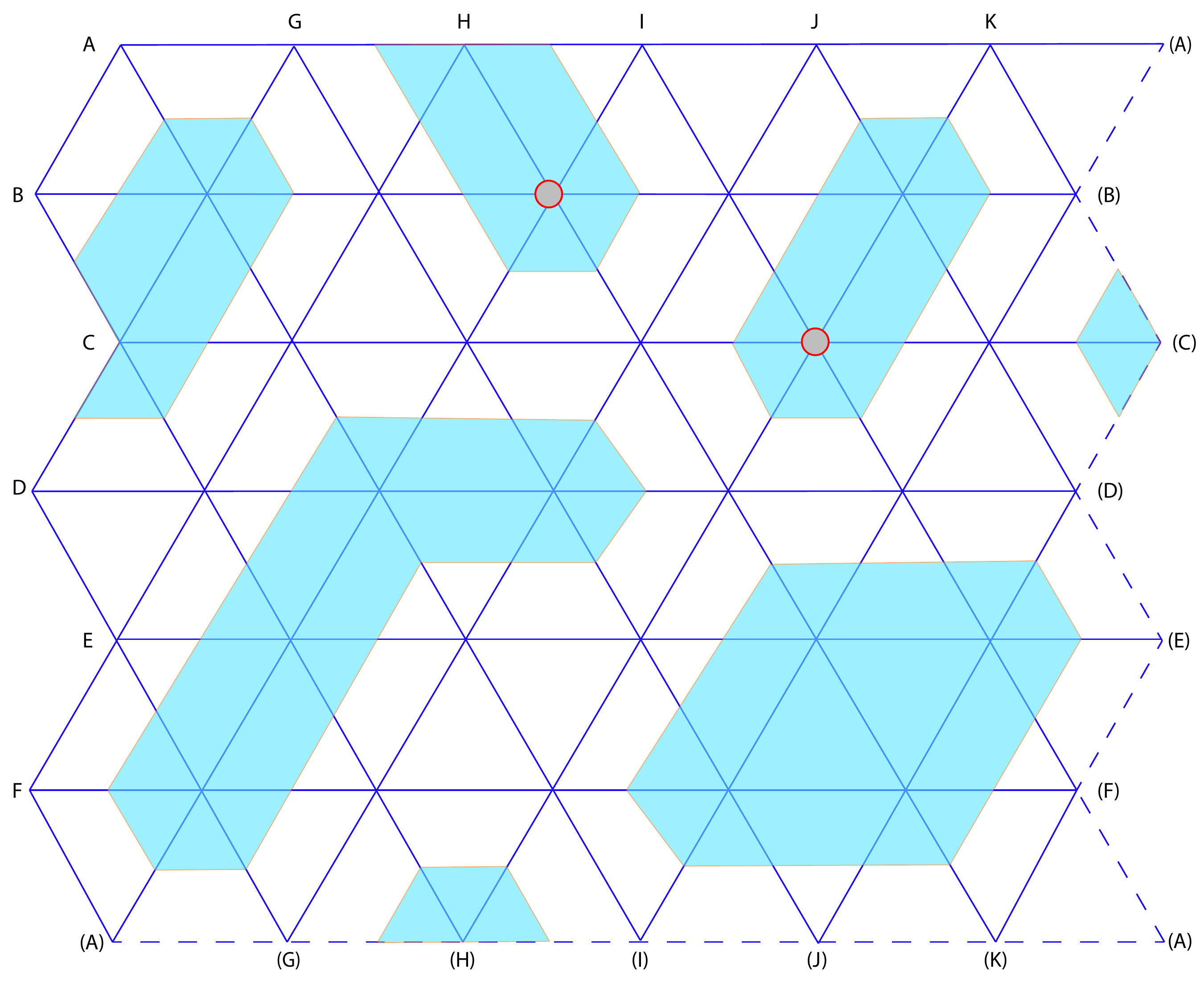}

In the next step, we need to call \texttt{grow(clusters, 1)} to grow each remaining cluster by one unit of lattice spacing in all possible directions (six directions for each vertex within the cluster), as indicated by the blue arrows in the figure below. As shown in the figure, the remaining two clusters (blue) grow to the larger clusters (green and orange): 

\includegraphics[width=1\columnwidth]{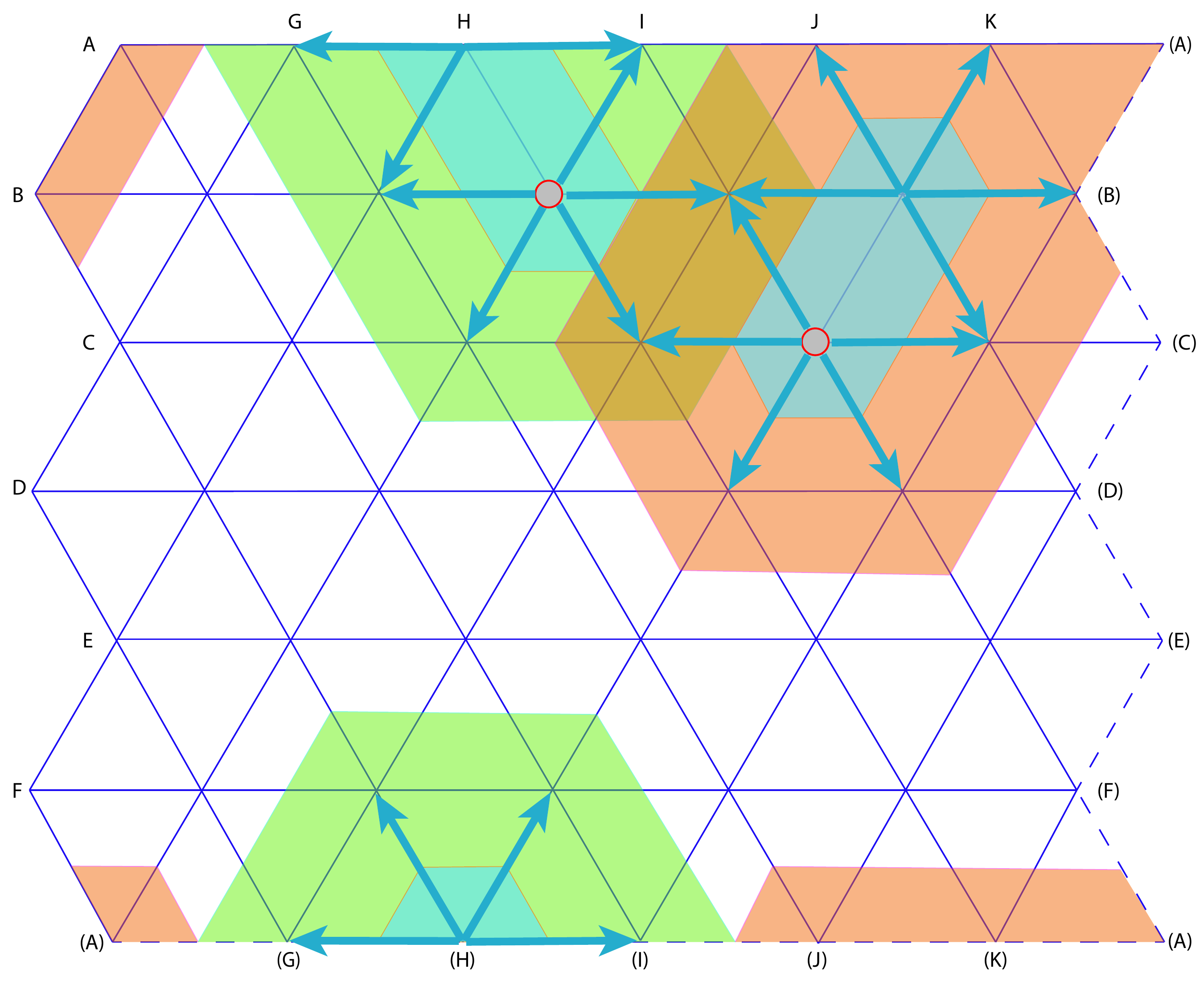}

Next, we call \texttt{join(clusters, 0)} to join overlapping clusters, i.e., any two clusters sharing a common set of vertices will be joined into a single one and this process is done iteratively until there are no overlapping clusters.   In the current example, the green and red clusters in the above figure are joined into a single bigger cluster (blue) in the following figure:  

\includegraphics[width=1\columnwidth]{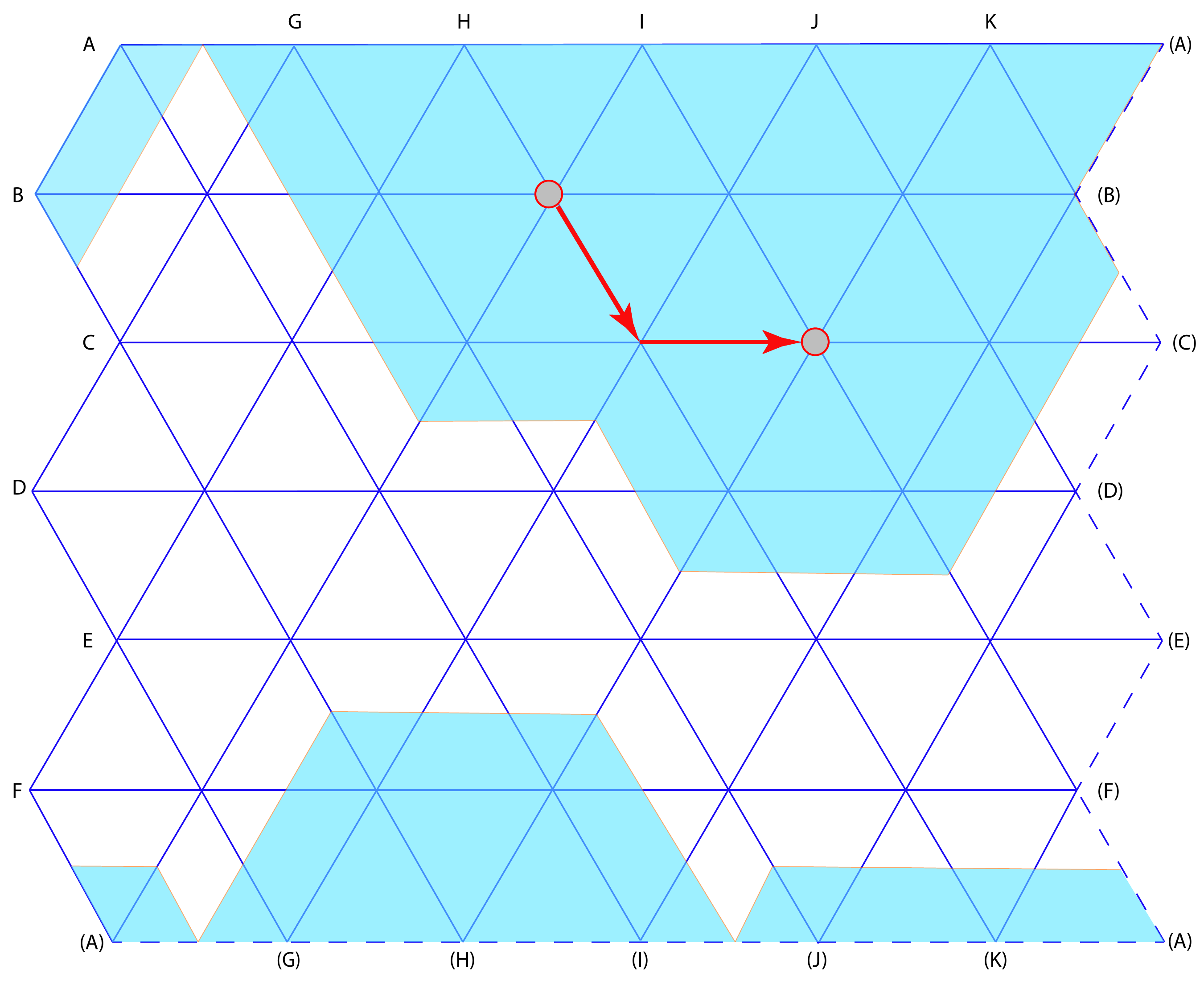}

After the merging of clusters, we repeat the above process, i.e., fuse all the anyons within each cluster (indicated by the red arrows in the above figure) and discard the empty cluster with trivial total charge $\mathbf{11}$ (vacuum), and then further grow and join the remaining clusters.  
 This iteration is stopped when we have zero or one remaining nontrivial anyon charge.  
 
After the end of the above iteration, we are at the final stage of our decoding algorithm.  If there is  no any remaining nontrivial anyon charge, i.e., 
\texttt{anyon\_charge == []},  we then have successfully corrected all the errors, and the decoder declares \texttt{success}. In the other situation with a single  nontrivial remaining anyon, we end up with a logical error, and will claim \texttt{failure} of the decoder.

\begin{figure}
		\centering
		\includegraphics[width=.8\columnwidth]{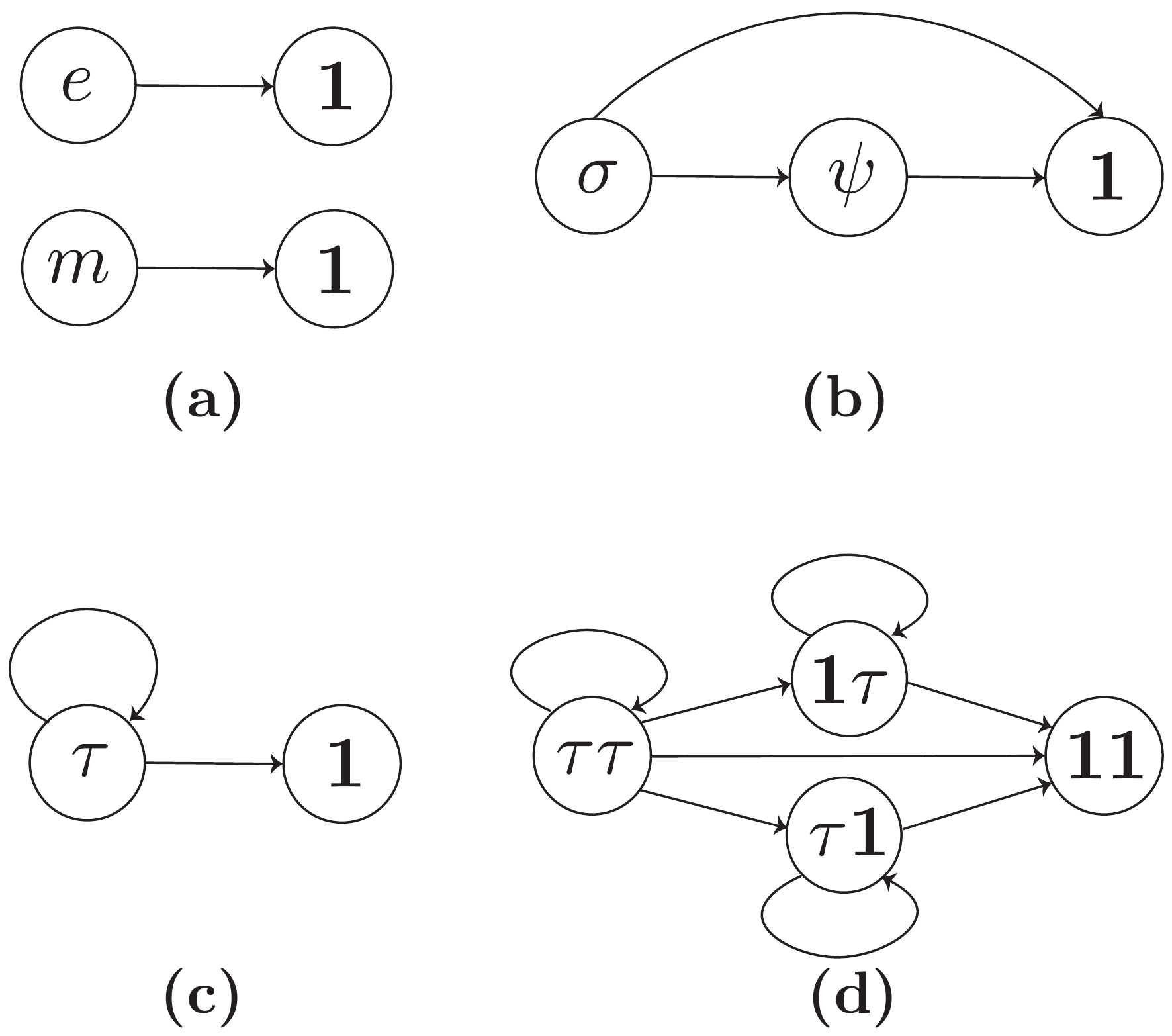}
		\caption{A graphic representation of the fusion rules (fusion graph) of (a) Toric code; (b) Ising category; (c) Fibonacci category (FIB); (d) doubled Fibonacci category (DFIB) corresponding to the Fibonacci Turaev-Viro code.  The fusion graph is a directed graph with the directed edge pointing from pair of incoming anyons represented by one vertex to the possible fusion outcome represented by another vertex. (a) and (b) have non-cyclic fusion structure, while (c) and (d) are cyclic. In (d), $\tau\tau$ can refer to either refer to $\tau\tau_\mathbf{1}$ or $\tau\tau_\tau$ }
		\label{fig:fusion_diagram}
	\end{figure}

\begin{figure*}
		\centering
		\includegraphics[width=2\columnwidth]{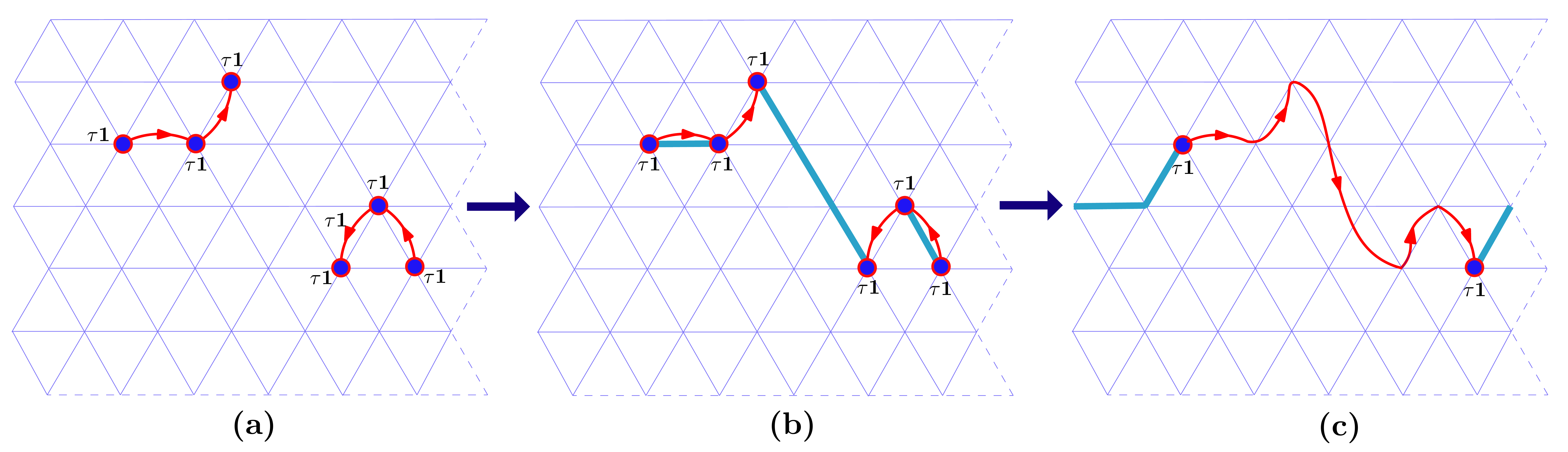}
		\caption{Illustration of the absence of threshold when applying the usual MWPM algorithm to the Fibonacci Turaev-Viro code. (a) Two triplets of $\mathbf{1}\tau$ anyons are created from the vacuum by $O(1)$ errors and are spatially separated apart at the order of code distance: $O(d)$. The underlying fusion structure can be seen from the curve diagrams, where the blue circles represent anyons with charge $\tau \mathbf{1}$ and the red directed edges specify the structure of the fusion basis. In particular, the labels on the circles specify the anyon charges. (b) When applying  the MWPM algorithm, two of the anyons in each triplet will be matched, while the remaining one in one triplet will be matched with the one in the other triplet. We show the result of matching by highlighted (light blue) edges on the decoding graph. For each pair of the matched anyons, one of the anyons (randomly chosen) will be transported along the highlighted path to fuse with the other. (c) After fusing the matched anyons, we have two remaining $\tau \mathbf{1}$ anyons, while the underlying curve diagram and the corresponding fusion trees of the ribbon graph states are merged. Now one further apply MWPM again to pair up the remaining anyons. A short path (indicated by the highlighted edges) will be chosen by the matching algorithm. One will fuse the anyons along the path selected by the matching algorithm and the anyon path will form a non-contractible cycle corresponding to a logical error. }
		\label{fig:MWPM_failure}
	\end{figure*}

\subsection{Fusion-aware iterative matching decoder}

The fusion-aware minimum-weight perfect matching (MWPM) decoder applies matching on the anyons with a preferred order of matching certain types of anyons according to the underlying structure in the anyon pair generation under the Pauli noise, as discussed in Sec.~\ref{sec:Pauli-noise}. 

As we have stated before, the non-Abelian Fibonacci fusion rule $\tau \times \tau = \mathbf{1} + \tau$ 
implies that a typical fusion of anyons in DFIB has probabilistic outcome into one of several fusion channels, e.g., $\tau \mathbf{1} \times \tau \tau = \mathbf{1} \tau + \tau \tau $. 
This is in stark contract to Abelian anyon models, where fusion processes are always deterministic. 
This difference can be seen using a graphical representation of the fusion rules (which we will call a \textit{fusion graph} from now on) as shown in Fig.~\ref{fig:fusion_diagram}(a,c,d). 
Since standard minimum-weight perfect matching (MWPM) decoder of the Abelian surface code is based on deterministically fusing pairs of Abelian anyons into the vacuum sector, this type of decoder will not work properly when applied to the Fibonacci string-net code.  

In Ref.~\cite{brell2014thermalization},  a MWPM decoder has been applied to the non-Abelian phenomenological Ising-anyon model, corresponding to the Ising category $\{\mathbf{1}, \sigma, \psi\}$. The corresponding fusion rules are: 
\be\label{Ising-fusion}
\psi \times \psi = \mathbf{1}, \quad \psi \times \sigma = \sigma,  \quad \sigma \times \sigma = \mathbf{1} + \psi,
\ee
as illustrated in the fusion graph  Fig.~\ref{fig:fusion_diagram}(b).  The basic strategy there is to apply MWPM to pair up and fuse a particular type of anyons: the $\sigma$-anyons. As indicated by the above fusion rule, the matched pair of anyons either fuse into the vacuum or the $\psi$-anyon.  When new $\psi$-anyons are generated from fusion, one further applies MWPM to pair and fuse all the pre-existing and newly generated $\psi$-anyons. Afterwards, one can clean up all the anyons and hence correct all the errors if the decoder succeeds.

For the Fibonacci Turaev-Viro code, we can adopt a similar strategy.   We first apply MWPM to pair up and fuse all the existing anyons, which will either fuse to vacuum or a certain type of anyons.   We then apply MWPM again to match the newly-generated anyons.  We iterate above process until all the anyons are annihilated into the vacuum and then declare \texttt{success} of the decoder if no logical error occurs. If there is any residual anyon or a logical error corresponding to  any anyon-path forming a non-contractible cycle, we claim \texttt{failure} of the decoder.  One can further introduce a particular order for merging different types  of anyons, similar to the above case in the phenomenological Ising-anyon model in Ref.~\cite{brell2014thermalization}.

\begin{figure*}[t]
		\centering
		\includegraphics[width=2\columnwidth]{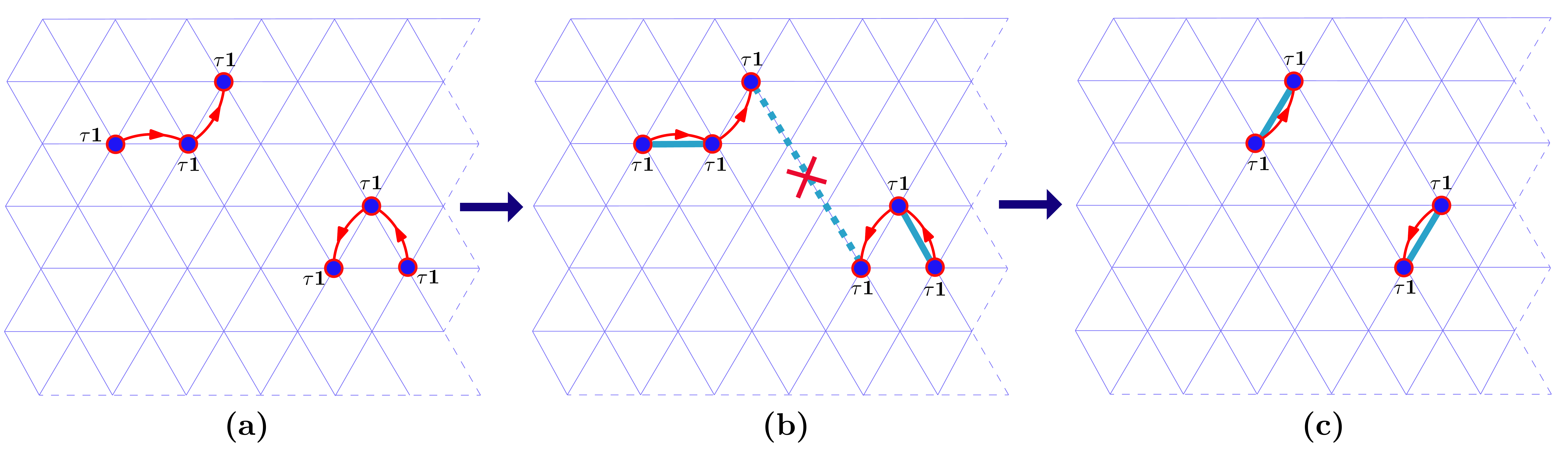}
		\caption{Resolve the issue in the previous example in Fig.~\ref{fig:MWPM_failure} by applying the iterative MWPM algorithm. In panel (b), one applies MWPM algorithm to match three pairs of anyons.  The matched pair involving anyons from different triplet clusters has a graph  distance 3, exceeding the fusing radius $r=1$ in the first iteration.  Therefore, the fusion of this matched pair is rejected. In panel (c),  the two previously matched pairs were fused and there are four remaining anyons. Apply another MWPM will match two pairs of anyons within their own cluster and hence successfully correct the errors.}
		\label{fig:MWPM_iterative}
	\end{figure*}

However, this type of MWPM decoder does not have a threshold.  The underlying reason is related to the particular fusion structure of the Fibonacci and doubled-Fibonacci categories. As shown in Fig.~\ref{fig:fusion_diagram}, the Ising category has a \textit{non-cyclic} fusion structure \cite{dauphinais2017fault} such that the fusion graph does not have any cycle (an edge pointing from and to the same vertex, i.e., the same anyon type). In contrast, both the Fibonacci category (FIB) and the doubled Fibonacci category (DFIB) have a \textit{cyclic} structure, i.e., containing at least a cycle in the fusion graph, which complicates the fusion process.  As we can see, the Ising fusion rule naturally implies a pair creation process: a pair of anyons, either two $\sigma$-anyons or two $\psi$-anyons are generated from the vacuum $\mathbf{1}$. Applying the matching can immediately annihilate these pairs into the vacuum. If one of the $\sigma$-anyon might further split into a $\sigma$-anyon and a $\psi$-anyon, we get a triplet $(\sigma, \sigma, \psi)$ which still contains a pair of $\sigma$. When applying the above MWPM algorithm, two $\sigma$'s will be paired up and must generate a $\psi$-particle if the anyons still stay close by. A further matching with the other $\psi$ can annihilate the pair of $\psi$'s back to the vacuum $\mathbf{1}$. 

As we can see, such pair creation mechanism naturally fits the above MWPM algorithm.  However, the Fibonacci category (FIB) and its doubled version (DFIB) does not have such simple pair creation structure.  In FIB,  two $\tau$-anyons can fuse into $\tau$ as well due to the cyclic structure, which can further fuse with another $\tau$ into the vacuum $\mathbf{1}$.  This means that not only a pair of $\tau$ can be generated from the vacuum $\mathbf{1}$, but also a triplet $(\tau, \tau, \tau)$. In fact, due to the cyclic structure, any number of $\tau$ can be created from the vacuum.  The absence of a pair generation structure is in contrast to the situation in the Ising category where only pairs of $\sigma$ can be generated from the vacuum.  Similarly, in DFIB, two $\tau \mathbf{1}$ can fuse into $\tau \mathbf{1}$, which means a triplet $(\tau \mathbf{1},\tau \mathbf{1},\tau \mathbf{1})$ or more $\tau \mathbf{1}$ can be generated from the vacuum $\mathbf{11}$.  Similar argument applies to triplets of $ \mathbf{1}\tau$ and $\tau\tau$.
    In the example of Pauli noise, we can see from Fig.~\ref{fig:x_error_basis_convention} that a single Pauli-X error can create 2, 3 or 4 anyons, without a clear pair creation mechanism like the $\sigma$-anyons in the Ising category.

Now we can give a proof that the above MWPM decoder does not have a threshold for the Fibonacci Turaev-Viro code (DFIB). Consider a counterexample shown in Fig.~\ref{fig:MWPM_failure}, where two triplets of $\tau \mathbf{1}$ anyons are created by $O(1)$ errors from the vacuum and spatially separated by $O(d)$, i.e., the order of the code distance. According to the algorithm stated above, we will first apply MWPM to pair up all the present anyons \footnote{If the total number of anyons is odd, then there will be a single anyon which is not paired up with any other anyon.}.  As indicated by the highlighted edges in Fig.~\ref{fig:MWPM_failure}(b), two of the anyons within each triplet are matched, while one anyon within one cluster is matched with the one in another triplet.  We then fuse the matched anyon pair by transporting one anyon (randomly chosen) along the highlighted path to fuse with the other anyon and ends up with the configuration in Fig.~\ref{fig:MWPM_failure}(c), where the curve diagrams are merged.    We then apply MWPM to pair up the two remaining anyons. A short path is preferred by the matching algorithm as shown by the highlighted edge.  When fusing the anyons along this short path, the anyon path forms a non-contractible (homologically nontrivial) cycle, which equivalently applies a chain of errors of length $O(d)$ and hence induces a logical error.  In summary, when applying the above MWPM algorithm, even $O(1)$ errors may not be corrected, meaning the effective code distance in this decoding scheme is only $O(1)$ instead of $O(d)$.  Therefore, the logical error rate is not exponentially suppressed  as a function of code distance $d$, and the decoder does not have a threshold. As one can easily see, such an issue also exists in the case of single copy of the Fibonacci category. Nevertheless, we note that such problem is not present in the  case of the triplet $(\sigma, \sigma, \psi)$ in Ising anyon models mentioned above. Using the MWPM algorithm for the Ising-anyon model,  one will first apply matching only to the $\sigma$-anyons within each triplet, which fuse into a $\psi$-particle, and then apply matching to the $\psi$-anyons, which fuse the two remaining $\psi$-anyons within each cluster.  The non-cyclic feature of Ising fusion rule avoids the above issue.  

Due to this issue, we need to modify the MWPM algorithm for FIB and DFIB. As we have seen, the main issue is that sometimes the matching algorithm pairs up anyons which are far apart and do not belong to the same anyon cluster generated from the vacuum.  Therefore, we propose an iterative MWPM algorithm.  In each iteration, we apply the MWPM algorithm, but only fuse the matched anyon pairs which are within a certain fusing radius $r$ (starting at one lattice spacing in the first iteration).  When fusing a certain anyon pair, we may generate some new anyons. Therefore,  we match and fuse anyons until all possible anyon pairs within the fusing radius $r$  have been fused and begin the next iteration by increasing the fusing radius by one lattice spacing, i.e., $r \rightarrow r+1$.  We keep doing this until all the anyons are  annihilated.  During this process, if any anyon path forms a non-contractible cycle, we claim \texttt{failure}.   If there is a single remaining anyon in the end of the algorithm, we also claim \texttt{failure}. Otherwise, we declare \texttt{success}.  As we will see in the next section, the numerical simulation suggests that such an iterative MWPM algorithm does actually  have a threshold.  

The above iterative MWPM algorithm is summarized below:

\noindent\makebox[\linewidth]{\rule{\columnwidth}{0.4pt}}

\textbf{Algorithm (iterative MWPM)}

\noindent\makebox[\linewidth]{\rule{\columnwidth}{0.4pt}}

\begin{small}
\nin \textit{\# Measure the anyon charge of each plaquette on the tailed trivalent lattice;  store the nontrivial anyon charge in a list ``anyon\_charge"}

\nin \texttt{anyon\_charge = get\_syndrome(state);}

\

\nin \textit{\# Initialize the fusing distance at one lattice spacing}

\nin \texttt{r=1;}

\

\nin \textit{\# While there is more than one anyon left}

\nin \textbf{\texttt{while}} \texttt{size(anyon\_charge) > 1}

\

\nin \qquad \textit{\# Apply the matching algorithm; only record anyon pairs with }

\nin \qquad \textit{maximal distance \texttt{r} and the corresponding path connecting }

\nin \qquad \textit{them}

\nin \qquad \texttt{[pairs, paths]=MWPM(anyon\_charge, r);}

\

\nin \qquad \textit{\# If no anyon pair within the fusing radius is matched}

\nin \qquad \textbf{\texttt{if}} \texttt{pairs == []}

\nin \quad \qquad \textit{\# Increase the fusion radius by one lattice spacing}

\nin \qquad \qquad \texttt{r = r+1};

\

\nin \qquad \texttt{\textbf{else}}

\nin \qquad \textit{\# Fuse the anyons along the paths given by MWPM; update}

\nin \qquad \textit{the list ``anyon\_charge" with the fused anyon charge; drop the }

\nin \qquad \textit{vacuum charge $\mathbf{11}$ from the list ``anyon\_charge"}

\nin \qquad \qquad \texttt{anyon\_charge = fuse\_anyons(pairs, }

\nin \qquad \qquad \texttt{paths);}

\

\nin \qquad \qquad \textit{\# Check (in the simulation) whether any anyon path }

\nin \qquad \qquad \ \ \ \textit{becomes  non-contractible, i.e., homologically nontrivial }

\nin \qquad \qquad \ \ \ \textit{after the  fusion process}

\nin \qquad \qquad \texttt{\textbf{if} anyon\_path == non-contractible}

\nin \qquad  \qquad \qquad  \textit{\# The decoder claims failure}

\nin \qquad  \qquad \qquad  \texttt{return} \textbf{\texttt{Failure}}

\

\nin \textbf{\texttt{end}}

\

\nin \textit{\# If there is no anyon excitations left}

\nin \texttt{\textbf{if} anyon\_charge == []}

\nin \qquad \textit{\# The decoder declares success}

\nin \qquad \texttt{return} \textbf{\texttt{Success}}

\nin \textit{\# If there is a single anyon left}

\nin \texttt{\textbf{else}}

\nin \qquad \texttt{return} \textbf{\texttt{Failure}}

\end{small}

\noindent\makebox[\linewidth]{\rule{\columnwidth}{0.4pt}}

\vspace{0.1 in}

As we can see, the iterative MWPM algorithm shares a similar hierarchical structure with the clustering algorithm where the cluster is grown by one  lattice space in each iteration.  When we apply this algorithm to the previous triplet example, we see that the previous issue is resolved, as illustrated in Fig.~\ref{fig:MWPM_iterative}.  In the first iteration, we have fusing radius $r=1$.  So when we apply matching, the  fusion of the matched pair with a long separation (three lattice spacing) is rejected by the decoder since the separation exceeds the fusing radius.  When we do the matching again, these anyons get matched and fused with their neighboring $\tau \mathbf{1}$ anyons generated from the last fusion. Therefore, all the fusion still occur within each triplet cluster and all the anyons are fused back into the vacuum.

\begin{figure*}[t]
		\centering
		\includegraphics[width=2\columnwidth]{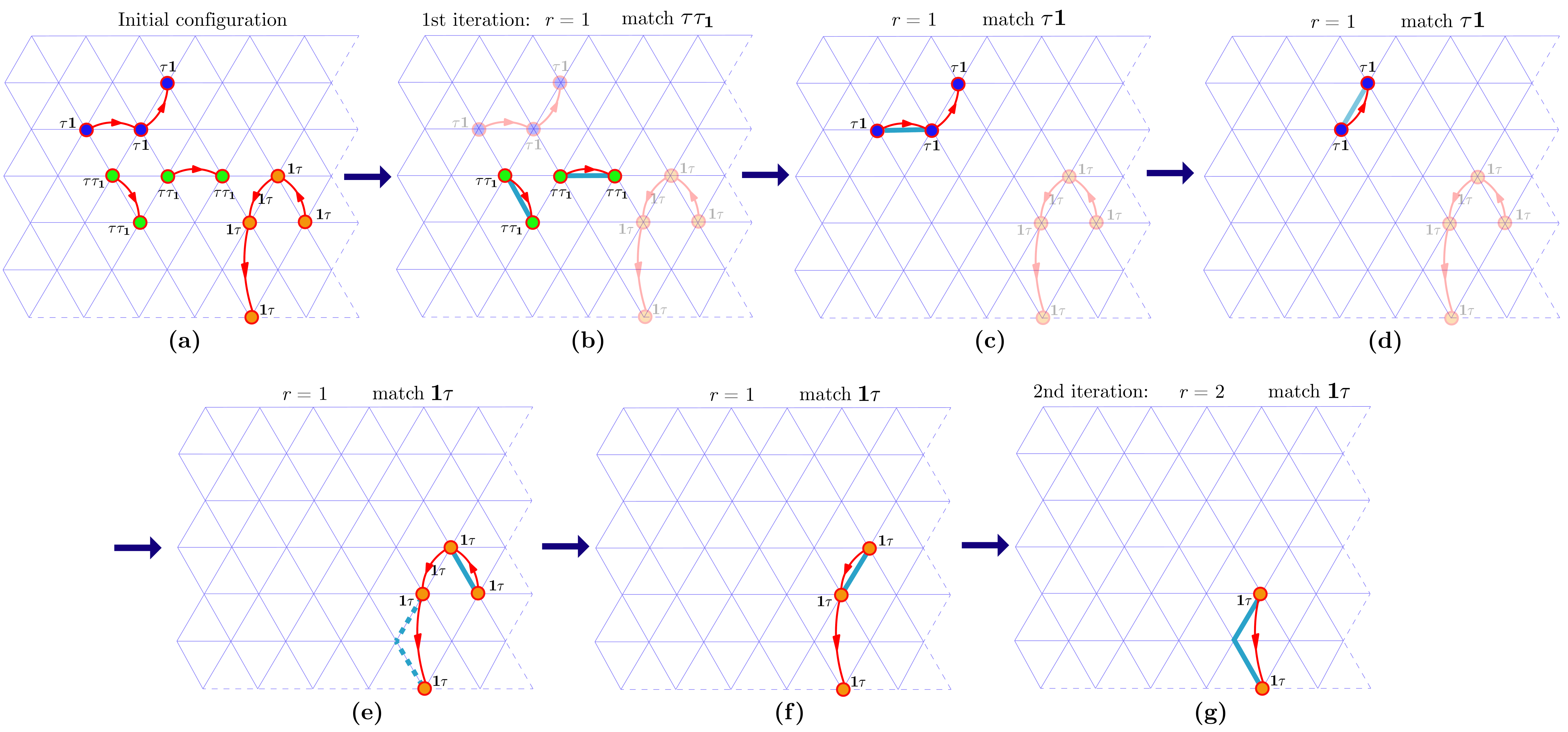}
		\caption{Illustration of the fusion-aware iterative MWPM decoder with an example of initial anyon configuration. (a) Four anyons clusters are generated from the vacuum sector. The anyon charges are shown on the nodes and branches of the curve diagram (see the branch label on the cluster with four $1\tau$ anyons). The two $\tau\tau_\mathbf{1}$ clusters are created by $\sigma_z$-errors.  (b) Match and fuse $\tau\tau_\mathbf{1}$ anyons only with fusing distance $r=1$.  (c, d) Match and fuse $\tau\mathbf{1}$ anyons only with fusing distance $r=1$. (e) Match and fuse $\mathbf{1}\tau$ anyons only with fusing distance $r=1$. The matched path (dashed lines) between the lower two anyons has  length 2 and hence exceeds the fusing distance $r=1$. The fusion between these two anyons is hence rejected.  (f) Continue to match and fuse $\mathbf{1}\tau$ anyons with fusing distance $r=2$. (g) In the second iteration, the decoder increases the fusing distance to $r=2$, and hence is able to match and fuse the remaining pair of $\mathbf{1}\tau$ anyons.}
		\label{fig:MWPM_iterative_aware}
	\end{figure*}

In the above iterative MWPM algorithm, we have not used any detailed syndrome information about the anyon types, similar to the case in the clustering algorithm. We call such an algorithm 
\textit{`blind'}. On the other hand, there are definitely certain features and patterns in the anyon creation process.   Therefore, we expect that an iterative MWPM decoder which is fusion-aware, i.e., has access to the detailed information of anyon type, should have a better performance.  In particular, in the context of depolarizing noise, a single $\sigma_z$ error generates a pair of anyons with charge $\tau\tau_\mathbf{1}$, as discussed in Sec.~\ref{sec:Pauli-Z} and shown in Fig.~\ref{fig:z_error_basis_convention}.  This is a distinct feature of $\sigma_z$ error when the system is exposed to all three types of noise ($\sigma_x, \sigma_y, \sigma_z$). In order to exploit such error feature, we require the decoder to have the following behavior: when the fusion-aware MWPM decoder sees a pair of $\tau\tau_\mathbf{1}$ anyons, it will prefer to match and fuse them first, rather than to pair any of them with other types of anyons.  This choice will tend to clean up the anyon clusters generated by $\sigma_z$ errors first, and avoids to fuse anyons belonging to different anyon clusters which will join these two clusters.

We hence propose the following modification to obtain the fusion-aware iterative MWPM decoder.  In each iteration, we start matching and fusing specific type of anyons in the following order [$\tau\tau_\mathbf{1}$, $\tau \mathbf{1}$, $ \mathbf{1}\tau$, $\tau\tau_\tau$, $\tau\tau_\mathbf{1}$, $\tau \mathbf{1}$, $ \mathbf{1}\tau$]. We only start fusing the next type once the current type of anyon pairs within the fusion radius have all been fused. After we exhausted all the types, we match and fuse the remaining anyons `blindly' within the fusing radius $r$, i.e., ignoring the detailed anyon charge type. In this case, one is allowed to match and fuse anyons with different anyon types. We then increase the fusing radius by one lattice spacing ($r \rightarrow r+1$) and continue to the next iteration.  

Now we explain the reason of choosing the above order of fusion.  According to the fusion graph in Fig.~\ref{fig:fusion_diagram}(d), one can choose the fusion order from left to right such that the anyon types generated in the fusion process are never on the left of the current type on the fusion  graph.  In that case, we can choose either the fusion order [$\tau\tau$, $\tau \mathbf{1}$, $ \mathbf{1}\tau$] or [$\tau\tau$, $ \mathbf{1}\tau$, $\tau \mathbf{1}$].  Note that here we do not distinguish $\tau\tau_\mathbf{1}$ and $\tau\tau_\tau$ and they make no difference in the fusion graph. However, due to the structure of $\sigma_z$-noise as mentioned above, we prefer to match and fuse $\tau\tau_\mathbf{1}$ first. On the other hand, $\tau\tau_\tau$-anyons are less likely to be produced, so we fuse them after $\tau \mathbf{1}$ and $ \mathbf{1}\tau$. Then when we start fusing $\tau\tau_\tau$-anyons, we are again on the very left of the fusion graph and will again produce all types of anyons ($\tau\tau_\mathbf{1}$, $\tau \mathbf{1}$, $ \mathbf{1}\tau$, and $\tau\tau_\tau$).  Therefore, after exhausting fusion of $\tau\tau_\tau$, we still need to start fusing $\tau\tau_\mathbf{1}$, $\tau \mathbf{1}$ and $ \mathbf{1}\tau$ in turn.  

We illustrate this algorithm with a specific example in Fig.~\ref{fig:MWPM_iterative_aware}. As we can see that the initial configuration in Fig.~\ref{fig:MWPM_iterative_aware}(a) has several anyon clusters generated from the vacuum sector by local noise, and each cluster has a total vacuum charge $\mathbf{11}$. In particular, two clusters  of $\tau\tau_\mathbf{1}$ anyons (green) are created by $\sigma_z$ errors.  The anyon clusters are all connected together, which poses significant challenge for the decoder.  Now the decoder starts with fusing radius $r=1$ and only perform matching on the $\tau\tau_\mathbf{1}$ anyons as shown in Fig.~\ref{fig:MWPM_iterative_aware}(b). As we see, the two pairs of $\tau\tau_\mathbf{1}$ are immediately matched and fused into the vacuum sector, and we are left with two separate anyon clusters with $\tau\mathbf{1}$ (blue) and $\mathbf{1}\tau$ (orange) anyons.  Due to the significant separation, decoding becomes much easier now. The decoder continues to match and fuse only the $\tau\mathbf{1}$ anyons, which takes two steps to annihilate all the $\tau\mathbf{1}$ anyons, as shown in Fig.~\ref{fig:MWPM_iterative_aware}(c, d).   The decoder now switches to match the $\mathbf{1}\tau$ anyons.  It performs two steps of matching and fusion with fusing radius $r=1$ in (e,f), and has exhausted the matching of all possible anyon pairs with a separation of one lattice spacing.   Now the decoder increases the fusing distance to $r=2$, and matches the remaining pair of  $\mathbf{1}\tau$ anyons separated with two lattice spacing in (g), which are then fused into the vacuum.  We can conclude that in this case, the fusion awareness makes the decoding much easier than the blind MWPM decoder which could prefer to join neighboring anyon clusters into a larger cluster and hence increases the chance of failure.

As we will see in the numerical results in the next section, this fusion-aware iterative MWPM decoder does give an overall improvement in logical fidelity over the `blind' iterative MWPM decoder.  

We note that Ref.~\cite{brell2014thermalization} has also adopted such fusion-aware strategies to the clustering decoder in the context of Ising anyon model, i.e., clustering different types of anyons separately in a certain chosen order.  We have also adopted this strategy to the Fibonacci Turaev-Viro code (DFIB). However, we do not observe any advantage compared to the clustering decoder described in Sec.~\ref{sec:clustering_decoder} which does not use the detailed syndrome information of anyon types.  This is potentially related to the sophisticated cyclic fusion structure of Fibonacci anyon as opposed to the non-cyclic fusion structure of Ising anyons.

%% file: sections/results.tex
\section{Numerical results}\label{sec:numerics}

    The Monte Carlo simulations described in Chapter~\ref{sec:simulation} were performed for the three different decoders described in Chapter~\ref{sec:decoding}. 
    Individual Pauli errors were picked using relative probabilities corresponding to the following noise models:
    \begin{itemize}
    	\item  depolarizing noise: \qquad $ \gamma_x = \gamma_y = \gamma_z = \frac{1}{3}\,, $
    	\item dephasing noise: \qquad \quad  $ \gamma_x = \gamma_y = 0 \,, \quad \gamma_z = 1\,, $
    	\item bit-flip noise:	\qquad \qquad  \, $ \gamma_x = 1 \,, \quad  \gamma_y = \gamma_z = 0\,. $
    \end{itemize} 

    Simulations were performed for a wide range of physical error rates.
	For depolarizing noise and pure bit-flip noise, we considered the linear system sizes $ L = 10,12,14,16,18 $. For dephasing noise the values $ L = 12,14,16,18,20,22 $ were used. \\     

    The logical failure rate for each $(p,L) $-pair was computed by averaging over  $ 10^{5} $ Monte Carlo samples. For the lowest error rates $ p = 0.01 $ (or $ p = 0.02 $ in case of dephasing noise), $ 10^{6} $ Monte Carlo samples were used in order to improve the accuracy of our results.
    As visible in the results below, this is ample to guarantee sufficiently small (95\%) confidence intervals for the average logical failure rates. 

    All simulations were done with the following values for the cutoff parameters:
    \begin{align*}
    	\mathtt{ N_{max} } &= 27\,,\\
    	\mathtt{ V_{max} } & = 2.5 \cdot 10^{7}\,.
    \end{align*}
    For depolarizing noise and bit-flip noise with $ L=18 $ the ratios of aborted Monte Carlo samples near the observed thresholds are shown in the table below. 
    The corresponding ratios of aborted failures are indicated between parentheses.
   \begin{center}
	\begin{tabular}{|c|c|c|c|}
	 	\hline
	 	& Clustering &  \begin{tabular}{@{}c@{}}Fusion-aware \\ MWPM\end{tabular} & Blind MWPM \\
	 	\hline
	 	Depolarizing & 6.1\%   & 0.6\%  & 0.8\%  \\
	 	noise & (17.4\%) & (2.1\%) & (2.9\%)\\  
	 	\hline
	 	Bit-flip  & 4.6\%  & 0.3\%  & 0.4\%  \\
	 	noise & (15.1\%) & (1.3\%) & (1.7\%) \\
	 	\hline
	\end{tabular}
	\end{center}
	We found that the ratio of aborted Monte Carlo samples drops rapidly below the threshold. For instance, at $ p = 0.04 $ less than 1.1\% of Monte Carlo samples (7.4\% of reported failures) were aborted for depolarizing noise with $ L=18 $.
	For dephasing noise, the ratio of aborted iterations is negligible for all system sizes and noise strengths we studied.\\

\subsubsection*{Determining the threshold}
    The error correction threshold is given by the critical value $ p_c $ below which the logical failure rate $ P_L $ is exponentially suppressed in terms of the system size $ L $. 
    For large system sizes, the threshold manifests itself as the physical error rate for which the logical failure rates of all system sizes coincide. 
    Hence, a rough estimate of the threshold can be obtained by plotting $ P_L $ in function of $ p $ for different system sizes and finding the error rate at which the various curves intersect. 
    
    A more accurate estimate for the error correction threshold can be obtained using the critical exponent method of Ref.~\cite{wang2003confinement}. 
    This method was introduced in the context of the toric code, where an exact mapping to a statistical model is known \cite{Dennis:2002ds, wang2003confinement}. 
    This model, the 2-dimensional random-bond Ising model (RBIM), undergoes a phase transition from an ordered to a disordered phase as the parameter corresponding to the physical error rate increases.  
    This implies a phase transition in the logical failure rate of the toric code. 
    Wang et al. demonstrated that in the regime $ L \gg | p - p_c|^{-\nu}$, where $ \nu $ is the critical exponent for the correlation length in the RBIM, the logical failure rate $ P_L $ depends only on the dimensionless ratio $ L ( p - p_c)^{\nu} $. 
    
    While the statistical model corresponding to the Fibonacci Turaev-Viro code is not known, it is expected that a similar scale invariant behavior occurs near the threshold here as well. 
    Specifically, for sufficiently large system sizes, we define the rescaled variable 
    \begin{equation}\label{eq:x}
    	x = (p - p_c) L^{1/\nu}\,,
    \end{equation}
    where $ \nu $ is some critical exponent, such that the logical failure rate $ P_L $ as a function of $ x $ is explicitly scale invariant. 
    We can find the correct values for $ p_c $ and $ \nu $ by fitting the values of $ P_L $ to the quadratic ansatz
    \begin{equation}\label{eq:ansatz_fit}
    	P_L(x) = A + B x + C x^2\,,
    \end{equation}
    originating from a truncated Taylor expansion in the neighborhood of $ x=0 $ ($ p = p_c $).
    We perform this fit explicitly with the data obtained for the clustering decoder with depolarizing noise and dephasing noise. 

\subsection{Clustering decoder}
\begin{figure}[h]
    	\centering
    	\begin{subfigure}[b]{0.4\textwidth}
    		\centering
    		\includegraphics[trim={1.2cm 7.3cm .5cm 6cm},clip,width=\linewidth] {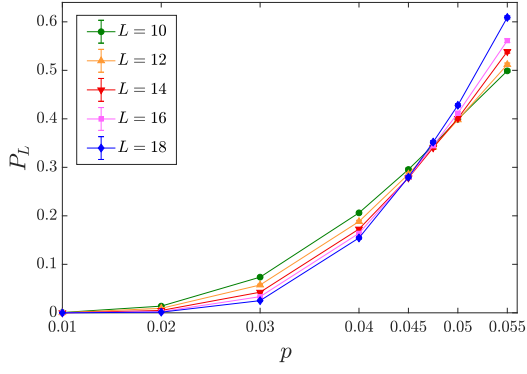}
    		\caption{}
    		\label{fig:cluster_simple_all}
    	\end{subfigure}
    	\begin{subfigure}[b]{0.4\textwidth}
    		\centering
    		\includegraphics[trim={1.2cm 7.3cm .5cm 6cm},clip,width=\linewidth] {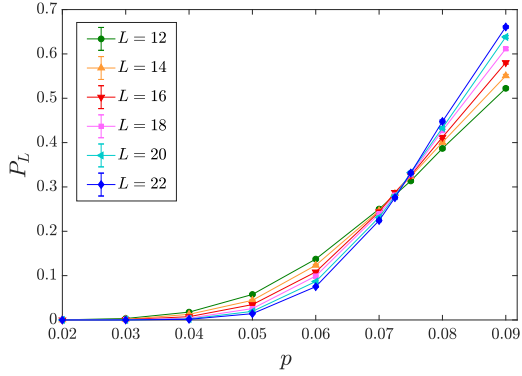}
    		\caption{}
    		\label{fig:cluster_simple_z}
    	\end{subfigure}
    	\begin{subfigure}[b]{0.4\textwidth}
    		\centering
    		\includegraphics[trim={1.2cm 7.3cm .5cm 6cm},clip,width=\linewidth] {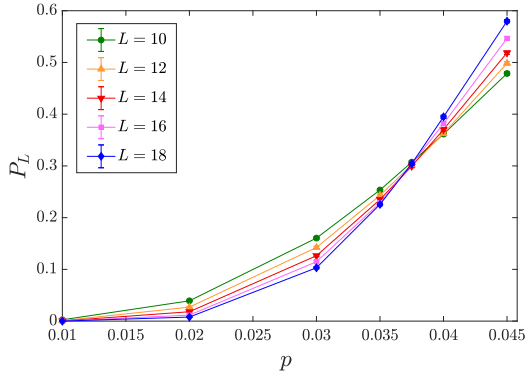}
    		\caption{}
    		\label{fig:cluster_simple_x}
    	\end{subfigure}
    	\caption{Logical failure rate $P_L$ as a function of the physical error rate $ p $ for the clustering decoder with (a) depolarizing noise, (b) pure dephasing noise, and (c) pure bit-flip noise.}
    	\label{fig:cluster}
    \end{figure}

    The logical failure rate $P_L $ of the clustering decoder in function of the noise strength $ p $, are shown in \figref{fig:cluster_simple_all}, \figref{fig:cluster_simple_z} and \figref{fig:cluster_simple_x} for depolarizing, dephasing noise and bit-flip noise, respectively. 
    These results clearly manifest threshold behavior.
    For pure bit-flip noise, the threshold can be estimated from the corresponding plot as $  p_c \approx 0.0375 \pm 0.0025 $. A more precise estimation, based on the finite-size ansatz discussed above, was made for depolarizing noise and dephasing noise.
    
\begin{figure}[h]
		\centering
		\begin{subfigure}{.45\textwidth}
			\centering
			\includegraphics[trim={.8cm 7.3cm .5cm 6cm},clip,width=1.0\linewidth]{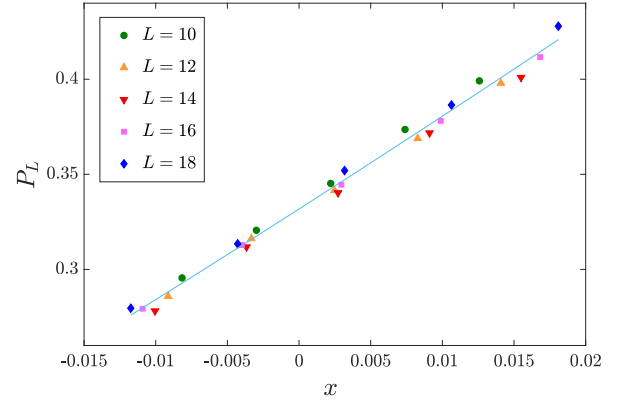}
			\caption{}
			\label{fig:scaling_all}
		\end{subfigure}
		\begin{subfigure}{.45\textwidth}
			\centering
			\includegraphics[trim={.8cm 7.3cm .5cm 6cm},clip,width=1.0\linewidth]{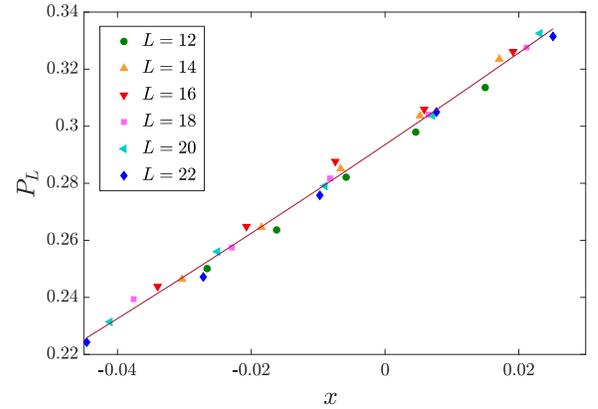}
			\caption{}
			\label{fig:scaling_z}
		\end{subfigure}
		\caption{Logical failure rate $ P_L$ as a function of the rescaled error rate $ x= (p - p_c) L^{(1/\nu)} $ for (a) depolarizing noise and (b) dephasing noise. The solid line represents the best fit of the model $ P_L = A + B x + C x^2 $.}
		\label{fig:scaling}
	\end{figure}

    For depolarizing noise the finite-size scaling ansatz Eq.~\eqref{eq:ansatz_fit} was fitted to the logical failure rates for $ p $ ranging from 0.045 to 0.05 in increments of 0.00125.  
   	The following values were found using a non-linear least squares fit: 
    \begin{align*} 
    	p_c & = 0.0470 \pm 0.0011\,, \\ 
    	\nu & =  1.62 \pm 0.33\,.
    \end{align*}	
	For dephasing noise, the ansatz was fitted to the logical failure rates obtained for $ p $ ranging from 0.07 to 0.075 in increments of 0.00125.  With this data, we found
    \begin{align}
    	p_c & = 0.0732 \pm 0.0006\,, \\ 
    	\nu & =  1.17 \pm 0.08\,.
    \end{align}
	The confidence intervals were estimated using the jackknife resampling method. 
	In both cases the obtained threshold is compatible with the rough estimate based on the crossing of the curves in \figref{fig:cluster}.
    The logical failure rates in terms of rescaled error rate $ x $ defined in Eq.~\eqref{eq:x} are shown in Figs.~\ref{fig:scaling_all} and \ref{fig:scaling_z} for depolarizing noise and dephasing noise,  respectively. 
    One can see that the obtained parameters do indeed result in a clear ``collapse'' of the data, as predicted by the finite-size scaling hypothesis.

    
    It is no surprise that the highest threshold is found for dephasing noise:
    no more than two anyonic excitations can be created by a single $ \sigma_z $ error, while up to four anyons can be created by a single $ \sigma_x $ or $ \sigma_y $ error. 
    Hence, the average number of new anyonic excitations created in each time step is the lowest for dephasing noise and the highest for pure bit-flip noise, with the value for depolarizing noise lying somewhere in between these two extremes. 
    The discrepancy between the thresholds for dephasing and bit-flip noise indicates that for biased noise, there is a preferred choice for the computational basis of the physical qubits.  
    
\subsection{Iterative matching decoders}
\begin{figure}[h]
    	\centering
    	\includegraphics[trim={1.2cm 7.3cm .5cm 5.59cm},clip,width=1\linewidth]{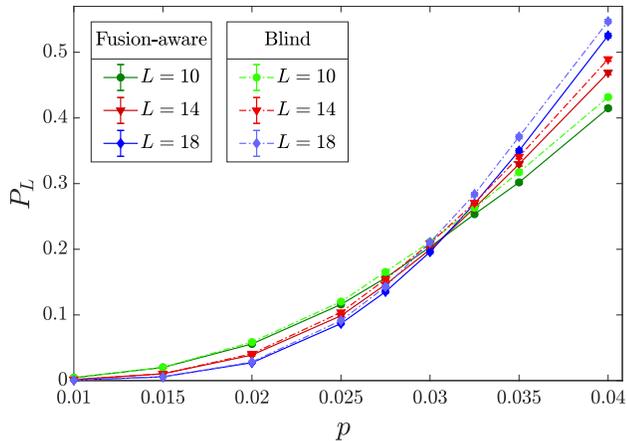}
    	\caption{Comparison between the logical failure rates for the fusion-aware and the blind iterative MWPM decoders under depolarizing noise.}
    	\label{fig:mwpm_comparison_all}
    \end{figure}

\begin{figure*}[h]
    	\centering
    	\begin{subfigure}[b]{0.4\textwidth}
    		\centering
    		\includegraphics[trim={1.2cm 7.3cm .5cm 6cm},clip,width=\linewidth] {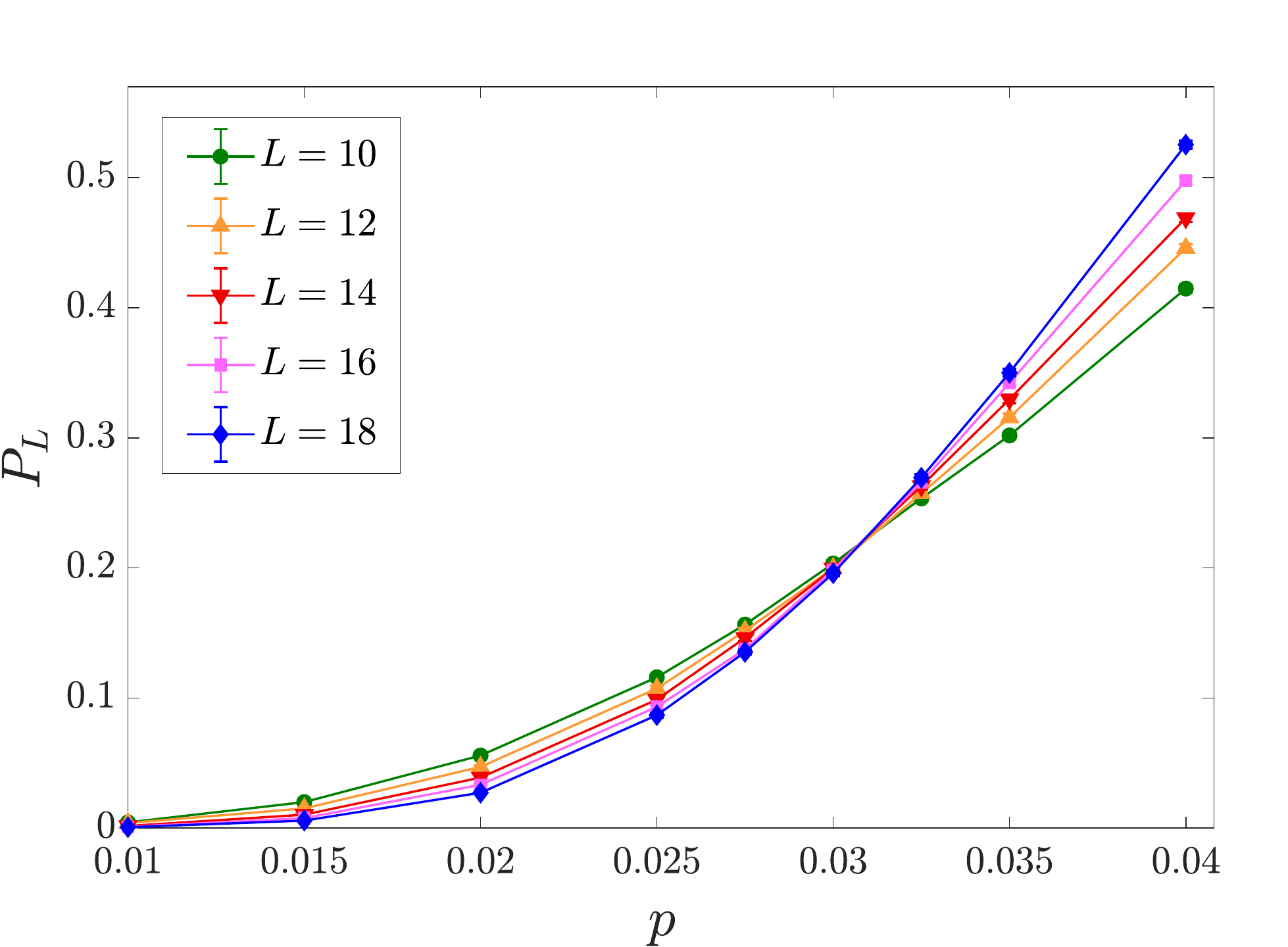}
    		\caption{}
    		\label{fig:mwpm_all}
    	\end{subfigure}
    	\begin{subfigure}[b]{0.4\textwidth}
    		\centering
    		\includegraphics[trim={1.2cm 7.3cm .5cm 6cm},clip,width=\linewidth] {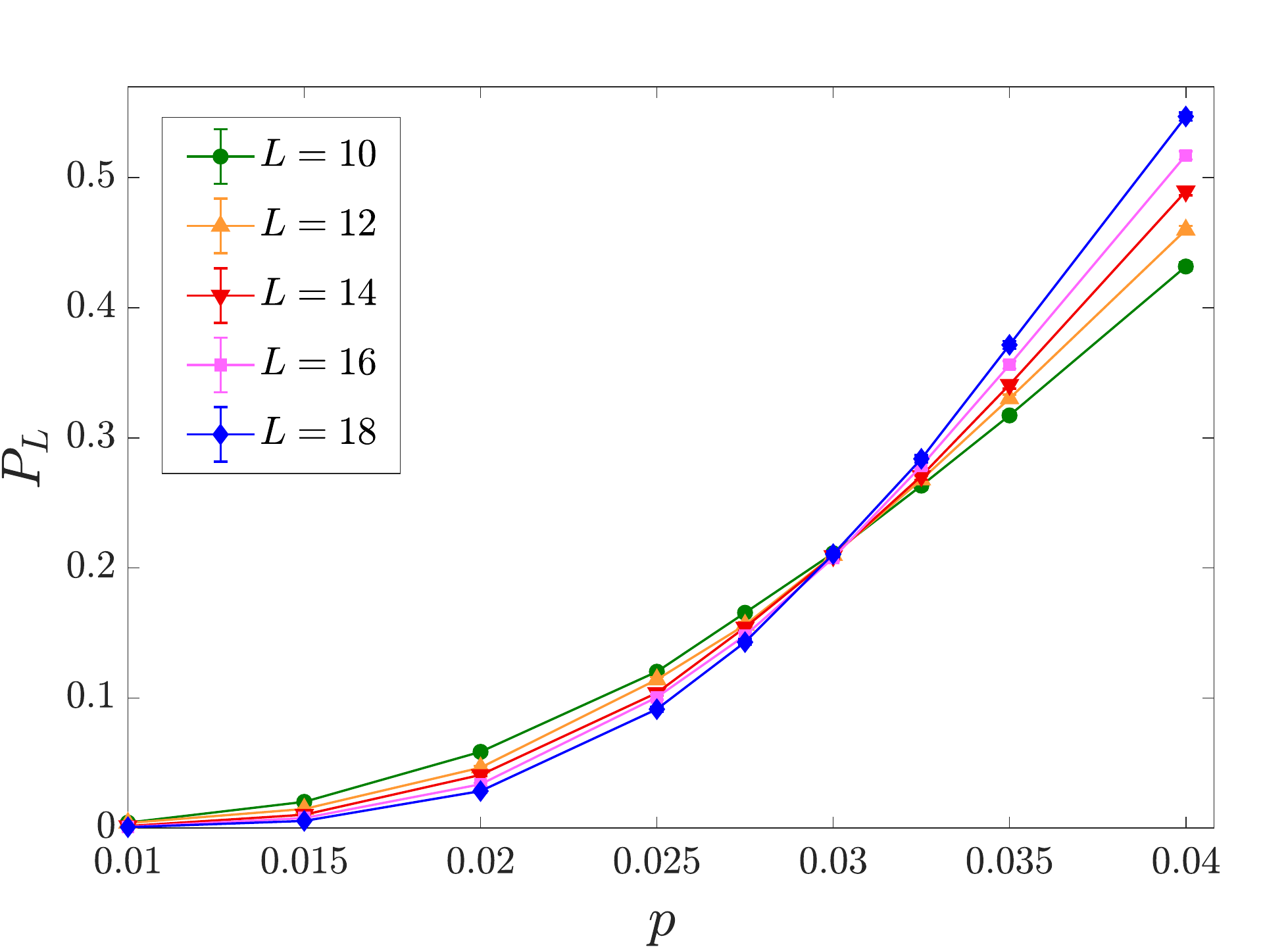}
    		\caption{}
    		\label{fig:mwpm_blind_all}
    	\end{subfigure}
    	\begin{subfigure}[b]{0.4\textwidth}
    		\centering
    		\includegraphics[trim={1.2cm 7.3cm .5cm 6cm},clip,width=\linewidth] {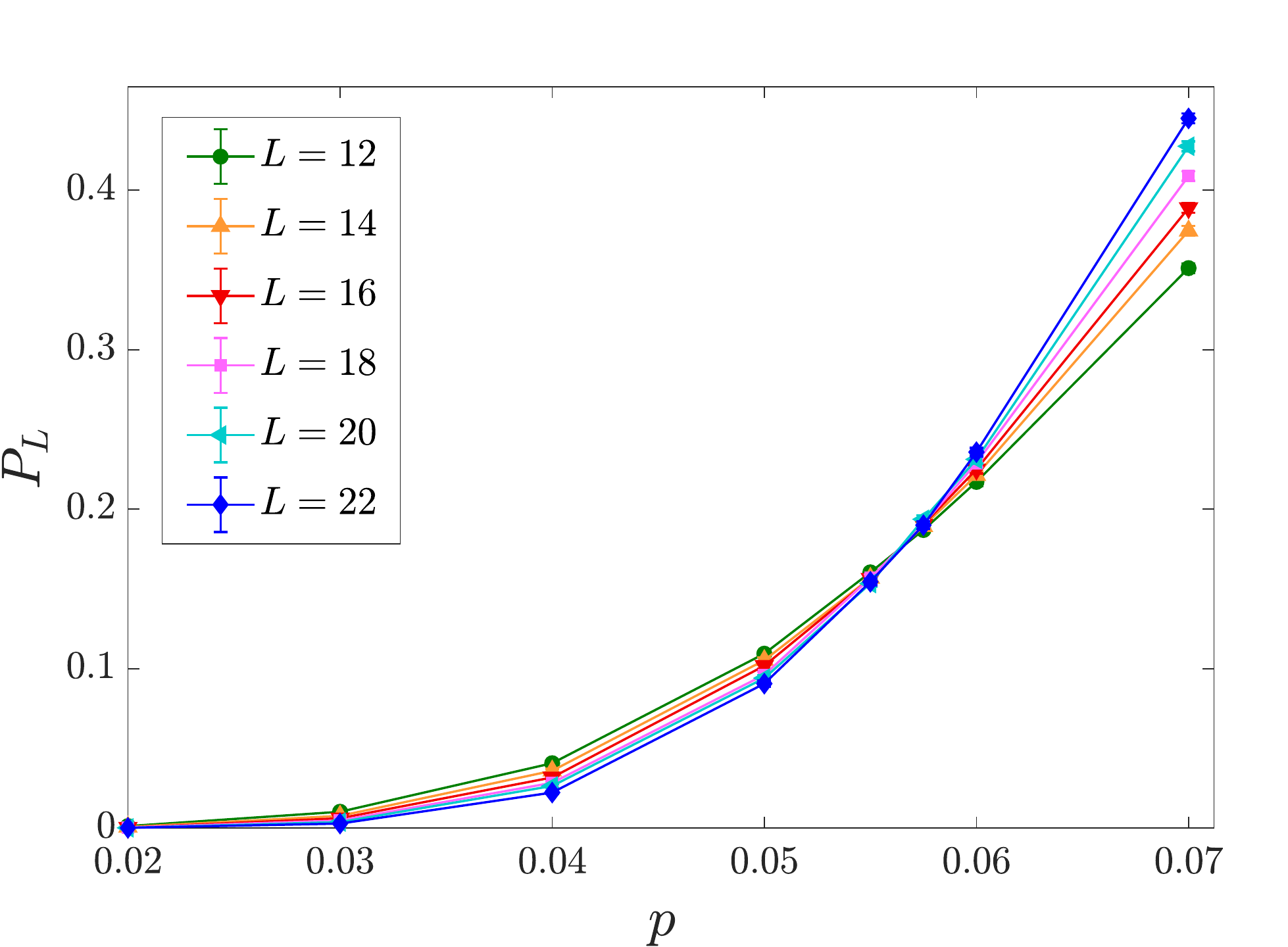}
    		\caption{}
    		\label{fig:mwpm_z}
    	\end{subfigure}
    	\begin{subfigure}[b]{0.4\textwidth}
    		\centering
    		\includegraphics[trim={1.2cm 7.3cm .5cm 6cm},clip,width=\linewidth] {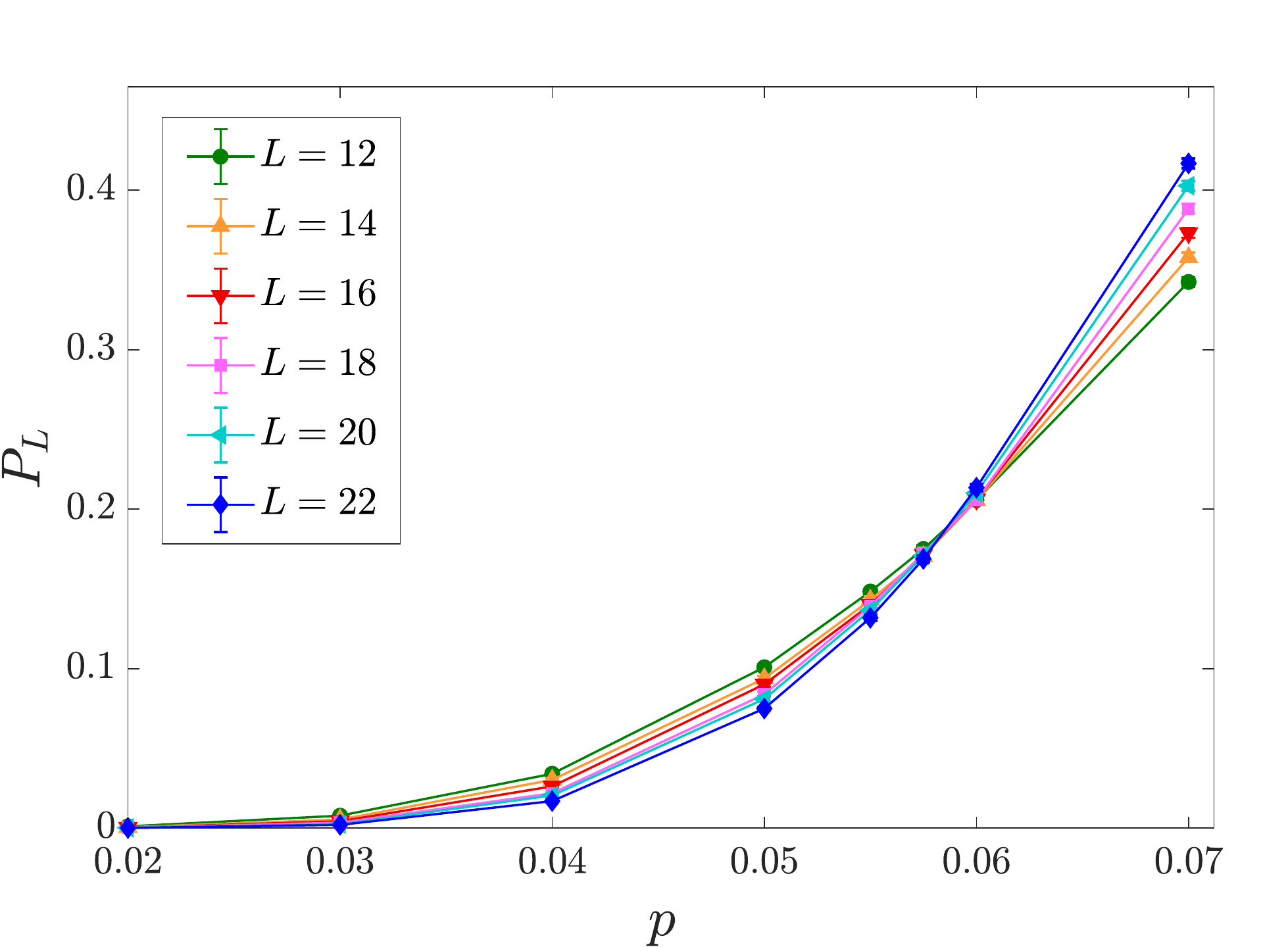}
    		\caption{}
    		\label{fig:mwpm_blind_z}
    	\end{subfigure}
       	\begin{subfigure}[b]{0.4\textwidth}
	    	\centering
	    	\includegraphics[trim={1.2cm 7.3cm .5cm 6cm},clip,width=\linewidth] {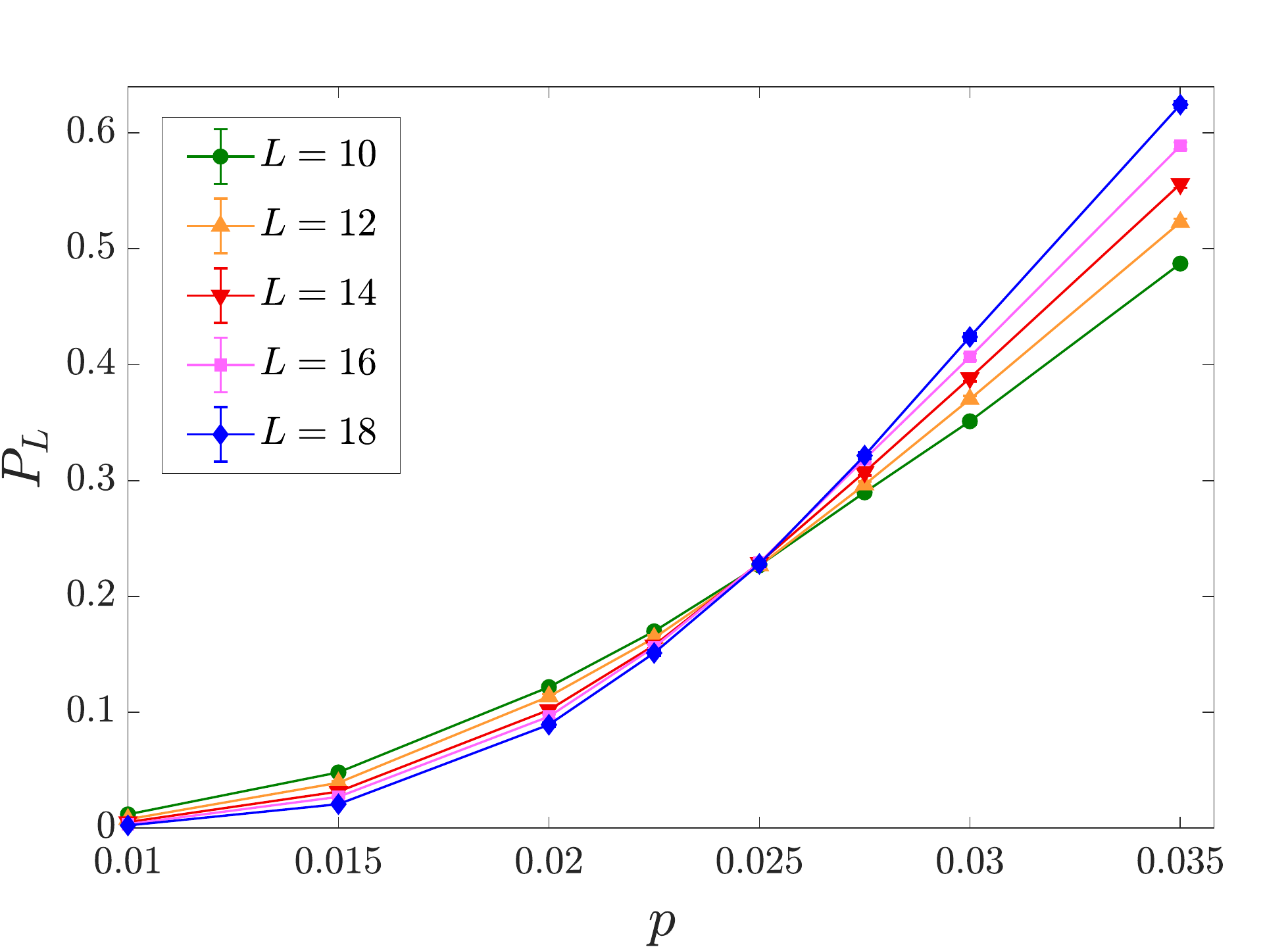}
	    	\caption{}
	    	\label{fig:mwpm_x}
    	\end{subfigure}
	    \begin{subfigure}[b]{0.4\textwidth}
	    	\centering
	    	\includegraphics[trim={1.2cm 7.3cm .5cm 6cm},clip,width=\linewidth] {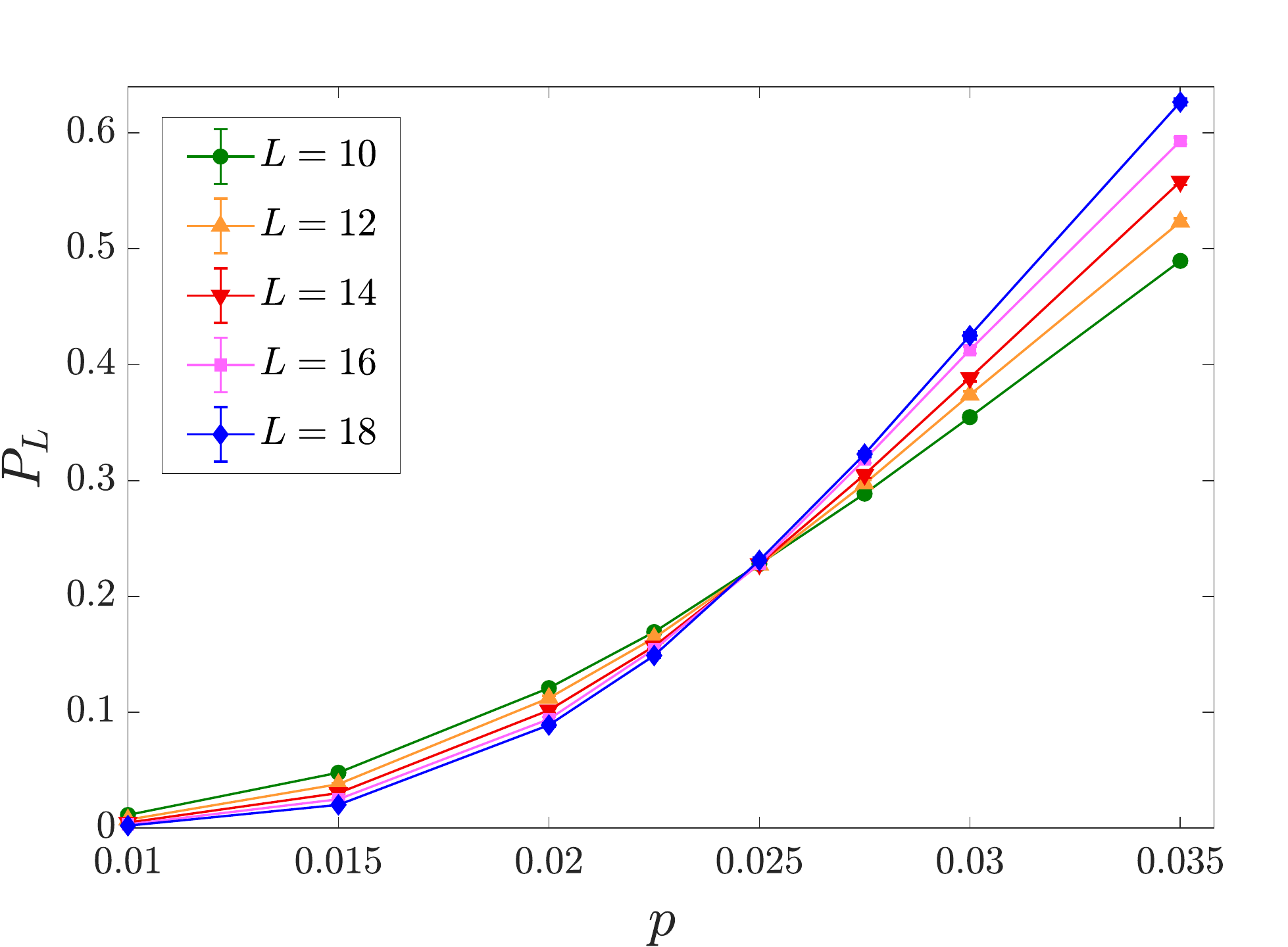}
	    	\caption{}
	    	\label{fig:mwpm_blind_x}
	    \end{subfigure}
    	\caption{(a,c,e) Logical failure rate $P_L$ as a function of the physical error rate $ p $ for the fusion-aware iterative MWPM decoder with (a) depolarizing noise, (c) pure dephasing noise, and (e) pure bit-flip noise.\\
    	(b,d,e) Logical failure rate $P_L$ as a function of the physical error rate $ p $ for the blind iterative MWPM decoder with (b) depolarizing noise, (d) pure dephasing noise, and (e) pure bit-flip noise.}
    	\label{fig:mwpm}
    \end{figure*}

    We consider two types of iterative MWPM decoders: a fusion-aware one and a blind one (see Sec.~\ref{sec:decoding}).
    The performance of these decoders under both types of noise are shown in~\figref{fig:mwpm}. 
    Note that the threshold obtained with these two types of decoders are very close.
    Under depolarizing noise, both decoders exhibit a threshold around $ p_c \approx 0.0300 \pm 0.0025$. 
    Under dephasing noise and bit-flip noise, respectively,  we find $ p_c \approx 0.0600 \pm 0.0025$ and $ p_c \approx 0.0250 \pm 0.0025$ for both decoders. 
    However, when closely comparing the results, as in \figref{fig:mwpm_comparison_all}, one finds a slight overall advantage for the fusion-aware decoder in the sense that its failure rates are lower than those obtained with the blind decoder.
    
    On the other hand, we see that both matching decoders have worse performance and lower thresholds compared to the clustering decoder which has not used the detailed syndrome information of anyon types. This is related to the fact that clustering decoder seems to be more natural for the Fibonacci code than the matching decoder. It remains an open question whether the optimal decoder for the Fibonacci Turaev-Viro code is fusion-aware.

\subsection{Discussion}
    The results above are expressed in terms of the average qubit error rate $ p $ in the fixed-rate sampling noise model described in Sec.~\ref{sec:noise_model}. 
    It is therefore not straightforward to compare these results to those of Abelian models such as the surface code, which are typically expressed in terms of an independent and identically distributed binomial noise strength $p_\text{i.i.d.} $. 
    However, for Abelian codes, both noise models are equally valid and their respective noise strengths can be compared using the relation between the i.i.d.~noise strength $ p_\text{i.i.d.} $ and the error rate $ p $ of a fixed-rate sampling noise model derived in App.~\ref{sec:relation_iid_noise}.
    
    Using this relation, one finds that the thresholds obtained for the clustering decoder under depolarizing ($p_c \approx 0.047$) and dephasing ($p_c \approx 0.073$) noise correspond to i.i.d~noise strengths of $4.6\%$ and $7.0\%$,  respectively.
    Remarkably, despite the complexity of the extended string-net code and the fact that it is not known whether or not the clustering decoder is optimal for this this code, these values compare very favorably with the optimal thresholds for the surface code\footnote{The results mentioned here were in fact obtained for the toric code (i.e., with periodic boundary conditions). It is likely that the true thresholds for surface codes are slightly lower \cite{fowler2013accurate}.} (under the assumption of perfect measurements), which are $18.9\%$ for depolarizing noise \cite{bombin2012strong} and around $10\%$ for dephasing noise \cite{Dennis:2002ds, wang2003confinement}.
    

%% file: sections/conclusion.tex
\section{Conclusion and outlook}\label{sec:discussion}

In order to estimate the non-Abelian error threshold, we have combined concepts and techniques from three seemingly distant fields: quantum error correction, topological quantum field theory, and tensor networks. In particular, we have developed a complete error correction scheme and decoding protocols for the  Fibonacci Turaev-Viro code, which supports a universal logical gate set via braiding or Dehn twists in two dimension.  Making use of the framework of tensor networks and tube algebra, we were able to estimate the code-capacity error correction threshold using a clustering decoder and a fusion-aware iterative matching decoder. The threshold of 4.75\% obtained for the clustering decoder is comparable to the code-capacity error threshold of the Abelian surface code in 2D, which is around $10\%$ \cite{Dennis:2002ds, yoder2017surface}. 

The main conceptual difference of our work and previous works which simulate the third spatial dimension of a 3D color code or 3D surface codes with the time dimension using a just-in-time decoder in a 2D measurement-based quantum architecture \cite{Bombin:2018wj, Browneaay4929} is that the computational power in our case comes from the 2D code space instead of the 3D code space, and no additional code switching or gauge fixing procedure is needed. Practically, the logical gates in our case can be implemented by braiding via  continuous code deformation, therefore the error threshold and logical error rate for fault-tolerant logical gates is expected to be the same as the fault-tolerant threshold for memory storage. Therefore, no extra decrease of the error threshold (compared to the storage threshold) due to the implementation of non-Clifford transversal gates, code switching or gauge fixing with 3D surface codes or color codes \cite{Bombin:2015hia, vasmer_three-dimensional_2019,  Bombin:2018wj, Browneaay4929}, as well as the just-in-time decoding is present in our case.  Another both fundamental and practical difference is that our scheme can also be implemented in a hybrid approach of active and passive topological  protection, where the majority of noise source are passively protected by a 2D Hamiltonian while only the thermal noise will need to be corrected via active error correction \cite{Kapit:2015cy}. This hybrid approach may greatly reduce the overhead of active error correction.

A natural future extension of the work presented here will be the adaptation of our error correction protocols to take into account measurement errors, and to determine the error threshold of the code in the presence of both measurement and circuit-level noise. In this setting of full fault-tolerant error correction, our measurement scheme which allows to distinguish the charges of different anyonic excitations may prove useful, since this information could help in identifying measurement errors when performing repeated syndrome measurements through a consistency check. Thus, while it is still an open question what the optimal decoder for the Fibonacci Turaev-Viro code is and whether it would make use of the detailed charge information in the error syndrome, it is likely that this charge information would yield a notable advantage in the presence of measurement errors.    More generally, the tensor-network representation used in the current work can also be further used to simulate coherent noise in these non-Abelian codes as well as in the usual surface code.

Another direction to explore is the application of our techniques to different models.  A first interesting route would be to investigate other types of Turaev-Viro codes, such as the Ising Turaev-Viro code, which has doubled Ising anyons as excitations. While not universal for quantum computation by braiding itself, extra topologicla charge measurement and Dehn twists permit a universal set of fault-tolerant logical gates. More importantly, this code would have a simpler non-cyclic fusion rule structure which could lead to a higher threshold, especially in the presence of measurement noise. In this context, our measurement scheme to extract the specific anyon charges would be particularly useful, since it was shown in Ref.~\cite{brell2014thermalization} that fusion-aware decoders can yield a significant advantage for Ising-type anyons. A second important direction here will be to investigate non-Abelian codes with a simpler structure, such as lower-weight syndrome operator and lower-depth measurement circuits. The Levin-Wen string-net models are sophisticated in the sense that their plaquette syndrome operator has weight 16. On the  other hand, Kitaev's non-Abelian quantum double models \cite{kitaev2003fault} can have a weight-4 syndrome operator, and could possibly be analyzed using an adaptation of the techniques presented in this work. It would therefore be interesting to further explore the error threshold of these alternative models which could be more practical in terms of an experimental implementation.

A different avenue of further research would be the investigation of planar string-net codes with suitable gapped boundary conditions, where information is then encoded in the fusion state of a number of well separated anyons rather than in the ground state degeneracy associated to a closed manifold with a nontrivial topology (high-genus surface). Alternatively, the logical information can also be encoded in the boundary degeneracy of an open manifold corresponding to a planar geometry in analogy with the Abelian surface code with $e$ and $m$ boundaries. This matter is of significant interest, since a planar geometry is highly attractive regarding experimental realization. The classification of these gapped boundaries has been performed in the language of (bi)module category theory \cite{kitaev_models_2012}, and recent progress has been made in capturing this formalism in terms of tensor network representations \cite{lootens_matrix_2020}. While it does not seem that suitable boundaries for a Fibonacci Turaev-Viro can be constructed in this language, the investigation of these concepts in the context of error correction using different non-Abelian codes would be of great interest.

Besides the study of the quantum memory property of the non-Abelian codes, an important direction is to study and simulate the detailed implementation of a universal set of logical gates in these codes.   Besides the approach of doing braiding and Dehn twists \cite{freedman2002modular, konig2010quantum, PhysRevLett.125.050502, PhysRevB.102.075105}, one can also perform transversal gates on a folded non-Abelian code equivalent to elements in the mapping class group \cite{PhysRevResearch.2.013285}.  An additional advantage of non-Abelian codes appears when they are placed on a hyperbolic surface, which admits both constant-rate encoding ($O(1)$ space overhead) and parallel universal logical gates via constant-depth circuits \cite{Lavasani:2019}. A promising direction is to explore non-Abelian codes in higher spatial dimension or on an expander graph. Along this direction, the ultimate goal is to explore and achieve the fundamental limit of space-time overhead.  This is because in higher dimension such as 4D, one can obtain a self-correcting quantum memory. In that case, local errors created when implementing a logical gate with a constant depth circuit \cite{PhysRevLett.125.050502, PhysRevB.102.075105, Lavasani:2019} can be corrected locally  in O(1) time. Eventually, this direction could evolve into a flourishing interface between quantum information and quantum topology.

An important aspect in terms of experimental implementation of the Fibonacci Turaev-Viro code is the realization of multi-controlled-Z (multi-qubit Toffoli) gates. Therefore, it would be interesting to further explore hardware-efficient implementations of multi-qubit gates, instead of always decomposing these into a longer sequence of two-qubit gates. Such multi-qubit gates are widely studied in Rydberg-atom and ion-trap systems, and it would be useful to further develop these gates in superconducting qubit systems as well.

Finally, besides the application to quantum error correction, the scheme developed in this paper also paves the way for quantum simulation of topological quantum field theory on a near-term quantum computer.  In particular, the measurement and correction schemes will be a crucial ingredient for the state preparation of the TQFT wave functions.  

%% file: sections/acknowledgment.tex
\section*{Acknowledgment}
The authors would like to thank Dominic Williamson, Volkher Scholz, and Nick Bultinck for their suggestions during the early stages of this project. Special thanks goes to Bram Vanhecke and Laurens Lootens for their valuable advice and the countless discussions on various aspects of the code. We also appreciate Steve Flammia for the discussion on decoding and classical simulation of Fibonacci anyons. 
The computational resources (Stevin Supercomputer Infrastructure) and services used in this work were provided by the Flemish Supercomputer Center (VSC), funded by Ghent University, the Research Foundation Flanders (FWO), and the Flemish Government.
AS, LB and FV were by supported by FWO, and ERC Grant No. QUTE (647905).

%% file: sections/ribbon_graphs.tex
\section{The Turaev-Viro TQFT and the extended Levin-Wen string-net model} \label{sec:extended_LW_and_TQFT}

The extended string-net model defined in Sec.~\ref{sec:model} is best introduced in the context of topological quantum field theory. 
In particular, it can be understood as a local Hamiltonian realization of the Turaev-Viro TQFT. 
Hence, the associated error correcting code should be thought of as a microscopic realization of a Turaev-Viro code \cite{konig2010quantum} on a system of qudits. 
In this appendix, we introduce the ribbon graph Hilbert space originating from the Turaev Viro TQFT. We introduce a basis of this Hilbert space which allows us to interpret ribbon graph configurations in terms of anyonic fusion states. 
We then discuss the tube algebra which emerges naturally in this context.
We conclude by illustrating how the (extended) Levin-Wen string-net model emerges naturally when constructing a lattice realization of the ribbon graph Hilbert space. 


\subsection{Category theory primer} \label{sec:category}

	The mathematical underpinning op topological order and anyon models, is category theory.
	Category theory in itself is a broad field of mathematics, which has been studied intensively for several decades,
	For the purpose of this work however, we do not need the full mathematical machinery of category theory. 
	Instead, we will give a condensed overview of the algebraic data of unitary fusion categories and unitary modular tensor categories, which are needed for defining our model in the remainder of this section.
	
	A \emph{unitary fusion category} (UFC) $ \C $ is defined by a finite set of simple objects  
	\begin{equation}\label{key}
		\{a_1, a_2, \dots a_N \},
	\end{equation}
	 and a collection of algebraic data for this set.
	The core of this data is formed by the fusion algebra
	\begin{equation}\label{eq:fusion_algebra}
		a \times b = \sum_{c} N_{ab}^c c\,,
	\end{equation}
	where $ N_{ab}^c \in \NN $ are called the fusion coefficients. 
	The fusion algebra satisfies the following associativity condition:
	\begin{equation}\label{eq:fusion_algebra_associativity}
		\sum_{e} N_{ab}^e N_{ec}^d = \sum_{f} N_{bc}^f N_{af}^d\,.
	\end{equation}
	We define
	\begin{equation}\label{eq:fusion_rules}
		\delta_{ab}^c =
		\begin{cases}
			0 \quad \text{if } N_{ab}^c = 0	\,,\\
			1 \quad \text{otherwise}.
		\end{cases}
	\end{equation}
	
	Among the simple objects, there must be a unique unit element $ \1 $, satisfying $ N_{\1a}^b =  N_{a\1}^b =\delta_{a,b} $ for all $ a,b \in \C $. 
	For each $ a \in \C $, there is a unique element $ a^* \in \C $ satisfying $ (a^*)^*  = a $, and $ N_{ab}^\1 = N_{ba}^\1 =  \delta_{a^*, b} $. 
		 
	A UFC associates to every fusion (resp. splitting) vertex, an $ N_{ab}^c $-dimensional vector space $ V_{ab}^c $ (resp. $ V^{ab}_c $) over $ \CC $.
	For simplicity, we will restrict to the multiplicity-free case, i.e.: $N_{ab}^c = \delta_{ab}^c$.
	
	The fusion associativity condition \eqref{eq:fusion_algebra_associativity} then implies that the fusion or splitting vector spaces corresponding to different orderings of fusion or splitting, are isomorphic:
	\begin{equation}\label{eq:fusion_associativity}
		\bigoplus_{e} V_{ab}^e \otimes V_{ec}^d \simeq \bigoplus_{f} V_{bc}^f \otimes V_{af}^d\,.
	\end{equation}
	The linear map between those two vector spaces is given by a 6-index object called the $F$-symbol:
	\begin{equation}\label{eq:F_category}
		\raisebox{-.9cm}{\includegraphics[scale=.40]{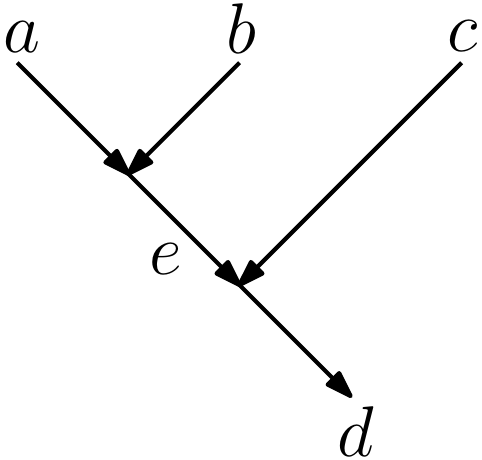}}
		\; =
		\sum_f
		F^{abe^*}_{cd^*f} 
		\;
		\raisebox{-.9cm}{\includegraphics[scale=.40]{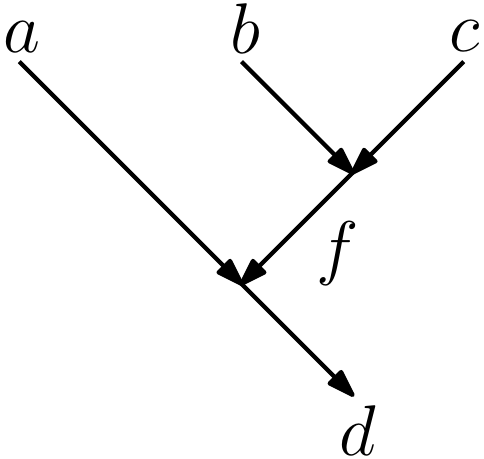}}.
	\end{equation}
	The $ F $-symbol is only defined for allowed fusion vertices. We use the convention that it is zero outside this subspace:
	\begin{equation}\label{eq:F_physicality}
		F^{abe}_{cdf} = \delta_{abe}\delta_{e^*cd}\delta_{adf}\delta_{bcf^*} F^{abe}_{cdf}\,,
	\end{equation}
	where $ \delta_{abc} = \delta_{ab}^{c^*} $.
	The $ F $-symbol must satisfy the following consistency condition, called the \emph{pentagon equation}:
	\begin{equation}\label{eq:pentagon_equation}
		F^{c f g^*}_{e^* d l^*} F^{b a f^*}_{e^* l k^*} = \sum_h
		F^{b a f^*}_{g^* c h^*} F^{h a g^*}_{e^* d k^*} F^{c b h^*}_{k^* d l^*}
		\,.
	\end{equation}

	Each simple object $ a \in \C $, we define a constant $ d_a \in \RR $ called the \emph{quantum dimension}:
	\begin{equation}\label{eq:quantdim}
		d_a = \dfrac{1}{\abs{F^{a a^* \1}_{a a^* \1}}}\,.
	\end{equation}
	The quantum dimensions satisfy
	\begin{equation}\label{eq:quantdim_cond}
		d_a d_b = \sum_k N_{ab}^c d_c \,, \quad \quad	d_\1 = 1\,, \quad \text{and} \quad d_{a^*} = d_{a}\,.
	\end{equation}
	The \emph{total quantum dimension} $ \D $ is defined as
	\begin{equation}\label{eq:tot_quantdim}
		\D = \sqrt{\sum_a d^2_a}\,.
	\end{equation}

	Finally, the $ F $-symbol, when viewed as a matrix $ [F^{ab}_{cd}]_{fe}  = F^{abe}_{cdf}$, must be unitary on the subspace on which it is defined:
	\begin{align}\label{eq:F_unitarity}
		 [(F^{ab}_{cd})^{-1}]_{ef} = [(F^{ab}_{cd})^{\dagger}]_{ef} = \left( [F^{ab}_{cd}]_{fe} \right)^*\,, \\
		 \sum_f \left(F^{abe'}_{cdf}\right)^* F^{abe}_{cdf}  = \delta_{e,e'} \delta_{abe}\delta_{e^*cd}\,. \nonumber
	\end{align}
	
	A \emph{unitary ribbon category} (URC) is obtained by adding the notion of braiding and twists to a UFC.
	Braiding is a unitary operation between fusion spaces that corresponds to a counterclockwise exchange:
	\begin{equation}\label{eq:R_move_ccw}
		R^{ab}: \;  V_{ab}^c \rightarrow V_{ba}^c \;: \;\; \raisebox{-.9cm}{\includegraphics[scale=.40]{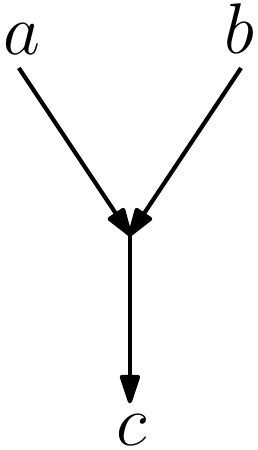}} \mapsto
		\raisebox{-.9cm}{\includegraphics[scale=.40]{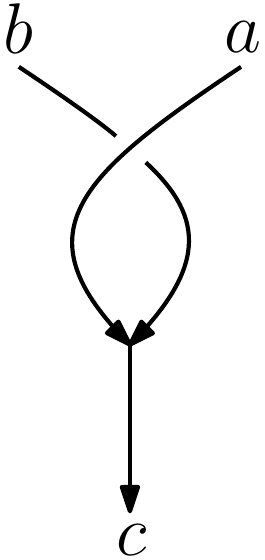}} \; = R^{ab}_c \;
		\raisebox{-.9cm}{\includegraphics[scale=.40]{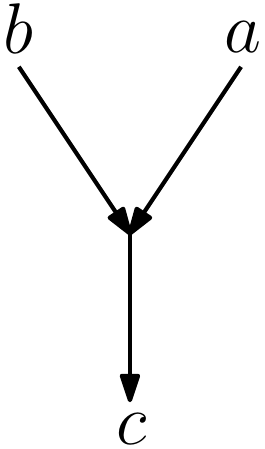}}. 
	\end{equation}
	Note that since we are working in the multiplicity-free case, $ R^{ab}_c $ is simply a complex phase.
	It's inverse corresponds to a clockwise exchange:
	\begin{equation}\label{eq:R_move_cw}
		(R^{ba})^{-1}: \;  V_{ab}^c \rightarrow V_{ba}^c \; : \;\; 
		 \raisebox{-.9cm}{\includegraphics[scale=.40]{fig/R_move_left.pdf}} \mapsto
		\raisebox{-.9cm}{\includegraphics[scale=.40]{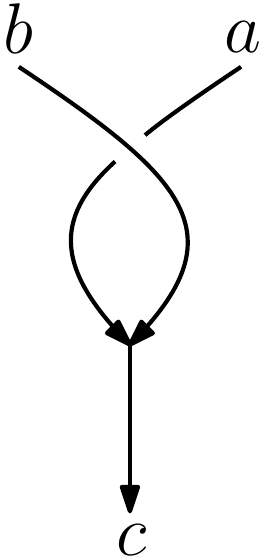}} \; = 
		\left(R^{ba}_{c}\right)^* \;
		\raisebox{-.9cm}{\includegraphics[scale=.40]{fig/R_move_right.pdf}}.
	\end{equation}
	Similar to the pentagon equation \eqref{eq:pentagon_equation} for the $ F $-symbol, the $ R $-symbol must obey consistency conditions called the \emph{hexagon equations}:
	\begin{align}
		R^{ca}_{e} F^{c^*a^*e}_{db^* g} R^{cb}_{g} &= \sum_{f} F^{a^*c^*e}_{db^*f} R^{cf}_{d} F^{b^*a^*f}_{dc^*g}\,, \label{eq:hexagon_equation1} \\
		\big(R^{ac}_{e}\big)^* F^{c^*a^*e}_{db^* g} \big(R^{bc}_{g}\big)^* &= \sum_{f} F^{a^*c^*e}_{db^*f} \big(R^{fc}_{d}\big)^* F^{b^*a^*f}_{dc^*g} \label{eq:hexagon_equation2}\,.
	\end{align}
	
	In a URC, every simple object $ a $ is assigned a \emph{topological phase} $ \theta_a $, used to ``untwist'' a ribbon:
	\begin{equation}\label{eq:topological_spin}
		\raisebox{-.6cm}{\includegraphics[scale=.40]{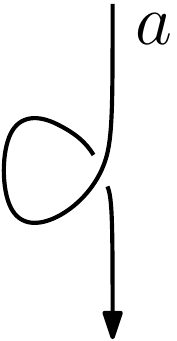}} \; = 
		\theta_a \;
		\raisebox{-.6cm}{\includegraphics[scale=.40]{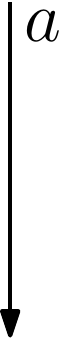}}
		\,, \qquad \quad
		\raisebox{-.6cm}{\includegraphics[scale=.40]{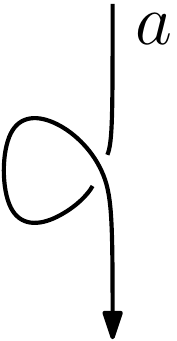}} \; = 
		\left(\theta_a\right)^* \;
		\raisebox{-.6cm}{\includegraphics[scale=.40]{fig/topological_phase_middle.pdf}}\,.
		\,
	\end{equation}
	The topological phase is related to the $R$-symbol as follows:
	\begin{equation}\label{eq:topological_spin_R_symbol}
		\theta_a = \left(R^{aa^*}_\1\right)^* .
	\end{equation}
	
	A \emph{Unitary Modular Tensor Category}, is a unitary ribbon category for which the $R$-matrices satisfy a certain nondegeneracy condition. We will not need the technical definition of modularity. For more details, we refer the reader to Ref.~\cite{wang2010topological}.
	
	\subsubsection{The Fibonacci category} \label{sec:FIB}
		Throughout this paper, we will place a special focus on the Fibonacci category. 
		This category contains two simple objects: 
		\begin{equation}
			 \{\1,  \tau \}\,,
		\end{equation}
	 	with only one nontrivial fusion rule
		\begin{equation}\label{eq:fib_fusion}
			\tau \times \tau = \1 + \tau\,.
		\end{equation}
		The pentagon equation \eqref{eq:pentagon_equation} with these fusion rules only has one solution that satisfies the unitarity condition \eqref{eq:F_unitarity}. 
		The only nontrivial $F$-matrix is
		\begin{equation}\label{eq:fib_F}
			[F^{\tau\tau}_{\tau\tau}] = 
			\begin{pmatrix}
				\phi^{-1} & \phi^{-\frac{1}{2}}  \\
				\phi^{-\frac{1}{2}}&  -\phi^{-1}
			\end{pmatrix},
		\end{equation}
		where $\phi=\frac{\sqrt{5}+1}{2}$ is the golden ratio. 
		For all other combinations of indices, $ F^{ijm}_{kln} $ is either 1 or 0, depending on whether or not the corresponding indices in Eq.~\eqref{eq:F_physicality} satisfy the branching rules.
		
		The quantum dimensions are
		\begin{equation}\label{eq:fib_qd}
			d_\mathbf{1} = 1\,, \qquad d_\tau = \phi\,,
		\end{equation}
		and the total quantum dimension is then given by 
		\begin{equation}\label{eq:fib_D}
			\D = \sqrt{1 + \phi^2} = \sqrt{2 + \phi} = \sqrt{ \sqrt{5} \phi}\,.
		\end{equation}
		The hexagon equation admits two solutions, which are related by complex conjugation. We will use the following:
		\begin{equation}\label{eq:fib_R}
			R^{\tau\tau}_{\mathbf{1}} = \e^{\frac{4\pi\ii}{5}} \, ,
			 \qquad R^{\tau\tau}_{\tau} = \e^{\frac{-3\pi\ii}{5}}\,,
		\end{equation}
		the other $R$-symbols allowed by the fusion rules are $ R^{\mathbf{1} a}_a = R^{a \mathbf{1}}_a = 1$ for any $ a $.
		
		Finally, the  modular $\mathcal{S}$-matrix is 
		\begin{equation}\label{eq:fib_S}
			\mathcal{S}=\frac{1}{\sqrt{2+\phi}}
			\begin{pmatrix}
				1 & \phi  \\
				\phi &  1
			\end{pmatrix}\,.
		\end{equation}
			
\subsection{The ribbon graph Hilbert space}\label{sec:ribbon_graph}
	The \emph{ribbon graph Hilbert space} is the Hilbert space associated to a surface by the Turaev-Viro TQFT defined by a unitary fusion category $ \C $ \cite{konig2010quantum}.  
	Even though it is defined for any unitary fusion category, we consider the special case where the input category $ \C $ is a unitary modular tensor category.
	Furthermore, in	addition to the conditions \eqref{eq:F_physicality}, \eqref{eq:pentagon_equation} and \eqref{eq:F_unitarity} that are satisfied for every UMTC, we impose the following three conditions for the $ F $-symbols: 
	\begin{align}
		F^{ijm}_{kln} = F^{jim}_{lkn^*} &= F^{lkm^*}_{jin} = F^{imj}_{k^*nl}\frac{v_m v_n}{v_j v_l}\,, \label{eq:F_tetrahedral} \\
		F^{ii^*1}_{j^*jk} &= \frac{v_k}{v_i v_j}\, \delta_{ijk}\,,  \label{eq:F_normalization} \\ 
		\left(F^{ijm}_{kln}\right)^* &= F^{lin}_{jkm^*} \,.	\label{eq:F_unitarity_2} 
	\end{align}
	The first of these conditions is known as tetrahedral symmetry, the second one is a normalization condition and the third condition is an alternative formulation of the unitarity condition Eq.~\eqref{eq:F_unitarity}.	\\   
	
	Let $ \Sigma $ be a compact orientable surface with a boundary, where we place a single marked point on each connected component of the boundary.
	A ribbon graph is a labeled, directed graph embedded into $ \Sigma $, with internal vertices of degree two and three, and a vertex of degree one at each marked boundary point of $ \Sigma $.
	The ribbons are labeled with the labels of $ \C $. Reversing the direction of an edge in the ribbon graph corresponds to conjugating the edge label. Edges assigned the trivial label $ 1 $ can be added to or removed from a given ribbon graph, as the vacuum label denotes the absence of a ribbon. We require that each internal vertex satisfies the fusion rules, i.e.: that $ \delta_{ijk} = 1 $ for every vertex with incoming lines $ i $, $ j $ and $ k $.
	
	The \emph{ribbon graph Hilbert space} $ \H_{\Sigma} $ is the space of formal linear combinations of ribbon graphs, modulo the following relations:
	\begin{align} 
		\raisebox{-0.4cm}{\includegraphics[scale=.40]{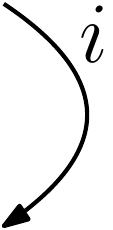}} \;\;
		&= \;\;
		\raisebox{-.4 cm}{\includegraphics[scale=.40]{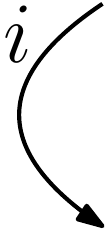}}
		\label{eq:rib_graph_eq_relations1} \;,\\
		\raisebox{-0.35cm}{\includegraphics[scale=.40]{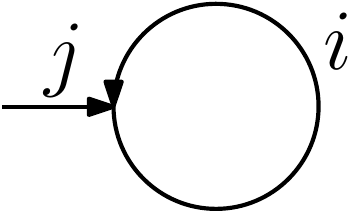}} \;\;
		&= \delta_{j1} \, d_i
		\label{eq:rib_graph_tadpole} \,,\\
		\raisebox{-0.45cm}{\includegraphics[scale=.40]{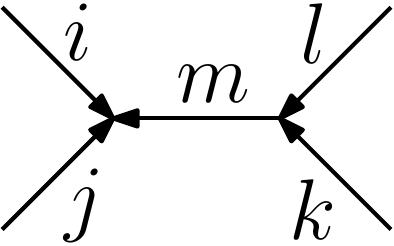}} \;
		&= 
		\sum_{n} \; F^{ijm}_{kln} \;\;
		\raisebox{-.75 cm}{\includegraphics[scale=.40]{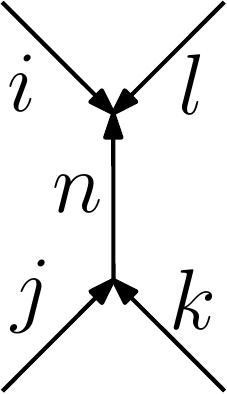}} \,. \label{eq:F-move} 
	\end{align}

	A convenient relation, known as the \emph{bubble bursting} equation, follows from Eqs.~ \eqref{eq:F-move} and Eq.~\eqref{eq:F_normalization}:
	\begin{equation}\label{eq:bubble_bursting}
		\raisebox{-1.cm}{\includegraphics[scale=.40]{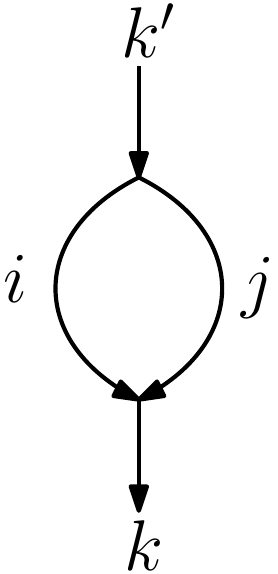}} \; = \frac{v_i v_j}{v_k} \, \delta_{k k'} \;\; \raisebox{-.8cm}{\includegraphics[scale=.40]{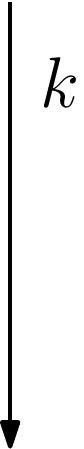}}\,.
	\end{equation}
	Another convenient relation that follows from Eq.~\eqref{eq:F_normalization} is
	\begin{equation}\label{eq:double_line}
		\raisebox{-.9cm}{\includegraphics[scale=.40]{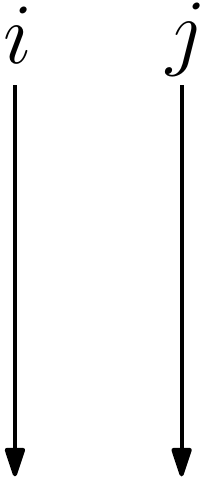}}\; = \sum_k \frac{v_k}{v_i v_j} \delta_{ijk^*} \raisebox{-1.15cm}{\includegraphics[scale=.40]{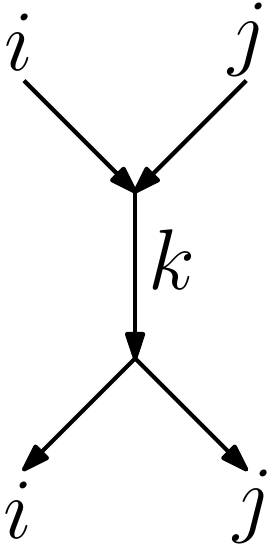}}\,.
	\end{equation}

	From here on, in order to simplify the notation and the graphical representations, we will assume that $ \C $ is self-dual, meaning $ i = i^* $ for all labels $ i \in \C$. This allows us to drop the orientations in the ribbon diagrams. The generalization to non-self-dual categories is straight-forward. 
	A much more elaborate treatment of the ribbon graph Hilbert space, including the non-self-dual case, can be found in Ref.~\cite{konig2010quantum}. 
	However, for our purpose, this simplified and condensed summary is sufficient.\\  
	
	We now consider the case where $ \Sigma $ is the $ n $-punctured sphere, $ \Sigma_n = S^2 \setminus (A_1 \cup \cdots \cup A_n) $, where each $ A_i $ represents a disk.
	To each hole we assign a single marked boundary point $ p \in \partial A_i $. A labeling $ \ell $ of $ \Sigma_n $ associates to each marked boundary point $ p $ a label $ \ell(p) \in \C $. The ribbon graph Hilbert space $ \H_{\Sigma_n} $ can then be decomposed as
	\begin{equation}\label{eq:rib_graph_hspace_decomp}
	\H_{\Sigma_n} = \bigoplus_\ell \H_{\Sigma_n}^\ell \,,
	\end{equation}
	where $ \H_{\Sigma_n}^\ell $ is the subspace spanned by ribbon graphs with an edge labeled by $ \ell(p) $ connected to $ p $, for every marked boundary point $ p $.
	
\begin{figure}[ht]
		\centering
		\includegraphics[scale=.35]{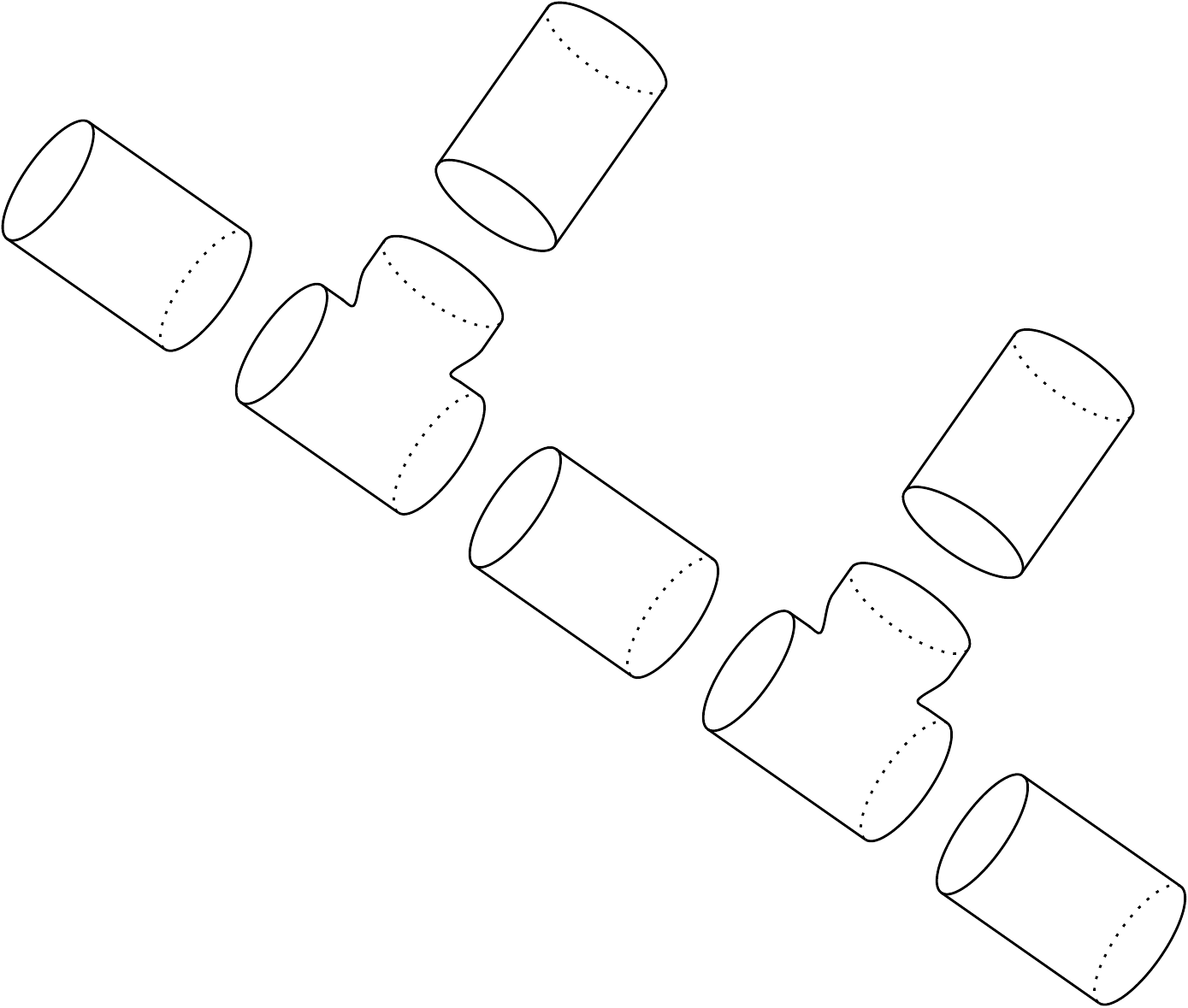}
		\caption{A standard pants decomposition for the sphere with four punctures $ \Sigma_4 $.}
		\label{fig:pants_decomposition_example}
	\end{figure}

	We can define a basis for $ \H_{\Sigma_n} $ based on a \emph{pants decomposition} of the surface $ \Sigma_n  $.
	A pants decomposition is associated to a rooted binary tree with $ n-1 $ leaves, where the root and leaves each correspond to a hole in $ \Sigma_n $. Its internal vertices of degree three correspond to \emph{pants segments}, each isomorphic with $ \Sigma_3 $, while its edges correspond to \emph{cylindrical segments}, each isomorphic with $ \Sigma_2 $. An example of a pants decomposition of $ \Sigma_4 $ is depicted in \figref{fig:pants_decomposition_example}.
	For a given pants decomposition, a basis for $ \H_{\Sigma_n} $ is obtained by fixing the ribbon graph on each leaf to
	\begin{equation}\label{eq:computational_basis_leaf}
		\raisebox{-.9cm}{\includegraphics[scale=.5]{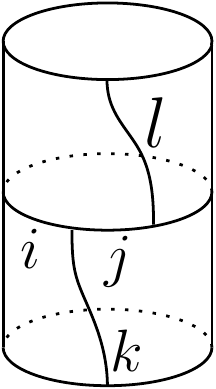}}
		\quad \equiv  \;
		\raisebox{-.9cm}{\includegraphics[scale=.46]{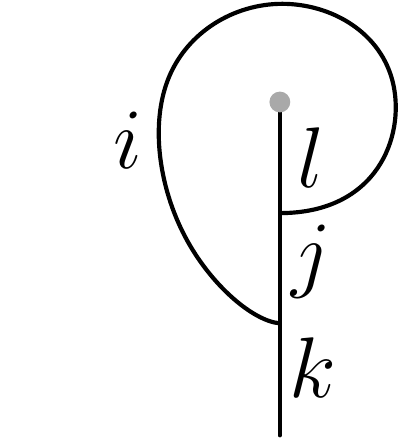}}
		\;,
	\end{equation}
	on each internal segment and on the root to
	\begin{equation}\label{eq:computational_basis_root}
		\raisebox{-.9cm}{\includegraphics[scale=.5]{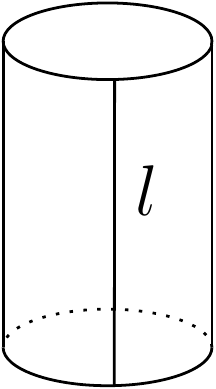}}
		\quad \equiv  \;
		\raisebox{-.9cm}{\includegraphics[scale=.46]{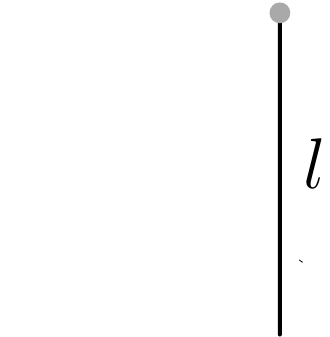}}
		\;,
	\end{equation}
	and finally connecting the ribbons in each pant segment.
	The set of all allowed label assignments of this ribbon graph, constitutes a basis for $ \H_{\Sigma_n} $. An inner product on $ \H_{\Sigma} $ can now be defined by setting these basis states to be orthonormal.	
	This basis is called a \emph{computational basis}, since it provides a way of encoding $ \H_{\Sigma_n} $ into $ 5n - 6 $ qudits. 
	A more general construction, based on specific triangulations of $ \Sigma $, exists and gives rise to an entire family of computational bases for $ \H_{\Sigma_n} $ \cite{konig2010quantum}. These different computational bases can be related using the equivalence rules \eqref{eq:rib_graph_eq_relations1}, \eqref{eq:rib_graph_tadpole} and \eqref{eq:F-move}, ensuring that the inner product they define is unique.
	
\subsection{Dehn twists and braid moves on $ \H_{\Sigma} $}
	Consider the set $ Q \in \Sigma $ of all marked boundary points.
	Diffeomorphisms $f: \Sigma \rightarrow \Sigma$ that map the $ Q $ onto itself, define an action  on $ \H_{\Sigma} $.
	Two special types of such transformations, known is Dehn twists and braid moves, are of particular interests to us, because of their relation with the topological properties of anyon models (i.e.: the $R$-matrices and topological phases appearing in App.~\ref{sec:category}.
			
	A \emph{Dehn twist} corresponds to a $ 2\pi $-counterclockwise twist along a simple closed curve on $ \Sigma_n $. If we consider the non-contractible curve $ \gamma $,
	\begin{equation}\label{eq:non_contractible_curve}
		\raisebox{-.5 cm}{\includegraphics[scale=.5]{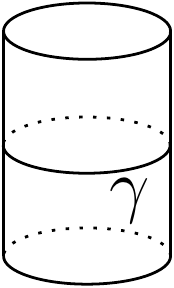}} \quad ,
	\end{equation}
	then a Dehn twist $ D(\gamma) $ along this curve acts as
	\begin{equation}\label{eq:dehn_twist}
		D(\gamma): \;
		\raisebox{-.5cm}{\includegraphics[scale=.5]{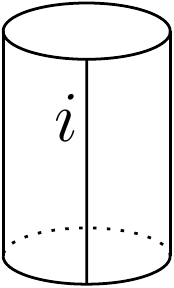}}
		\quad  \mapsto \quad
		\raisebox{-.5cm}{\includegraphics[scale=.5]{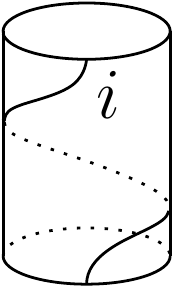}}\;,		
	\end{equation}
	or, schematically
	\begin{equation*}
			D(\gamma): \; \raisebox{-.55cm}{\includegraphics[scale=.5]{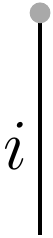}}
		\quad  \mapsto \quad
		\raisebox{-.55cm}{\includegraphics[scale=.5]{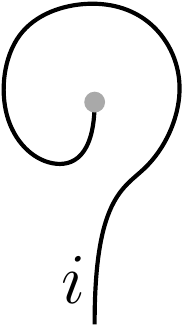}}
		\;.
	\end{equation*}

	A \emph{braid move} acts on two holes of a pair of pants in $ \Sigma_n $ and is defined as a $ \pi $-counterclockwise twist along a simple closed curve on $ \Sigma_n $ enclosing the two holes, followed by a $ \pi $-twist in the opposite direction on each of the legs corresponding to the two holes,
	\begin{equation}\label{eq:braid_move}
		\raisebox{-.7 cm}{\includegraphics[scale=.5]{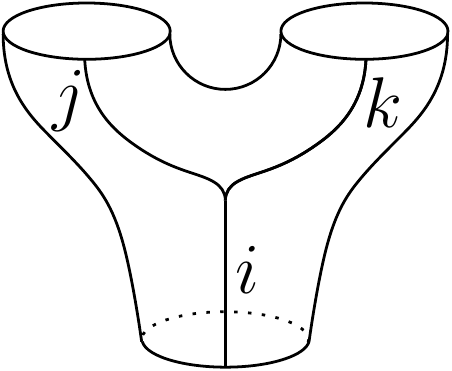}}
		\quad \mapsto \quad
		\raisebox{-.7 cm}{\includegraphics[scale=.5]{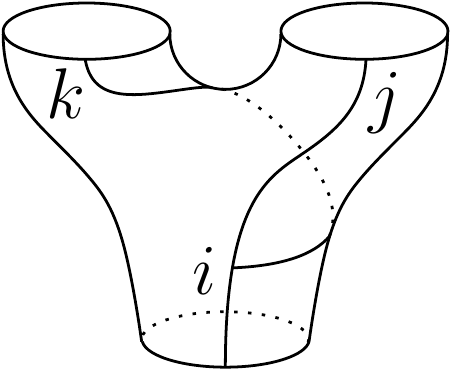}}\;,
	\end{equation}
	or, schematically
	\begin{equation*}
		\raisebox{-.6cm}{\includegraphics[scale=.5]{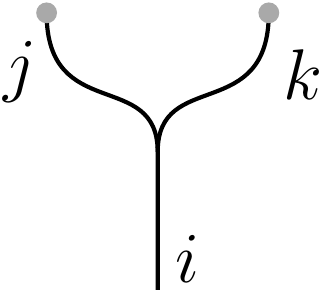}}
		\quad  \mapsto \quad
		\raisebox{-.6cm}{\includegraphics[scale=.5]{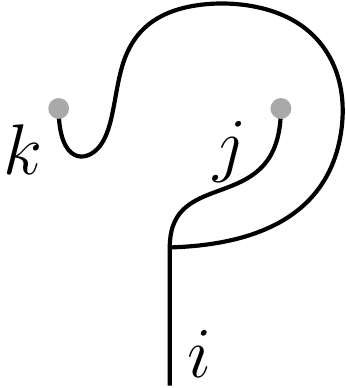}}\;.
	\end{equation*} 

	The action of both Dehn twists and braid moves on $ \H_{\Sigma} $ follows by linearly extending their action on the computational basis ribbon graphs to the full Hilbert space.
	
\subsection{The anyonic fusion basis} \label{sec:anyonic_fusion_basis}
	States described using the computational basis constructed above, do not transform cleanly under the action of the mapping class group generators, making them inconvenient operationally. 
	Below, we will define an orthonormal basis that simultaneously diagonalizes all Dehn twists and behaves nicely under braid moves. While the construction itself is tedious, this basis will reveal much more of the mathematical structure of $ \H_{\Sigma} $, and will prove invaluable for the rest of this work.  
	In particular, this construction will reveal the relation between states in $ \H_{\Sigma} $ and the 
	Drinfeld center or categorical double $\mathcal{D} \mathcal{C}$ of the input category $\mathcal{C}$. 	
	
	It is important to note that the categorical double of a unitary fusion category is always braided, meaning that one does not need to specify braiding properties for the input category in order to have emergent anyon statistics.
	In the special case we are considering however, that where the input category $ \C $ is itself modular, the Drinfeld center has a special structure.	
	For a modular input category $ \C $ the categorical double $ \D\C $ is isomorphic to the direct product of $ \C $ and its reverse category $ \bar{\C}$: $ \D\C \cong \C \otimes \bar{\C}$ \cite{etingof2016tensor}.
	The reverse category $ \bar{\C}$ is obtained from $ \C $ as follows: for each label $ a \in \C $, there is a corresponding label $ \bar{a} \in \bar{C}$ which is the same as $ a $, except for its chirality, which is opposite: $ \theta_{\bar{a}} = (\theta_{a})^*$. Furthermore the $R$-matrix is replaced by	$ R^{\bar{b}\bar{c}}_{\bar{a}}  = (R^{cb}_a)^* $, which one can interpret as swapping the meaning of clockwise and counterclockwise. 
	The doubled category $ \D\C \cong \C \otimes \bar{C}$ then has elements $ \{a \otimes \bar{b} \mid a \in \C ,\, \bar{b}\in \bar{\C}\} $, which we will denote as $ a \bar{b} \equiv a \otimes \bar{b} $, fusion rules \
	\begin{equation}
		\delta_{a\bar{a'} b\bar{b'} c\bar{c'}} = \delta_{abc} \, \delta_{\bar{a'} \bar{b'} \bar{c'}} = \delta_{abc} \, \delta_{a'b'c'}\,,
	\end{equation}
	and numerical data
	\begin{align}
		\theta_{a \bar{a'}} &= \theta_a \theta_{\bar{a'}} = \theta_a (\theta_{a'})^* \,, \label{eq:doubled_topological_spin} \\
		R^{b \bar{b'} \, c \bar{c'}}_{a \bar{a'}} &= R^{b c }_{a} R^{ \bar{b'} \bar{c'}}_{\bar{a'}} = R^{b c}_{a} (R^{ c' b'}_{a'})^*\,,	\label{eq:doubled_R} \\
		F^{a \bar{a'}\, b \bar{b'}\, e \bar{e'}}_{c \bar{c'}\, d \bar{d'}\, f \bar{f'}}	&= F^{a b e}_{c d f}	F^{a' b' e'}_{c' d' f'}	\,. \label{eq:doubled_F}
	\end{align}
	In the following, we will exploit this structure, in order to make the connection between $ \H_{\Sigma} $ and the Drinfield center of $ \C $ explicit. \\

	A \emph{vacuum line}, denoted by a dashed line, is defined as a weighted superposition of ribbons with different labels,
	\begin{equation}\label{eq:vacuum_line}
	\raisebox{-0.4cm}{\includegraphics[scale=.40]{fig/vacuum_line.pdf}} \quad
	= 
	\frac{1}{\D} \sum_{i} d_i  \,\,\,
	\raisebox{-.4 cm}{\includegraphics[scale=.40]{fig/line_i.pdf}} \,.
	\end{equation}
	On can easily show that vacuum loops have the property that all other ribbons can freely pass over them:
	\begin{equation}\label{eq:vacuum_line_pulling_through}
	\raisebox{-.85cm}{\includegraphics[scale=.45]{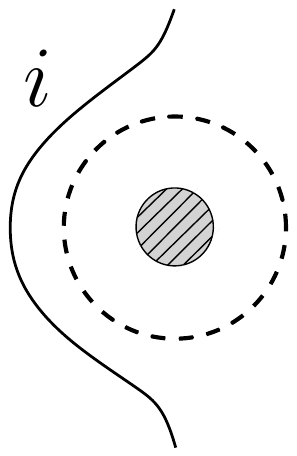}} \quad
	=  \quad
	\raisebox{-.85cm}{\includegraphics[scale=.45]{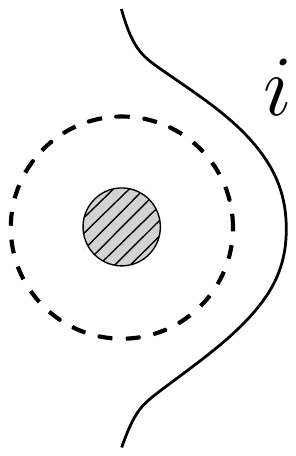}} \; ,
	\end{equation}
	where the hashed area inside the vacuum loop denotes a generic ribbon graph configuration. This represents a local identity, where the ribbon graphs inside the hashed area and outside the diagram are assumed to be the same for the left- and right-hand side. 
	Intuitively, vacuum loops render the region they contain invisible from ribbon graph configurations outside.
	Another useful identity is
	\begin{equation}\label{eq:vacuum_loop_doubling}	
		\raisebox{-.72cm}{\includegraphics[scale=.45]{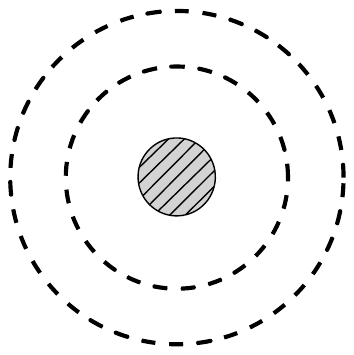}} \;
		=  \D \;\;
		\raisebox{-.47cm}{\includegraphics[scale=.45]{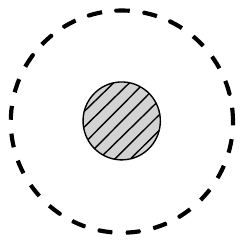}} \; ,
	\end{equation}
	which means we can pull one vacuum loop out of another. \\
	
	Fix a pants decomposition for $ \Sigma_n $ associated to some rooted binary tree $ T $, and fix a labeling $ \ell $ of the marked boundary points like in Eq.~\eqref{eq:rib_graph_hspace_decomp}. 
	A labeling of $ T $ assigns an anyon label from the input category $ \C $ to every edge of $ T $. Such a labeling of $ T $ is said to be fusion-consistent if the fusion rules of $ \C $ are satisfied at every internal vertex of $ T $. A pair of labelings of $ T $ is said to be boundary-consistent with $ \ell $ if for each marked boundary point $ p $, the labels $ a $ and $ b $ assigned to the corresponding leaf of $ T $ by the respective labelings of $ T $ satisfy
	\begin{equation}\label{eq:fusion_basis_boundary_condition}
		\delta_{a b\,\ell(p)} = 1.
	\end{equation}
	A pair of fusion-consistent labelings $ d $ of $ T $ that is boundary-consistent with $ \ell $ is called a \emph{$ \ell $-consistent doubled anyonic fusion diagram}. 
	
	The \emph{anyonic fusion basis} for $ \H_{\Sigma_n}^\ell $ is then an orthonormal basis indexed by $ \ell $-consistent doubled anyonic fusion diagrams.  We still haven't defined the ribbon graph states $ \ket{\ell,d} \in \H_{\Sigma_n}^\ell $, that correspond to the $ \ell $-consistent doubled anyonic fusion diagram $ d $. We will do this by first constructing a three-dimensional ribbon graph on the thickened surface $ \Sigma_n \times [-1,1] $, and then reducing it to a regular ribbon graph on $ \Sigma_n $.
	
	Think of each of the labelings in $ d $ as a ribbon graph, and embed these in the two-dimensional slices $ \Sigma_n \times \{1\} $ and $ \Sigma_n \times \{-1\} $ of the thickened surface, respectively. Place the marked boundary points of $ \Sigma_n $ in $ \Sigma_n \times \{0\} $, and add a vacuum loop in $ \Sigma_n \times \{0\} $ around each of the boundary components. Then close off the diagram by adding a ribbon graph edge in $ \Sigma_n \times \{0\} $ labeled by $ \ell(p) $ connected to every boundary point $ p $, and connect the corresponding edges of the graphs in $ \Sigma_n \times \{1\} $ and $ \Sigma_n \times \{-1\} $ to this new edge, \emph{inside} the vacuum loop around $ p $,
	\begin{align}\label{eq:an_fus_basis_closing_off}
		\raisebox{-1.1cm}{\includegraphics[scale=.52]{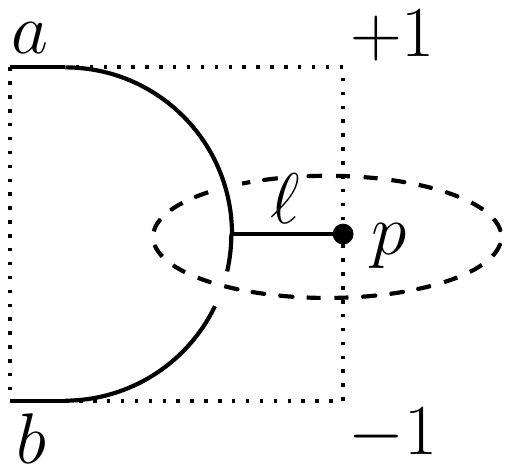}}
		\;\; ,
	\end{align}
	where the vertical direction represents the additional coordinate. Boundary consistency of $ d $ ensures that the vertex created by this closing off of the diagram is a valid ribbon graph vertex. 
	Finally, in order to reduce the resulting diagram to a state in $ \H_{\Sigma_n}^\ell $, visualize it as a two-dimensional ribbon graph with crossings in $ \Sigma_n $ (this requires slightly offsetting $ \Sigma_n \times \{1\} $ and $ \Sigma_n \times \{-1\} $, the convention for the offset only affects the phase of $ \ket{\ell,d} $ ). These crossings can be resolved using the $R$-symbol of the input category by combining  \eqref{eq:R_move_ccw}, \eqref{eq:F-move} and \eqref{eq:F_normalization} into
	\begin{align}\label{eq:resolve_crossing}
		\raisebox{-.6cm}{\includegraphics[scale=.40]{fig/uncrossing_left.pdf}}
		\quad = \sum_{k} \frac{v_k}{v_i v_j} \, R^{ij}_k \,\,\,
		\raisebox{-.6cm}{\includegraphics[scale=.40]{fig/uncrossing_right.pdf}}\;.
	\end{align}
	The state $ \ket{\ell,d} \in \H_{\Sigma_n}^\ell $ is defined as the resulting superposition of ribbon graphs in $ \Sigma_n $ that is obtained by resolving all crossings in this way. The three-dimensional ribbon graphs introduced here can be manipulated in the same way as regular ribbon graphs in $ \Sigma_n $, and it is sometimes useful to perform manipulations in the three-dimensional picture before reducing the ribbon graph back to two dimensions. It was shown in Appendix A of Ref.~\cite{konig2010quantum} that the anyonic fusion basis is a complete orthonormal basis for $ \H_{\Sigma_n}^\ell $. A basis for the entire Hilbert space $ \H_{\Sigma_n} $ is then obtained by taking the union of the bases for all values of $ \ell $ according to \eqref{eq:rib_graph_hspace_decomp}.
	
	The anyonic fusion basis states of $ \H_{\Sigma_2} $ take the form
	\begin{equation}\label{eq:anyonic_fusion_basis_states_2}
		\ket{k, \ell, a, b} = \; \raisebox{-.58cm}{\includegraphics[scale=.4]{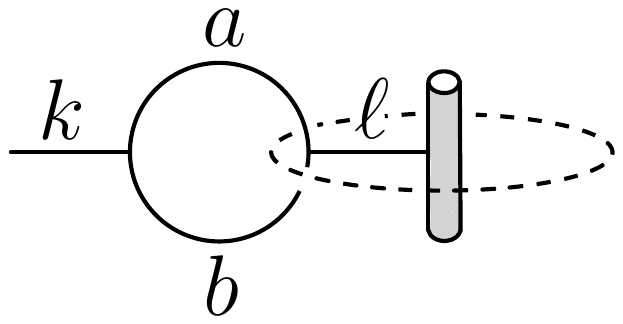}} \; \equiv \; \raisebox{-.8cm}{\includegraphics[scale=.4]{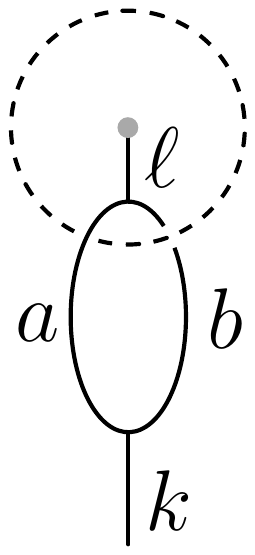}} \;.
	\end{equation}
	\begin{widetext}
	Reducing this to a two-dimensional ribbon graph yields 
	\begin{align}\label{eq:doubled_leaf_segment_red}
		\raisebox{-.8cm}{\includegraphics[scale=.40]{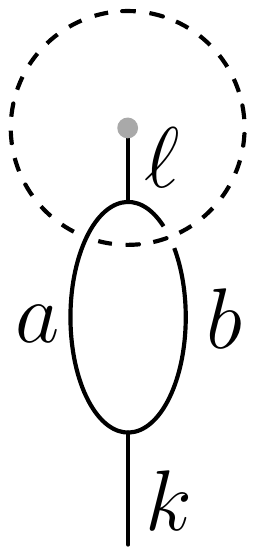}}
		\quad = 
		\frac{1}{\D} v_a v_b \sum_{\alpha, \beta} v_\alpha v_\beta \sum_{\gamma, \delta} d_\gamma d_\delta R^{a \alpha}_{\gamma} R^{\alpha b}_{\delta} G_{\alpha a b}^{k \delta \gamma} G_{b \alpha \ell}^{\beta a \delta} G_{a \gamma \delta}^{k \beta \alpha}
		\raisebox{-.8cm}{\includegraphics[scale=.40]{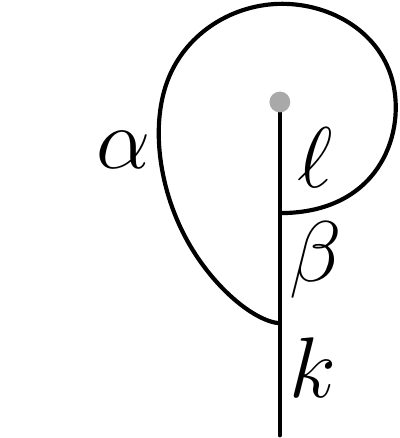}} \; ,
	\end{align} 
	where the six-index $ G $-symbol is defined as
	\begin{align}\label{eq:Gsymbol_def}
		\raisebox{-0.62cm}{\includegraphics[scale=.40]{fig/vertex_loop.pdf}} = v_\lambda v_\mu v_\nu G^{i j k}_{\lambda \mu \nu} \raisebox{-0.42cm}{\includegraphics[scale=.40]{fig/vertex.pdf}}\,.
	\end{align}
	From Eqs.~\eqref{eq:F_tetrahedral}, \eqref{eq:F_unitarity_2}, \eqref{eq:F-move} and \eqref{eq:bubble_bursting}, it then follows that, for an input category describing a self-dual anyon model,
	\begin{equation}\label{eq:Gsymbol_value}
		G^{i j k}_{\lambda \mu \nu} = \frac{1}{v_k v_\nu} F^{i j k}_{\lambda \mu \nu} \,.
	\end{equation}
	
	For the general case of $ \H_{\Sigma_n} $ with a standard pants decomposition we get anyonic fusion basis states of the form
	\begin{align}\label{eq:anyonic_fusion_basis_states_n}
		\hspace{-10pt}
		\ket{\vec{\ell}, \vec{a}, \vec{b}, \vec{c}, \vec{d}} =
		\hspace{-.5cm} \raisebox{-1.8cm}{\includegraphics[scale=.5]{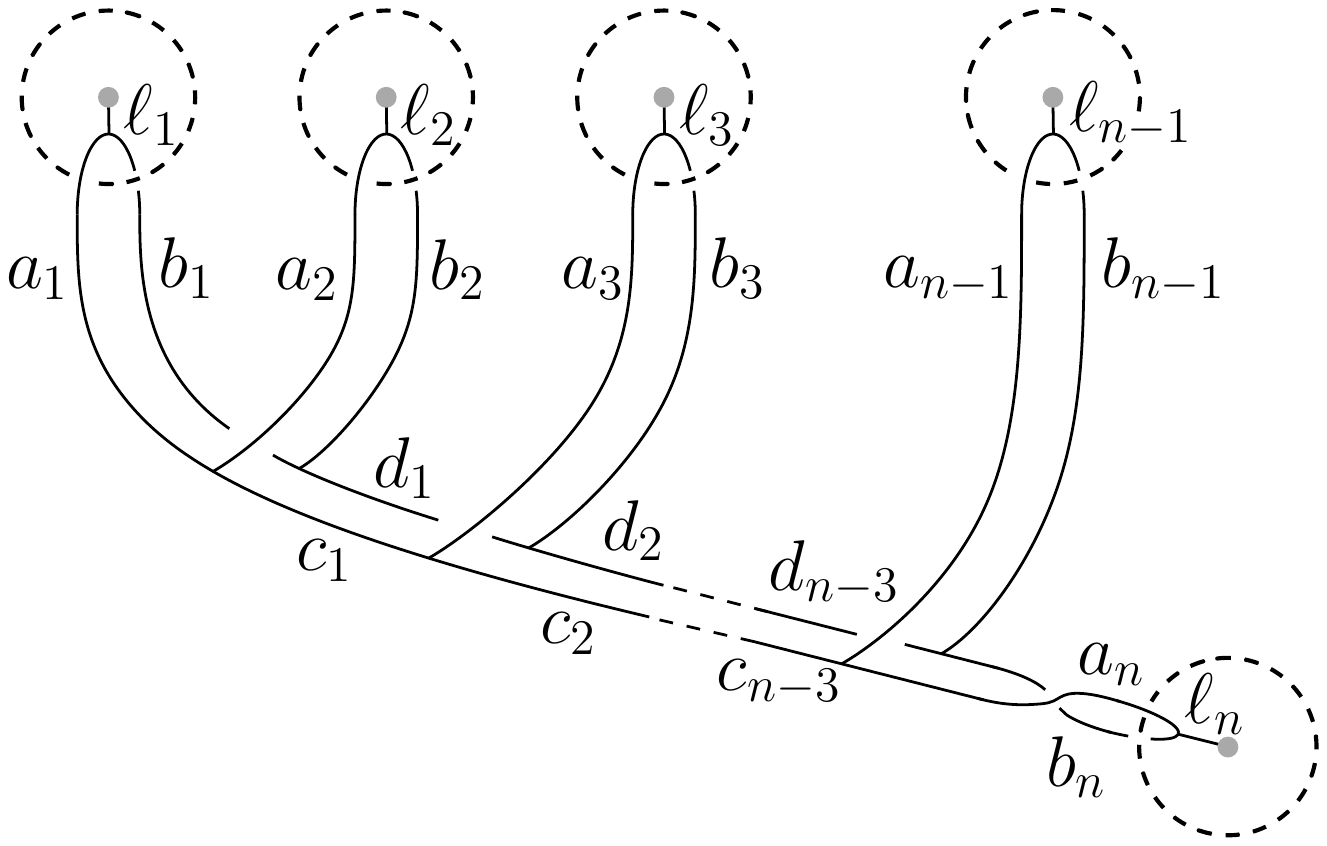}}  \hspace{-.8cm}.
	\end{align}
	Note that the doubled ribbons were crossed before attaching them to the root $ \ell_n $. This is a matter of convention, and was done to give it the same geometry as the other leaves in the pants decomposition.  
	The vacuum loop around the root puncture of the pants decomposition is unnecessary in principle, but was added here to further symmetrize the expression between all punctures\footnote{The inclusion is this additional vacuum loop should be accompanied by a factor $ 1/D$ in order to ensure proper normalization. We choose not to write it explicitly to avoid additional clutter. Unless mentioned otherwise, proper normalization is always assumed for anyonic fusion basis states.}.
	These choices are particularly convenient for the case of self-dual input categories that we are considering here.
	
	Reducing basis state \eqref{eq:anyonic_fusion_basis_states_n} to a two-dimensional ribbon graph, is done by first using Eqs.~\eqref{eq:double_line} and \eqref{eq:R_move_cw} to reduce all doubled interior lines to single lines, which gives 
	\begin{equation}\label{eq:anyonic_fusion_basis_red1}
			\ket{\vec{\ell}, \vec{a}, \vec{b}, \vec{c}, \vec{d}} = \quad \sum_{\vec{k},\vec{l}}
		\left(\prod_{x=1}^{n} \frac{v_{k_{x}}}{v_{a_x}v_{b_x}} \right) \left(\prod_{y=1}^{n-3} \frac{v_{l_{y}}}{v_{c_y}v_{d_y}} \right) \left(R^{b_n a_n}_{k_n}\right)^* \hspace{-.3cm}
		\raisebox{-1.8cm}{\includegraphics[scale=.5]{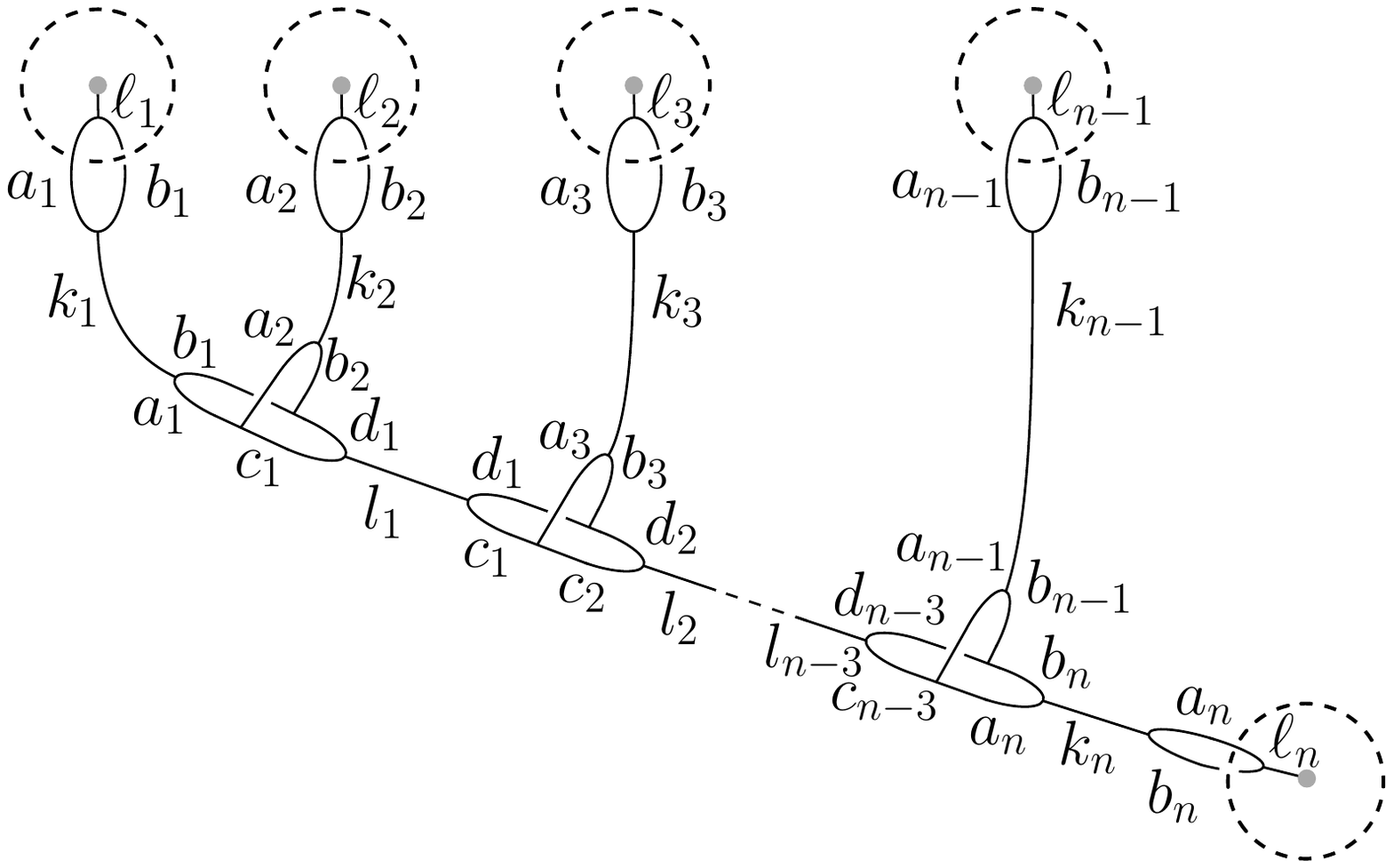}} \hspace{-.7cm}.
	\end{equation}

	The leaf segments can then be reduced as in \eqref{eq:doubled_leaf_segment_red}, while the pants segments in \eqref{eq:anyonic_fusion_basis_red1} can be reduced as follows: 
	\begin{equation}\label{eq:doubled_pants_segment_red}
		\raisebox{-.9cm}{\includegraphics[scale=.4]{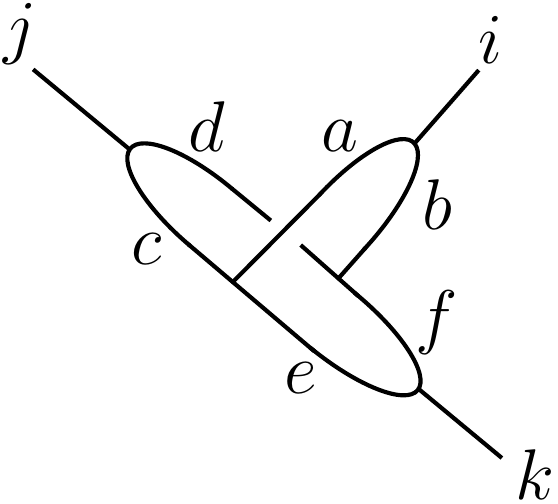}}
		\quad = 
		v_a v_b v_c v_d v_e v_f \sum_{\gamma, \delta} d_\gamma d_\delta R^{a d}_{\gamma} G_{def}^{k b \delta} G_{d a e}^{c \delta \gamma} G_{\gamma d c}^{j \delta a} G_{\delta b a}^{i j k}
		\raisebox{-.75cm}{\includegraphics[scale=.4]{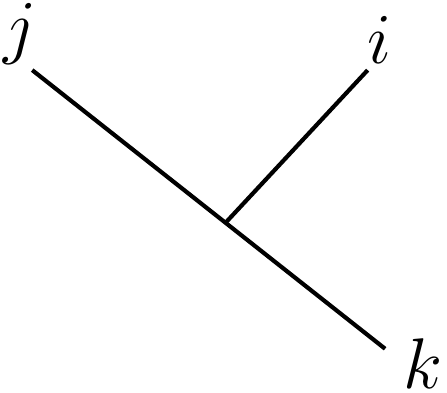}} \,.
	\end{equation}

	By combining all of this, one finds 
	\begin{equation}\label{eq:anyonic_fusion_basis_red2}
		\ket{\vec{\ell}, \vec{a}, \vec{b}, \vec{c}, \vec{d}} =  	\sum_{\vec{\alpha}, \vec{\beta}, \vec{k}} X^{\vec{a}, \vec{b}, \vec{c}, \vec{d}, \vec{\ell}}_{\vec{\alpha}, \vec{\beta}, \vec{k}, \vec{l}} 
		\hspace{-1.2cm}
		\raisebox{-2.7cm}{\includegraphics[scale=.5]{fig/anyonic_fus_bas_state_n_red2.pdf}} \,,
	\end{equation}
	where the coefficients $  X^{(\vec{a}, \vec{b}, \vec{c}, \vec{d}, \vec{\ell})}_{\vec{\alpha}, \vec{\beta}, \vec{k}, \vec{l}} $ follow from Eqs.~\eqref{eq:anyonic_fusion_basis_red1}, \eqref{eq:doubled_leaf_segment_red} and \eqref{eq:doubled_pants_segment_red}. 
	\end{widetext}

	The basis states of the anyonic fusion basis for a given pants decomposition, can be interpreted as fusion states of anyons from the doubled category $ \D\C $. 
	Indeed, one can easily verify that the topological phases and the $R$-matrix match with \eqref{eq:doubled_topological_spin} and \eqref{eq:doubled_fusion_basis_stacking}, respectively, by computing the action of a Dehn twist and of a braid move. Furthermore the bases corresponding to different pants decompositions are related by the correct doubled $F$-symbol \eqref{eq:doubled_F}. 
	These calculations are performed in detail in Ref.~\cite{konig2010quantum}. For completeness, we include them (for the self-dual case) below.
	
    \subsubsection{The action of Dehn twists, braiding, and recoupling on the anyonic fusion basis}
    	Consider a cylindrical segment in the pants decomposition (for which we have constructed the anyonic fusion basis), with labels $ a $ and $ b $ assigned to it by an anyonic fusion basis state $ \ket{\ell,d}$. The three-dimensional ribbon graph in $ \Sigma_2 \times [-1,1] $ corresponding to this cylindrical segment, can be represented as the thickened disk
    	\begin{equation}\label{eq:doubled_fusion_cyl_segment}
			\raisebox{-1cm}{\includegraphics[scale=.4]{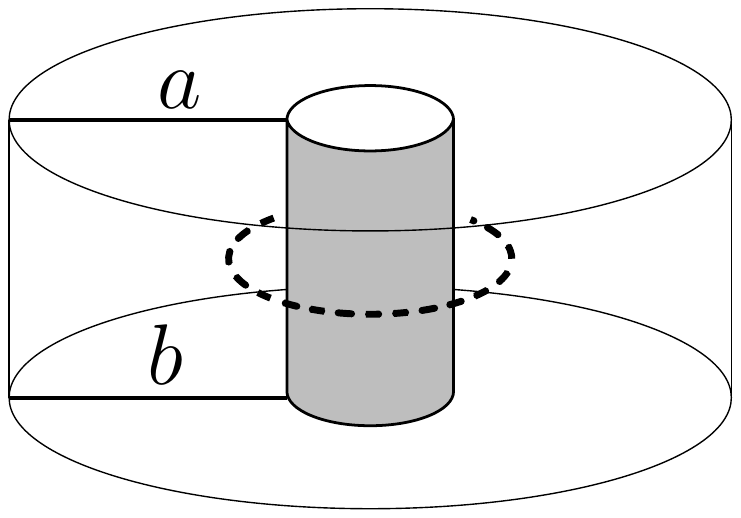}}\quad,
		\end{equation}
		where the shaded region denotes the inner component of $ (\partial \Sigma_2) \times [-1,1] $. 
		Note that we have applied \eqref{eq:vacuum_line_pulling_through} in order to pull a vacuum loop from the shaded region. 
		This vacuum loop can be thought of as originating at one of the boundary components inside the shaded region, using \eqref{eq:vacuum_loop_doubling}.
		For simplicity we will stop drawing the outer boundary of the cylindrical segment. 
		A Dehn-twist along a non-contractible curve $ \gamma $ around the cylinder acts on this three-dimensional ribbon graph in exactly the same way as it would on a regular ribbon graph \eqref{eq:dehn_twist}, giving
		\begin{align}\label{eq:doubled_fusion_cyl_segment_twisted}
			D(\gamma)\; &
			\raisebox{-.7cm}{\includegraphics[scale=.4]{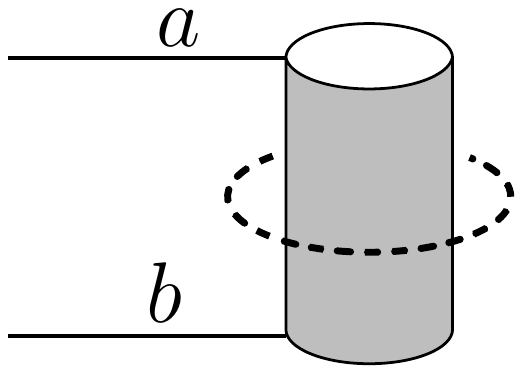}} \; = \;
			\raisebox{-.9cm}{\includegraphics[scale=.4]{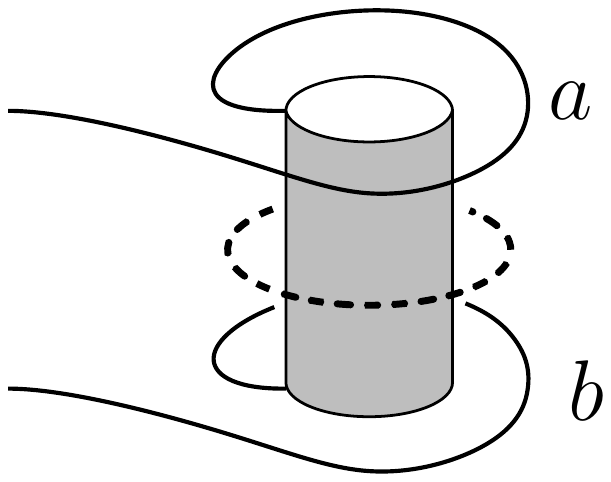}} \; \nonumber \\
			& = \;
			\raisebox{-.73cm}{\includegraphics[scale=.4]{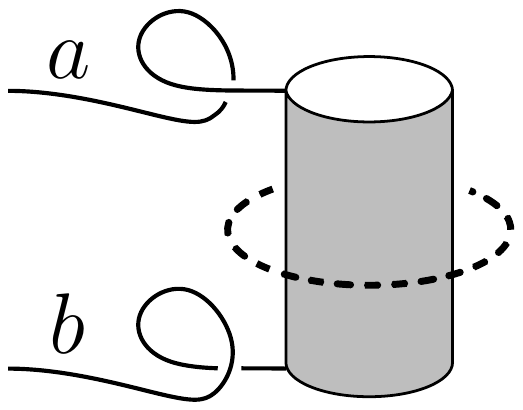}} \; 
			= \theta_a \theta_b^* \;
			\raisebox{-.7cm}{\includegraphics[scale=.4]{fig/3D_cylinder_segment_straight.pdf}}\;,
		\end{align}
		where the resulting state was simplified by moving the ribbons to $ \Sigma_n \times \{0\} $ and pulling them through the vacuum loop, and using \eqref{eq:topological_spin} to remove the resulting kinks.
		The above identity shows that the anyonic fusion basis diagonalizes Dehn twists, and that the corresponding eigenvalues are the topological spins of the doubled anyons given in Eq.~\eqref{eq:doubled_topological_spin}.
		
		Now consider a pants segment, with corresponding labels $ a, b $ and $ c $, and $ a', b' $ and $ c' $ in $ \ket{\ell, d} $. The resulting three-dimensional ribbon graph is 
		\begin{equation}\label{eq:doubled_fusion_pants_segment}
			\raisebox{-1cm}{\includegraphics[scale=.5]{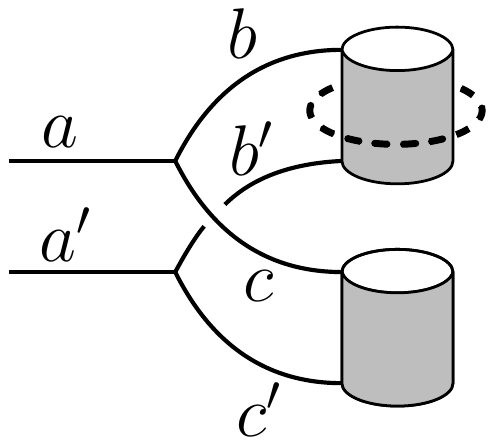}}\;,
		\end{equation}
		where we have again pulled a vacuum loop from one of the shaded regions.
		A braid move on the two interior holes acts in the same way it would on a regular ribbon graph \eqref{eq:braid_move}, giving
		\begin{widetext}
			\begin{equation}\label{eq:doubled_fusion_pants_segment_braided}
				\raisebox{-1cm}{\includegraphics[scale=.5]{fig/3D_pants_segment_straight.pdf}} \;\; \mapsto \quad
				\raisebox{-1.2cm}{\includegraphics[scale=.5]{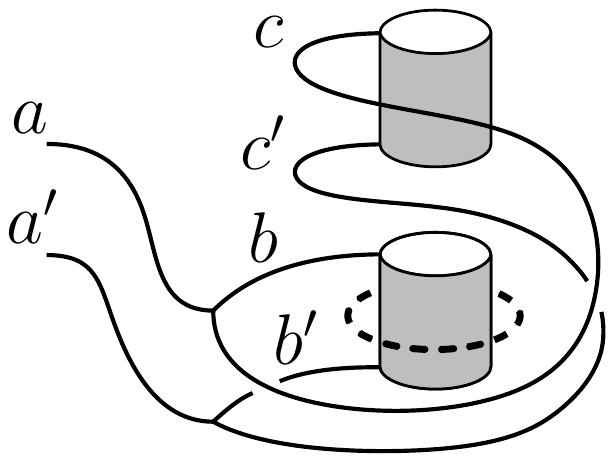}} \quad = \quad
				\raisebox{-1cm}{\includegraphics[scale=.5]{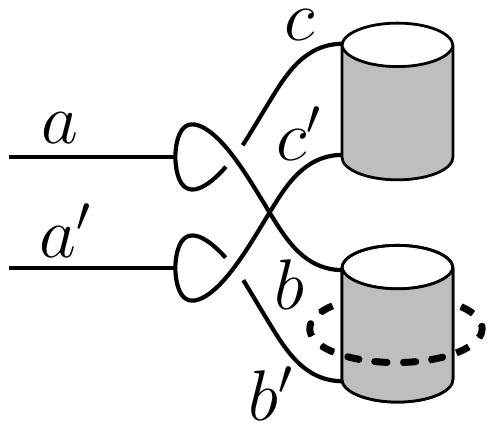}}
				\quad = R^{bc}_a \left(R^{c'b'}_{a'}\right)^* \;
				\raisebox{-1cm}{\includegraphics[scale=.5]{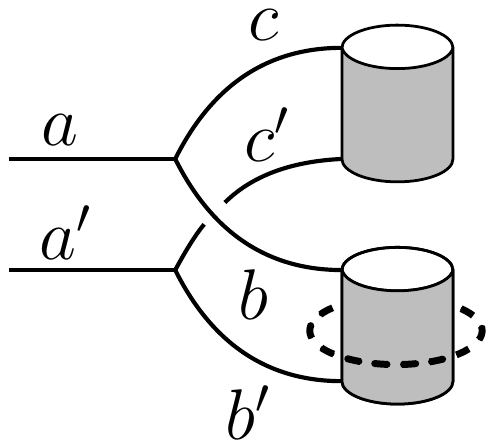}}\;,
			\end{equation}
		\end{widetext}
		where we have again pulled the upper and lower ribbons through the vacuum loop, and have used \eqref{eq:R_move_ccw} and \eqref{eq:R_move_cw} to remove the resulting kinks.
		A quick glance at \eqref{eq:doubled_R} confirms that the is again consistent with the interpretation of fusion basis state as fusion states of doubled anyons.
		
		Finally we investigate the relation between anyonic fusion basis states associated with different pants decompositions of $ \Sigma_n $. For this we can simply use Eq.~\eqref{eq:F-move} on both the top and the bottom ribbon graph independently in the three-dimensional picture. For $ \Sigma_4 $ this gives the relation
		\begin{widetext}
			\begin{equation}\label{eq:doubled_F_move}
				\raisebox{-1.7cm}{\includegraphics[scale=.52]{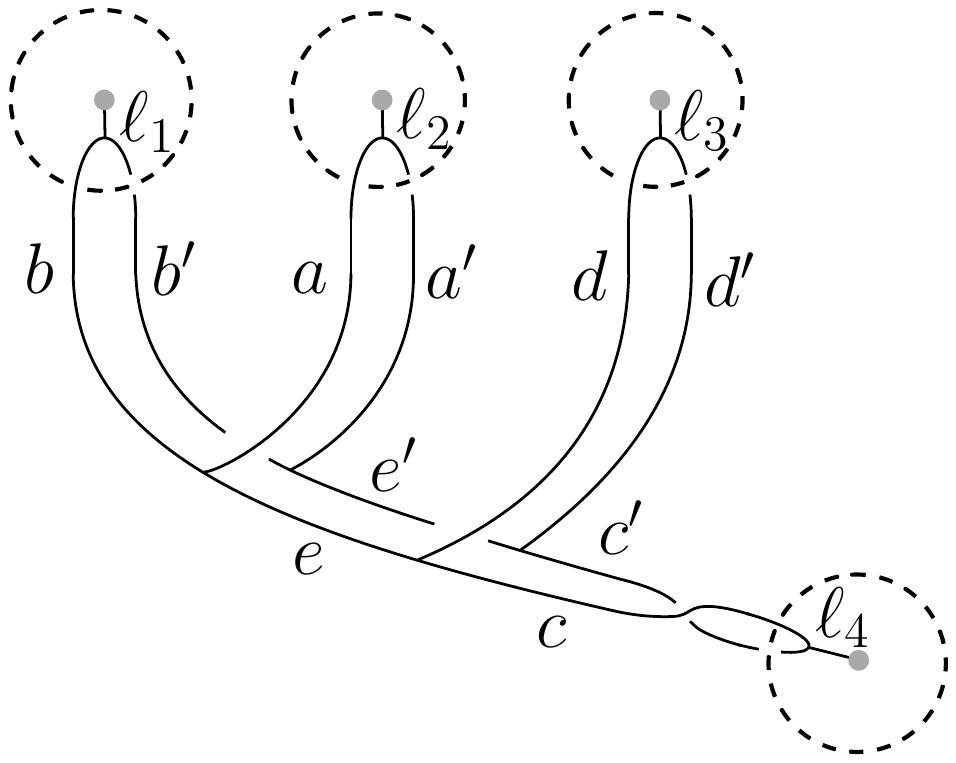}}
				\quad = \sum_{f, f'} F^{abe}_{cdf} F^{a'b'e'}_{c'd'f'}\quad
				\raisebox{-1.7cm}{\includegraphics[scale=.52]{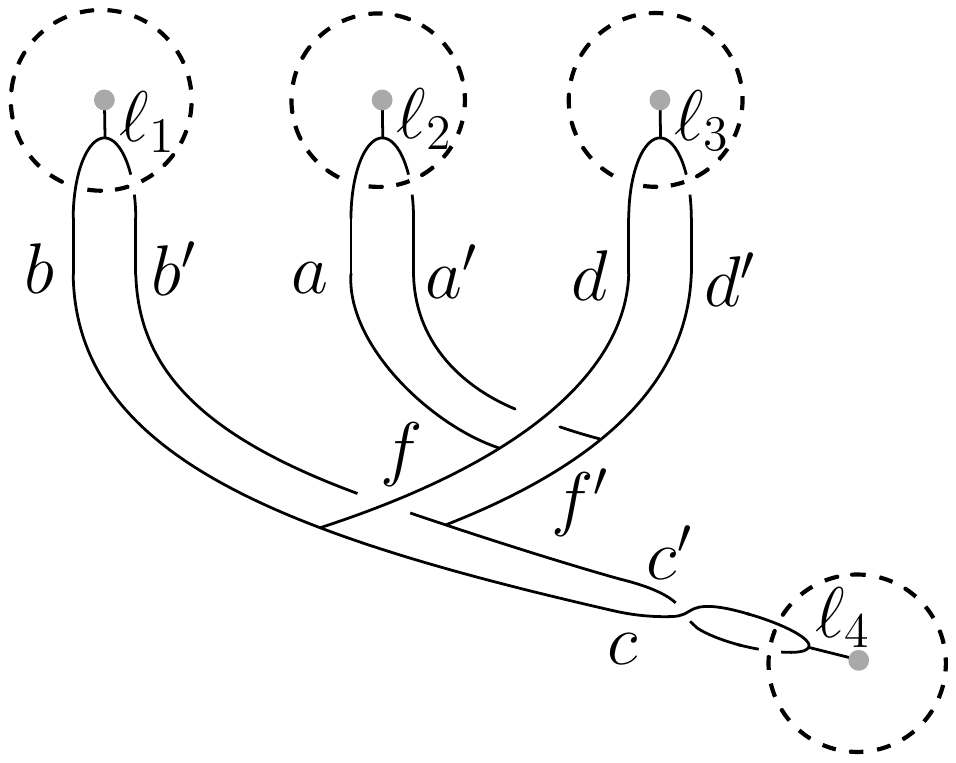}} \hspace{-1cm},
			\end{equation}
		\end{widetext}
		which is indeed the correct $F$-matrix for of the doubled theory $ \D\C $ given in Eq.~\eqref{eq:doubled_F}.
		
	\subsubsection{Higher-genus surfaces} \label{sec:higher-genus}
		The definition of the anyonic fusion basis can be extended to surfaces of a higher genus. 
		Consider a surface $ \Pi_n $ with genus $ g $, and $ n $ punctures. We start by noting that $\Pi_n$ homeomorphic to the $ (n+g) $-puncuted sphere $\Sigma_{n+g}$ with a handle (a punctured torus) glued $ g $ punctures. This is known as a handle decomposition.
		
		As before, we start by fixing a pants decomposition of $ \Sigma_{n+g} $ corresponding to a rooted binary tree $ T $ of our choosing.
		Anyonic fusion basis states of $ \H_{\Pi_n} $ are then labeled with a labeling $ \ell $ of the $ n $ marked boundary points, a doubled anyonic fusion diagram $ d $ of $ n+g $ anyons (including the root), and a set $ \jmath  $ of $ g $ doubled anyon labels which we will call the \emph{handle labels}. 
		As before, the doubled anyonic fusion diagram $ d $ must be $ \ell $-consistent. 
		In addition to this, it must also be consistent with the handle labels $ \jmath $, by which we mean that for a handle $ h \in \{1,\dots, g\}$ with handle label $\jmath(h) = b_+\overline{b_-}$, the doubled anyonic label $ a_+ \overline{a_-} $ of the corresponding leaf of $ T $ must satisfy $ \delta_{a_+ b_+ b_+} = \delta_{a_- b_- b_-} = 1 $
		
		The state $ \ket{\ell, d, \jmath} $ $ \H_{\Pi_n} $ is then constructed similarly as the anyonic fusion basis states on the punctured sphere.
		Start by constructing the three-dimensional ribbon graph corresponding to the doubled fusion diagram $ d $ on  $ \Sigma_{n+g} \times [-1, 1]$. In the $ n $ regular punctures, the diagram is closed off according to the boundary labeling $ \ell $ using \eqref{eq:an_fus_basis_closing_off}. 
		In the remaining $ g $ punctures, the diagram is closed off using one of the following ribbon configurations:
		\begin{align}\label{eq:handle_diagrams}
			\raisebox{-1.1cm}{\includegraphics[scale=.5]{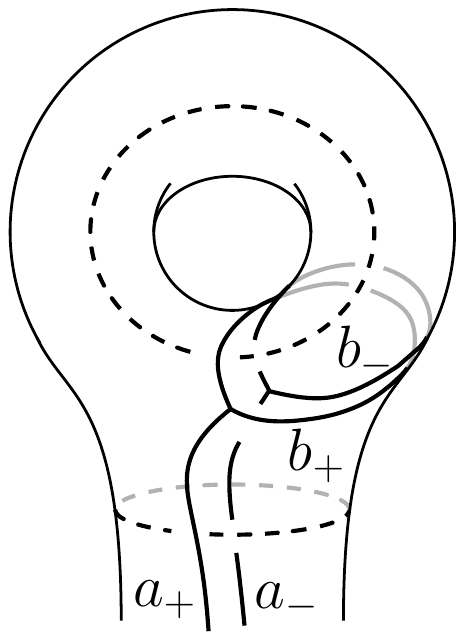}}\;, \qquad \qquad 
			\raisebox{-1.1cm}{\includegraphics[scale=.5]{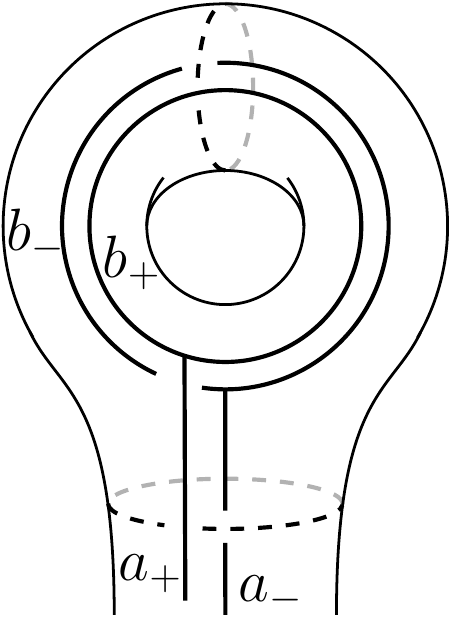}}\;,
		\end{align}
		where $ a_+ $ and $ a_- $ are the labels of the corresponding edge in the doubled fusion diagram, and $ b_+ $ and $ b_- $ are the handle labels of the corresponding handle $ h $.
		
		Finally, this three-dimensional ribbon graph in $ \Pi_{n} \times [-1, 1]$ is reduced to a regular ribbon graph in $ \Pi_{n} $ as before, using Eq.~\eqref{eq:resolve_crossing} to remove any crossings.
		The two possible choices for the handle configuration in \eqref{eq:handle_diagrams} result in two nonequivalent bases. The unitary operator relating them can be found in Ref.~\cite{konig2010quantum}. For $ a_+ = a_- = \mathbf{1} $, it is given by the S-matrix of the doubled category $ \C \otimes \bar{\C}$.
		Note that Eqs.~\eqref{eq:doubled_fusion_cyl_segment_twisted} and \eqref{eq:doubled_fusion_pants_segment_braided} are still satisfied, meaning that the resulting ribbon graph states indeed transform appropriately under the action of the mapping class group.
		
		On a torus, the anyonic fusion basis states (with the first handle choice and $ \jmath = e\bar{f} $) take the following form:
		\begin{widetext}	
		\begin{equation}\label{eq:anyonic_fusion_basis_states_torus}
			 \ket{\vec{\ell}, \vec{a}, \vec{b}, \vec{c}, \vec{d}, e,f}  = \quad
			 \raisebox{-2.1cm}{\includegraphics[scale=.55]{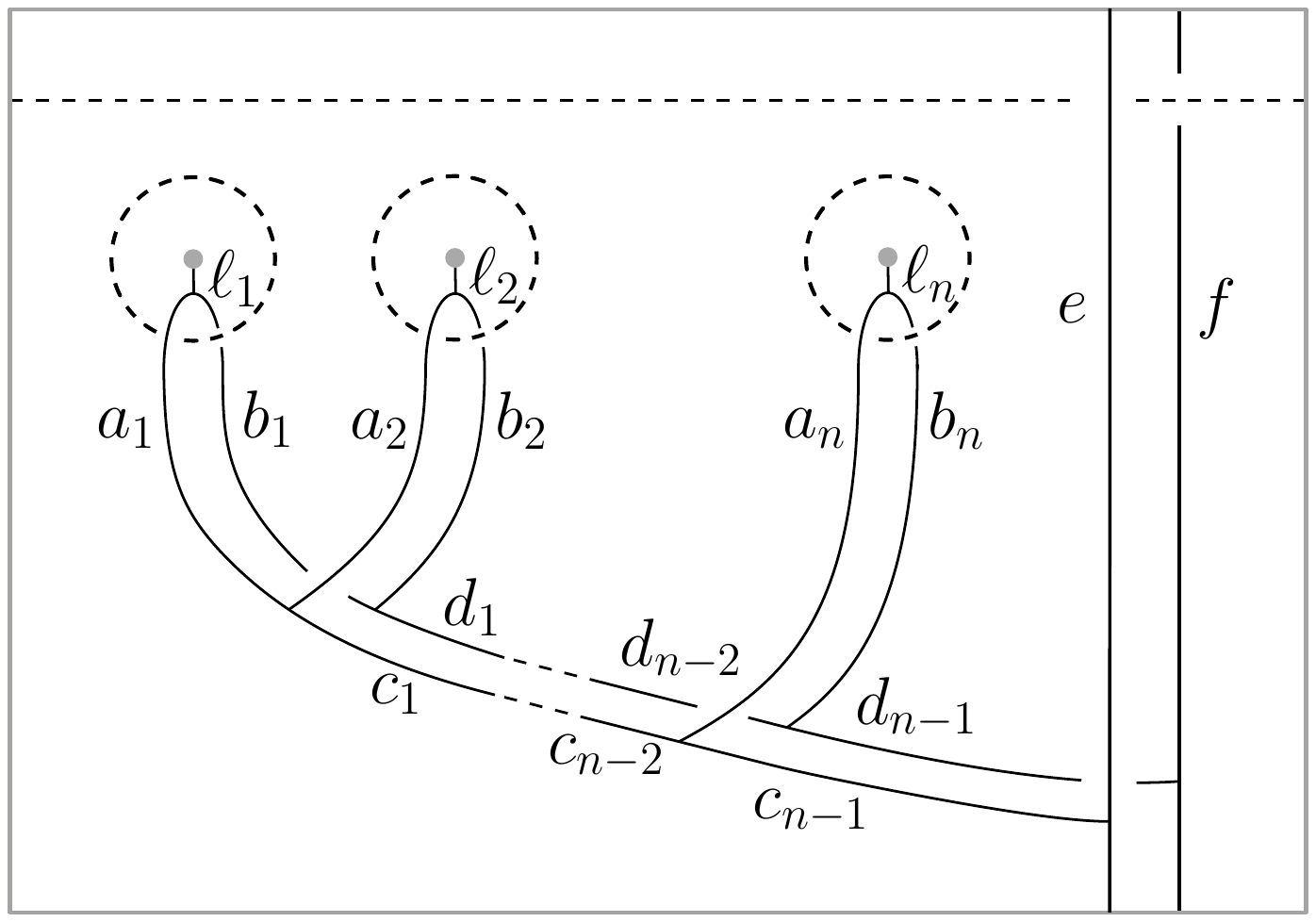}} \; ,
		\end{equation}
		\end{widetext}
		where the gray rectangle represents the periodic boundary conditions of a torus.
		
		The subspace of $ \H_{\Pi_n} $ where all punctures have the doubled anyon label $ 1\bar{1} $ is called the \emph{anyonic vacuum}, and is isomorphic to $ \H_{\Pi_{n=0}} $.
		It follows from \eqref{eq:handle_diagrams}, that the anyonic vacuum on a surface with nonzero genus is degenerate.
		For a torus, this degeneracy is precisely the number of anyon labels in $ \C \otimes \bar{\C}$, as can be deducted from  \eqref{eq:anyonic_fusion_basis_states_torus}.
		The vacuum subspace on a torus is spanned by the different possible handle labels. These states are locally indistinguishable since the loops around the handle can be continuously deformed in an arbitrary way without changing the resulting state.
		
		Note that in the case where the doubled anyons associated to the punctures of the
		torus have a trivial total charge (corresponding to $ c_{n-2} = d_{n-2} = \mathbf{1} $ above), the handle labels in the corresponding anyonic fusion states have no influence on braiding operations on these anyons, since the loops around the handle ( the lines labeled with $e$ and $f$ above) can always be pulled through a group of anyons with a trivial total charge without changing the state.
		In practice, when working with states with a trivial total charge, one can largely ignore the precise value of the handle loop labels, since they can not be affected by local operations on $ \H_{\Pi}\ $.
		The only operations that can modify the handle labels are those involving paths that are homotopically nonequivalent to the ribbons connected to the punctures.  An example of such an operation is a clockwise exchange of two punctures along a path that is homotopically nonequivalent to the ribbons connecting these punctures in the anyonic fusion basis [e.g.: when using the basis in \eqref{eq:anyonic_fusion_basis_states_torus}, exchanging the first two punctures along a path that crosses the vertical gray boundary].
		Subspaces with fixed handle loop labels $ \jmath = e\bar{f} $, are not invariant under such operators. Indeed, describing the outcome of this exchange (in the original basis), will yield a superposition in the loop labels $ e $ and $ f $.	
		An extensive treatment of fusion spaces of anyons on a torus can be found in Ref.~\cite{pfeifer2012translation},  the ``outside'' and ``inside'' bases introduced there, correspond to the first and second handle choices [depicted in \eqref{eq:handle_diagrams}],  respectively.

\subsection{The tube algebra} \label{sec:tube_algebra}
	Let $ \Sigma_A $ and $ \Sigma_B $ be two surfaces, and let $ \Sigma $ be the surface obtained by gluing these surfaces together along one or more boundary components, in such a way that the marked boundary points are matched. 
	Ribbon graphs on $ \sigma_A $ and $ \Sigma_B $ with matching edge-labels in the glued boundary components, can then be glued together to form a ribbon graph on $ \Sigma $. 
	In this way, we can associate to every element $ s \in \H_{\Sigma_B} $ a linear map $ \hat{s} : \H_{\Sigma_A} \rightarrow \H_{\Sigma}$. The result of acting with $ \hat{s} $ on state $ t \in \H_{\Sigma_B} $ is a linear combination of ribbon graphs comprising $ s $ and $ t $ that have matching boundary labels.
	The map 
	\begin{equation} 
		\widehat{} \;: \H_{\Sigma_B} \rightarrow \widehat{\H_{\Sigma_B} }
	\end{equation} 
	is an isomorphism between vector spaces, hence we can interpret ribbon graphs as both states and operators interchangeably. 
	Since the composition of operators in $ \widehat{\H_{\Sigma_2}} $ is again an operator in $ \widehat{\H_{\Sigma_2}} $, they define an operator algebra, known as Ocneanu's \emph{tube algebra} \cite{ocneanu2001operator}. 
	The computational basis of $ \H_{\Sigma_2} $ then corresponds to the basis $ \{ O_{k l \alpha \beta} \} $ of $ \widehat{\H_{\Sigma_2} }$ with
	\begin{equation}\label{eq:tube_operator}
		O_{k l \alpha \beta} = \raisebox{-.8cm}{\includegraphics[scale=.4]{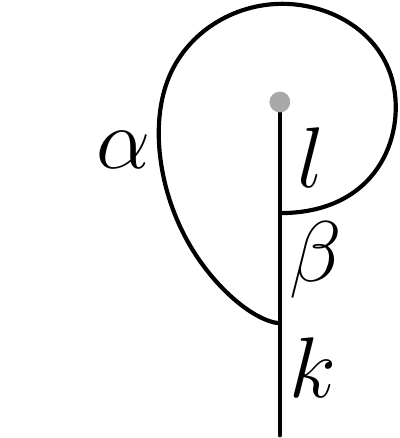}} \; .
	\end{equation}
	
	Consider the operators
	\begin{equation}
		\cP^{a\bar{b}}_{kl} \equiv  \frac{1}{\D} \frac{v_a v_b}{v_k} \quad \raisebox{-.8cm}{\includegraphics[scale=.4]{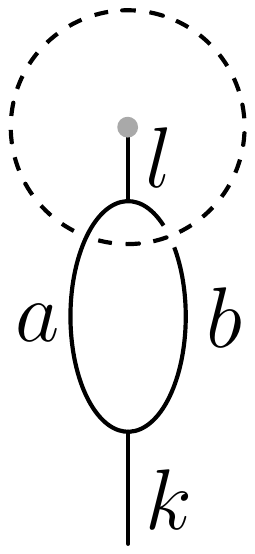}} \; ,
	\end{equation}
	corresponding to (a rescaling of) the anyonic fusion basis states \eqref{eq:anyonic_fusion_basis_states_2} of  $ \widehat{\Sigma_2} $.
	One can show that stacking two such anyonic fusion basis states yields
	\begin{align}
		&\raisebox{-.58cm}{\includegraphics[scale=.4]{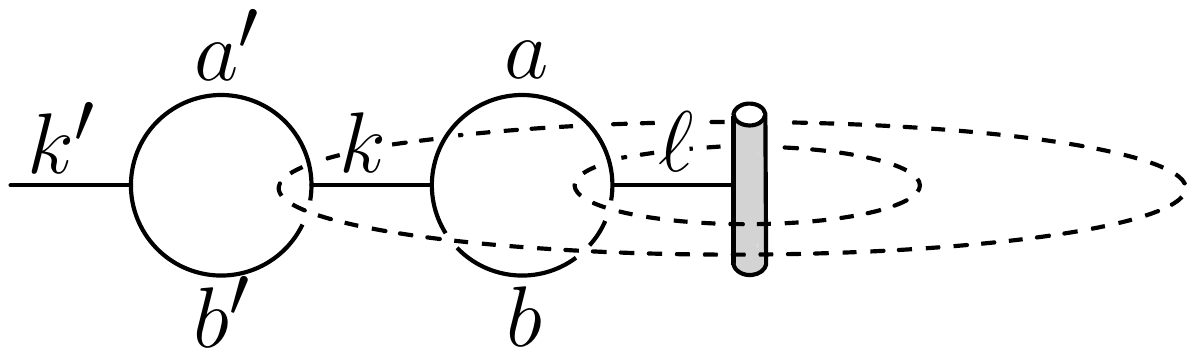}} \nonumber \\
		\quad &= \D \, \delta_{aa'} \delta_{bb'} \frac{v_k}{v_a v_b} \quad
		\raisebox{-.58cm}{\includegraphics[scale=.4]{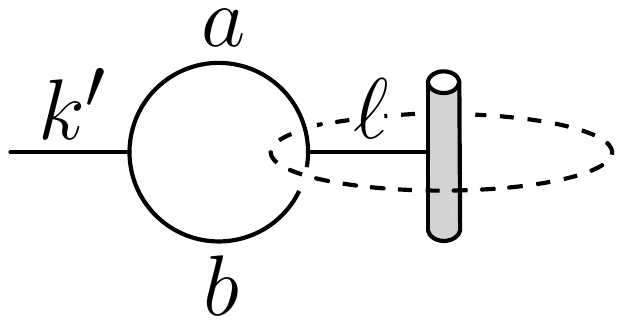}} \;. \label{eq:doubled_fusion_basis_stacking} 
	\end{align}
	Hence, the operators $ \cP^{a\bar{b}}_l \equiv \cP^{a\bar{b}}_{ll} $ are the simple idempotents of the tube algebra:
	\begin{equation} \label{eq:tube_algebra_idempotents}
		\cP^{a\bar{b}}_{l} \equiv  \frac{1}{\D} \frac{v_a v_b}{v_l} \quad \raisebox{-.8cm}{\includegraphics[scale=.4]{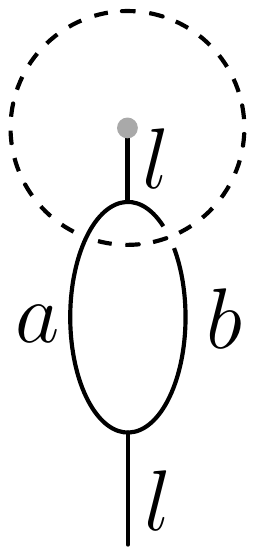}} \; .
	\end{equation}
	Similarly, the operators $ 	\cP^{a\bar{b}}_{kl}  $ with $ k \ne l $ are nilpotent.

	In general the simple idempotents Eq.~\eqref{eq:tube_algebra_idempotents} do not commute with all elements from the tube algebra. 
	In particular, if any nilpotents $ 	\cP^{a\bar{b}}_{kl}  $  ($ k\neq l $) exist for some pair of labels $ a\bar{b} $, then these will not commute with the simple idempotents $ \cP^{a\bar{b}}_{l} $ since Eq.~\eqref{eq:doubled_fusion_basis_stacking} implies $ \cP^{a\bar{b}}_{l} \cP^{a\bar{b}}_{kl}  = \cP^{a\bar{b}}_{kl}   $ and $ \cP^{a\bar{b}}_{kl} \cP^{a\bar{b}}_{l} = 0  $.	
	For every pair of labels $ a\bar{b} $, one can construct a unique central idempotent as follows: 
	\begin{equation}\label{eq:tube_algebra_central_idempotents}
		\cP^{a\bar{b}}  =  \sum_{l\, |\, \delta_{abl} = 1} \cP^{a\bar{b}}_l \,.
	\end{equation}
	These central idempotents project on the different anyon superselection sectors in a puncture: for a state $ \psi \in \H_{\Sigma_n} $ that is a superposition of basis states \eqref{eq:anyonic_fusion_basis_states_n}, stacking the central idempotent $ \cP^{a\bar{b}} $ onto the $ i^{\text{th}} $ hole results in the state where we only keep those basis states in the superposition with anyon label $ a\bar{b} $ associated to the $ i^{\text{th}} $ hole.
	Note that since these projectors commute with all elements of the tube algebra, the anyon label of a puncture is stable against local deformations of the ribbon graph (obtained by locally acting with tube algebra elements). 
	Stacking the irreducible idempotent $ \cP^{a\bar{b}}_l $ on the  $ i^{\text{th}} $ hole instead, further limits the resulting superposition to states with boundary label $ l $ in the $ i^{\text{th}} $ hole.
	
	The simple idempotents can be expressed in the basis of basic tube operators Eq.~\eqref{eq:tube_operator}:
	\begin{equation}\label{eq:idempotent_decomp}
		\cP^{a\bar{b}}_l = \sum_{\alpha \beta} P^{(abl)}_{\alpha \beta} O_{l l \alpha \beta}\,,
	\end{equation}
	where the coefficients $ P^{(abl)}_{\alpha \beta} $ follow from \eqref{eq:doubled_leaf_segment_red}
	\begin{equation}\label{eq:idempotent_decomp_coeff}
		P^{(abl)}_{\alpha \beta} = \frac{1}{\D^2} \frac{d_a d_b}{v_l} v_\alpha v_\beta \sum_{\gamma, \delta} d_\gamma d_\delta R^{a \alpha}_{\gamma} R^{\alpha b}_{\delta} G_{\alpha a b}^{l \delta \gamma} G_{b \alpha l}^{\beta a \delta} G_{a \gamma \delta}^{l \beta \alpha} \,.
	\end{equation}
		
	For the Fibonacci input category (FIB), the doubled category is known as the doubled Fibonacci category (DFIB): 
	\begin{equation}\label{eq:DFIB}
		\text{DFIB} = 	\mathrm{FIB} \times \mathrm{FIB}^* = \{\1, \tau \} \times \{\bar{\1}, \bar{\tau }\}\,.
	\end{equation}	
	The central idempotents corresponding to these 4 anyon types are given by
	\begin{align}
		\cP^{\1\bar{\1}} &= \dfrac{1}{\D^2} \left(O_{\1\1\1\1} + \phi O_{\1 \1 \tau \tau} \right ), \label{eq:P11} \\
		\cP^{\1\bar{\tau}} &= \dfrac{1}{\D^2} \left(O_{\tau \tau \1 \tau}  + \e^{4\pi \ii/5} O_{\tau\tau \tau 1}  + \sqrt{\phi} \e^{-3\pi \ii/5} O_{\tau\tau\tau\tau} \right ), \label{eq:P12} \\
		\cP^{\tau \bar{\1}} &= \dfrac{1}{\D^2} \left (O_{\tau  \tau \1 \tau}  + \e^{-4\pi \ii/5} O_{\tau\tau\tau \1} + \sqrt{\phi} \e^{3\pi \ii/5} O_{\tau\tau\tau\tau} \right ), \label{eq:P21}\\
		\cP^{\tau\bar{\tau}} &= \frac{1}{\D^2} \bigg(\phi^2 O_{\1\1\1\1} - \phi O_{\1 \1 \tau \tau} +   \phi O_{\tau \tau \1 \tau}  + \phi O_{\tau\tau\tau \1} \nonumber \\
		& \quad + \dfrac{1}{\sqrt{\phi}} O_{\tau\tau\tau\tau}  \bigg) \label{eq:P22}.
	\end{align}
	where $ \phi = (1+\sqrt{5})/2 $ is again the golden ratio. 
	The last entry decomposes into two simple idempotents: $ \cP^{\tau \bar{\tau}} = \cP^{\tau \bar{\tau}}_\1 + \cP^{\tau \bar{\tau}}_\tau $ with		
	\begin{align}
		\cP^{\tau\bar{\tau}}_\1 &= \dfrac{1}{\D^2} (\phi^2 O_{\1\1\1\1} - \phi O_{\1\1\tau \tau}), \label{eq:P221}\\
		\cP^{\tau\bar{\tau}}_\tau &= \dfrac{1}{\D^2}  \left(\phi O_{\tau \tau \1  \tau}  + \phi O_{\tau\tau\tau \1} + \dfrac{1}{\sqrt{\phi}} O_{\tau\tau\tau\tau} \right). \label{eq:P222}
	\end{align}
		
	It is important to note that both the +1 eigenstates of $ \cP^{\tau\bar{\tau}}_\1 $ and those of $ \cP^{\tau\bar{\tau}}_\tau $, should be interpreted as containing a $ \tau\bar{\tau} $ anyon. 
	It follows from Eq.~\eqref{eq:doubled_fusion_basis_stacking} that their respective +1 eigenspaces are related through the nilpotent operators
	\begin{align} 
		 \cP^{\tau\bar{\tau}}_{\1\tau}  = \e^{-3\pi \ii/10} \dfrac{\phi  }{\D} O_{\1\tau\tau\tau}\,, \\
		 \cP^{\tau\bar{\tau}}_{\tau\1}  =  \e^{3\pi \ii/10} \dfrac{\sqrt{\phi}}{\D} O_{\tau\1\tau\tau}\,,
	\end{align}
	as follows
	\begin{align} 
		\cP^{\tau\bar{\tau}}_{\1\tau} \cP^{\tau\bar{\tau}}_\1 = \cP^{\tau\bar{\tau}}_\tau,
		\qquad \quad
		\cP^{\tau\bar{\tau}}_{\tau\1} \cP^{\tau\bar{\tau}}_\tau = \cP^{\tau\bar{\tau}}_\1 .	
	\end{align}

\subsection{The Levin-Wen string-net model as a lattice realization of $ \H_{\Sigma}\ $} \label{sec:lattice}
	We now turn to the realization of the ribbon graph Hilbert space $ \H_{\Sigma_n} $ as the ground space of a lattice spin model.
	We will consider surfaces $  \Sigma_n $ containing  $ n $ punctures and no other boundary components.
	
	Let $ \mathcal{T} $ be a triangulation of $  \Sigma_n $, and denote its dual graph by $ \Lambda $. Then $ \Lambda $ is a connected graph with vertices of degree three (each corresponding to a triangle in $ \mathcal{T} $). 
	Note that we take $ \Lambda $ to contain boundary edges for each hole, corresponding to edges in $ \mathcal{T} $ that lie along the boundary of the respective holes.
	We now modify the $ \Lambda $ as follows: for each hole, remove all but one of the boundary edges, and then remove the resulting vertices of degree two by identifying the two edges for each of them.
	The are then left with a lattice containing exactly one boundary edge for each hole, and vertices of degree 3.
	
	A qudit with local Hilbert space $ \H_e = \CC^{N}$ is placed on each edge $ e $ of $ \Lambda $, where $ N $ is the number of anyon labels in the input category $ \C $.
	We choose an orthonormal basis $ \{\ket{i}_e\} $ for this local Hilbert space, where each basis element is labeled by an anyon type of $ \C $. This gives a lattice spin system with a Hilbert space $ \H  \equiv \left(\H_e \right)^{\otimes E} = (\CC^{N})^{\otimes E}$, where $ E $ is the number of edges in $ \Lambda $.
	
	For every vertex $ v $ in  $ \Lambda $, we define the following vertex operator:
	\begin{align}\label{eq:Qv}
		Q_v  = \sum_{i j k} \delta_{ijk} \ket{ijk} \bra{ijk}\,,
	\end{align}
	where $ i $, $ j $ and $ k $ are the labels of the qubits associated the three edges connected to vertex $ v $. One can easily see that all $ Q_v $ operators commute with one another, meaning that they can be simultaneously diagonalized. 
	We denote the simultaneous +1 eigenspace of all $ Q_v $ by $ \H_\text{s.n.} \equiv \{ \ket{\psi} \in \H \,|\, \forall v : Q_v \ket{\psi} = \ket{\psi} \} $, and will refer to it as the \emph{string-net subspace} of $ \H $. 
	Let $ P $ be the number of plaquettes in $ \Lambda $, and let $ \Delta \simeq \Sigma_{n+P} $ be the surface obtained by placing a puncture at the center of each plaquette in $ \Sigma_n $. The states in $  \H_\text{s.n.} $ can then be regarded as (superpositions of) ribbon graphs on $ \Delta $ by embedding the lattice $ \Lambda$ in surface $ \Delta $. More precisely, one can show that the string-net subspace is isomorphic to  $ \bigoplus_\ell \H_{\Sigma_{n+P}}^{(\ell, 1^P)} $, which is the subspace of $ \H_{\Delta}$ defined by the condition that all $ P $ holes corresponding to plaquettes in $ \Lambda $ must have a trivial boundary label.

	For each plaquette $ p $ of the lattice $ \Lambda $, we now introduce a plaquette operator which corresponds to adding a vacuum loop (devided by the total quantum dimension) around the puncture in $ \Sigma_{n+P}$, that corresponds to the said plaquette:
	\begin{align}\label{eq:ribbon_graph_plaquette_operator}
		B_p \Bigg |  \raisebox{-1.1cm}{\includegraphics[scale=.32]{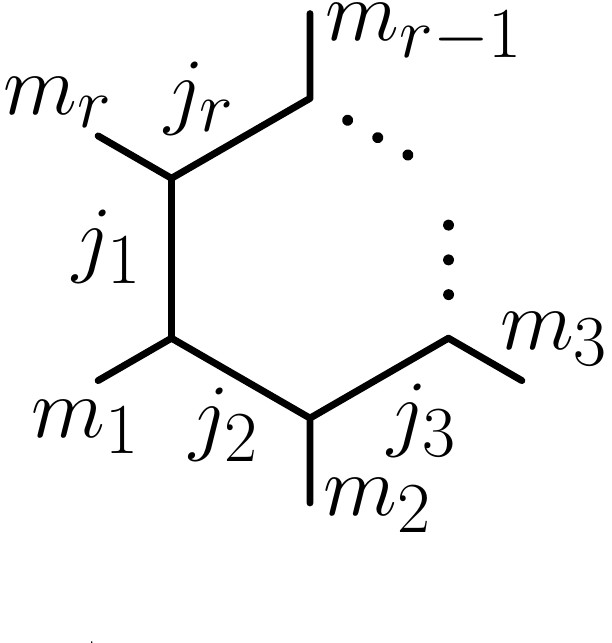}} \Bigg \rangle &= 
		\frac{1}{\D} \Bigg |
		\raisebox{-1.1cm}{\includegraphics[scale=.32]{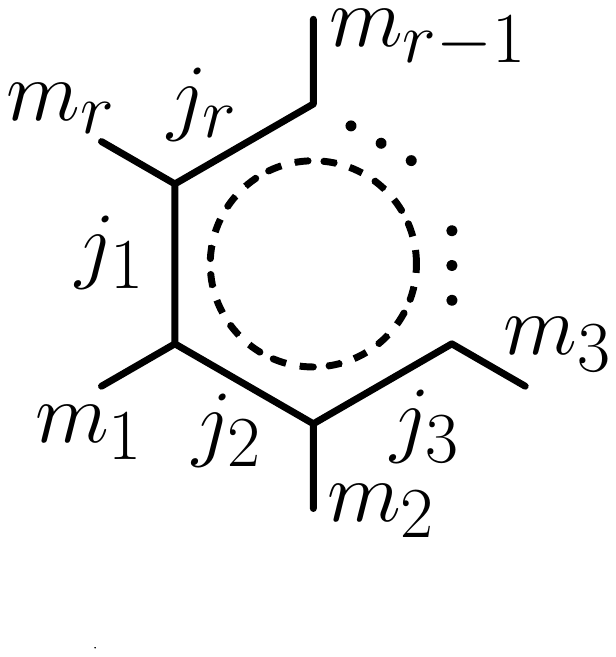}} \Bigg \rangle \nonumber \\
		& = 
		\frac{1}{\D^2} \sum_{s} d_s \Bigg |
		\raisebox{-1.1cm}{\includegraphics[scale=.32]{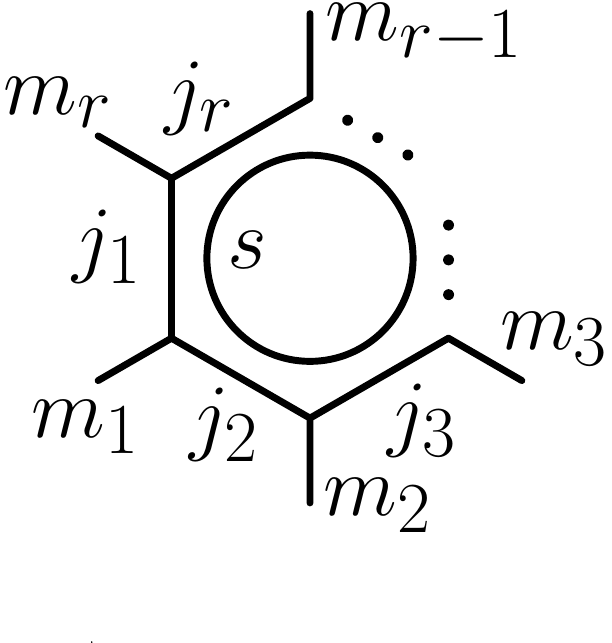}} \Bigg \rangle \,.
	\end{align}
	Its action on $ \H_{\text{s.n.}} $ can be determined by repeatedly using Eq.~\eqref{eq:F-move} to pull the interior loop onto the lattice, giving
	\begin{equation}\label{eq:ribbon_graph_Bp}
		B_p = \dfrac{1}{\D^2} \sum_s d_s O_p^s\,,
	\end{equation}
	with 
	\begin{widetext}
	\begin{equation}\label{eq:ribbon_graph_plaquette_operator_action}
	    O_p^s  \Bigg |  	\raisebox{-1.1cm}{\includegraphics[scale=.32]{fig/plaquette_1.pdf}} \Bigg \rangle 
	    =  \sum_{k_1,..,k_r}\left(\prod_{\nu = 1}^r F^{m_\nu j_{\nu+1} j_{\nu}}_{s k_{\nu} k_{\nu+1}}\right)
	    \Bigg |
        \raisebox{-1.1cm}{\includegraphics[scale=.32]{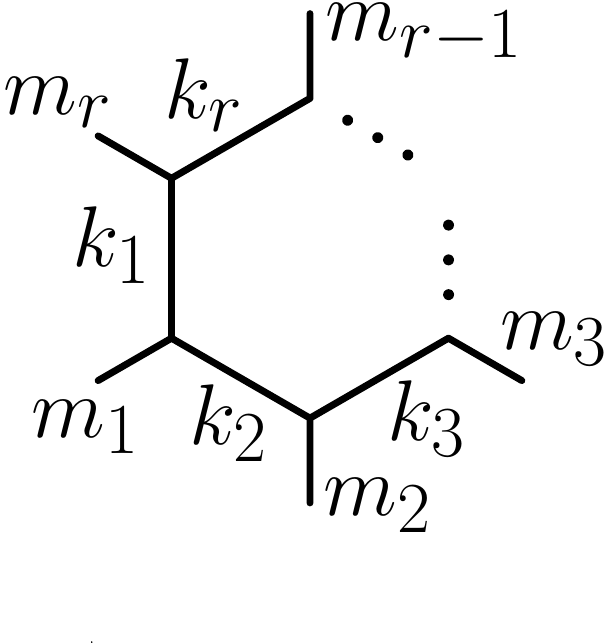}} \Bigg \rangle \,.
	\end{equation}
	\end{widetext}
	Note that it follows form the $F$-symbol condition \eqref{eq:F_physicality} that $ B_p = 0 $ outside the subspace where all involved vertices $ v $ satisfy the vertex condition imposed by $ Q_v $, meaning $ B_p $ is properly defined on the entire Hilbert space $ \H $.
	
	Using the identity \eqref{eq:vacuum_loop_doubling}, one can see that the operators $ B_p $ are in fact projectors. They are in fact the lattice realization of the central idempotent $ \cP^{\1\bar{\1}} $ of the tube algebra defined in Eq.~\eqref{eq:tube_algebra_central_idempotents}.
	Furthermore, because $ B_p $ corresponds to adding a vacuum loop around a puncture inside plaquette $ p $, all $ B_p $ operators commute. Hence, $ \{Q_v\} \cup \{B_p\} $ is a set of commuting projectors.
	
	We can now introduce the following Hamiltonian on $ \H $:
	\begin{equation}\label{eq:ribbon_graph_on_lattice_hamiltonian}
		H_{\Lambda} = - \sum_v Q_v - \sum_p B_p\,,
	\end{equation}
	for which the ground space $ \H_{\Lambda} $ is precisely the simultaneous +1 eigenspace of all $ Q_v $ and $ B_p $ operators. 
	We claim that $ \H_{\Lambda}$ is isomorphic to $ \H_{\Sigma_n} $.
	Indeed, since adding a vacuum loop around a puncture allows any ribbons to be pulled across it ( effectively hiding the presence of the said puncture), the +1 eigenspace of $ B_p $ inside $ \H_\text{s.n.} $ is isomorphic to $ \bigoplus_\ell \H_{\Sigma_{n+P-1}}^{(\ell, 1^{P-1})} $. 
	By this reasoning, one finds that the simultaneous +1 eigenspace of all $ Q_v $ and $ B_p $ operators is isomorphic to $ \H_{\Sigma_n} $.

	A state in the ribbon graph Hilbert space $ \H_{\Sigma_n} $ can be explicitly mapped to the ground space $ \H_{\Lambda} $ of $ H_{\Lambda} $ with the map
	\begin{equation*}\label{eq:map_ribbon_to_lattice}
		\Gamma_{\Lambda} \, : \,\H_{\Sigma_n} \, \rightarrow \, \H_{\Lambda}
	\end{equation*}
	whose action is defined as follows:
	\begin{itemize}
		\item Given a state in $ \H_{\Sigma_{n}} $, continuously deform the ribbons in an arbitrary way to avoid all (imaginary) punctures at the centers of the plaquettes of $ \Lambda $. This deformation yields a state in $\bigoplus_\ell \H_{\Delta}^{(\ell, 1^P)} $, which can be identified with a lattice state in the string-net subspace. This lattice state is then an eigenstate with eigenvalue +1 of all the vertex operators $ Q_v $.
		\item Apply the total projector $ B  = \prod_p B_p $ to this lattice state to map it to the ground space of the string-net Hamiltonian $ H_{\Lambda} $.
	\end{itemize}
	It was shown in Ref.~\cite{konig2010quantum} that this map defines an isomorphism between the ribbon graph Hilbert space $ \H_{\Sigma_{n}} $ and the string-net ground space $ \H_{\Lambda} $, which confirms our claim that  $ \H_{\Lambda}  \simeq \H_{\Sigma_{n}}$.
	
	For the case where where $ n=0 $, \eqref{eq:ribbon_graph_on_lattice_hamiltonian} is precisely the Levin-Wen Hamiltonian introduced in Ref.~\cite{levin2005string}, which is conjectured to describe all doubled topological phases.
	It's important to note that the braiding properties of the input category $ \C $ are not required to define the Levin-Wen model, which is defined for any unitary fusion category satisfying the additional rules \eqref{eq:F_normalization}, \eqref{eq:F_tetrahedral} and \eqref{eq:F_unitarity_2}.
	In fact, the original construction has since been generalized in order to drop these additional requirements, for example see Ref.~\cite{hahn2020generalized}. 
	
\subsection{The extended string-net model} \label{sec:extended_sn}
	In the lattice realization of the ribbon graph Hilbert space above, we found that the string-net Hilbert space of the model is isomorphic to $ \bigoplus_\ell \H_{\Delta}^{(\ell, 1^P)} $. 
	We will now introduce a small modification in Hamiltonian \eqref{eq:ribbon_graph_on_lattice_hamiltonian}, such that the string-net subspace $ \H_\text{s.n.} $ will now be isomorphic to $ \H_{\Delta} $ instead. 
	This is achieved by modifying the lattice on which the model is defined, in order to introduce additional degrees of freedom for each plaquette.
	Such a modification was first proposed in Ref.~\cite{hu2018full}, with the goal of incorporating charged and dyonic excitations in (doubled) topological phases. 
	Here, we will refer to this modified Levin-Wen model as the \emph{extended Levin-Wen model}, or the \emph{extended string-net model}.
	
\begin{figure}[h]
		\centering
		\includegraphics[width=0.5\linewidth]{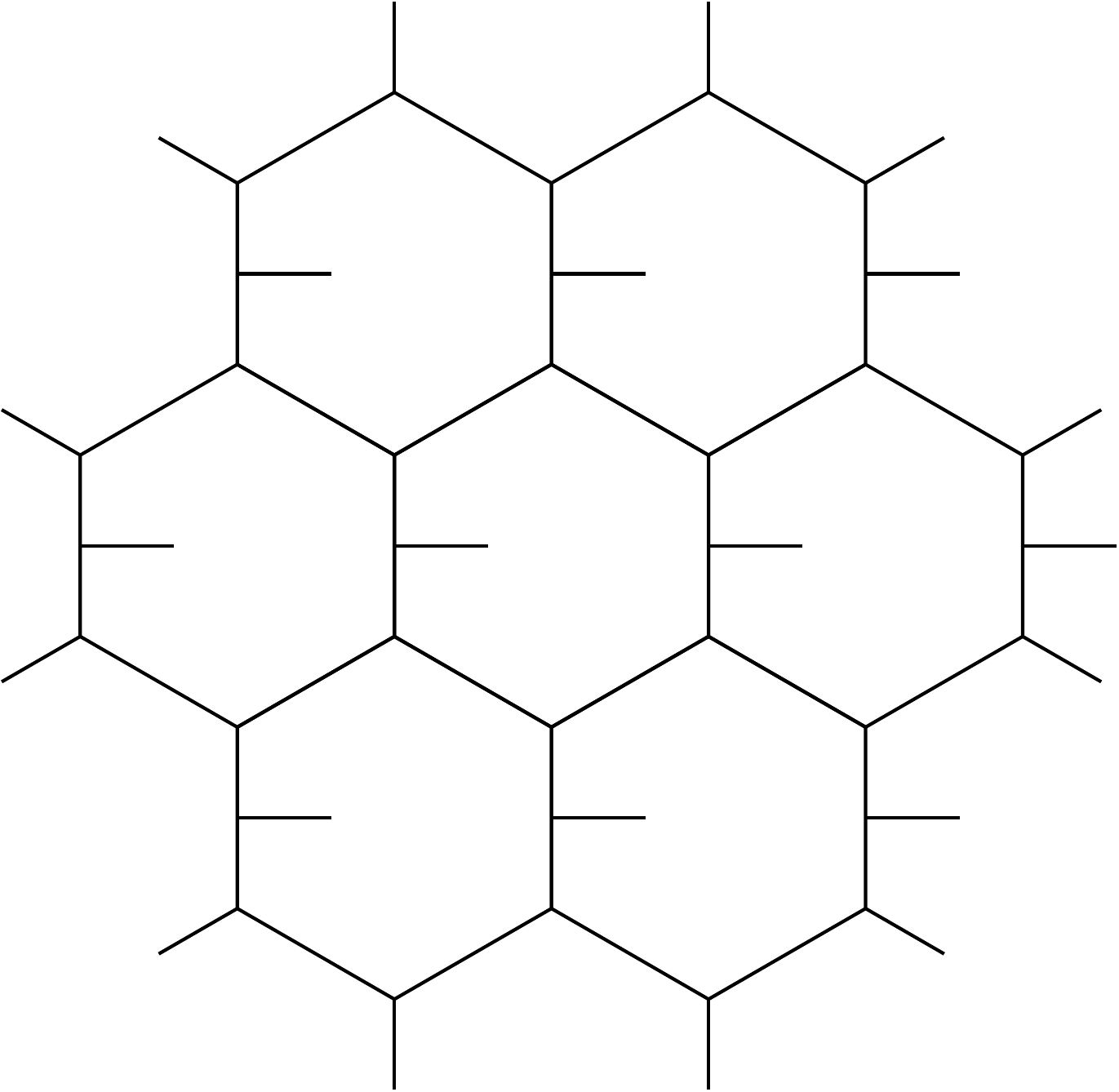}
		\caption{The tailed honeycomb lattice.}
		\label{fig:tailed_lattice}
	\end{figure}

	The extended string-net model is defined on a tailed lattice, which is obtained by adding a ``tail'' edge inside of each plaquette, as is depicted for the honeycomb lattice in \figref{fig:tailed_lattice}. This additional edge will allow for strings to end inside plaquettes, which corresponds to allowing any boundary labels for the $ P $ holes corresponding to the plaquettes in $ \Delta $, as opposed to fixing them to $ 1 $ as before.
	
	The modifications that must be made to \eqref{eq:ribbon_graph_on_lattice_hamiltonian} are straightforward: the sum over all vertices is extended to include the additional $ P $ vertices connected to the tail edges, and the plaquette operator $ B_p $ must be replaced by a modified operator $ \tilde{B}_p $ to accommodate the additional qudits:
	\begin{equation}\label{eq:extended_sn_hamiltonian}
		H = - \sum_v Q_v - \sum_p \tilde{B}_p\,.	
	\end{equation}
	
	The action of the new plaquette operator $ \tilde{B}_p $ is to project the tail qudit of plaquette $ p $ onto the local base state $ \ket{\1} $ associated to the vacuum label of $ \C $, and subsequently add a vacuum loop (divided by $ \D $) in the interior of $ p $. 
	Informally, one could write this as $ \tilde{B}_p = B_p \ket{\1}_t\bra{\1}_t$, where $ t $ stands for the tail qudit.
	The operators $ \tilde{B}_p $ thus essentially removes the tail edges from the picture, reducing the extended model to the regular string-net model. This means that the ground space of the Hamiltonian \eqref{eq:code_hamiltonian} on the tailed lattice is equivalent to the ground space of the regular string-net model defined on $ \Sigma $. 
	
	Since the string-net subspace of the extended string-net model is isomorphic to $ \H_{\Sigma_{n+P}} $, we are now capable of defining the action of the tube algebra (introduced in App.~\ref{sec:tube_algebra}) on any plaquette in $ \H_\text{s.n.} $.
	The action of the tube operator $ O_{x y \alpha \beta} $ on a plaquette $ p $, corresponds to  gluing the tube
	\begin{equation}\label{eq:tube_1}
		\includegraphics[scale=.4]{fig/tube_1.pdf}
	\end{equation}
	inside the puncture in $ \Sigma_{n+P} $ associated with the said plaquette. 
	Its action on the lattice string-net Hilbert space $ \H_\text{s.n.} $, can be found by resolving the resulting ribbon graph into the lattice using a sequence of 2-2 Pachner moves ($F$-moves) followed by a 1-3 Pachner move, as shown in Fig.~\ref{fig:tube_derivation} in Sec.~\ref{sec:model}. 
	The resulting expression is:
	\begin{widetext}
		\begin{equation} \label{eq:tube_operator_lattice}
		O_{x y \alpha \beta } \, \Bigg |
		\raisebox{-1.1cm}{\includegraphics[scale=.32]{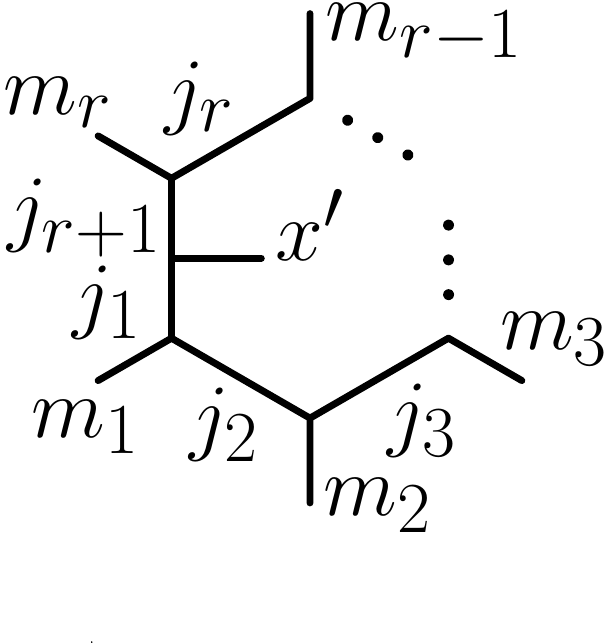}}
		\Bigg \rangle
		= \delta_{x,x'} \Bigg |
		\raisebox{-1.1cm}{\includegraphics[scale=.32]{fig/plaquette_tube.pdf}}
		\Bigg \rangle
		= \delta_{x,x'} \dfrac{v_\alpha v_\beta}{v_y}\!\!
		\sum_{k_1,\ldots,k_{r+1}} 
		\!\!\!\!\! F^{j_1 j_{r+1} x}_{\alpha \beta k_{r+1}} \bigg( \! \prod_{\nu = 1}^{r} F^{m_\nu j_{\nu} j_{\nu+1}}_{\alpha k_{\nu+1} k_{\nu} } \! \bigg) F^{ k_{r+1} k_1 y}_{\alpha j_1 \beta}
		\Bigg |
		\raisebox{-1.1cm}{\includegraphics[scale=.32]{fig/plaquette_stick_2.pdf}}
		\Bigg \rangle \,.
		\end{equation}
	\end{widetext}
	
	
	The precise definition of $ \tilde{B}_p $ can now be formulated:
	\begin{equation}\label{eq:Bp}
		\tilde{B}_p = \cP^{1\bar{1}} = \dfrac{1}{\D^2} \sum_s d_s O_{\1 \1 s s}\,.
	\end{equation}
	This corresponds the central idempotent of the tube algebra that projects onto the trivial (doubled) anyon label. Hence, we can reformulate the action of the plaquette operator $ \tilde{B}_p $ as projecting onto the subspace where plaquette $ p $ has trivial anyon charge $ \1\bar{\1} $.
	The plaquette operators in the extended string-net Hamiltonian can therefore be considered as generalized stabilizer generators of an error correcting code, in the sense that they characterize the ground space $ \H_{\Lambda}^{\ell} $ in terms of the absence of anyonic charges.
	When restricted to the string-net subspace $ \H_{\text{s.n.}} $, excitations manifest themselves as states containing nontrivial anyons. The precise doubled anyon label of these local excitations can then be obtained through a projective measurement with the lattice realization of the central idempotents of the tube algebra.
	An important distinction between the extended and the regular Levin-Wen model, is the fact that the excitation spectrum of the extended Levin-Wen model inside $ \H_{\text{s.n.}} $ contains all anyon types in $ \D\C $, while only the doubled anyons compatible with a trivial boundary label can be realized with the regular Levin-Wen model.\\

\subsection{Realizing the anyonic fusion basis on the tailed lattice} \label{sec:anyonic_fusion_basis_lattice}
	The isomorphism between the ribbon graph Hilbert space for $ \Delta $ and  the string-net subspace $ \H_{\text{s.n.}} $ on the lattice, implies that we can obtain a basis for $ \H_{\text{s.n.}} $ by mapping the anyonic fusion basis states from $ \H_{\Delta} $ to $\H_{\text{s.n.}} $.
	A lattice realization of the anyonic fusion basis allows us to decompose any state inside the string-net subspace in
	terms of its anyon content and underlying doubled anyonic fusion tree. 
	We consider the case where $ \Delta $ has genus $ g = 0 $. Generalizing to surfaces with a nonzero genus requires us to include handles in the construction, the procedure for this is analogous to the one below.	
	
	In order to simplify the graphical notation, doubled ribbons will be represented using bold red lines, allowing to draw Eq.~\eqref{eq:anyonic_fusion_basis_states_n} somewhat more compactly as
	\begin{equation}\label{eq:anyonic_fusion_basis_comp}
		\ket{\vec{\bm{a}}, \vec{\bm{b}}} =	\raisebox{-1cm}{\includegraphics[scale=.5]{fig/anyonic_fus_bas_state_n_comp.pdf}}  .
	\end{equation}	
	Here the bold leaf labels $ \bm{a}_i = (a^+_i a^-_i)_{\ell_i}$ represent a doubled anyon label $ (a^+_i a^-_i) $ together with a compatible boundary label $ \ell_i $. The bold labels $ \boldsymbol{b} = b^+_j b^-_j $ represent doubled anyon labels by which we label the branches of the doubled fusion tree. 
	It will always be clear from the context whether or not a bold label represents a leaf or a branch label. 
	The closing off of the doubled ribbon graph inside the vacuum loop at the root and each leaf is always implied in the compact notation for the fusion basis states on the left hand side of this expression.
	
	Mapping the anyonic fusion state \eqref{eq:anyonic_fusion_basis_comp} to a lattice state is done using the isomorphism in Eq.~\eqref{eq:map_ribbon_to_lattice} (while treating the nontrivial plaquettes as punctures).
	One starts by embedding the fusion tree on the fattened lattice. (Note that this embedding is not arbitrary, but is fixed by the anyonic fusion basis on $ \H_{\Delta} $.)
	For example, when only 5 plaquettes carry a nontrivial anyon label, this embedding could look like
	\begin{equation}\label{eq:anyonic_fusion_basis_lattice1}
		\raisebox{-2cm}{\includegraphics[scale=.35]{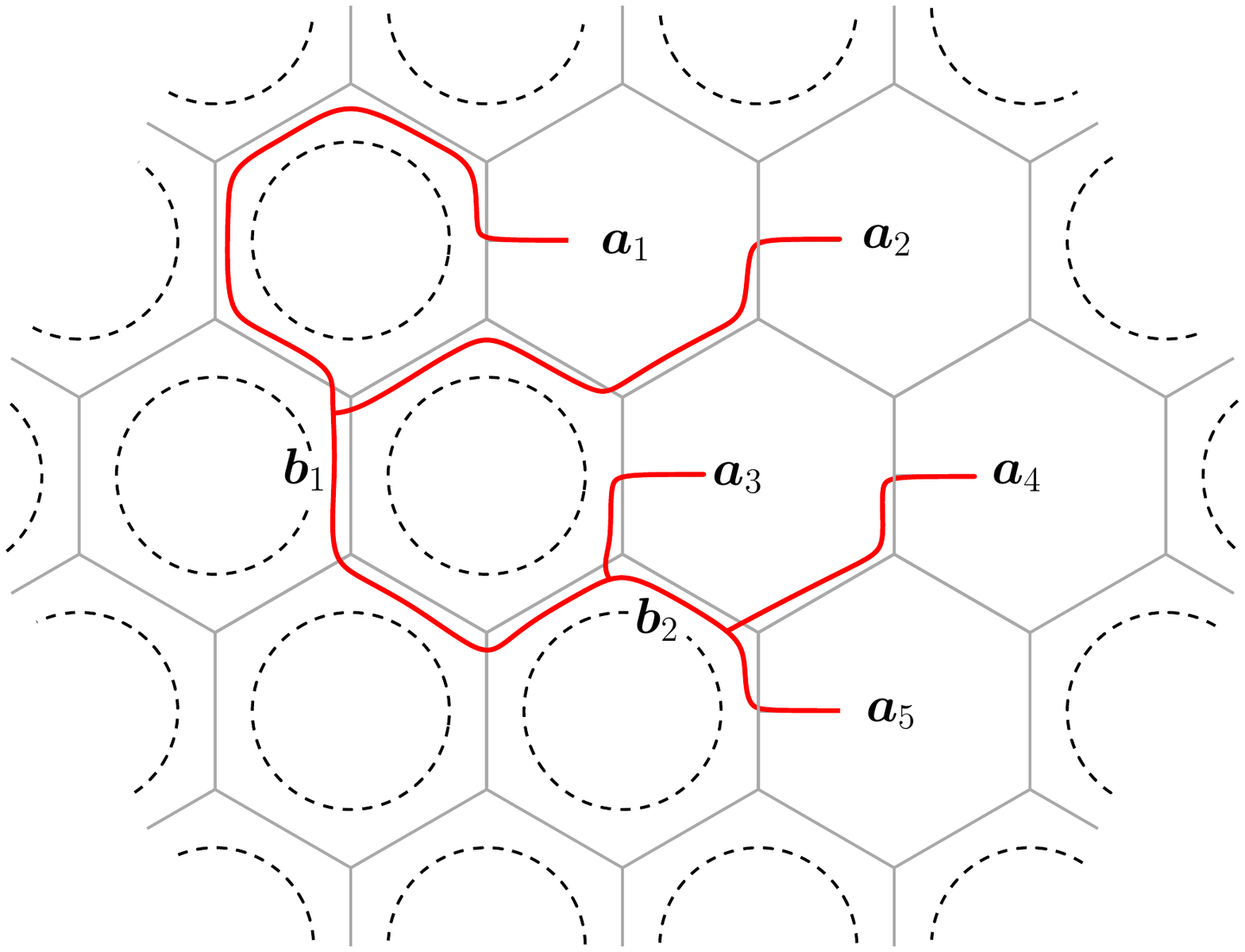}} \quad ,
	\end{equation}
	where we chose not to draw the leaves with vacuum labels explicitly, but included vacuum loops in the corresponding plaquettes instead.
	Using the reduction to two-dimensional ribbon graphs in \eqref{eq:anyonic_fusion_basis_red2}, \eqref{eq:anyonic_fusion_basis_lattice1} can be rewritten as
	\begin{equation}\label{eq:anyonic_fusion_basis_lattice2}
		\sum_{\vec{\alpha}, \vec{\beta}, \vec{k}} X^{(\vec{\boldsymbol{a}}, \vec{\boldsymbol{b}})}_{\vec{\alpha}, \vec{\beta}, \vec{k}, \vec{l}} \quad \raisebox{-2cm}{\includegraphics[scale=.35]{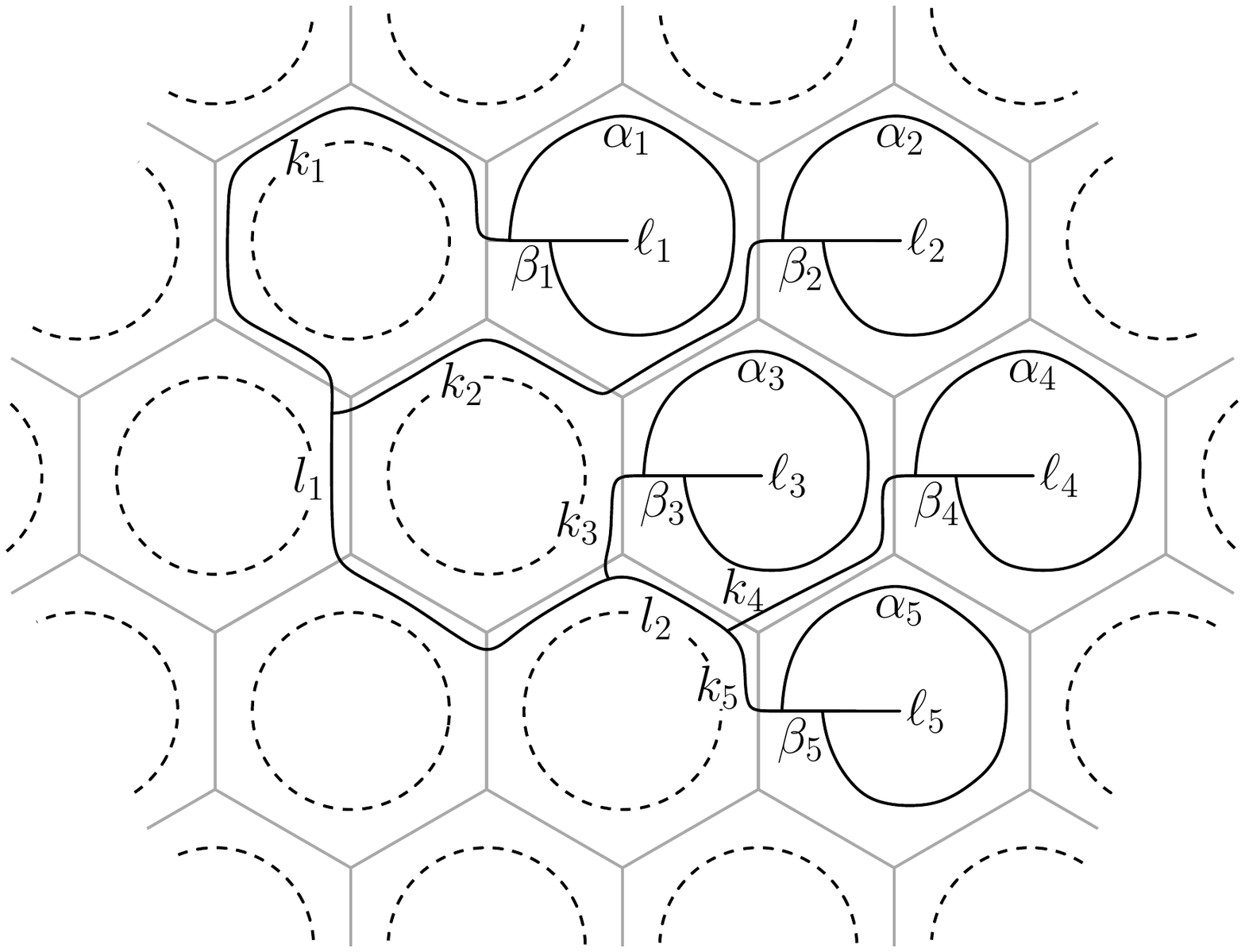}} \;\; .
	\end{equation}
	The actual lattice state is then obtained by using the definition \eqref{eq:vacuum_line} to write out the vacuum loops as superpositions of ribbon loops, repeatedly using Eq.~\eqref{eq:F-move} to pull all ribbons onto the lattice edges, and then finally including the all missing (trivial) tail qudits.
	
	It should be apparent by now that even though we are considering a rather simple fusion state with only 5 nontrivial charges, this corresponds to an exceedingly complex superposition of lattice qubit states. 
	In App.~\ref{sec:TN} we will see that tensor network states provide the perfect framework for describing the anyonic fusion states in terms of local entanglement degrees of freedom.

%% file: sections/tensor_networks.tex
\section{Tensor network representations} \label{sec:TN}
	During the previous two decades, tensor networks states \cite{verstraete2008matrix, bridgeman2017hand, cirac2020matrix} have emerged as a vital tool for both the theoretical and numerical study of strongly correlated quantum many body systems. 
	Their strength originates in the fact that they can be used to describe low-energy states of interacting systems in therms of local entanglement degrees of freedom, which allows for a very efficient representation of these states despite the exponential scaling of the quantum many-body Hilbert space.
	
	It is well known that string-net ground states allow for a remarkably simple PEPS (projected entangled pair state) representation, constructed from the algebraic data of the input category \cite{gu2009tensor, buerschaper2009explicit}. 
	Below, we will illustrate how such a tensor network representation is obtained, and then expand on these known results by constructing an explicit PEPS representation of any anyonic fusion basis state on a tailed lattice.


\subsection{A tensor network representation for the string-net ground states}
	We start by constructing the PEPS representation of the ground state
	\begin{equation}\label{eq:gs}
		\mathcal{N} \prod_p B_p \left( \otimes_e \ket{\mathbf{1}}_e \right)\,,
	\end{equation}
	where $ \mathcal{N} $ is a normalization factor. 
	Graphically, (a patch of) this state can be represented as
	\begin{widetext}
	\begin{equation}\label{eq:vacuum_patch_lattice}
		 \raisebox{-1.5cm}{\includegraphics[scale=.34]{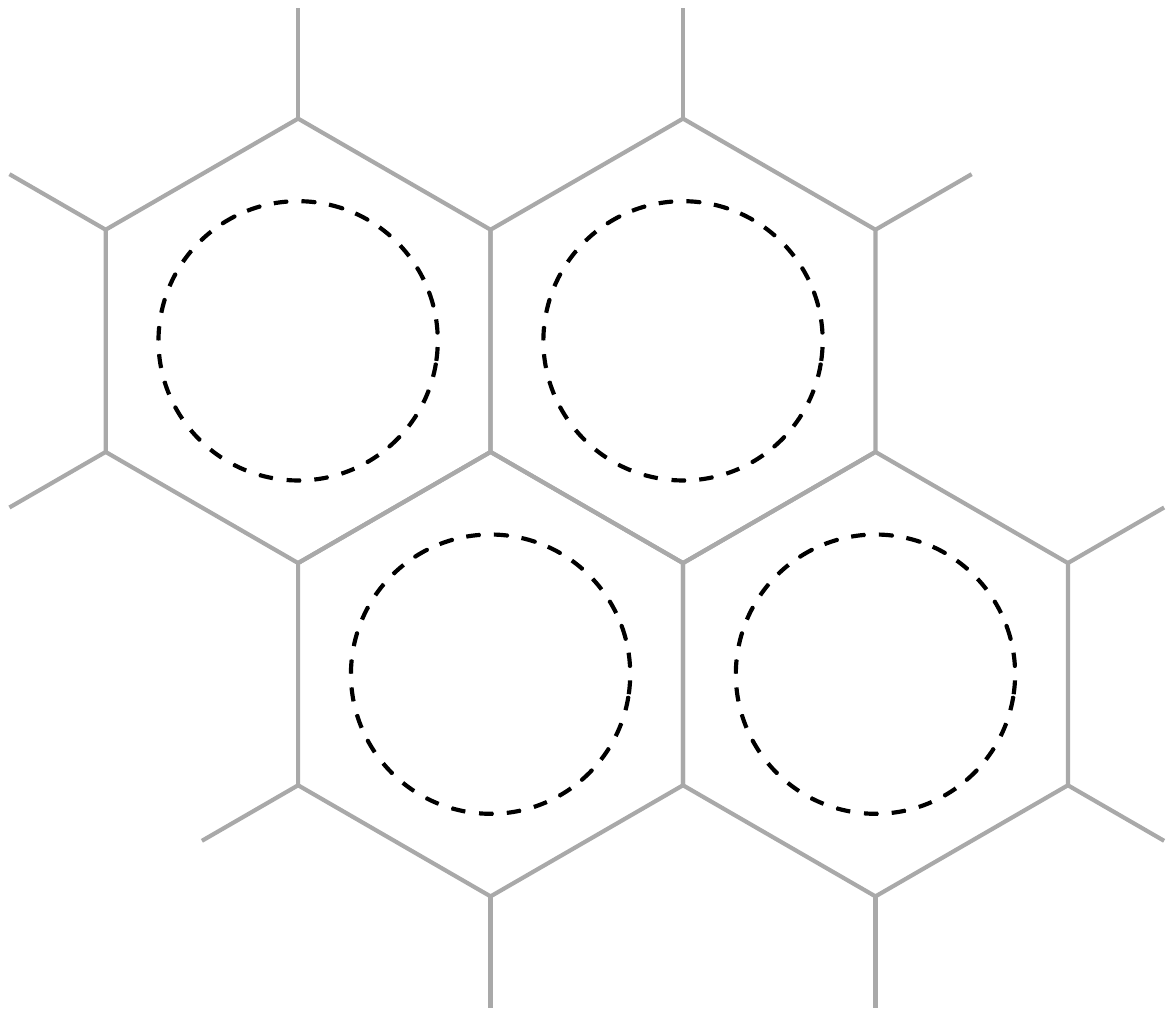}} \; 
		 = \;\mathcal{N} \sum_{\{ \mu\}} d_{\mu_1} d_{\mu_2} \cdots 
		\raisebox{-1.5cm}{\includegraphics[scale=.34]{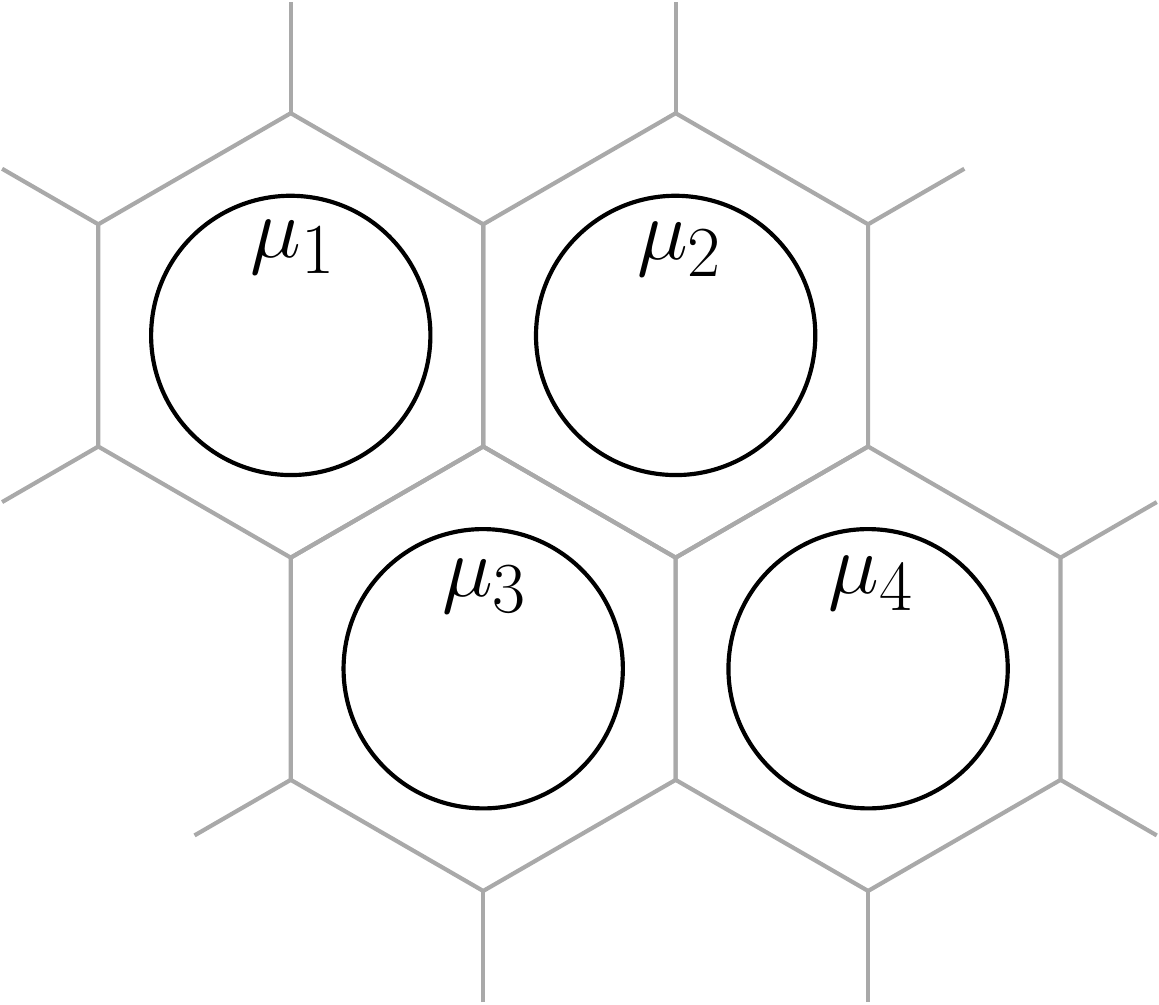}}
		\;\; , 
	\end{equation}
	\end{widetext}
	where qudits in the $ \ket{\mathbf{1}} $ state are represented by the gray edges.
	To find
	the state from this graphical notation, one has to resolve the loops appearing
	in Eq.~\eqref{eq:vacuum_patch_lattice} into the lattice. 
	This is done in two steps. 
	First we fuse the neighboring loops along each edge using Eq.~\eqref{eq:double_line}:
	\begin{equation}\label{eq:TN_pull_onto_edge}
		\raisebox{-0.4cm}{\includegraphics[scale=.5]{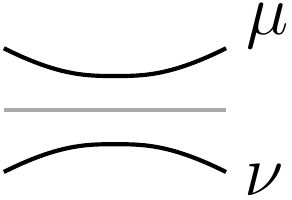}} =  \sum_{k} \frac{v_k}{v_\mu v_\nu} \delta_{\mu\nu k}\; \raisebox{-0.4cm}{\includegraphics[scale=.5]{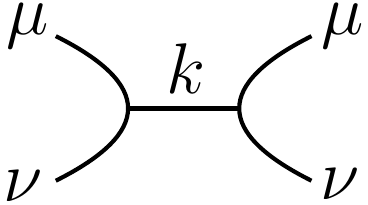}} \nonumber \,.
	\end{equation}
	Then, the resulting configuration at every vertex is resolved into the lattice using Eq.~\eqref{eq:Gsymbol_def}:
	\begin{equation}\label{eq:TN_vertex_config}
		\raisebox{-0.62cm}{\includegraphics[scale=.4]{fig/vertex_loop.pdf}} = v_\lambda v_\mu v_\nu G^{i j k}_{\lambda \mu \nu} \raisebox{-0.42cm}{\includegraphics[scale=.4]{fig/vertex.pdf}}\,.
	\end{equation}
	\vspace{1pt}
	The PEPS tensor is obtained by splitting the factor in Eq.~\eqref{eq:TN_pull_onto_edge} evenly
	between the two adjacent vertices. The result is
	\begin{align}\label{eq:TN_vertex_tensor}
		\raisebox{-1cm}{\includegraphics[scale=.46]{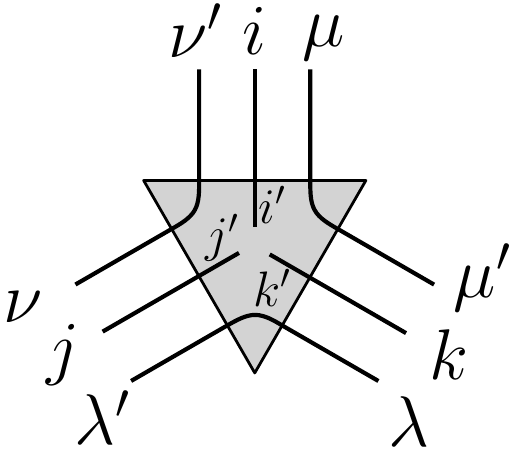}} \; =   \delta_{i i'} \delta_{j j'} \delta_{k k'} \delta_{\lambda \lambda'} \delta_{\mu \mu'} \delta_{\nu \nu'}\, \sqrt{v_i v_j v_k} \,G^{i j k}_{\lambda \mu \nu} \,,
	\end{align}
	where $ i' $ , $ j' $ and $ k' $ represent the physical degrees of freedom associated to the qudits on the 3 edges.
	Note that the physical degrees of freedom are doubled, since each of them is appearing at the two vertices connected to a given edge. The diagonal structure of Eq.~\eqref{eq:TN_vertex_tensor}, ensures that the values of the two physical indices representing the same qudit are always equal. 
	We impose the convention that for every closed loop with label $ \mu $ on the virtual level, a factor $ d_\mu $ is implied. This convention automatically takes care of the factors $ d_{\mu_i} $ appearing in Eq.~\eqref{eq:vacuum_patch_lattice}, which can then be represented as
	\vspace{1cm}
	\begin{align}\label{eq:TN_PEPS_GS}
		\raisebox{-2cm}{\includegraphics[scale=.5]{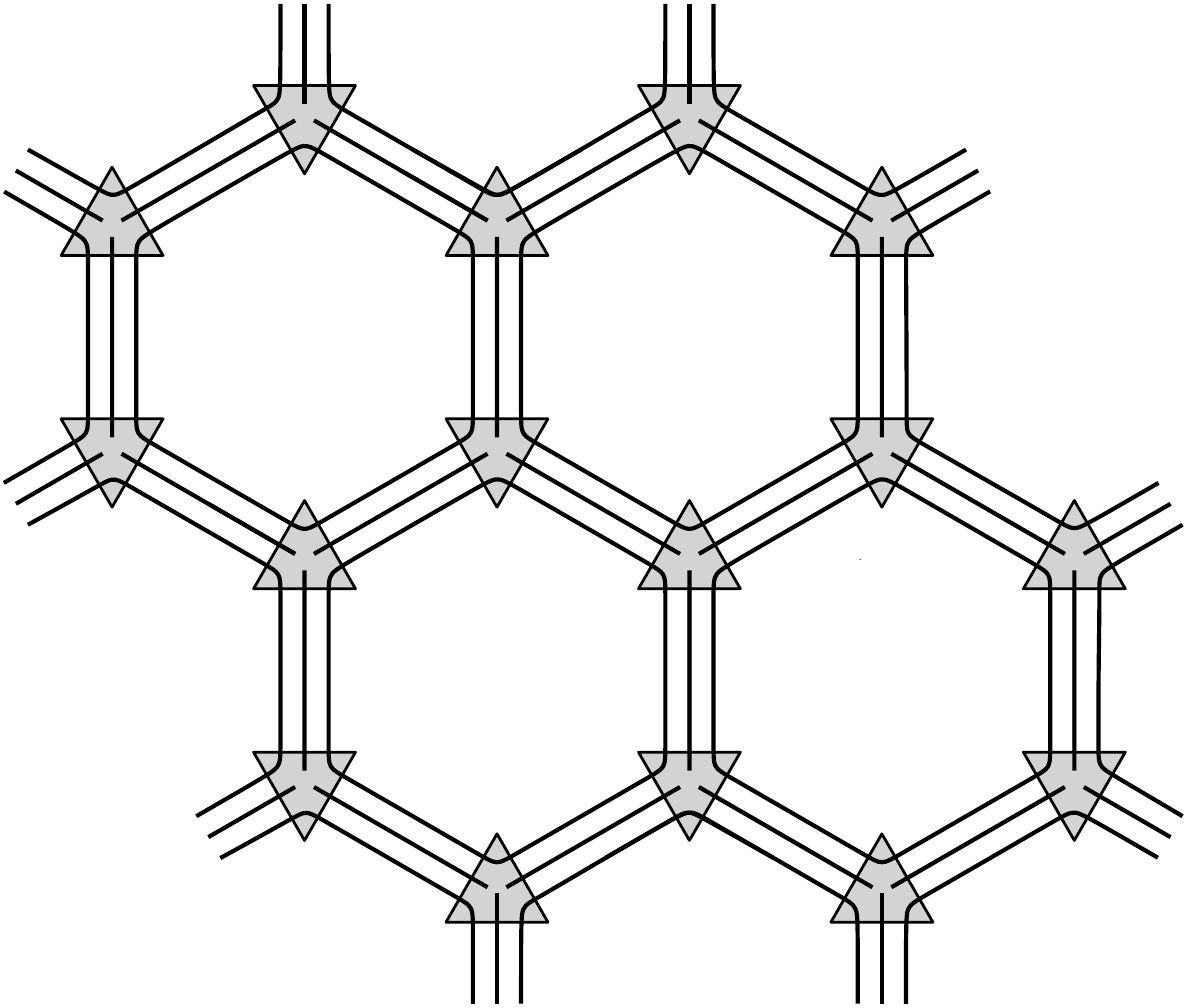}} \;\; .
	\end{align}
	
	To simplify the notation, we will follow the convention that a pair of legs crossing through a tensor represents a Kronecker delta between the indices on these legs, indicating that the tensor is diagonal in this pair of indices.
	
	The tensor network state constructed above corresponds specifically to the ground state Eq.~\eqref{eq:gs}. It can, however, be modified to represent any string-net ground state.
	To do this, first note that we can obtain any ground state by acting with the total projector $ B = \prod_p B_p $ on some state $ \ket{\phi} \in \H_{\text{s.n.}}$ corresponding to a configurations of strings on the lattice. 
	The tensor network state for $ B \ket{\phi} $, can then be obtained by modifying our original construction, to account for additional strings running between the vacuum loops in Eq.~\eqref{eq:vacuum_patch_lattice}.
	Locally, such a string-configuration can take two forms: a single string, or two strings fusing to a third one.
	
	The first configuration looks as follows: 
	\begin{widetext}
		\begin{equation}\label{eq:vacuum_patch_lattice_string}
			\sum_{\{ \mu\}}  d_{\mu_1} d_{\mu_2} d_{\mu_3} d_{\mu_4} \raisebox{-1.5cm}{\includegraphics[scale=.34]{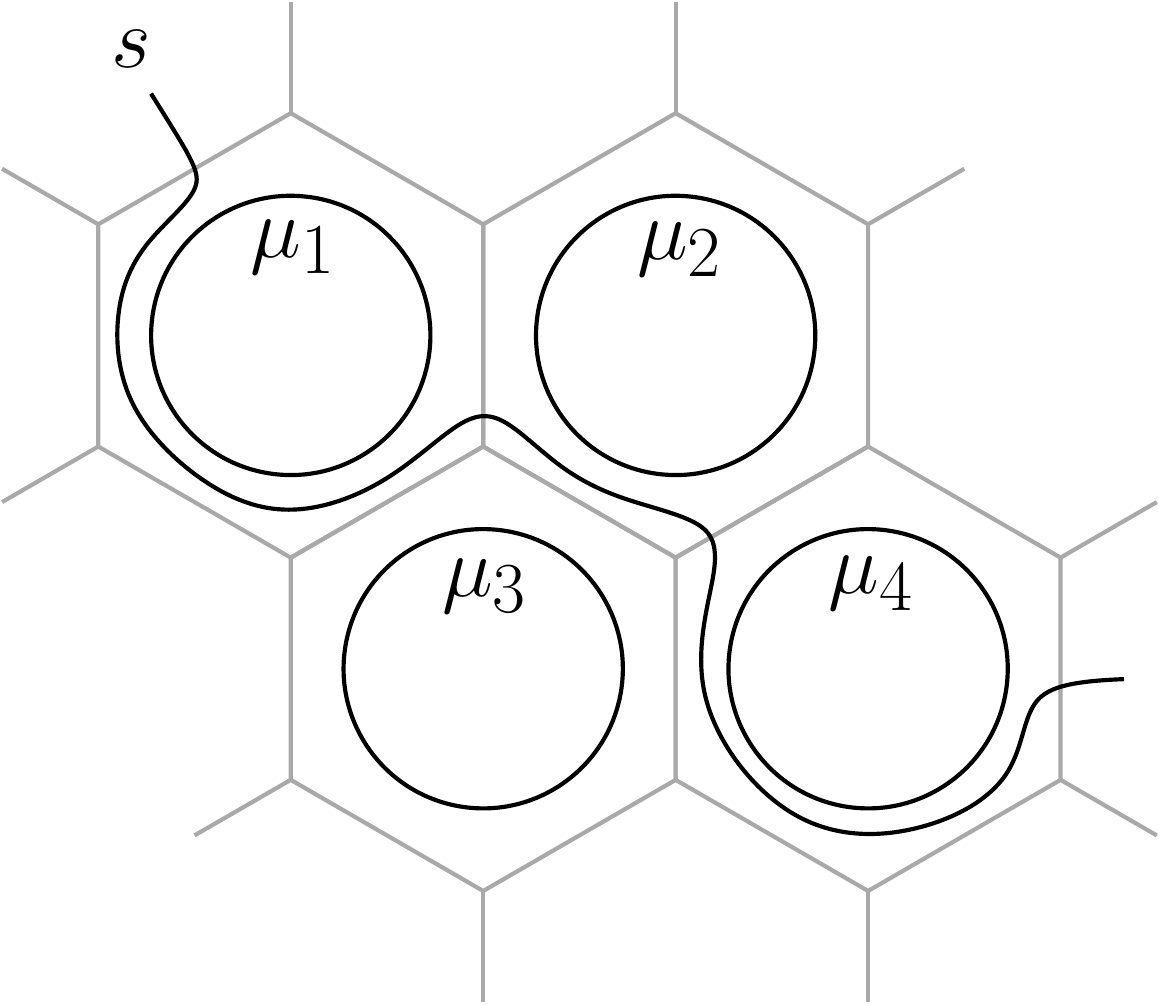}}
			\; =  \sum_{\{ \mu\}} \sum_{\mu_1',\mu_2',\mu_4'} d_{\mu_3} \frac{v_{\mu_1} v_{\mu_1'}}{v_s} \frac{v_{\mu_2} v_{\mu_2'}}{v_s} \frac{v_{\mu_4} v_{\mu_4'}}{v_s}
			\raisebox{-1.5cm}{\includegraphics[scale=.34]{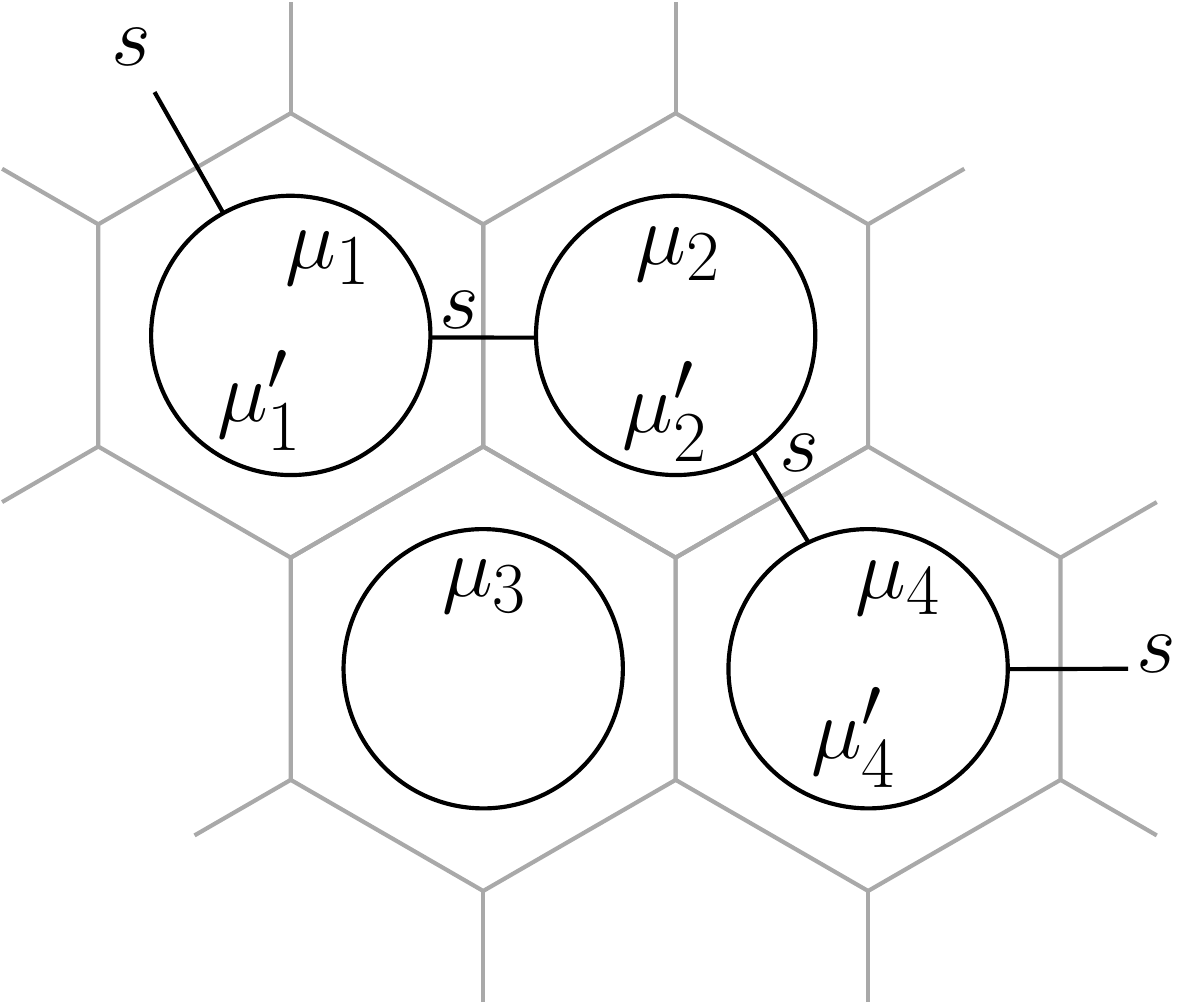}}
			\;\; ,
		\end{equation}
	where we have pulled the string $ s $ into the different loops along its path using Eq.~\eqref{eq:double_line}.
	Using $ F $-moves, we can again pull ribbons from neighboring plaquettes into each edges:
	\begin{align}\label{eq:TN_pull_onto_edge_string}
		\raisebox{-0.4cm}{\includegraphics[scale=.5]{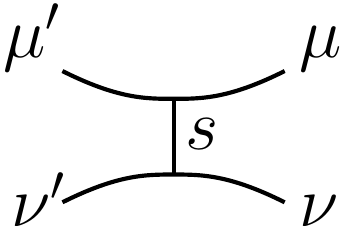}} = 
		\sum_{k} F^{\mu \mu' s}_{\nu' \nu k} \; \raisebox{-0.4cm}{\includegraphics[scale=.5]{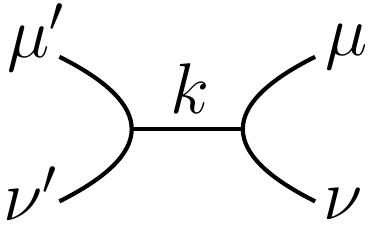}} 
		 & = 
		\sum_{k}  \sqrt{v_s v_{\mu} v_{\mu '}}  \sqrt{\frac{v_k}{v_\mu v_\nu}} \sqrt{\frac{v_k}{v_{\mu '}v_{\nu'}}} 
		\sqrt{v_s v_\nu v_{\nu '}} G^{\mu \mu' s}_{\nu' \nu k} \; \raisebox{-0.4cm}{\includegraphics[scale=.5]{fig/munus2.pdf}} \;. 
	\end{align}
	\end{widetext}
	The first two factors on the right hand side can be recognized as the symmetrized contribution to each vertex from Eq.~\eqref{eq:TN_pull_onto_edge}. The first factor on the right hand side of Eq.~\eqref{eq:TN_pull_onto_edge_string} will therefore contribute to the vertex to the left of the edge, while the second factor will contribute to the right vertex, ensuring that we can use the PEPS tensor Eq.~\eqref{eq:TN_vertex_tensor} on both vertices. The third and fourth factors on the right hand side will contribute to the upper and lower plaquette of the edge in Eq.~\eqref{eq:TN_pull_onto_edge_string}, respectively. When combined with the factors in Eq.~\eqref{eq:vacuum_patch_lattice_string}, these contributions result in a total factor of the form $ \dfrac{v_{\mu} v_{\mu'}}{v_s} \cdot v_s v_{\mu} v_{\mu '} = d_\mu d_{\mu'} $ for each plaquette separately [instead of a single quantum dimension factor like in Eq.~\eqref{eq:vacuum_patch_lattice}]. 
	
	If we now place the crossing tensor
	\begin{equation}\label{eq:TN_crossing_tensor}
		\raisebox{-1.cm}{\includegraphics[scale=.46]{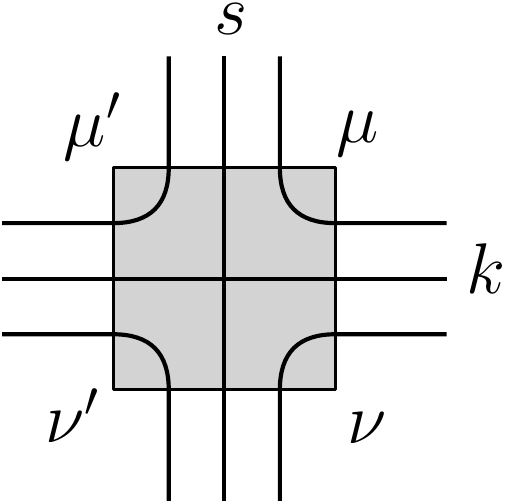}} \; =  G^{\mu \mu' s}_{\nu' \nu k} \,,
	\end{equation}
	in the PEPS at every crossing of the ribbon with label $ s $ with a lattice edge, this gives the correct superposition of qudit states.
	Note that the quantum dimension factors in each plaquette are again taken care of by the convention for closed loops at the virtual level.
	Eq.~\eqref{eq:vacuum_patch_lattice_string} can then be represented with the following tensor network state:
	\begin{align}\label{eq:TN_PEPS_string}
		\raisebox{-2cm}{\includegraphics[scale=.5]{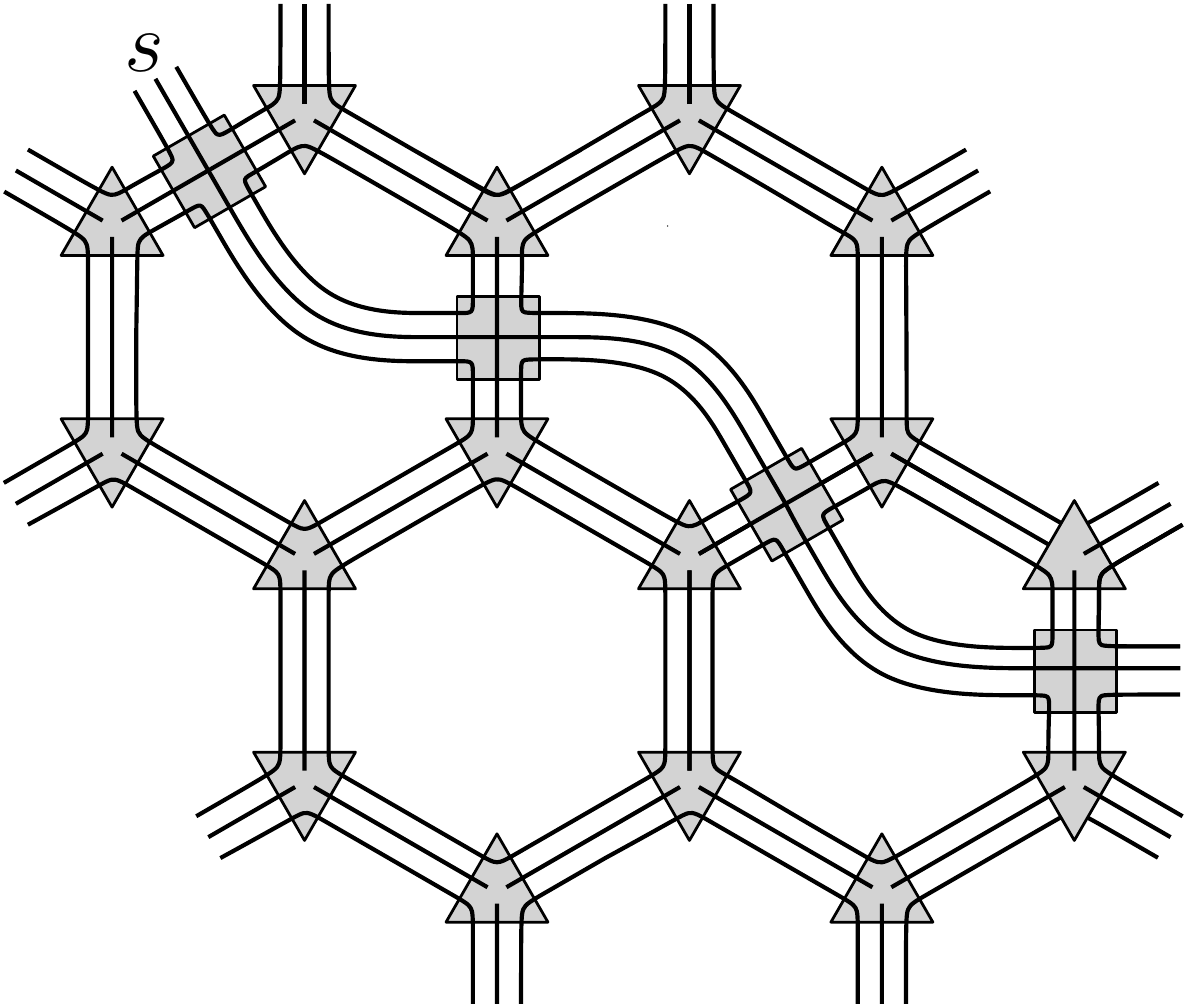}} \quad ,
	\end{align}
	where the additional loops in the plaquette correspond to the second summation on the right hand side of Eq.~\eqref{eq:vacuum_patch_lattice_string}.
	Note that in the tensor network representation above, one should ensure that the string-label is fixed to the correct string-label $ s $. 
	In case one were to use the tensors above to represent a closed string, an additional Kronecker-delta tensor must be included to avoid summing over all string-labels.
	
	The tensor network representation of a ground state obtained from a string-configuration containing fusing strings, is derived in a similar fashion.
	We again start by pulling the strings onto the various loops using Eqs.~\eqref{eq:double_line} and \eqref{eq:Gsymbol_def}:
		\begin{align}\label{eq:vacuum_patch_lattice_fusion}
			\sum_{\{ \mu\}}& \, d_{\mu_1} d_{\mu_2}	\raisebox{-1.5cm}{\includegraphics[scale=.34]{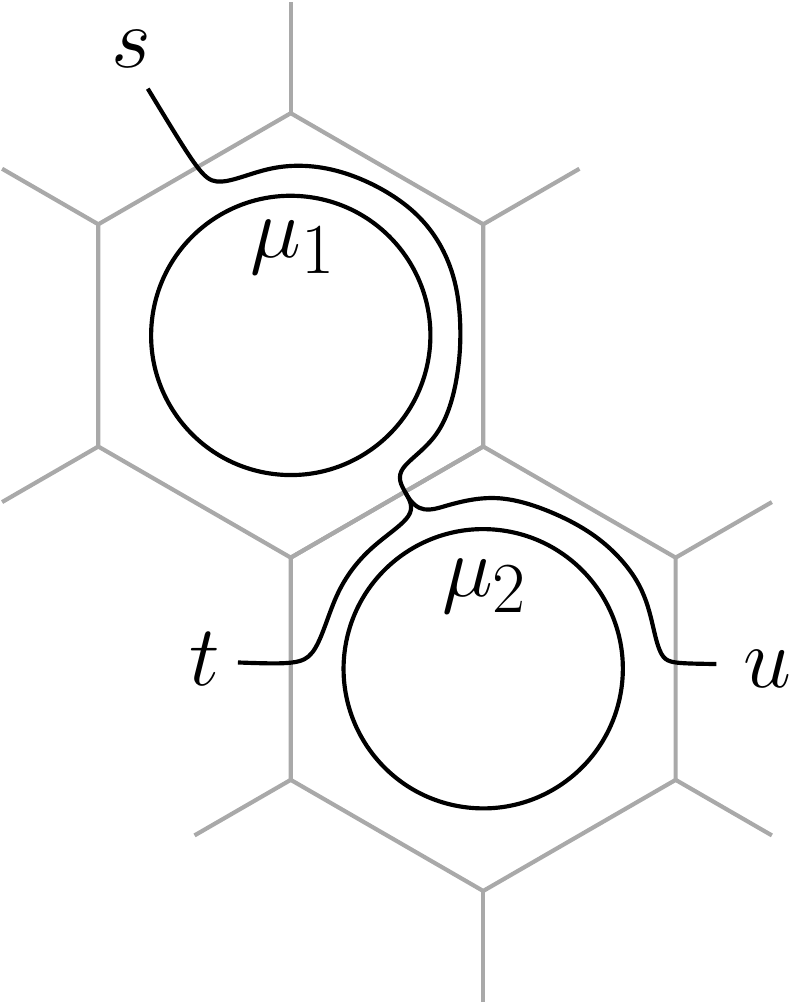}} \nonumber\\
			&=
			\sum_{\mu_1',\mu_2',\mu_2''} \frac{v_{\mu_1} v_{\mu_1'}}{v_s} \frac{ v_{\mu_2'} 	v_{\mu_2''}}{v_t v_u}
			\raisebox{-1.5cm}{\includegraphics[scale=.34]{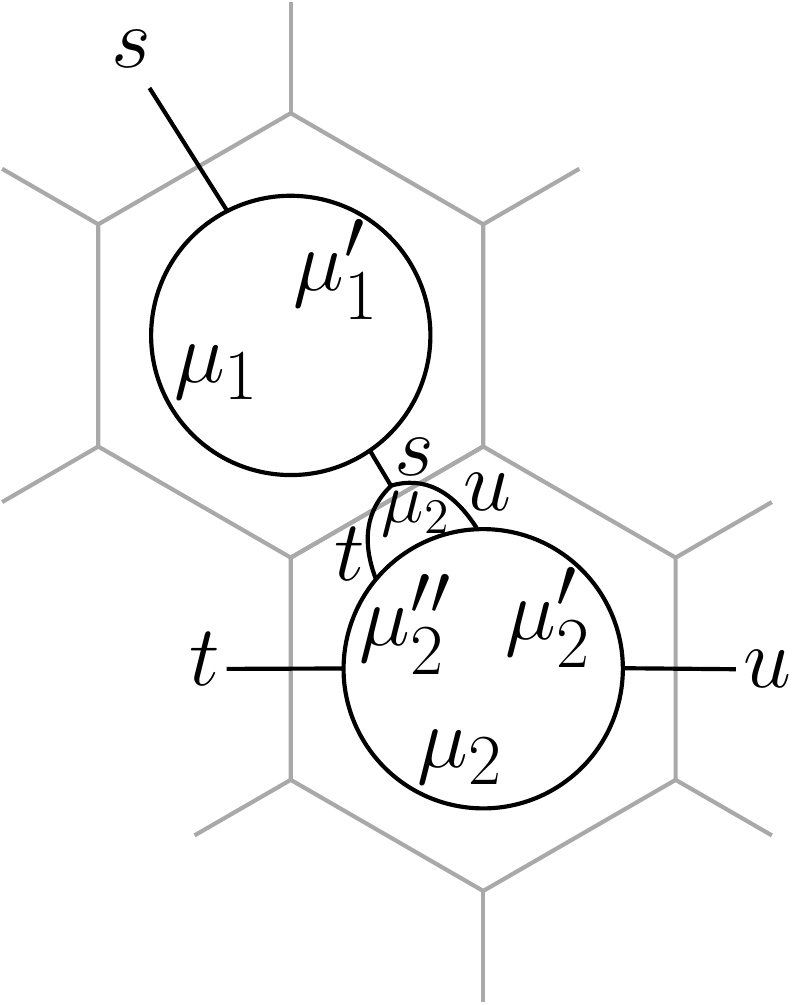}}  \\
			 &= 
			\sum_{\mu_1',\mu_2',\mu_2''} \frac{v_{\mu_1} v_{\mu_1'}}{v_s} v_{\mu_2} v_{\mu_2'} 	v_{\mu_2''}
			G^{s \mu_2'' \mu_2'}_{\mu_2 u t}
			\raisebox{-1.5cm}{\includegraphics[scale=.34]{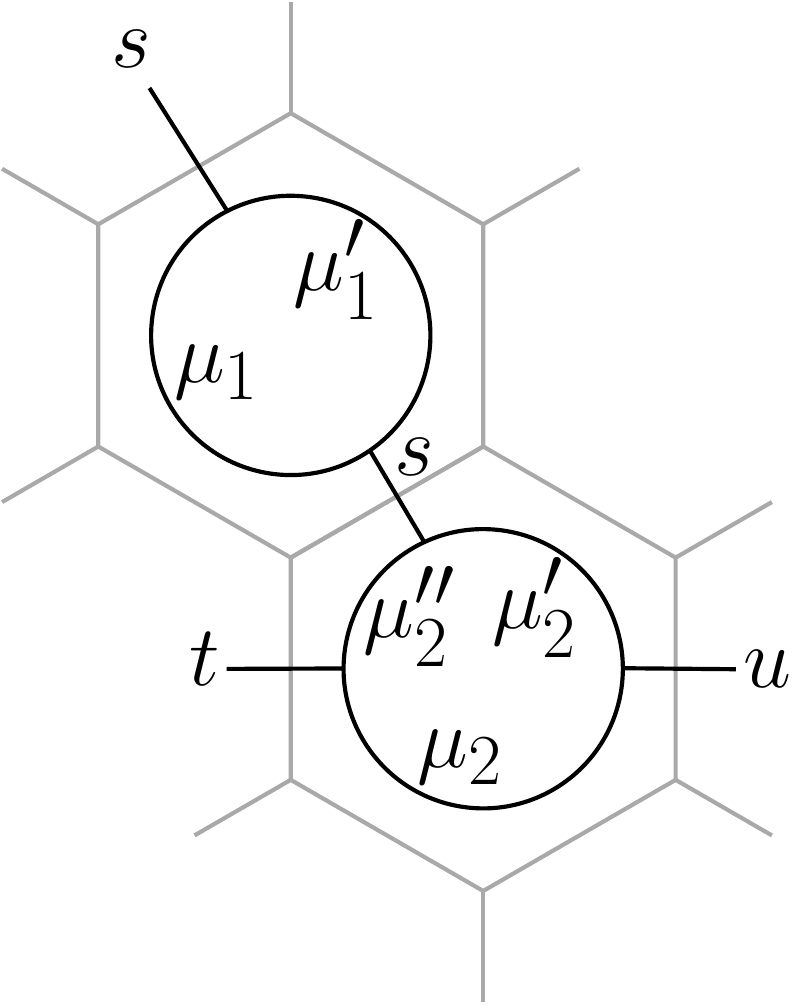}}
			\; . \nonumber
		\end{align}
	The edge crossings can again be resolved as in Eq.~\eqref{eq:TN_pull_onto_edge_string}, giving exactly the same situation as before for the upper plaquette. 
	For the lower plaquette, the combined contribution from of the last factor in Eq.~\eqref{eq:TN_pull_onto_edge_string} for the three edge crossings combines with the right hand side of Eq.~\eqref{eq:vacuum_patch_lattice_fusion} to a total factor of the form
	\begin{multline}\label{eq:TN_plaq_factor_fusion}
		\sqrt{v_s v_{\mu'} v_{\mu ''}} \sqrt{v_t v_{\mu} v_{\mu''}} \sqrt{v_u v_{\mu} v_{\mu '}} 	v_{\mu} v_{\mu'} v_{\mu''} 
		G^{s \mu'' \mu'}_{\mu u t} \\
		\qquad \qquad \qquad = d_{\mu} d_{\mu'} d_{\mu''} \sqrt{v_s v_t v_u} G^{s \mu'' 	\mu'}_{\mu u t} \,.
	\end{multline}
	We can now define the tensor
	\begin{align} 
		\raisebox{-.7cm}{\includegraphics[scale=.46]{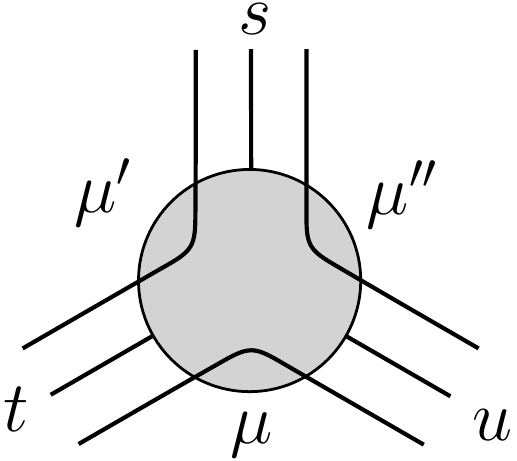}} \quad &=   \sqrt{v_s v_t v_u} G^{s \mu'' \mu'}_{\mu u t} \nonumber\\
		&= \sqrt{v_s v_t v_u} G^{s t u}_{\mu \mu'' \mu'} \label{eq:TN_fusion_tensor}\,,
	\end{align}
	to represent the fusion of strings on the virtual level.
	This tensor, along with the closed loop convention at the virtual level, takes care of all remaining factors in Eq.~\eqref{eq:TN_plaq_factor_fusion}, and gives the correct superposition of qudit states.
	A ground state obtained from a initial configuration containing fusing strings can then be represented as:
	\begin{align}\label{eq:TN_PEPS_fusion}
		\raisebox{-2cm}{\includegraphics[scale=.5]{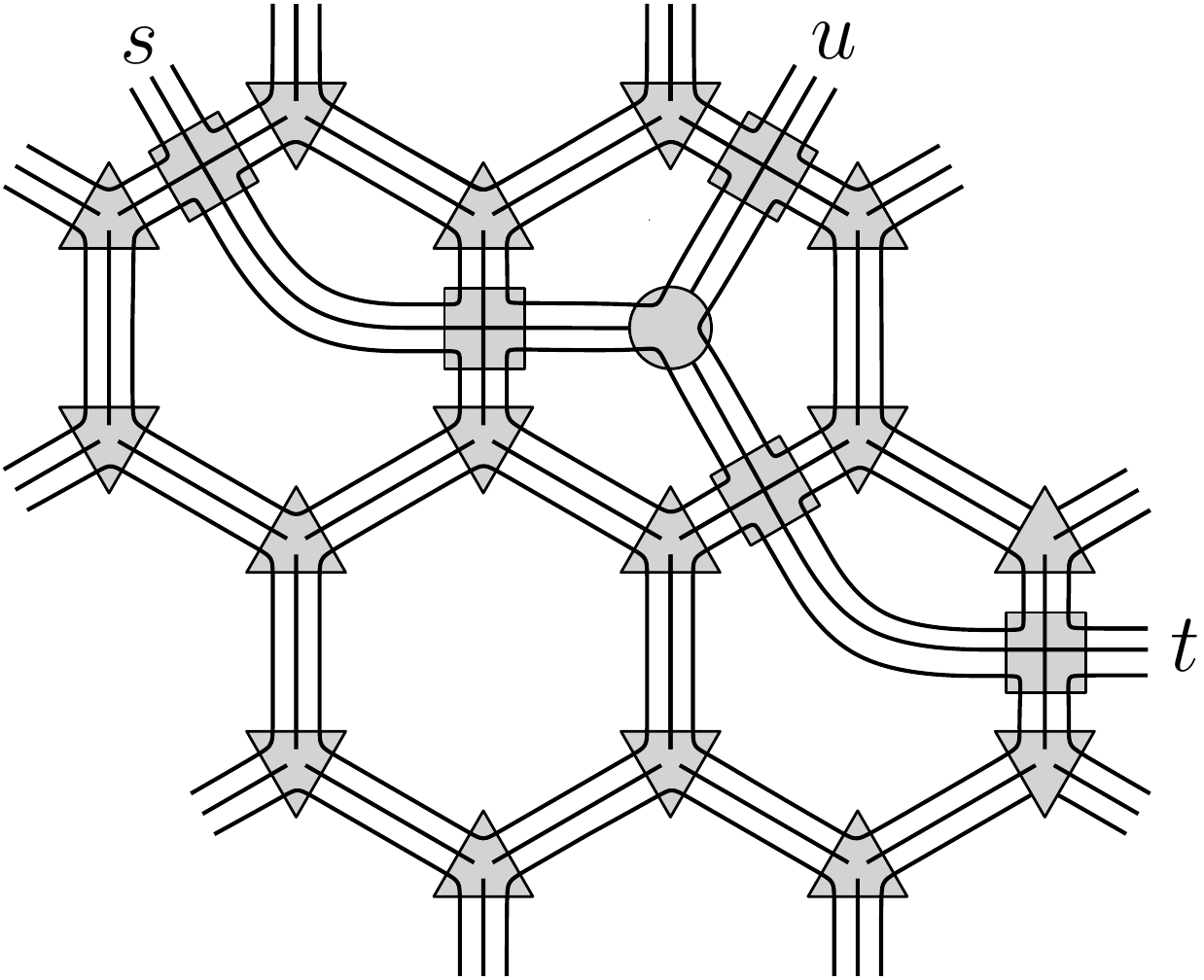}} \;\; ,
	\end{align}
	where one must again be cautious to fix the string-labels to the correct values when ``closing off'' this tensor network on the virtual level. 
	
	The tensors in Eqs.~\eqref{eq:TN_vertex_tensor}, \eqref{eq:TN_crossing_tensor} and \eqref{eq:TN_fusion_tensor} can be used to construct the tensor network representation of any ground state of the Levin-Wen model.
	Ground states of the extend Levin-Wen model can be constructed by adding the following tensor on ever vertical edge:
	\begin{align}\label{eq:TN_tail_tensor}
		\raisebox{-1.3cm}{\includegraphics[scale=.46]{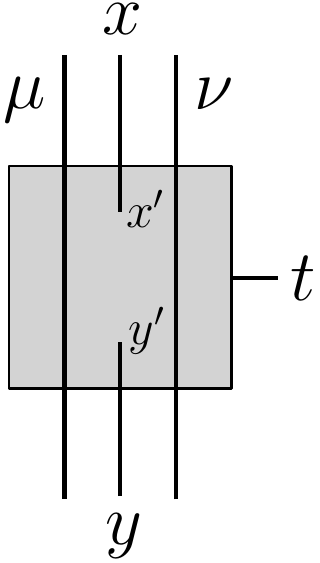}} \; =   \delta_{x x'} \delta_{y y'} \delta_{x'y'} \delta_{t \mathbf{1}}\,.
	\end{align}
	This tensor simply includes a trivial tail qudit and replaces the qudit on the vertical edge by two identical ones.
	Since its action is trivial, we deem it unnecessary to draw it explicitly.
	In the tensor network diagrams below, its pretense is implied in any plaquette where the tail qudit is not specified explicitly.

\subsection{Tensor network representations for anyonic fusion basis states}
	The construction above can be extended to a tensor network representation of any anyonic fusion basis state.
	An important ingredient that is missing is the tensor network representation of the following ribbon configuration:   
	\begin{multline}\label{eq:vacuum_patch_lattice_string_end}
		\sum_{\mu} d_{\mu} \raisebox{-1cm}{\includegraphics[scale=.34]{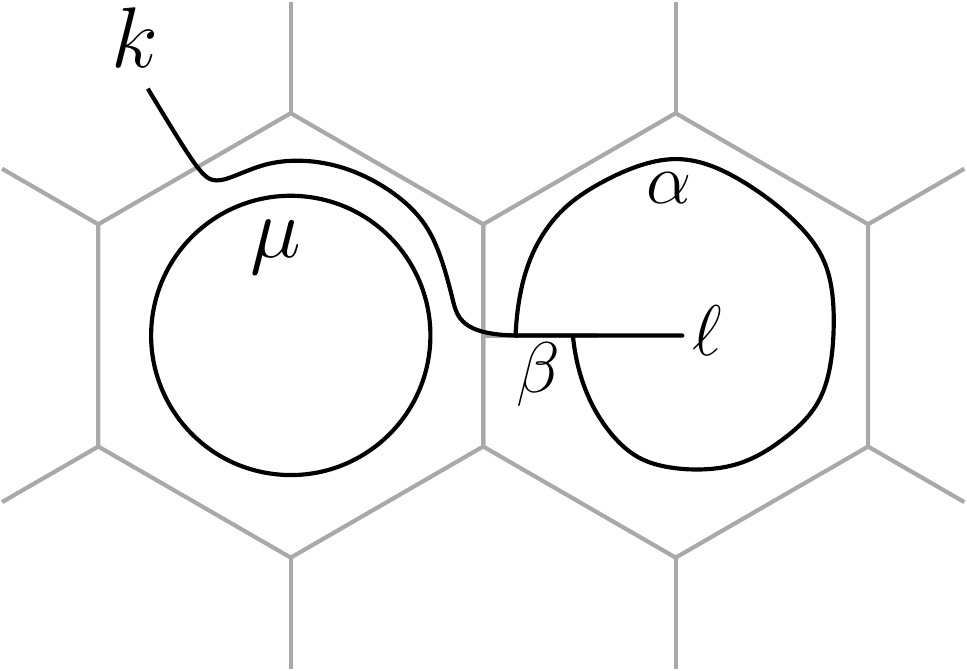}}
		\quad \\
		= \sum_{\mu} \sum_{\mu'} \frac{v_{\mu} v_{\mu'}}{v_k}
		\raisebox{-1cm}{\includegraphics[scale=.34]{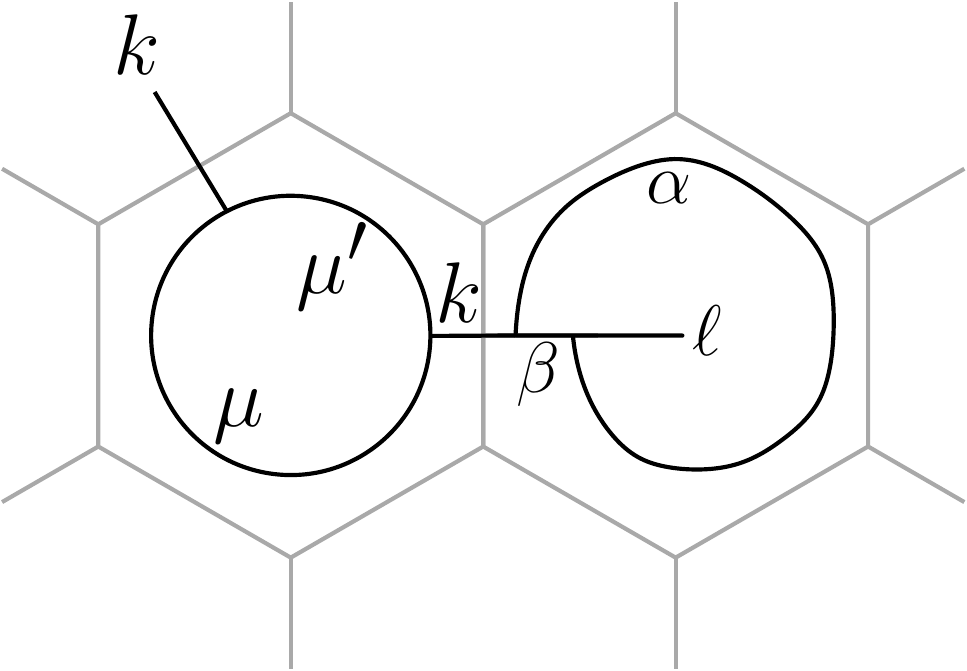}}
		\; .
	\end{multline}
	\vspace{.1cm}
	We again follow the same approach and pull the ribbons onto the physical lattice, which now has a non-trivial tail edge:
	\begin{widetext}
	\begin{align}\label{eq:TN_pull_onto_edge_string_end}
		\raisebox{-1.1cm}{\includegraphics[scale=.5]{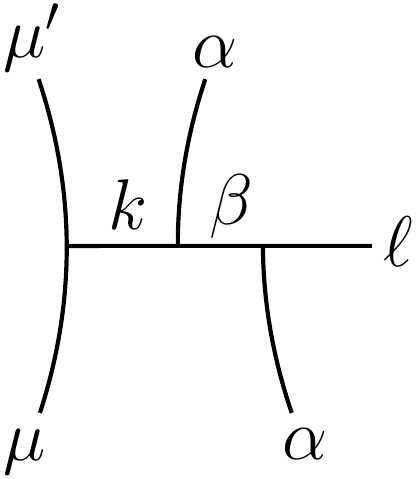}} \; = \sum_{x,y}  \sqrt{\frac{v_y}{v_{\mu}v_{\alpha}}}  \sqrt{\frac{v_x}{v_{\mu'} v_\alpha}}   \sqrt{v_k v_{\mu} v_{\mu '}} \sqrt{v_k v_x v_y} v_\alpha v_\beta G^{\mu \beta x}_{\alpha \mu' k} G^{\alpha y \mu}_{x \beta \ell} \; \raisebox{-1.1cm}{\includegraphics[scale=.5]{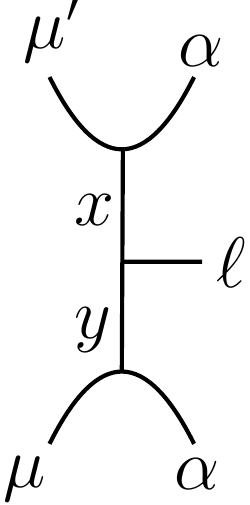}} \,,
	\end{align}
	\end{widetext}
	where the steps in the reduction are completely analogous to those in Eq.~\eqref{eq:TN_pull_onto_edge_string}. The first two factors on the right hand side will contribute to the bottom and top vertices,  respectively, ensuring that we can use the vertex tensor Eq.~\eqref{eq:TN_PEPS_GS} at both vertices. The third factor can be recognized from Eq.~\eqref{eq:TN_pull_onto_edge_string} as the factor contributing to the left plaquette. We then absorb the remaining factors into the definition of the string-end tensor
	\begin{equation}\label{eq:TN_string_end_tensor} 
		\raisebox{-.75cm}{\includegraphics[scale=.46]{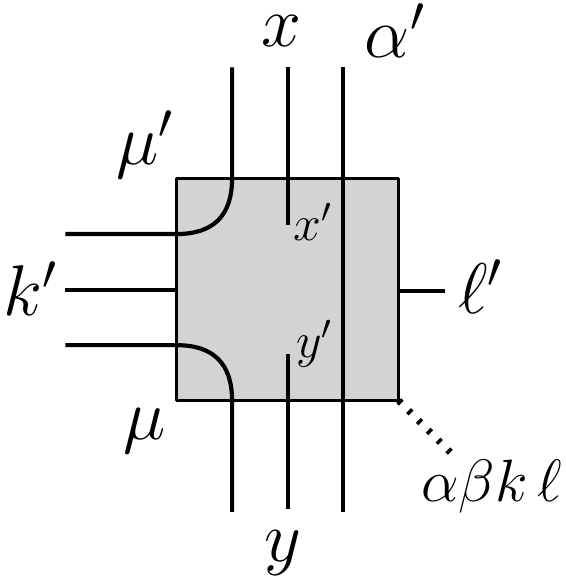}} \hspace{-2.7cm}
		\begin{split} \hspace{2.5cm}
        =& \, \delta_{x x'} \delta_{y y'} \delta_{k k'} \delta_{\ell \ell'} \delta_{\alpha \alpha'} \\ 
        & \qquad  \sqrt{v_k v_x v_y} \frac{v_\beta}{v_\alpha} G^{\mu \beta x}_{\alpha \mu' k} G^{\alpha y \mu}_{x \beta \ell} \,.
        \end{split} 
	\end{equation}
	This definition differs from the remaining factors on the right hand side of Eq.~\eqref{eq:TN_pull_onto_edge_string_end} by $ \frac{1}{d_\alpha} $, in order to counter the factor $ d_\alpha $ that arises from the closed loop convention on the virtual level. This tensor has three physical indices, $ x' $, $ y' $ and $ \ell' $, representing the physical state of the two qudits on the vertical edge and the connected tail qudit.
	The dotted leg with index $ \alpha \beta k \ell$ is included to avoid summing over different values of the string labels, as we are representing a state in which the labels $ \alpha $, $ \beta $  $ k $ and $ \ell $ are fixed.
	
\begin{figure}[ht]
		\centering
		\includegraphics[scale=.52]{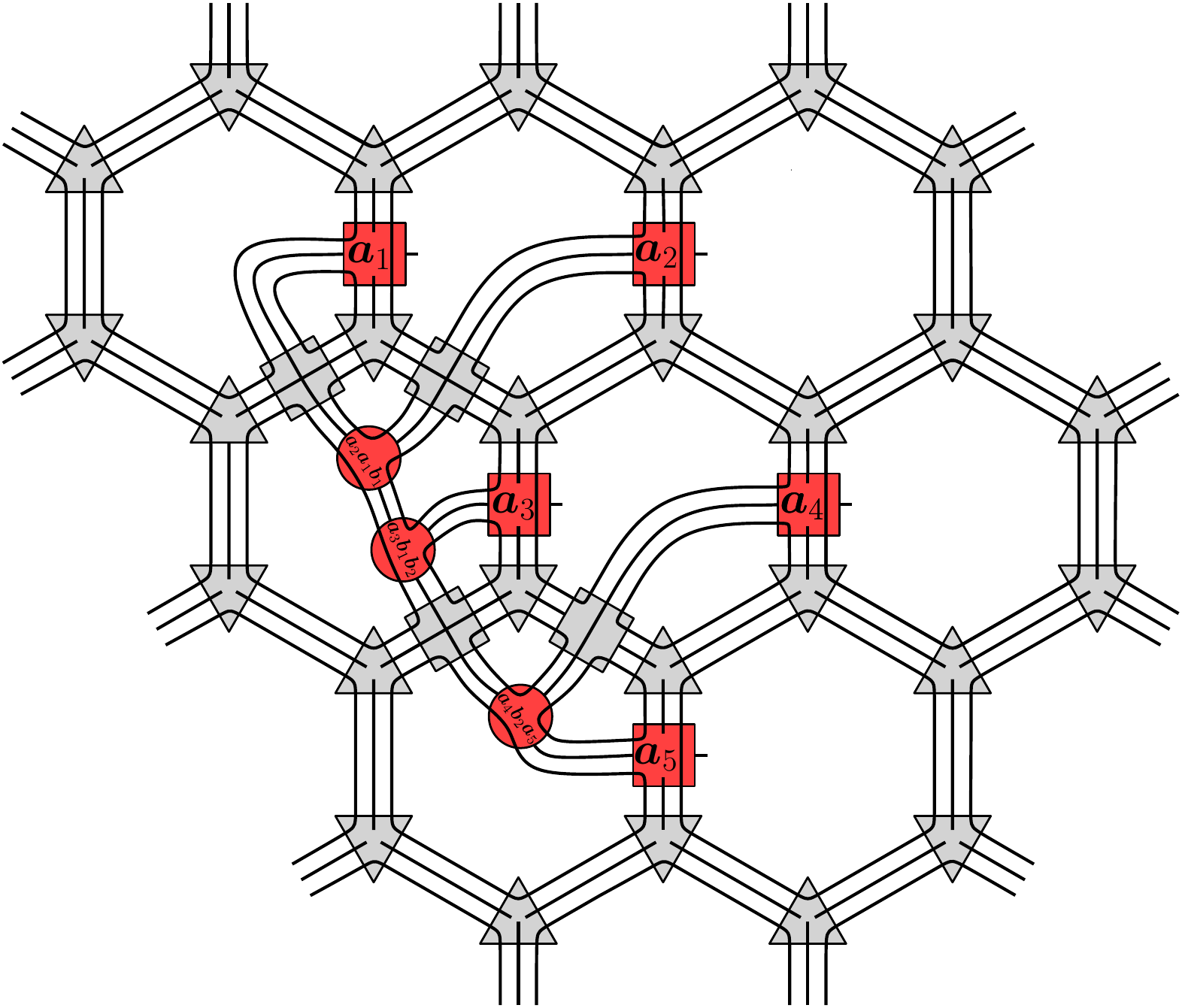}
		\caption{PEPS representation of the anyonic fusion basis state in Eq.~\eqref{eq:anyonic_fusion_basis_lattice1}.}
		\label{fig:PEPS_basis_state}
	\end{figure}

	We now have all the ingredients we need to construct a PEPS realization of an arbitrary anyonic fusion basis state on the tailed lattice. All that remains is to construct PEPS tensors representing the doubled leaf and pants segments appearing in Eq.~\eqref{eq:anyonic_fusion_basis_red1} by combining the string end and fusion tensors with the decompositions of the doubled ribbons to two-dimensional ribbon configurations. 
	The factors $ \frac{v_k}{v_a v_b} $ in Eq.~\eqref{eq:anyonic_fusion_basis_red1} that result from the reduction of the doubled ribbons in the interior branches of the fusion basis state are split evenly into both sides of the resulting single ribbon. 
	Each doubled leaf segment therefore receives an extra factor $ \sqrt{\frac{v_k}{v_a v_b}} $, which, when combined with the factors on the right hand side of Eq.~\eqref{eq:doubled_leaf_segment_red} yields the definition of the excitation tensor for a doubled anyon with a leaf label $ \boldsymbol{a} = a\bar{b}_\ell $:
	\vspace{-.3cm}
	\begin{widetext}
	\begin{equation}\label{eq:TN_excitation_tensor_def}
		\raisebox{-1.2cm}{\includegraphics[scale=.46]{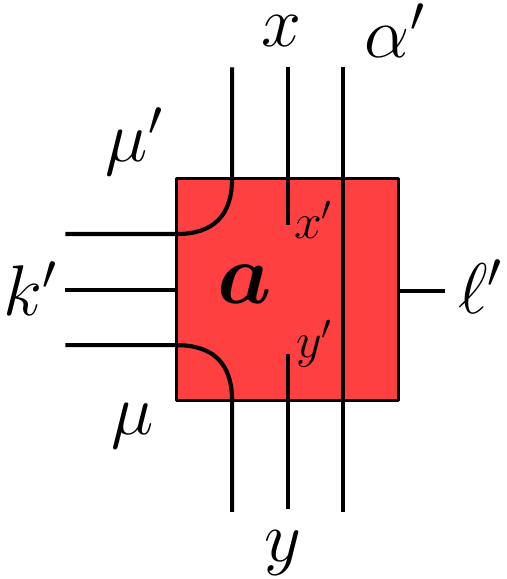}} \, = 
		\frac{\sqrt{v_a v_b}}{\D}  \sum_{\alpha \beta k} \sqrt{v_k} v_\alpha v_\beta \sum_{\gamma, \delta} d_\gamma d_\delta R^{a \alpha}_{\gamma} R^{\alpha b}_{\delta} G_{\alpha a b}^{k \delta \gamma} G_{b \alpha \ell}^{\beta a \delta} G_{a \gamma \delta}^{k \beta \alpha} \quad
		\raisebox{-1.2cm}{\includegraphics[scale=.46]{fig/PEPS_string_end_tensor.pdf}} ,
	\end{equation}
	or, equivalently,
	\begin{equation}\label{eq:TN_excitation_tensor_def2}
		\raisebox{-1.2cm}{\includegraphics[scale=.46]{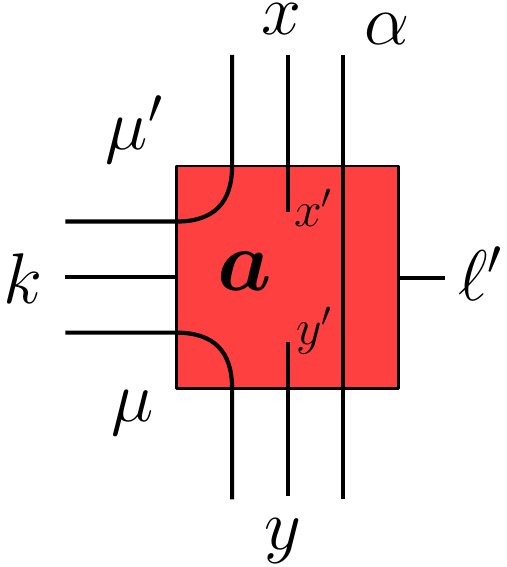}} \, = \delta_{x x'} \delta_{y y'} \delta_{\ell \ell'}
		\frac{\sqrt{v_a v_b}}{\D} \sqrt{v_x v_y} d_k \sum_\beta d_\beta 
		\sum_{\gamma, \delta} d_\gamma d_\delta R^{a \alpha}_{\gamma} R^{\alpha b}_{\delta} G_{\alpha a b}^{k \delta \gamma} G_{b \alpha \ell}^{\beta a \delta} G_{a \gamma \delta}^{k \beta \alpha}   G^{\mu \beta x}_{\alpha \mu' k} G^{\alpha y \mu}_{x \beta \ell} \,.
	\end{equation}
	For the tensor representing the root of the fusion tree, a factor $ \left(R^{b_5 a_5}_{\ell_5}\right)^* $ must be included , in accordance with Eq.~\eqref{eq:anyonic_fusion_basis_red1}. 
	We will not include it implicitly, but it is implied whenever a tensor is the root of the fusion tree.
	In a similar way, the doubled pants segment on the left hand side of Eq.~\eqref{eq:doubled_pants_segment_red} gets an additional factor $ \sqrt{\frac{v_i v_j v_k}{v_a v_b v_c v_d v_e v_f}} $ which, when combined with the reduction on the right hand side, leads to a doubled fusion tensor for three doubled Fibonacci anyons $ \boldsymbol{a} = a\bar{b} $, $ \boldsymbol{b} = c\bar{d} $ and $ \boldsymbol{c} = e\bar{f} $ of the form
	\vspace{-.1cm}
	\begin{equation}\label{eq:TN_doubled_fusion_tensor}
		\raisebox{-.8cm}{\includegraphics[scale=.46]{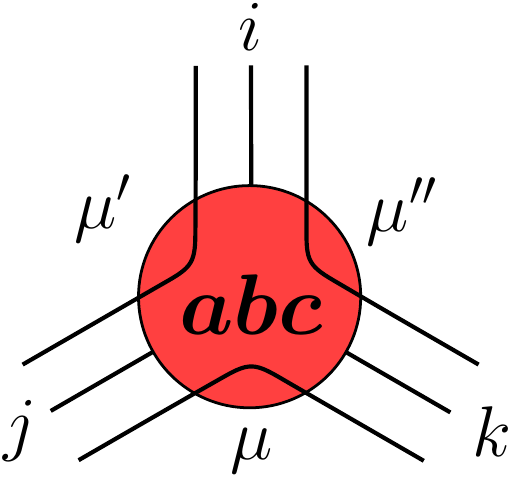}} \, = 
		\sqrt{v_i v_j v_k v_a v_b v_c v_d v_e v_f} \sum_{\gamma, \delta} d_\gamma d_\delta R^{a d}_{\gamma} G_{def}^{k b \delta} G_{d a e}^{c \delta \gamma} G_{\gamma d c}^{j \delta a} G_{\delta b a}^{i j k} \quad
		\raisebox{-.8cm}{\includegraphics[scale=.46]{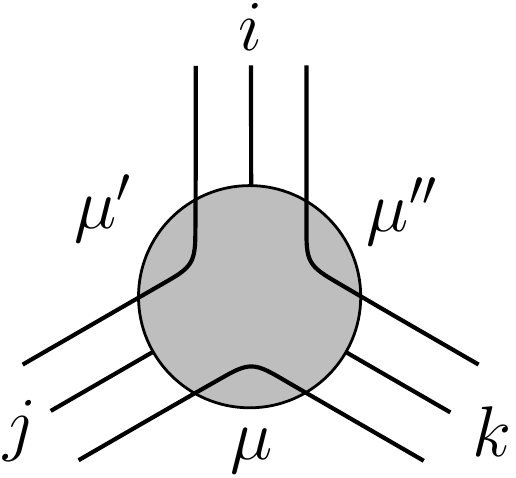}} \,.
	\end{equation}
	A PEPS representation of an arbitrary anyonic fusion basis state of the form \eqref{eq:anyonic_fusion_basis_comp} can then be constructed by applying the following correspondence rules:
	\vspace{-.3cm}
	\begin{equation}\label{TN_doubled_fusion_correspondence_rules}
		\raisebox{0cm}{\includegraphics[scale=.42]{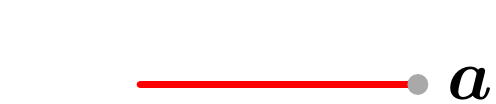}}
		\quad \rightarrow \quad
		\raisebox{-1.2cm}{\includegraphics[scale=.46]{fig/PEPS_excitation_tensor_bis.pdf}}
		\; ,  \qquad \qquad
		\raisebox{-.9cm}{\includegraphics[scale=.42]{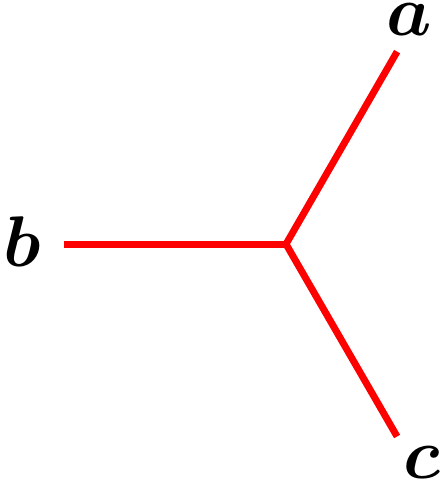}}
		\; \rightarrow \quad
		\raisebox{-1.cm}{\includegraphics[scale=.46]{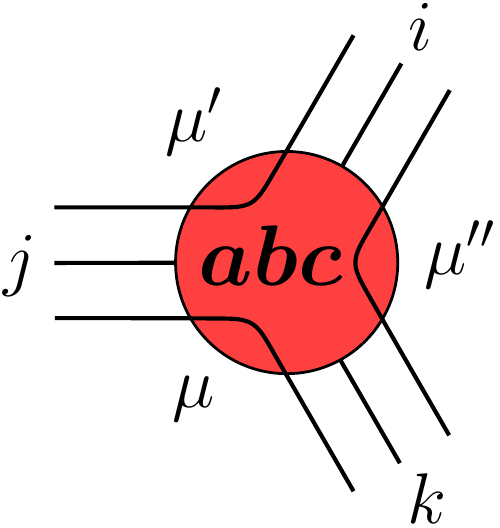}} \;.
	\end{equation}
	\vspace{-.3cm}
	\end{widetext}
	For the specific case of the fusion basis state in Eq.~\eqref{eq:anyonic_fusion_basis_lattice1}, the resulting PEPS is depicted in \figref{fig:PEPS_basis_state}. 

	The PEPS representation for anyonic fusion basis states we have just derived provides a powerful tool for numerical calculations with the relevant states in the physical subspace. 
	It was obtained here through the explicit reduction of the anyonic fusion basis states defined in App.~\ref{sec:anyonic_fusion_basis}. 
	Even though the graphical calculus on these three-dimensional ribbon configurations shows that they indeed behave as fusion states of doubled Fibonacci anyons under twists and braiding, it is not intuitively clear how this behavior translates once such a configuration has been reduced to a superposition of qudit states. It is therefore useful to explicitly verify that our PEPS ansatz itself has the desired behavior under relevant operations.
	We perform such checks explicitly in Sec.~\ref{sec:TN_consistency}.
	

\subsection{Square PEPS tensors} \label{sec:square_PEPS}
	In practice, it is often much more convenient to work with a square geometry for the PEPS. This can be achieved by contracting the tensors within each segment.
	In order to simplify the resulting square tensors, we combine each group of triple indices to a single index. Note that, since the definition of the tensor \eqref{eq:TN_vertex_tensor} requires that each set of tripled indices satisfies the Fibonacci fusion rules in order to yield a nonzero value, these regrouped virtual indices only need to have dimension 5, corresponding to the 5 combinations $abc$ satisfying $ \delta_{abc}  = 1$: $\{ \1\1\1,\, \1\tau\tau,\, \tau\1\tau, \,\tau\tau\1, \,\tau\tau\tau \}$.
	A similar trick can be used when grouping the physical indices, since these come in three sets of 3 labels that must satisfy the fusion condition in each vertex (furthermore, these sets share indices that where doubled by the tensor network construction).
	The resulting square tensors will be colored gray and red for segments of which the tails end inside plaquettes containing trivial and nontrivial DFIB charges, respectively:
	\begin{equation}\label{eq:PEPS_sqaure_tensors}
		\raisebox{-0.4cm}{\includegraphics[scale=.3]{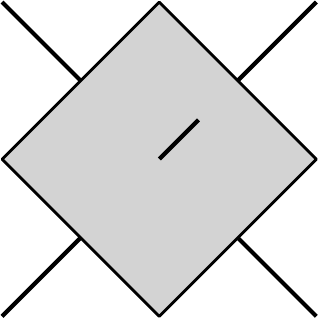}}
		\; \equiv \hspace{-.3cm}
		\raisebox{-1.4cm}{\includegraphics[scale=.36]{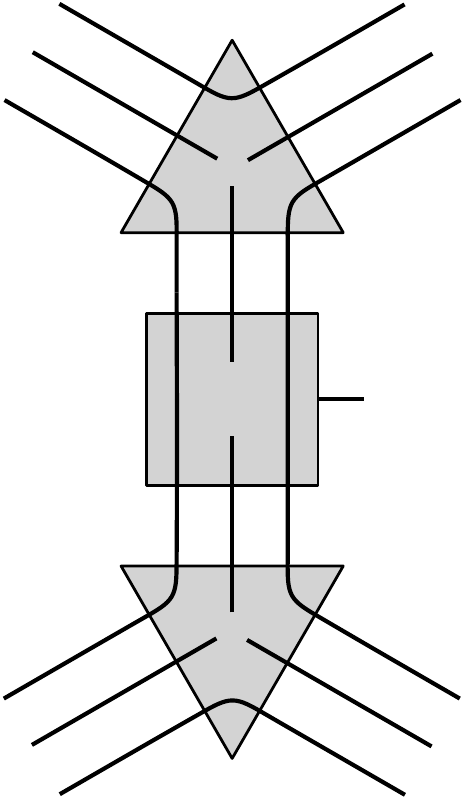}}   \hspace{-.1cm}, \qquad
		\raisebox{-0.4cm}{\includegraphics[scale=.3]{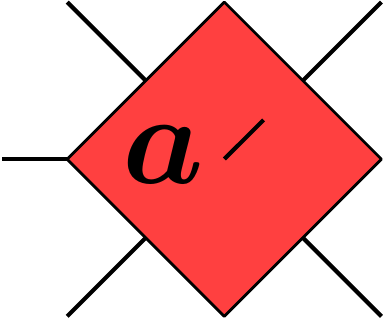}}
		\; \equiv 
		\raisebox{-1.4cm}{\includegraphics[scale=.36]{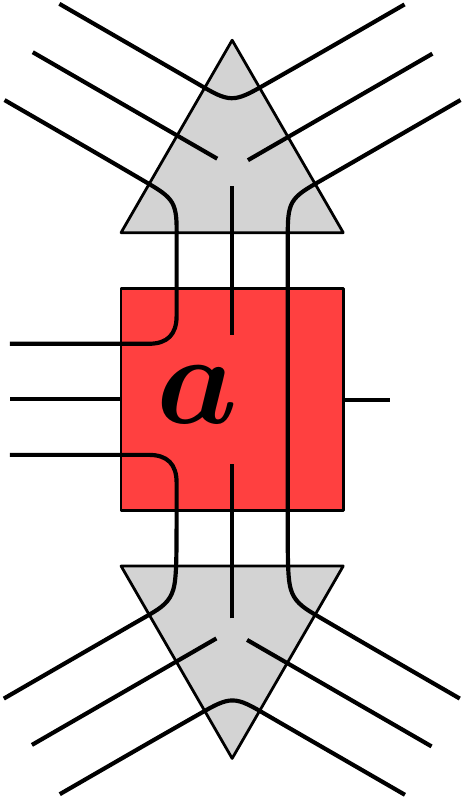}} .
	\end{equation}
	With this simplified notation, the fusion state depicted in \figref{fig:PEPS_basis_state} can  be written as
	\begin{equation}\label{eq:PEPS_squared}
		\raisebox{-1.9cm}{\includegraphics[scale=.5]{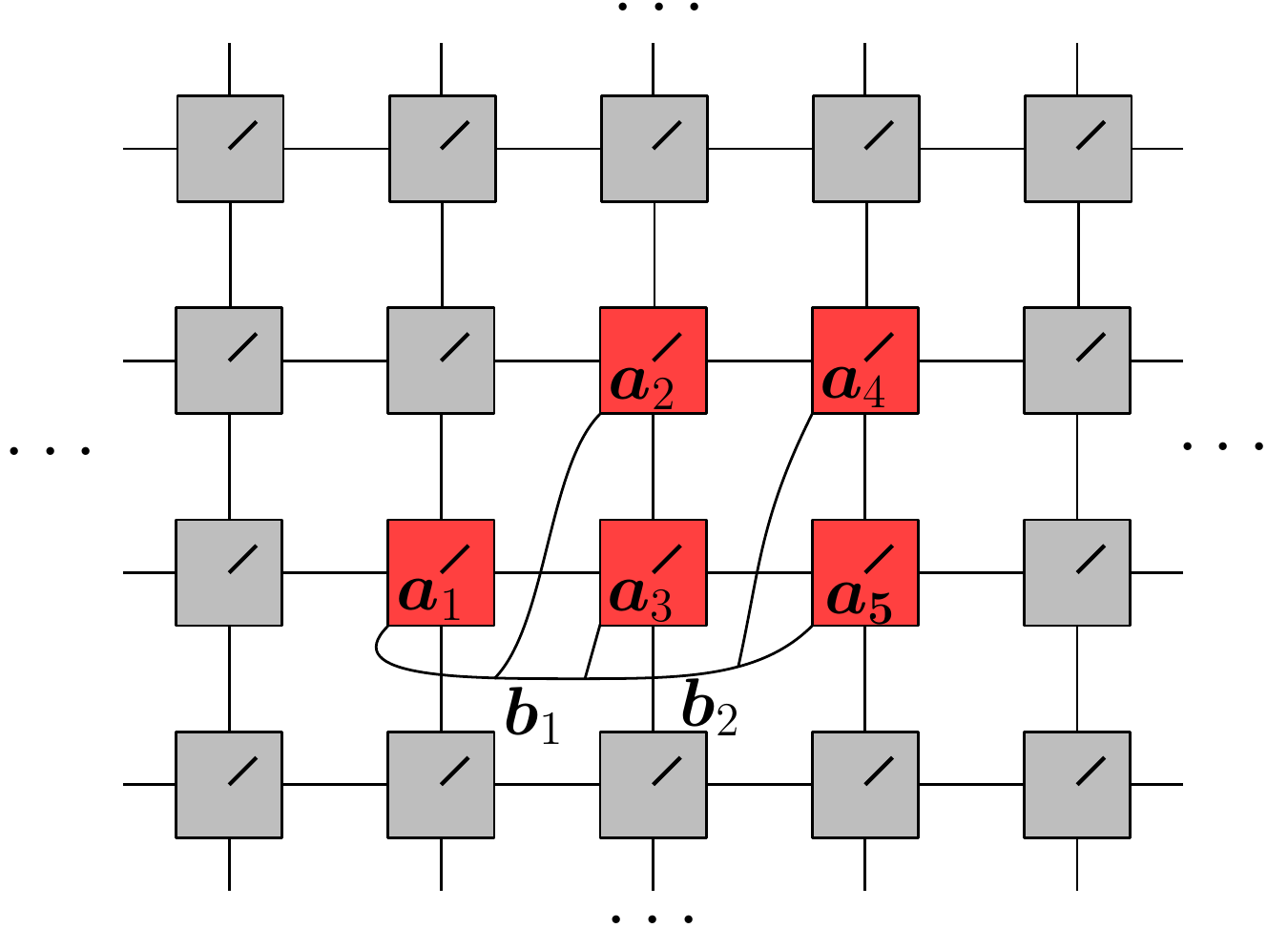}}\;,
	\end{equation}
	where we rotated the lattice counterclockwise by $ 60\degrees$ before obtaining the squared lattice.
	In this simplified notation, we	assume that the crossing tensors defined in Eq.~\eqref{eq:TN_crossing_tensor} are inserted at every edge crossing of a virtual string, and assume that the strings are fused using the appropriate doubled fusion tensors Eq.~\eqref{eq:TN_doubled_fusion_tensor}.	
	Note that when using the simplified tensors (without the triple indices), one must be careful to correctly enforce the closed loop convention adopted when defining the PEPS tensors on the honeycomb lattice.
	
\subsection{Consistency of the tensor network representation} \label{sec:TN_consistency}
	From the PMPO description of topological order \cite{bultinck2017anyons} it is known that an anyon ansatz in PEPS must satisfy certain consistency conditions. 
	It was shown through numerical calculations that our PEPS representation of the anyonic fusion basis states indeed satisfies all properties required of a tensor network description of anyonic excitations, providing an important consistency check for our framework. 	In particular, the braiding and fusion behavior of the DFIB excitation tensors was explicitly verified.
	Below we showcase some of these consistency checks, which were performed numerically with the Fibonacci input category, but should hold in general for any modular ribbon category satisfying Eqs.~\eqref{eq:F_tetrahedral}, \eqref{eq:F_normalization} and \eqref{eq:F_unitarity_2}.
	It is worth noting that these consistency conditions are in fact implied by the construction of the tensors. The numerical checks merely confirm that no mistakes where made in the derivation.
	
	
	
	We start by noting that the representation of ribbons on the virtual level as depicted in Eq.~\eqref{eq:TN_PEPS_string}, can be understood as a MPO operator acting on the virtual level of a PEPS.
	The MPO tensor itself is given by what we have previously called the crossing tenor,
	\begin{equation}\label{eq:MPO_tensor_fib}
		\raisebox{-1.cm}{\includegraphics[scale=.46]{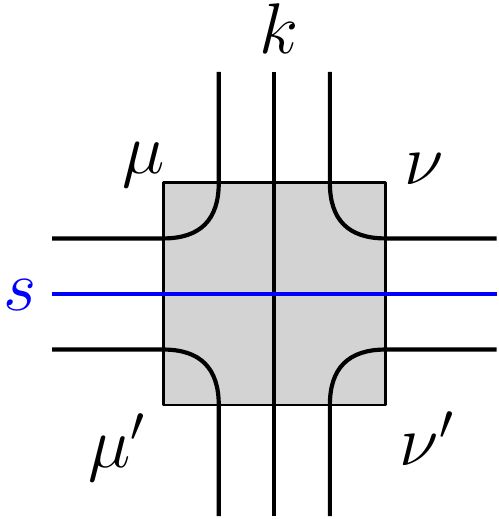}} \; =   G^{\mu'\nu' k}_{\nu\mu \, s} \,,
	\end{equation}
	where we have rotated the definition in Eq.~\eqref{eq:TN_crossing_tensor} over 90 degrees and we now denote the index that encodes the ribbon label $ s $ in blue. 
	We refer to this label as the \emph{block label} of the MPO tensor. 
	For notational convenience we will often assume that the tripled lines are grouped (as done in Sec.~\ref{sec:square_PEPS}), in which case we only explicitly write the block label for the grouped index.
	It should be emphasized however that we still maintain the closed loop convention introduced before, even when we depict indices as being grouped. 
	From the pentagon equation for the input category it follows that the MPOs can be moved freely through the ground state PEPS vertex tensors, which is referred to as the \emph{pulling through} property.
	This corresponds to the freedom to continuously deform the ribbons is App.~\ref{sec:ribbon_graph}, and the fact that the string-net ground state inherits this property.
	Adding the single block MPOs with a weight $ w_s = d_s / \D^2 $ for each block $ s $ and closing the resulting operator into a loop results in a projector MPO (PMPO) that acts trivially on any region of the PEPS with vacuum total charge. 
	This PMPO can be thought of as the virtual representation of a vacuum loop (divided by the total quantum dimension).
	
	The fusion tensors in Eq.~\eqref{eq:TN_fusion_tensor} that represent ribbon fusion on the virtual level can be interpreted as tensors that represent the fusion of MPOs of different blocks:
	\begin{equation}\label{eq:MPO_fusion}
		\raisebox{-1.cm}{\includegraphics[scale=.75]{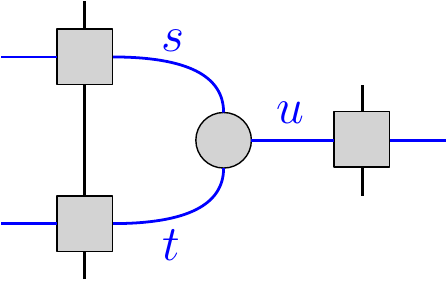}} \;.
	\end{equation}
	Using these fusion tensors we can build a virtual representation of the tube algebra introduced in App.~\ref{sec:tube_algebra}, which is then generated by elements of the form
	\begin{equation}\label{eq:MPO_tube}
		\raisebox{-1.2cm}{\includegraphics[scale=.5]{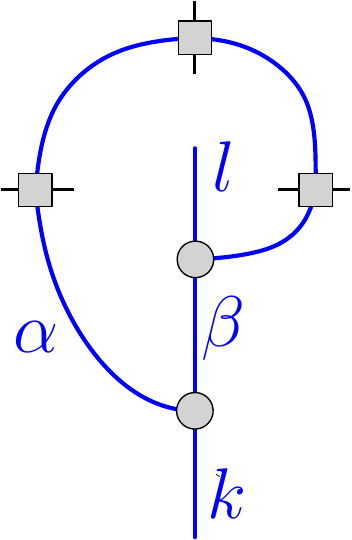}} \;.
	\end{equation}
	It was shown in Ref.~\cite{bultinck2017anyons} that these objects form a $ \C^* $-algebra whose central idempotents correspond to the topological superselection sectors of the theory, which corresponds to the results presented in App.~\ref{sec:tube_algebra} that were we was obtained using the graphical calculus of the ribbon graph Hilbert space. 
	In particular, we can derive the expression for the idempotents and nilpotents of the $ \C^* $-algebra by simply realizing Eq.~\eqref{eq:tube_algebra_idempotents} on the virtual level. 
	The operators of interest are the tube operators of the form $ \cP^{a\bar{b}}_{kl} $, that represent both the simple idempotents and nilpotents of the tube algebra. 
	Just as in App.~\ref{sec:tube_algebra} these can be decomposed into a superposition of the basis elements in Eq.~\eqref{eq:tube_operator} as
	\begin{equation}\label{eq:potent_decomp}
		\cP^{a\bar{b}}_{kl} = \sum_{\alpha \beta} P^{(abl)}_{\alpha \beta k} O_{k l \alpha \beta}\,,
	\end{equation}
	where the coefficients $ P^{(abl)}_{\alpha \beta k} $ are given by
	\begin{equation}\label{eq:potent_decomp_coeff}
		P^{(abl)}_{\alpha \beta k} = \frac{1}{\D^2} \frac{d_a d_b}{v_k} v_\alpha v_\beta \sum_{\gamma, \delta} d_\gamma d_\delta R^{a \alpha}_{\gamma} R^{\alpha b}_{\delta} G_{\alpha a b}^{k \delta \gamma} G_{b \alpha l}^{\beta a \delta} G_{a \gamma \delta}^{k \beta \alpha} \,.
	\end{equation}
	On the virtual level, these operators can be represented by the tensor
	\begin{align}\label{eq:MPO_potent_tensor}
		\raisebox{-1.2cm}{\includegraphics[scale=.45]{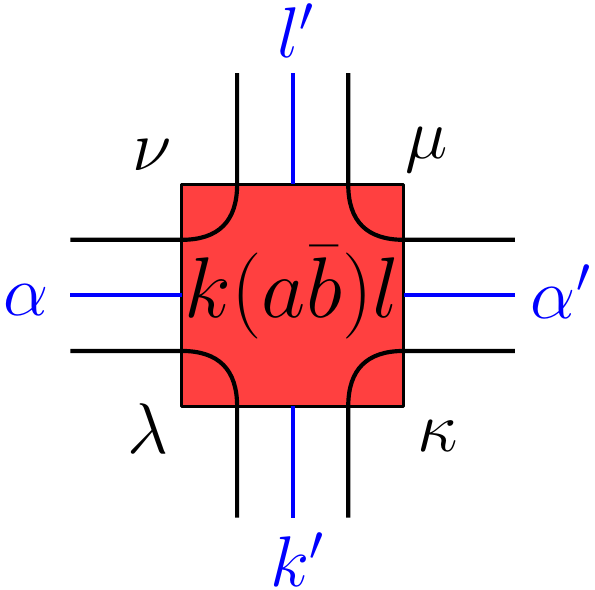}}
		\, =  \delta_{k k'} \delta_{l l'} \delta_{\alpha \alpha'} \sum_{\beta} P^{(abl)}_{\alpha \beta k} \hspace{-.6cm}
		\raisebox{-2.5cm}{\includegraphics[scale=.45]{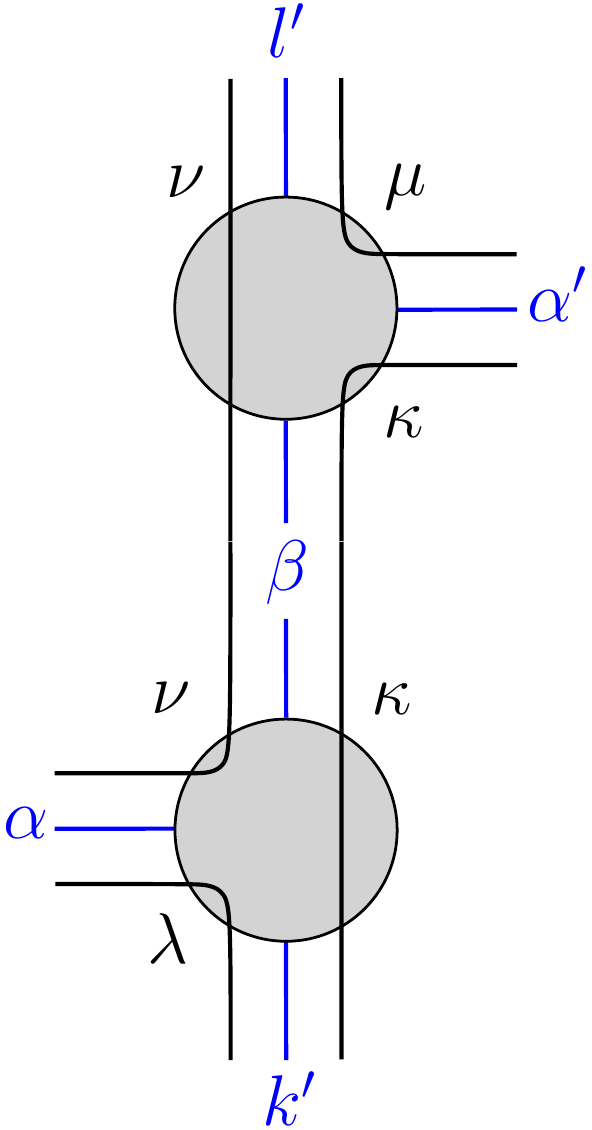}}
		\hspace{-.4cm},
	\end{align}
	giving rise to an MPO of the form
	\begin{equation}\label{eq:potent_MPO}
		\raisebox{-1.9cm}{\includegraphics[scale=.65]{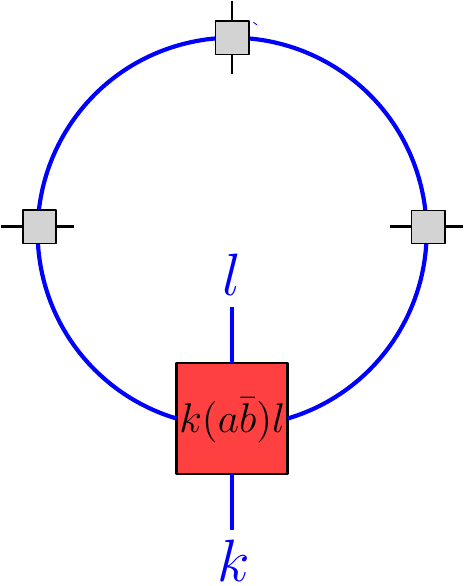}}\;.
	\end{equation}
	As the central idempotents $ \cP^{a\bar{b}} $ of the tube algebra can be constructed as $ \cP^{a\bar{b}} = \sum_{l} \cP^{a\bar{b}}_{ll} $, the tensors in Eq.~\eqref{eq:MPO_potent_tensor} can be combined to give the square tensors representing the central idempotents of the $ \C^* $-algebra:
	\begin{equation}\label{eq:MPO_idempotent_tensor}
		\raisebox{-.9cm}{\includegraphics[scale=.45]{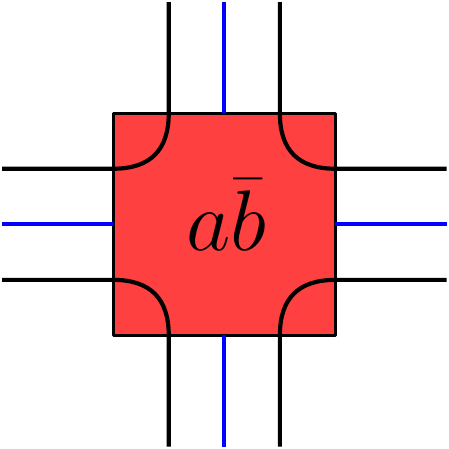}}
		\; =  \sum_{l}  \quad
		\raisebox{-.9cm}{\includegraphics[scale=.45]{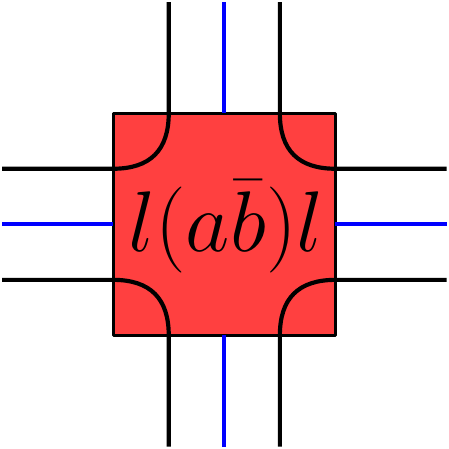}}
		\;,
	\end{equation}
	giving a representation for the central idempotents on the virtual level.
	
	Using these explicit expressions for the $ \cP^{a\bar{b}}_{kl} $ on the virtual level, all properties required of the anyon ansatz in our model can be explicitly verified. As a first check, the stacking behavior of the idempotents and nilpotents on the virtual level was verified by explicitly computing $ \cP^{a\bar{b}}_{kl} \cP^{a'\bar{b}'}_{k'l'} $, giving the expected results:
	\begin{align}\label{eq:MPO_potent_stacking}
	    \raisebox{-2.75cm}{\includegraphics[scale=.64]{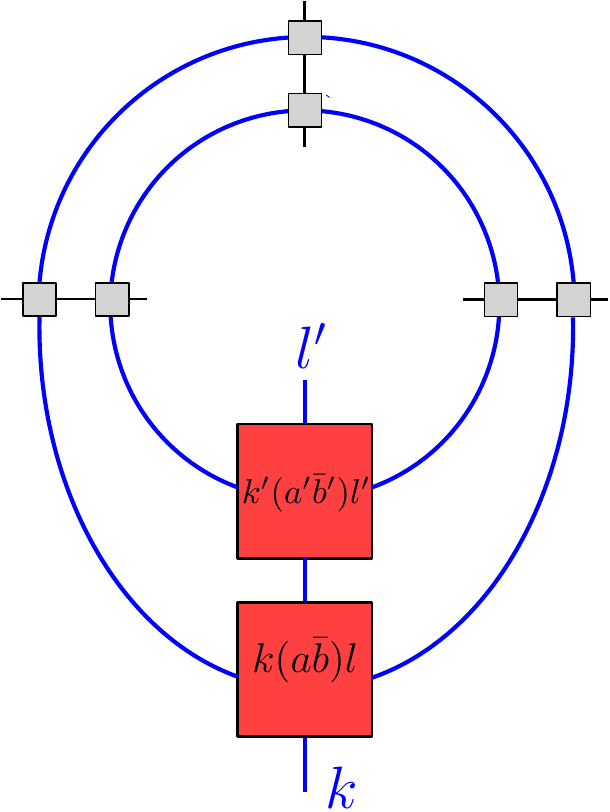}} \hspace{-.12cm}
		= \delta_{a a'} \delta_{b b'} \delta_{l k'} \hspace{-.2cm}
		\raisebox{-1.6cm}{\includegraphics[scale=.64]{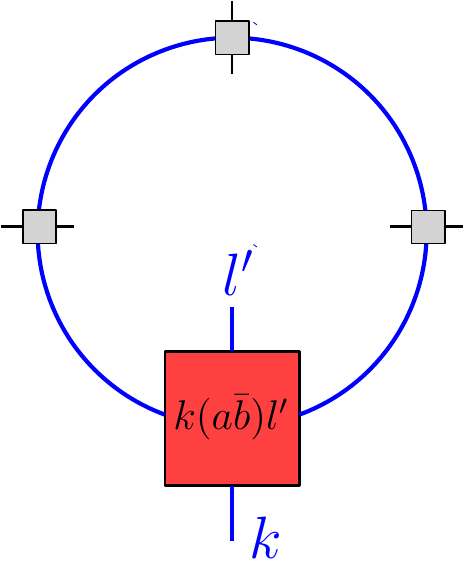}} \hspace{-.07cm},
	\end{align}
	This identity can be seen as the virtual representation of Eq.~\eqref{eq:doubled_fusion_basis_stacking}.
	Next it was explicitly verified that the excitation tensors defined in Eq.~\eqref{eq:TN_excitation_tensor_def} behave correctly under the virtual action of the tube algebra idempotent and nilpotent MPO operators in Eq.~\eqref{eq:MPO_potent_tensor}:
	\begin{align}
		 \raisebox{-1.9cm}{\includegraphics[scale=.65]{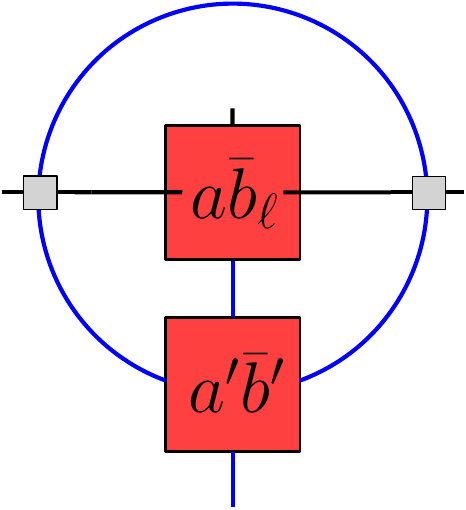}}
		\; = \delta_{a a'} \delta_{b b'} \quad
		\raisebox{-.7cm}{\includegraphics[scale=.65]{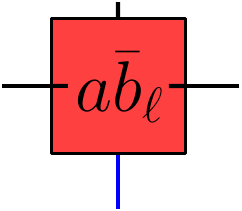}}
		\;,  \label{eq:excitation_tensor_central_idempotent}
	\end{align}
	\begin{align}
		 \raisebox{-3cm}{\includegraphics[scale=.65]{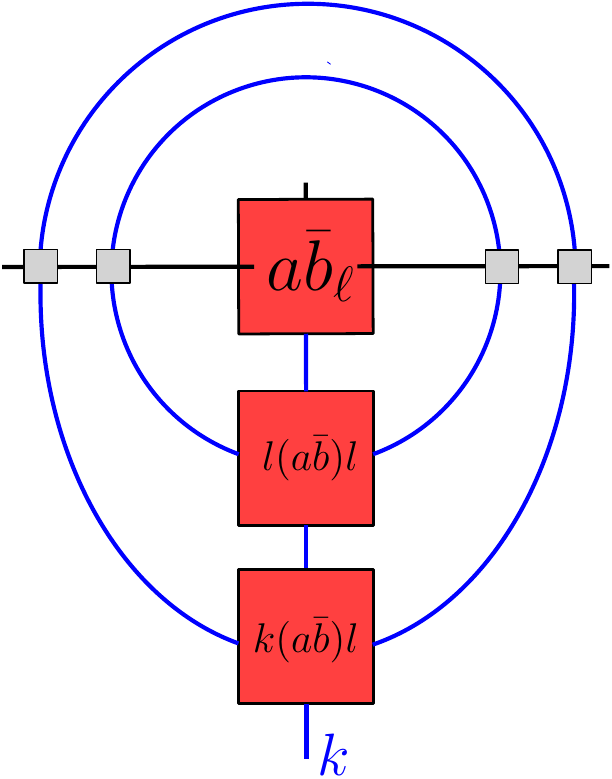}}
		\propto \
		\raisebox{-1.85cm}{\includegraphics[scale=.65]{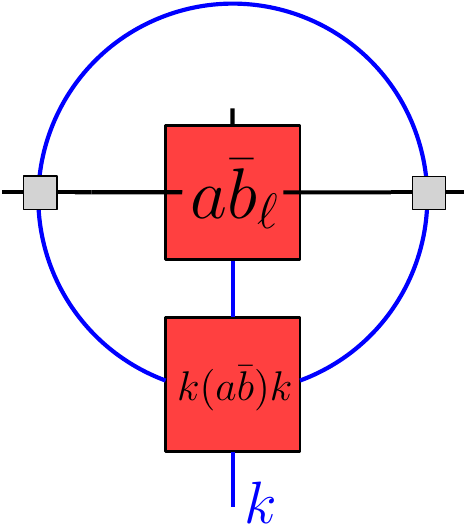}}
		\;. \label{eq:excitation_tensor_simple_idempotents}
	\end{align}
	\noindent In these expressions the excitation tensors \eqref{eq:TN_excitation_tensor_def} were rotated by 90 degrees counterclockwise. 
	For the specific case of DFIB excitations, \eqref{eq:excitation_tensor_simple_idempotents} shows that the $ \tau\bar{\tau}_1 $ and $ \tau\bar{\tau}_\tau $ excitations have \emph{virtual} support in both the $ \cP^{\tau\bar{\tau}}_{11} $ and $ \cP^{\tau\bar{\tau}}_{\tau\tau} $ simple idempotents. 
	It should be stressed that, even though we represent the diagrams in single line notation, the closed loop convention for the PEPS representation of string-net states must be used in the actual computations of all contractions.
	
	We conclude this section with the topological properties (fusion rules, braiding properties and topological spin) of our anyon ansatz.
	To simplify the expressions we again denote a DFIB charge and FIB tail label with a single bold label, $ \boldsymbol{a} = a_+\overline{a_-}_\ell $, or $ \boldsymbol{a} = a_+\overline{a_-} $ depending on the context. 
	It was verified that fusion of two excitations tensors using the doubled fusion tensor \eqref{eq:TN_doubled_fusion_tensor} only has virtual support in the correct MPO central idempotent: 
	\vspace{-1.1cm}
	\begin{widetext}
	\begin{align}\label{eq:excitation_doubled_fusion_check}
		\raisebox{-1.9cm}{\includegraphics[scale=.65]{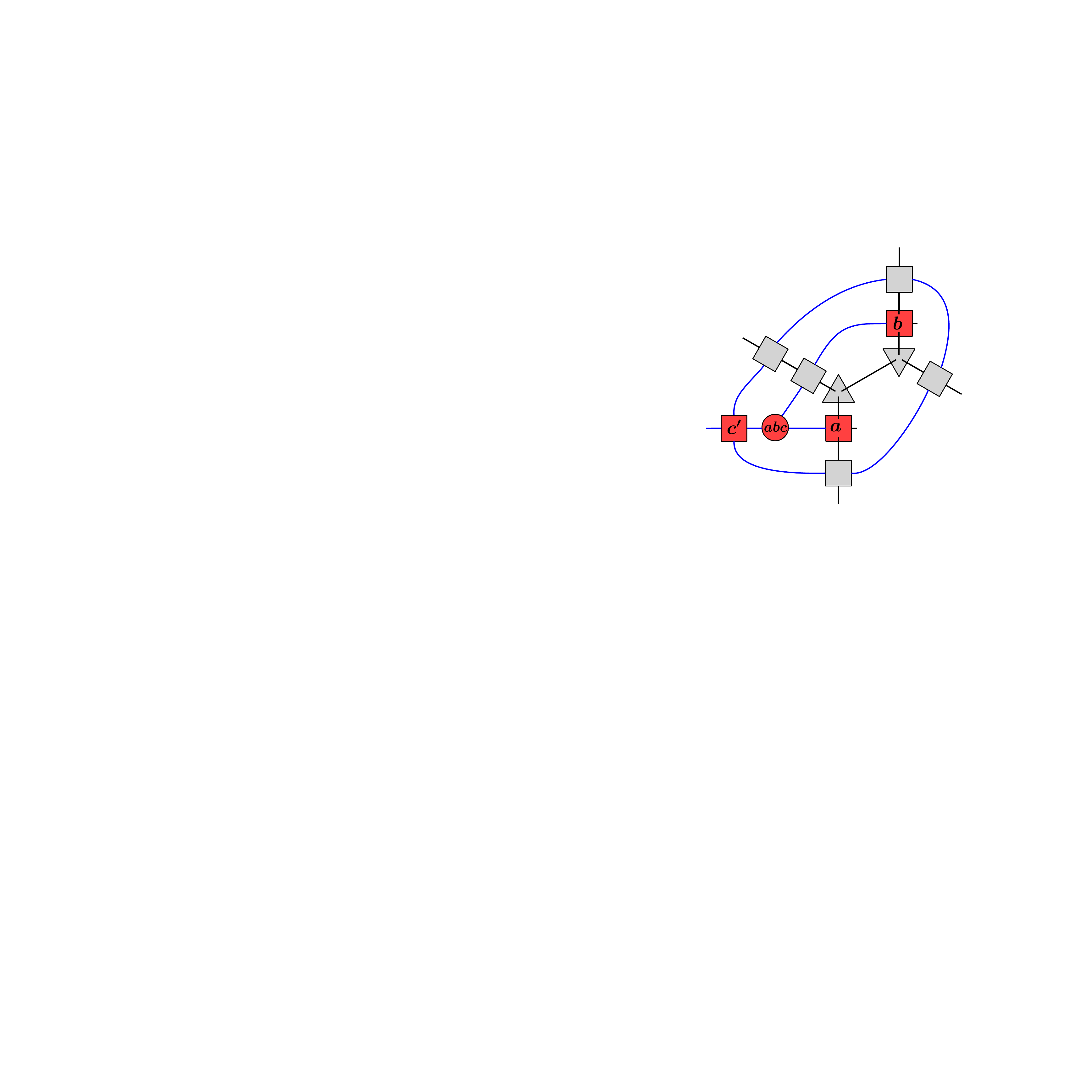}}
		\quad = \delta_{\boldsymbol{c}\boldsymbol{c}'} \quad
		\raisebox{-1.cm}{\includegraphics[scale=.65]{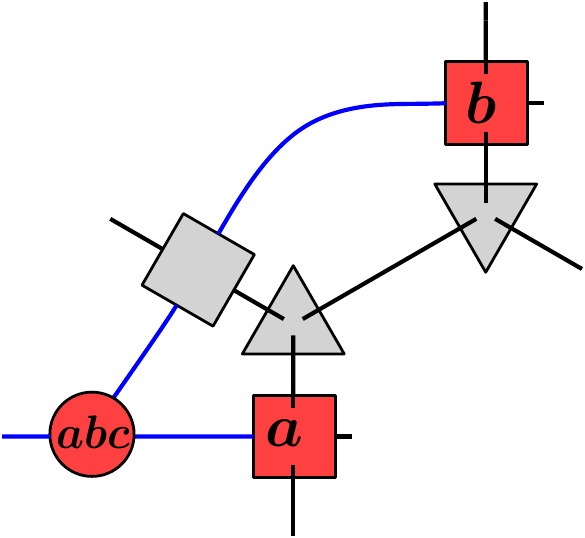}}
		\; .
	\end{align}
	The braiding properties of the doubled anyons, can be translated to the following relation on the virtual level:
	\begin{equation}\label{eq:excitation_braiding_check}
		\raisebox{-1.9cm}{\includegraphics[scale=.65]{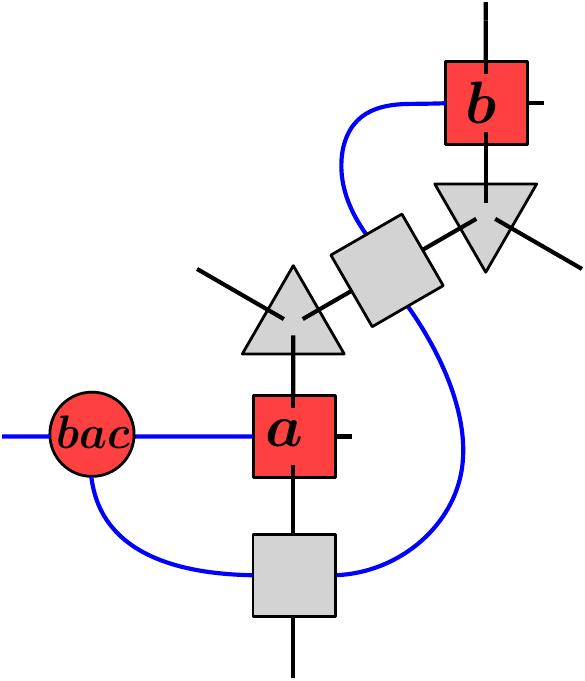}}
		\quad = R^{\boldsymbol{a}\boldsymbol{b}}_{\boldsymbol{c}} \quad
		\raisebox{-1.cm}{\includegraphics[scale=.65]{fig/MPO_doubled_fusion_right.pdf}}
		\; .
	\end{equation}
	\end{widetext}
	As for all other relations in this section, this was explicitly verified for the Fibonacci input category. This confirms that our ansatz does indeed possess the correct DFIB braiding behavior. 
	Finally, the topological spin of anyonic excitations, emerges on the virtual level as follows: 
	\begin{equation}\label{eq:excitation_topological_spin_check}
		\raisebox{-1.7cm}{\includegraphics[scale=.65]{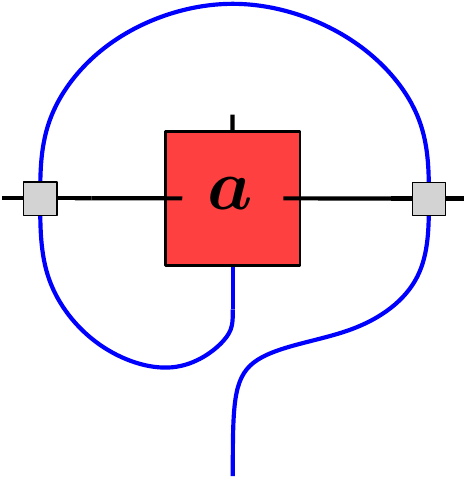}}
		= \theta_{\bm{a}} \quad
		\raisebox{-.9cm}{\includegraphics[scale=.65]{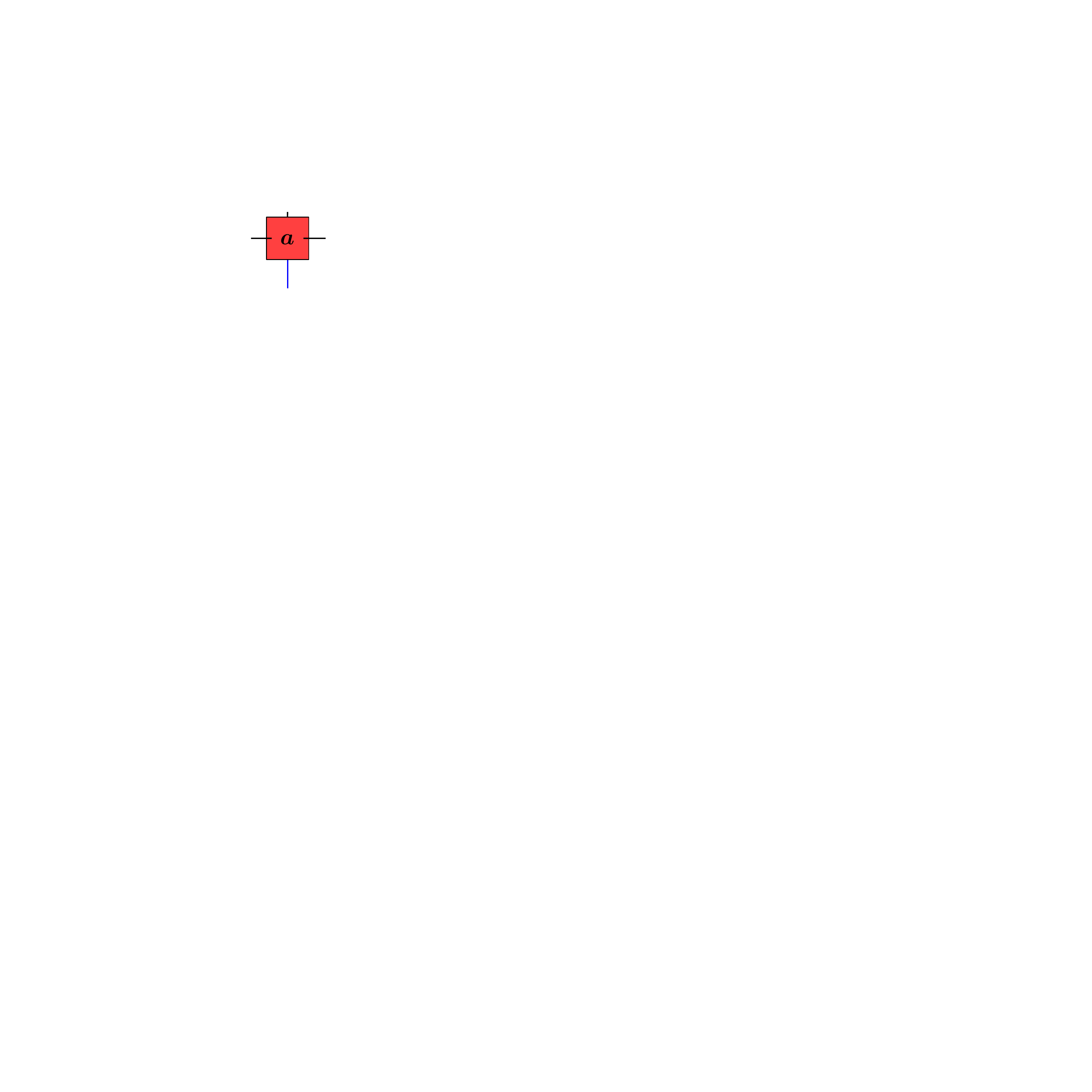}}
		\; ,
	\end{equation}
	where $ \theta_{\bm{a}} = \theta_{a^{+}} (\theta_{a^{-}})^*$ according to Eq.~\eqref{eq:doubled_topological_spin}.
	Alternatively, this can be expressed on the level of the central idempotents as 
	\begin{equation}\label{eq:excitation_topological_spin_idempotents_check}	\raisebox{-3cm}{\includegraphics[scale=.65]{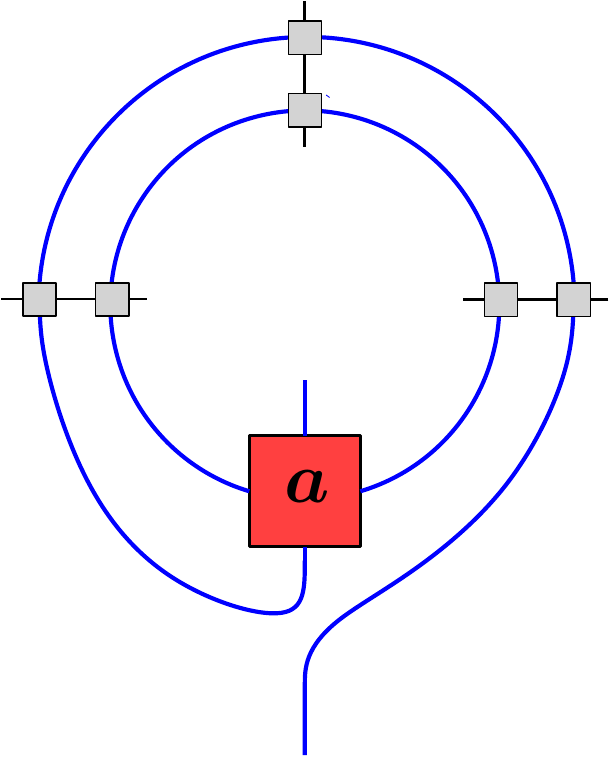}}
		\;= \theta_{\bm{a}} \;
		\raisebox{-1.9cm}{\includegraphics[scale=.65]{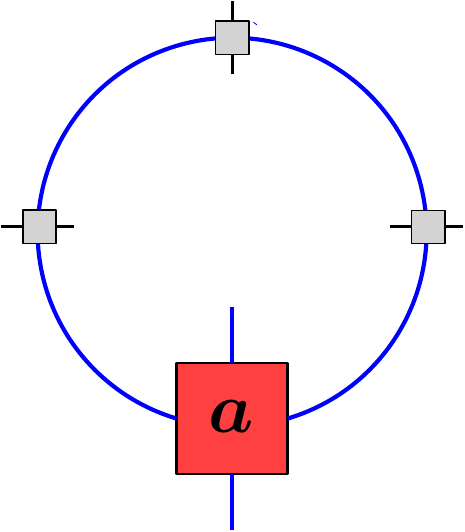}}
		\; .
	\end{equation}
	Both relations where verified for the Fibonacci input category, which ensures that the PEPS ansatz was indeed derived correctly.

%% file: sections/result_scaling.tex
\section{Scaling of largest connected group of anyons}\label{sec:scaling}
Below, we present the results concerning the scaling of the average size of the largest group of connected anyons created by the application of depolarizing noise.
These result represent a ``worst case scenario'' for the noise model (see Sec.~\ref{sec:noise_model}) used in our threshold simulations, in the sense that they correspond to the improbable case where charge measurements of the plaquettes affected by noise operators always yield a nontrivial outcome. 

\begin{figure}[h]
	\centering
	\includegraphics[trim={1cm 7.5cm 1cm 5cm},clip,width=\linewidth] {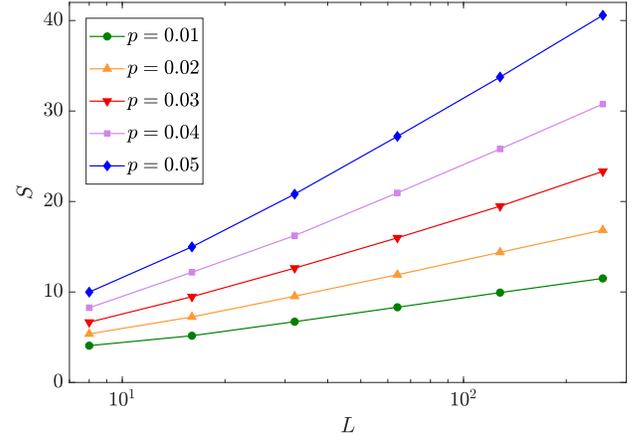}
	\caption{The worst case scenario scaling of the average size $ S $ of the largest connected group of anyons with linear system size $ L $, for the depolarizing noise model.}
	\label{fig:cluster_size}
\end{figure}

The average size $ S $ of the largest connected group of anyons after noise application as a function of the linear system size $ L $, was determined by simulating the (fixed-rate sampling) noise process for a total of $ 10^4 $ Monte Carlo samples with depolarizing noise.
The results for a range of noise strengths $ p $ are shown in \figref{fig:cluster_size}.

As expected, we find that $ S $ scales logarithmically with the system size $ L $. This confirms that classical simulation of the system subjected to depolarizing noise, is indeed possible for noise strengths up to at least $ p = 0.05$.

%% file: sections/relation_iid_noise.tex
\section{Relating fixed-rate sampling and i.i.d. noise}\label{sec:relation_iid_noise}

The results above were presented in terms of the average number of errors per qubit $ p $, which is the parameter characterizing the noise strength in our fixed-rate sampling noise model. This is a natural measure in the context of this work, as the non-Abelian nature of our quantum memory and the fact that it undergoes constant syndrome measurements mean that noise processes cannot be formulated in terms of an independent and identically distributed noise model. However, in order to compare our results to known error threshold results for Abelian codes, we require a way of relating $ p $ to the i.i.d. noise strength $ p_{\text{i.i.d.}} $ used in such models. To this end, we illustrate the relation between the Poisson and i.i.d. binomial noise models in Abelian stabilizer codes, for which both these noise models are equally valid, following the reasoning presented in \cite{brell2014thermalization}.

We again consider our Poisson noise model in which a total of $ T $ error operations are executed, where $ T $ is drawn from a Poisson distribution with mean $ T_0 = |E| p $ and $ |E| $ is the total number of edges, but now apply it in the context of an Abelian stabilizer code. For each individual error process, an edge $ e $ of the lattice is chosen at random and a Pauli operator $ \sigma_i $ is applied according to relative probabilities $\{ \gamma_x,\, \gamma_y,\, \gamma_z\} $. If we think of the global fixed-rate sampling noise model as arising from a superposition of fixed-rate sampling noise processes on each edge of the lattice individually, then we may assume that the latter can be approximately characterized as Poisson distributed events themselves. According to this reasoning, the noise model would then be captured by a superposition of three different Poisson distributions at the level of each individual edge, with means $ \gamma_x p $, $ \gamma_y p $ and $ \gamma_z p $ for $ \sigma_x $, $ \sigma_y $ and $ \sigma_z $ errors, respectively.

Since all Pauli errors acting on a given edge either commute or anticommute, we can compute the i.i.d probability for a net $ \sigma_x $ error on an individual edge, $ p_{\text{i.i.d.}}^x $, by adding the probabilities of all Poisson error processes where the number of $ \sigma_x $ errors acting on the edge is odd and the number of $ \sigma_y $ and $ \sigma_z $ errors acting on the edge are both even, or where the number of $ \sigma_x $ errors acting on the edge is even and the number of $ \sigma_y $ and $ \sigma_z $ errors acting on the edge are both odd. We therefore get:
\begin{widetext}
\begin{equation}
    \begin{split}
    p_{\text{i.i.d.}}^x &= 
    \left(\displaystyle\sum_{k=0}^{+\infty} \e^{-\gamma_x p} \frac{(\gamma_x p)^{(2k+1)}}{(2k+1)!}\right)
    \left(\displaystyle\sum_{k=0}^{+\infty} \e^{-\gamma_y p} \frac{(\gamma_y p)^{(2k)}}{(2k)!}\right)
    \left(\displaystyle\sum_{k=0}^{+\infty} \e^{-\gamma_z p} \frac{(\gamma_z p)^{(2k)}}{(2k)!}\right)
     \\
    & \qquad \quad  + \left(\displaystyle\sum_{k=0}^{+\infty} \e^{-\gamma_x p} \frac{(\gamma_x p)^{(2k)}}{(2k)!}\right)
    \left(\displaystyle\sum_{k=0}^{+\infty} \e^{-\gamma_y p} \frac{(\gamma_y p)^{(2k+1)}}{(2k+1)!}\right)
    \left(\displaystyle\sum_{k=0}^{+\infty} \e^{-\gamma_z p} \frac{(\gamma_z p)^{(2k+1)}}{(2k+1)!}\right)
    \\
    &= \frac{\e^{-p}}{4} \left(\e^{p} + \e^{(\gamma_x - \gamma_y - \gamma_z)p} - \e^{(-\gamma_x + \gamma_y - \gamma_z)p} - \e^{(-\gamma_x - \gamma_y + \gamma_z)p}\right).
    \end{split}
\end{equation}
Similar expressions can be derived in an analogous fashion for $ p_{\text{i.i.d.}}^y $, $ p_{\text{i.i.d.}}^z $ and $ p_{\text{i.i.d.}}^{\text{none}} $, which denote the i.i.d. error probability for a net $ \sigma_y $ error, net $ \sigma_z $ error or no net error at all on an individual edge, respectively:
\begin{align}
    p_{\text{i.i.d.}}^y &= \frac{\e^{-p}}{4} \left(\e^{p} + \e^{(-\gamma_x + \gamma_y - \gamma_z)p} - \e^{(\gamma_x - \gamma_y - \gamma_z)p} - \e^{(-\gamma_x - \gamma_y + \gamma_z)p}\right)\\
    p_{\text{i.i.d.}}^z &= \frac{\e^{-p}}{4} \left(\e^{p} + \e^{(-\gamma_x - \gamma_y + \gamma_z)p} - \e^{(\gamma_x - \gamma_y - \gamma_z)p} - \e^{(-\gamma_x + \gamma_y - \gamma_z)p}\right)\\
    p_{\text{i.i.d.}}^{\text{none}} &= \frac{\e^{-p}}{4} \left(\e^{p} + \e^{(\gamma_x - \gamma_y - \gamma_z)p} + \e^{(-\gamma_x + \gamma_y - \gamma_z)p} + \e^{(-\gamma_x - \gamma_y + \gamma_z)p}\right)\;.
\end{align}
\end{widetext}
Hence, for $ p \ll 1$ we have $ p_{\text{i.i.d.}}^x \approx \gamma_x p $, $ p_{\text{i.i.d.}}^y \approx \gamma_y p $, $ p_{\text{i.i.d.}}^z \approx \gamma_z p $ and $ p_{\text{i.i.d.}}^{\text{none}} \approx 1 - p $.

%% file: sections/paperclip_list.tex
\section{Paperclip configurations}\label{sec:paperclip_list}

%

The paperclip algorithm (see Sec.~\ref{sec:paperclip}) works by exploiting the fact that refactoring moves along the boundary of an individual tile, can be classified into 16 different paperclip configurations, and that these entirely determine the corresponding sequence of swaps.
These 16 configurations can be split into two groups corresponding to refactoring moves along or against the orientation of the curve diagram.
Each of these groups can itself be split into two groups, depending on whether the initial piece of curve containing the anyon has an incoming or outgoing orientation.
Within each of these groups, the different configurations can be distinguished using their respective turn numbers, as defined in Sec.~\ref{sec:paperclip}
Below, we list all possible paperclip configurations, and the corresponding sequence of swaps for each of them.

\begin{widetext}
	\subsection{Refactoring along the curve}	
	\begin{align*}
		& \boxed{-3}\,
		\raisebox{-1cm}{\includegraphics[scale=.45]{fig/paperclip_along_in_-3.pdf}}
		\rightarrow   		\raisebox{-.5cm}{\includegraphics[scale=.48]{fig/paperclip_along_in_-3_line.pdf}} 
		\;\;: \quad S^{-1}[B]	\\
		& \boxed{-6}\,   
		\raisebox{-1cm}{\includegraphics[scale=.45]{fig/paperclip_along_in_-6.pdf}}
		\rightarrow 		\raisebox{-.5cm}{\includegraphics[scale=.48]{fig/paperclip_along_in_-6_line.pdf}} 
		\;\;: \quad S^{-1}[B] \, S^{-1}[H] \, S^{-1}[H^r] \\
		& \boxed{+6}\,		\raisebox{-1.3cm}{\includegraphics[scale=.45]{fig/paperclip_along_in_+6.pdf}}
		\rightarrow 		\raisebox{-.5cm}{\includegraphics[scale=.48]{fig/paperclip_along_in_+6_line.pdf}} 
		\;\;: \quad S[T^r] \, S[T] \, S[B] \\
		& \boxed{+9}\, 		\raisebox{-1.3cm}{\includegraphics[scale=.45]{fig/paperclip_along_in_+9.pdf}}
		\rightarrow 		\raisebox{-.5cm}{\includegraphics[scale=.48]{fig/paperclip_along_in_+9_line.pdf}}
		\;\; : \quad S[T^r] \, S[T] \, S[B] \, S[H] \, S[H^r]\\
		 \\
		 \\
		 \\
	%
%
%
		& \boxed{+3}\,
		\raisebox{-1cm}{\includegraphics[scale=.45]{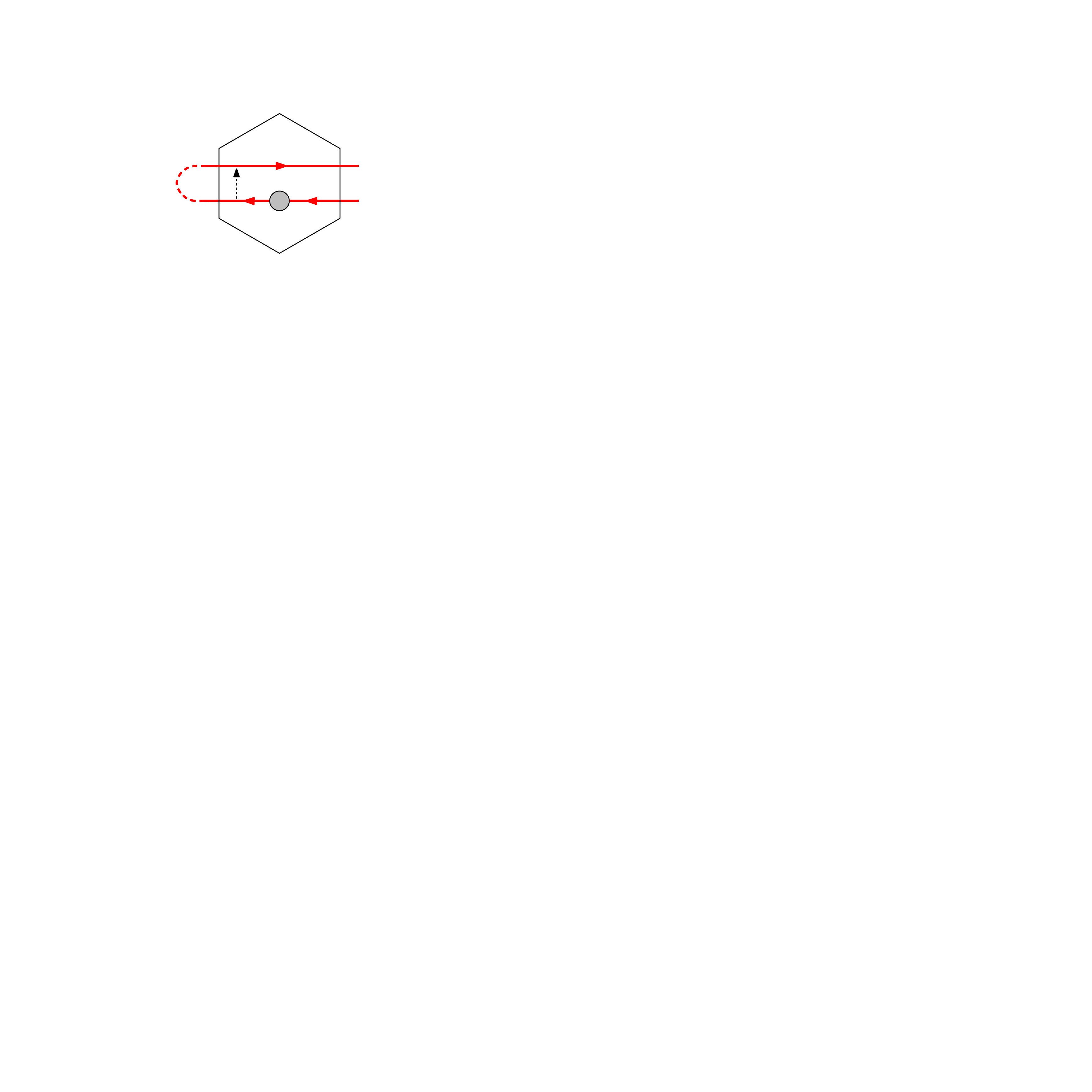}}
		\rightarrow   		\raisebox{-.5cm}{\includegraphics[scale=.48]{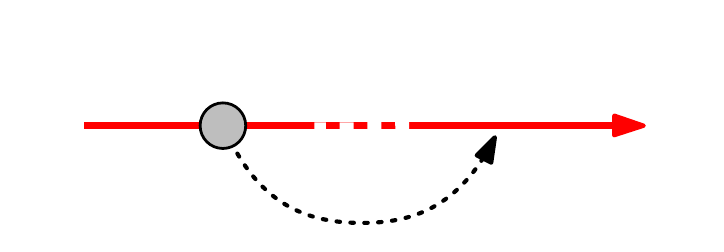}}
		\;\;: \quad S[B] \\
		& \boxed{+6}\,   
		\raisebox{-1cm}{\includegraphics[scale=.45]{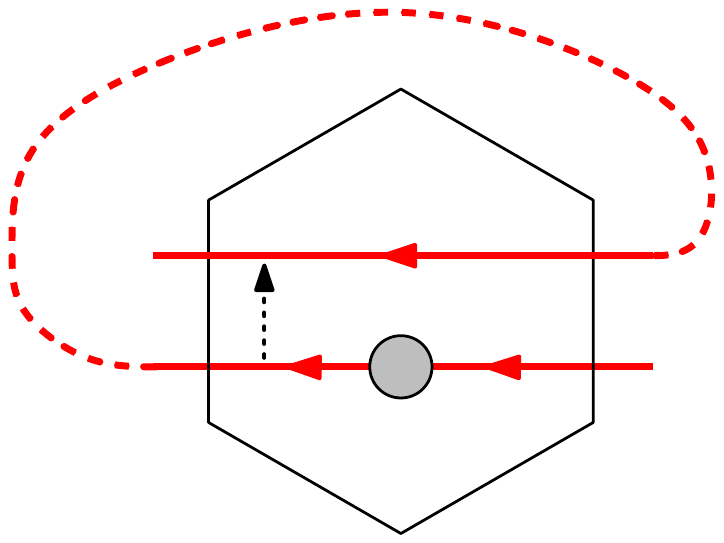}}
		\rightarrow 		\raisebox{-.5cm}{\includegraphics[scale=.48]{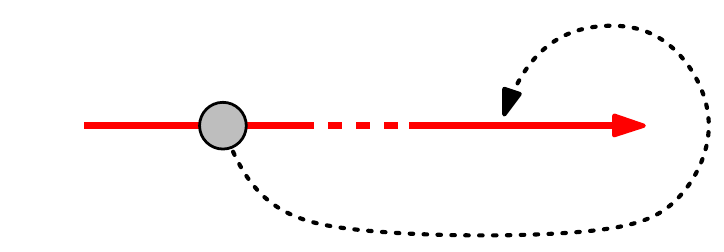}}
		\;\;: \quad  S[B] \, S[H] \, S[H^r]\\
		& \boxed{-6}\, 		\raisebox{-1.3cm}{\includegraphics[scale=.45]{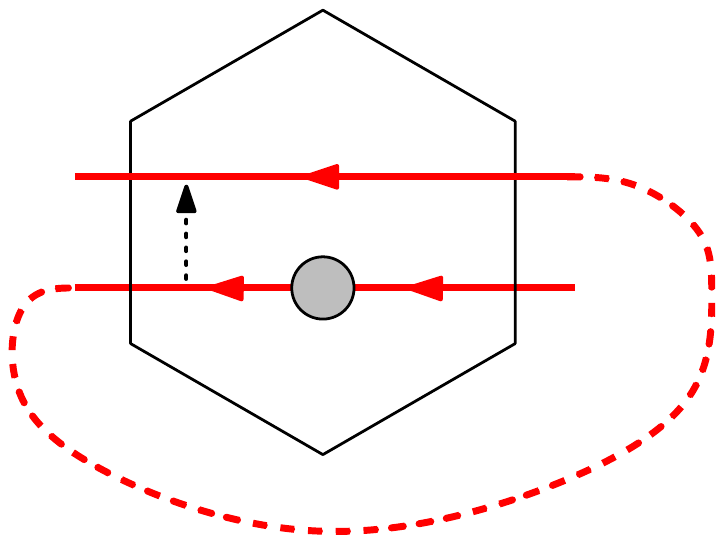}}
		\rightarrow 		\raisebox{-.5cm}{\includegraphics[scale=.48]{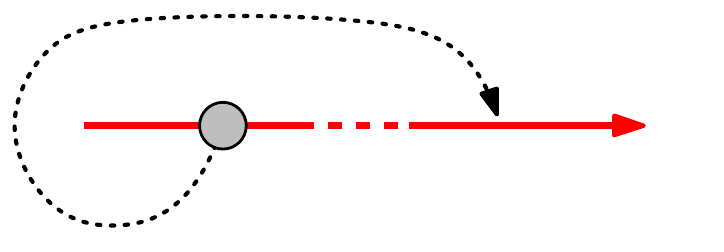}} 
		\;\;: \quad  S^{-1}[T^r] \, S^{-1}[T] \, S^{-1}[B] \\
		& \boxed{-9}\, 		\raisebox{-1.3cm}{\includegraphics[scale=.45]{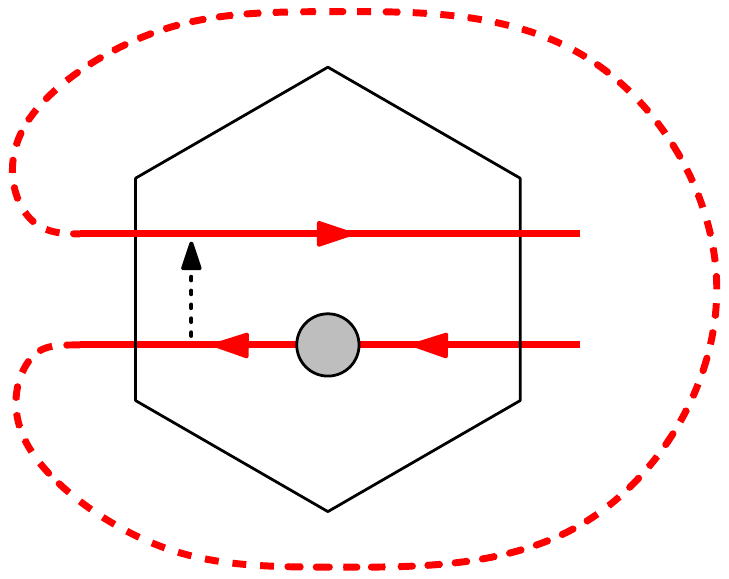}}
		\rightarrow 		\raisebox{-.5cm}{\includegraphics[scale=.48]{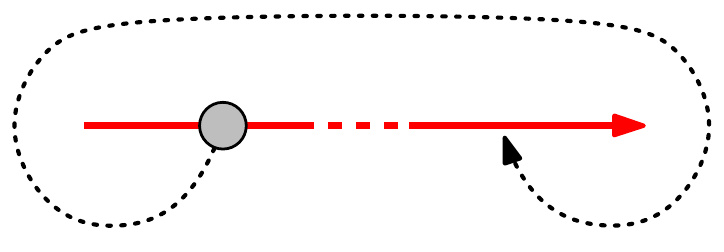}}
		\;\;: \quad S^{-1}[T^r] \, S^{-1}[T] \, S^{-1}[B] \, S^{-1}[H]
	\end{align*}

\subsection{Refactoring against the curve}
	\begin{align*}
		& \boxed{+3}\,
		\raisebox{-1cm}{\includegraphics[scale=.45]{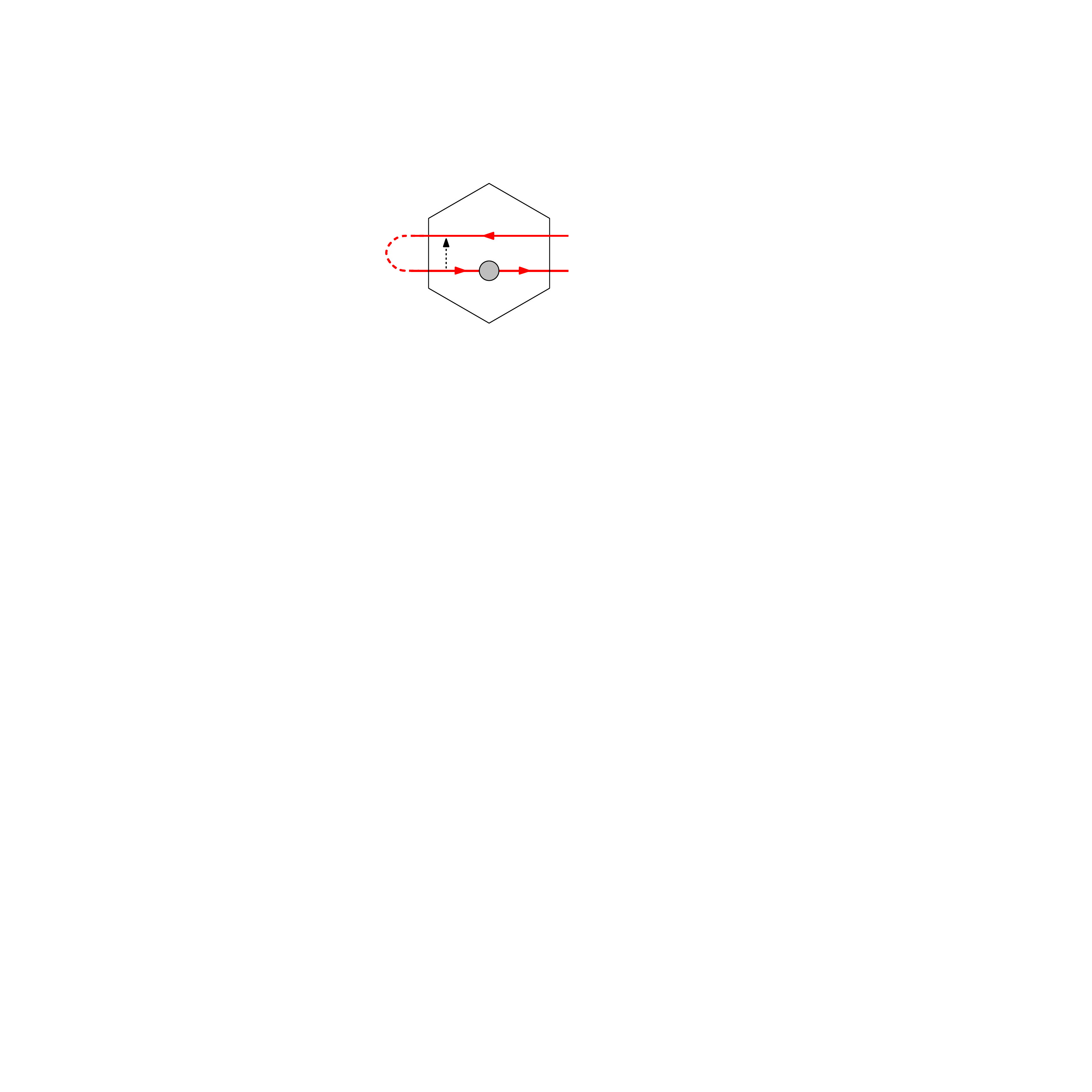}}
		\rightarrow   	\raisebox{-.5cm}{\includegraphics[scale=.48]{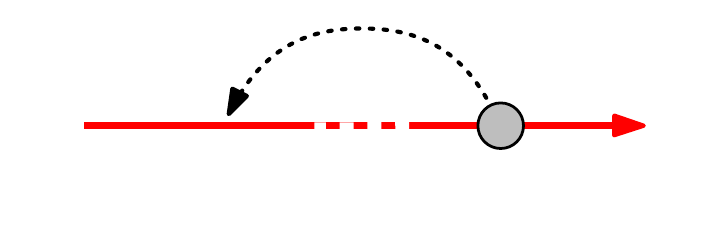}}
		\;\;: \quad \quad S[B^r]\\
		& \boxed{+6}\,   
		\raisebox{-1cm}{\includegraphics[scale=.45]{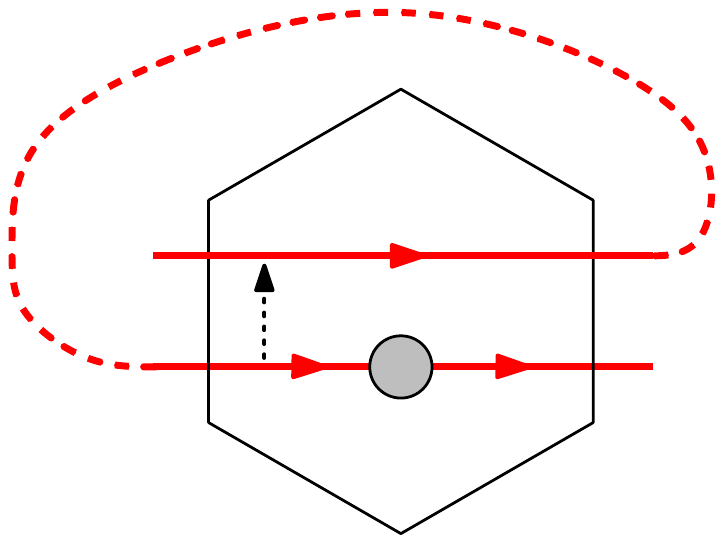}}
		\rightarrow 	\raisebox{-.5cm}{\includegraphics[scale=.48]{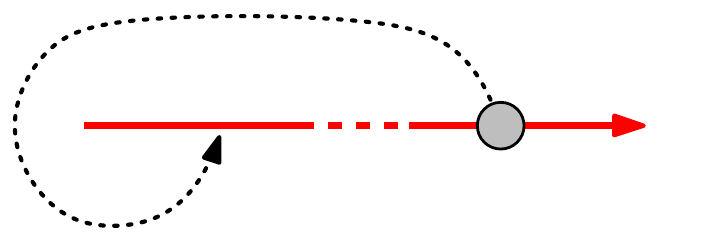}} 
		\;\;: \quad  S^{-1}[H]\, S^{-1}[H^r]\, S^{-1}[B]\\
		& \boxed{-6}\, 	\raisebox{-1.3cm}{\includegraphics[scale=.45]{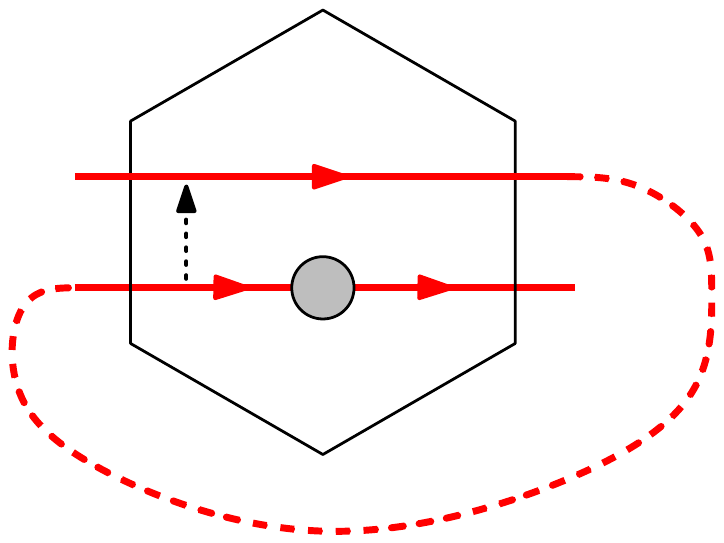}}
		\rightarrow 	\raisebox{-.5cm}{\includegraphics[scale=.48]{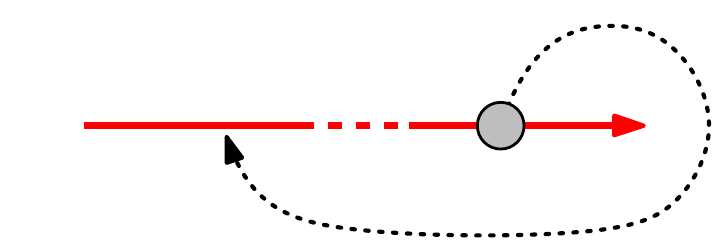}} 
		\;\;: \quad S[B]\, S[T^r]\, S[T]  \\
		& \boxed{-9}\, 		\raisebox{-1.3cm}{\includegraphics[scale=.45]{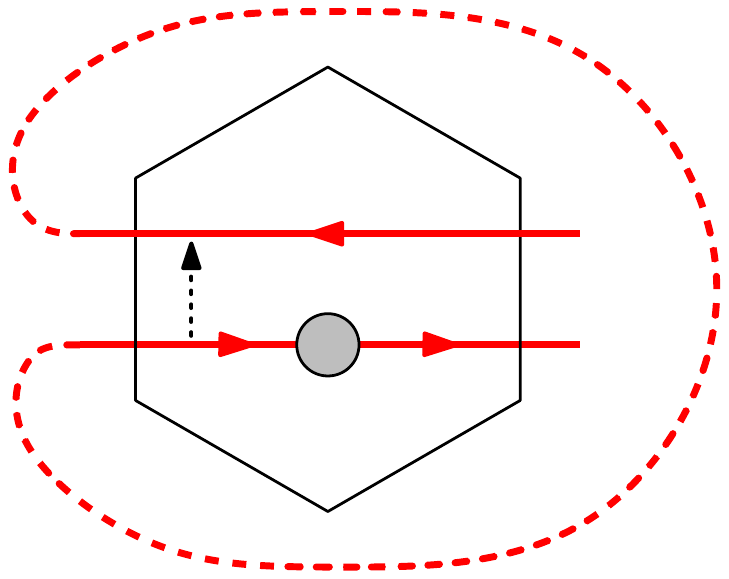}}
		\rightarrow 	\raisebox{-.5cm}{\includegraphics[scale=.48]{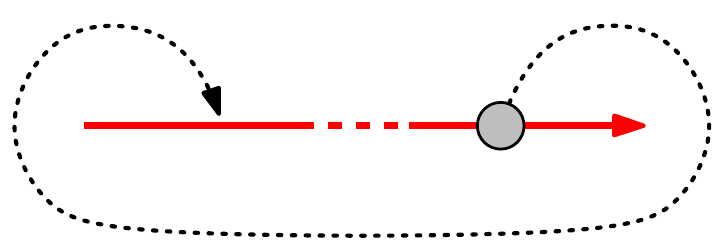}}
		\;\;: \quad  S^{-1}[H]\, S^{-1}[H^r]\, S^{-1}[B^r]\, S^{-1}[T^r]\, S^{-1}[T]\\
		\\
		\\
		\\
%
		& \boxed{-3}\,
		\raisebox{-1cm}{\includegraphics[scale=.45]{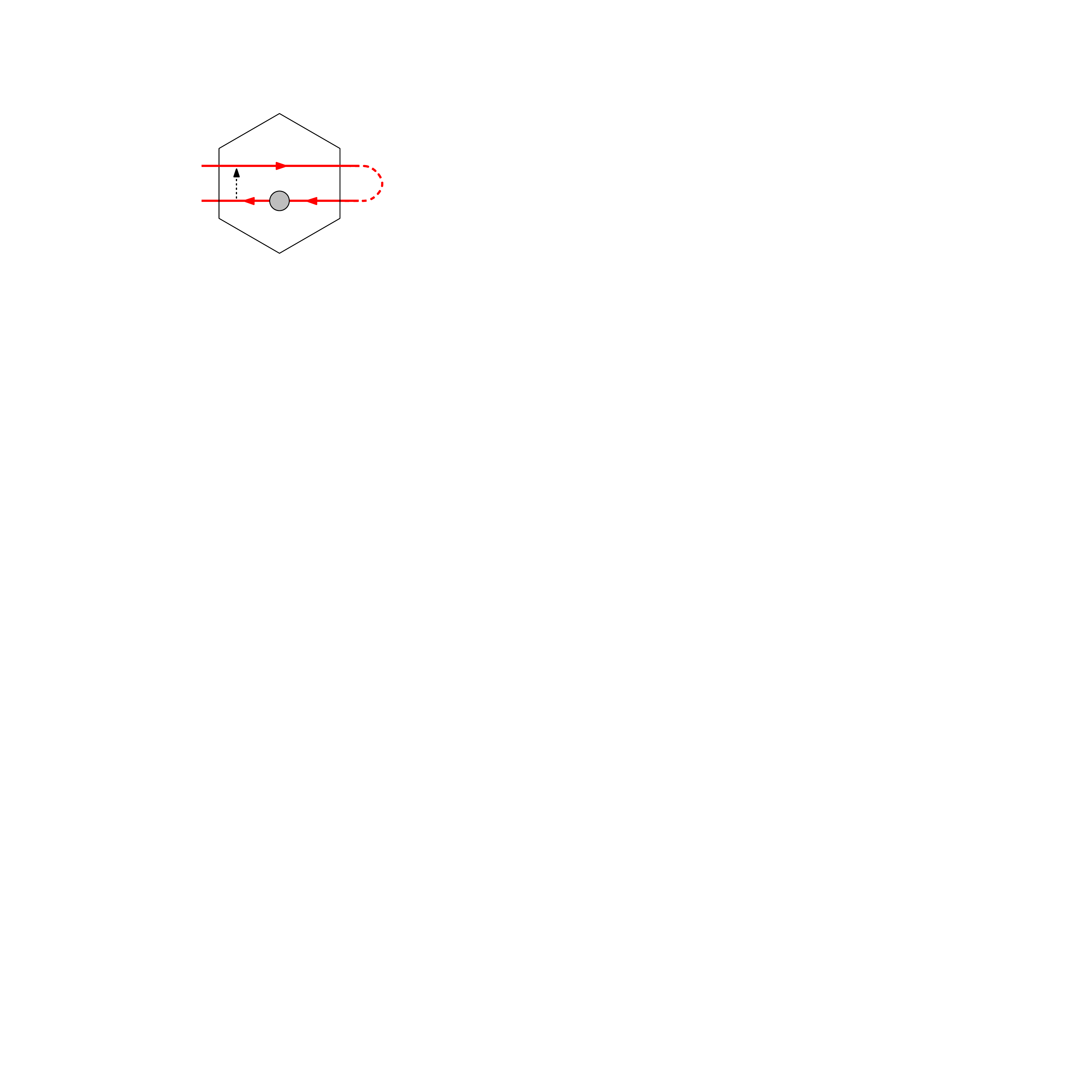}}
		\rightarrow   		\raisebox{-.5cm}{\includegraphics[scale=.48]{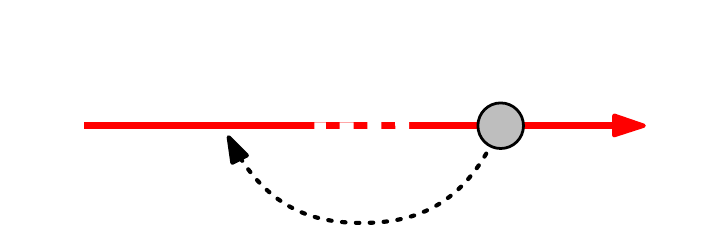}}
		\;\;: \quad  S^{-1}[B^r] \\
		& \boxed{-6}\,   
		\raisebox{-1cm}{\includegraphics[scale=.45]{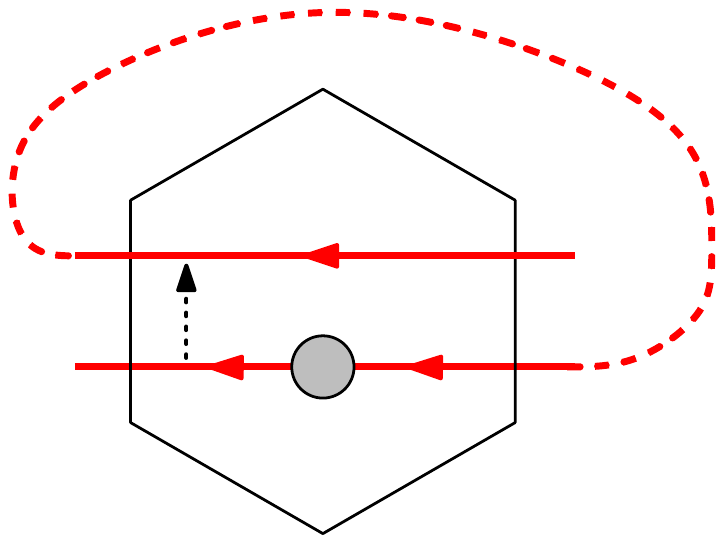}}
		\rightarrow 		\raisebox{-.5cm}{\includegraphics[scale=.48]{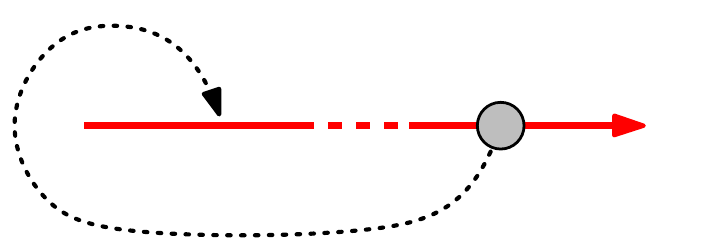}}
		\;\;: \quad  S[H]\, S[H^r]\, S[B] \\
		& \boxed{+6}\, 		\raisebox{-1.3cm}{\includegraphics[scale=.45]{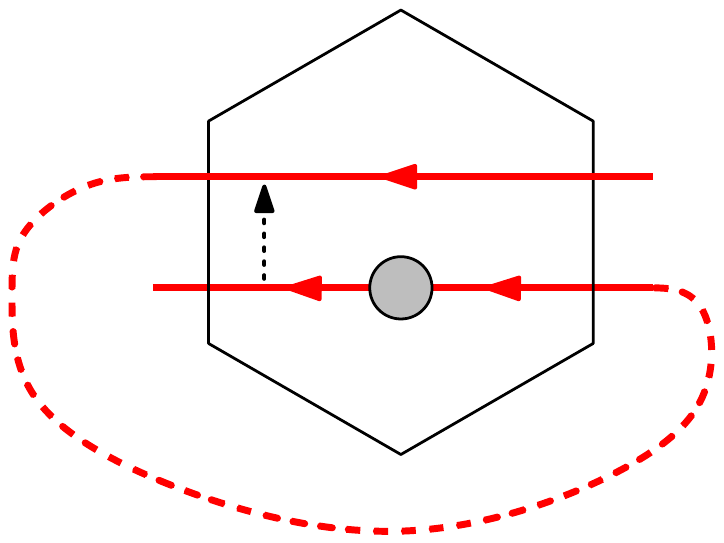}}
		\rightarrow 		\raisebox{-.5cm}{\includegraphics[scale=.48]{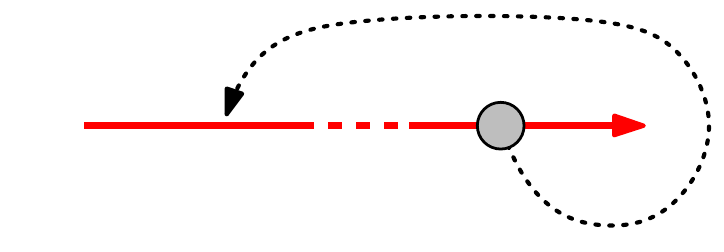}}
		\;\;: \quad  S^{-1}[B]\, S^{-1}[T^r]\, S^{-1}[T] \\
		& \boxed{+9}\, 		\raisebox{-1.3cm}{\includegraphics[scale=.45]{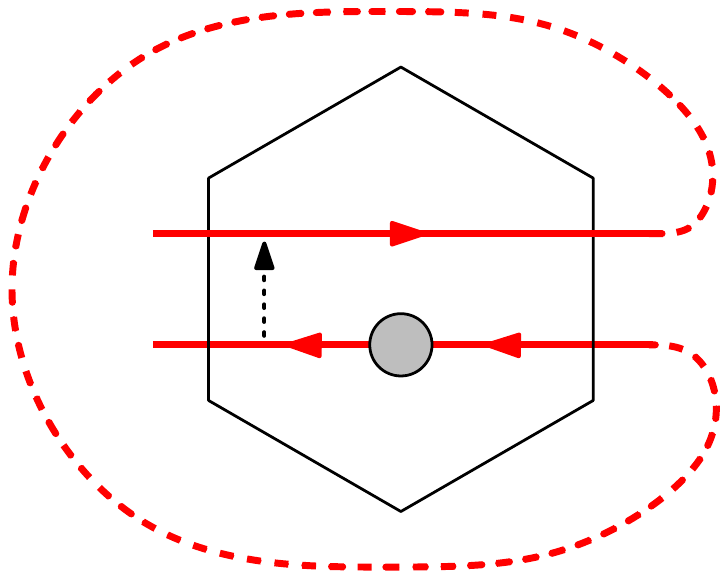}}
		\rightarrow 		\raisebox{-.5cm}{\includegraphics[scale=.48]{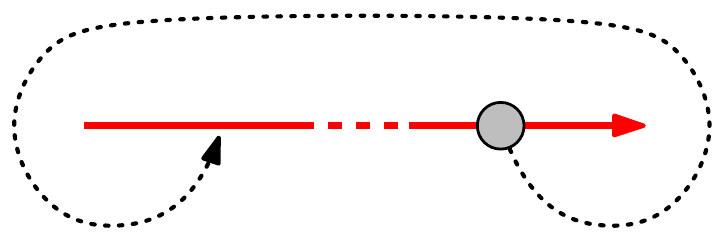}}
		\;\;: \quad  S[H]\, S[H^r]\, S[B^r]\, S[T^r]\, S[T]
	\end{align*}

\end{widetext}